\documentclass [11pt]{article}

\usepackage{jheppub}

\usepackage[utf8]{inputenc}

\usepackage{amsfonts}

\usepackage{amsmath,amssymb}

\usepackage{mathtools}

\usepackage{graphicx,subfigure}
\graphicspath{ {./figures/} }

\usepackage[space]{grffile}
\usepackage{mathtools}

\usepackage{simplewick}

\usepackage{array}

\usepackage{float}

\usepackage{color}

\usepackage{mathrsfs}

\usepackage{ytableau}

\usepackage{tikz}

\usepackage{slashed}

\usepackage{booktabs}
\usepackage{tabularx}

\usepackage[section]{placeins}

\def\nn{\nonumber}

\newcommand{\bfj}{{\bf j}} % bold j
\newcommand{\cO}{\mathcal{O}}
\newcommand{\cT}{\mathcal{T}}

\newcommand{\<}{\langle}
\renewcommand{\>}{\rangle}

\newcommand{\myRho}{\rho}

 \definecolor{verde}{rgb}{0,0.7,0.2}

\newcommand{\myP}{\boldsymbol{p}}

\newcommand{\tr}{\text{tr}}

\renewcommand{\Im}{\operatorname{Im}}

\newcommand{\pder}[1]{\partial_#1}

\newcommand{\al}{\alpha}
\newcommand{\be}{\beta}
\newcommand{\la}{\lambda}

\newcommand{\aldot}{\dot{\alpha}}
\newcommand{\bedot}{\dot{\beta}}

\newcommand{\plub}[1]{[#1]}
\newcommand{\minb}[1]{\langle #1 \rangle}
\newcommand{\Lmax}{L_{\text{max}}}
\newcommand{\Nmax}{N_{\text{max}}}

\newcommand{\sh}{\xi} %spinor helicity
\newcommand{\tsh}{\tilde{\xi}} %spinor helicity

\usetikzlibrary{calc,snakes, patterns,angles,quotes,arrows.meta,shapes.misc,decorations.pathmorphing,decorations.markings}

\title{\boldmath Bounds on photon scattering}

\author[ab]{Kelian H\"aring,}
\author[cb]{Aditya Hebbar,}
\author[d b]{Denis Karateev,}
\author[d b]{Marco Meineri,}
\author[b]{Jo\~ao Penedones}

\affiliation[a]{Theoretical Physics Department,
CERN, 1211 Geneva 23, Switzerland}

\affiliation[b]{Fields and Strings Laboratory, Institute of Physics\\ École Polytechnique Fédéral de Lausanne (EPFL)
	\\ Route de la Sorge, CH-1015 Lausanne, Switzerland}

\affiliation[c]{CPHT, CNRS, École Polytechnique\\Institut Polytechnique de Paris
	\\ Route de Saclay, 91128 Palaiseau, France}

\affiliation[d]{
	D\'epartment de Physique Th\'eorique, Universit\'e de Gen\`eve,\\
	24 quai Ernest-Ansermet, 1211 Gen\`eve 4, Switzerland}

\emailAdd{kelian.haring@epfl.ch}
\emailAdd{aditya.hebbar@polytechnique.edu}
\emailAdd{ denis.karateev@unige.ch}
\emailAdd{marco.meineri@unige.ch}
\emailAdd{joao.penedones@epfl.ch}

\abstract{
We study 2-to-2 scattering amplitudes of massless spin one particles in $d=4$ space-time dimensions, like real world photons. 
We define a set of non-perturbative observables (Wilson coefficients) which describe these amplitudes at low energies. 
We use full non-linear unitarity to construct various novel numerical bounds on these observables. 
For completeness, we also rederive some bounds using positivity only. 
We discover and explain why some of these Wilson coefficients cannot be bounded. 
}

\begin{document}
	\begin{flushleft}
		\hfill \parbox[c]{40mm}{CERN-TH-2022-161}
	\end{flushleft}
\maketitle

\section{Introduction and summary}

There has been a lot of new progress in the non-perturbative S-matrix bootstrap, driven by the development of efficient numerical tools, see \cite{Paulos:2016fap, Paulos:2016but, Paulos:2017fhb,Doroud:2018szp, He:2018uxa, Cordova:2018uop, Guerrieri:2018uew, Paulos:2018fym, Homrich:2019cbt, EliasMiro:2019kyf, Cordova:2019lot, Bercini:2019vme, Gabai:2019ryw,Bose:2020shm,Bose:2020cod,Correia:2020xtr,Kruczenski:2020ujw, Guerrieri:2020bto,Hebbar:2020ukp,Karateev:2019ymz,Karateev:2020axc,Guerrieri:2020kcs,Tourkine:2021fqh,Guerrieri:2021ivu,He:2021eqn,EliasMiro:2021nul,Guerrieri:2021tak,Chen:2021pgx,Cordova:2022pbl,Albert:2022oes,Sinha:2020win,Chowdhury:2021ynh,Karateev:2022jdb,Chen:2022nym}.\footnote{For an overview of recent results and discussion of some future directions, see \cite{Kruczenski:2022lot}.}

In this paper we focus on $4$ space-time dimensions.
We consider a class of theories
which have a single photon-like particle, namely a massless particle with spin one. We denote this particle by $\gamma$.
The goal of this paper is to study the 2-to-2 scattering amplitudes of the process
\begin{equation}
	\label{eq:scattering_under_consideration}
	\gamma(\la_1)\gamma(\la_2) \rightarrow \gamma(\la_3)\gamma(\la_4)\,,
\end{equation}
where $\lambda_i$ is the helicity of the in/out particles. 
We are agnostic about the high energy behaviour of the theory apart from the requirement that the 2-to-2 scattering amplitude must obey the usual S-matrix bootstrap principles: Lorentz invariance, unitarity and analyticity.

Scattering amplitudes of particles with spin in $4d$ was reviewed systematically in \cite{Hebbar:2020ukp}.\footnote{This problem was studied in the 60s by many authors \cite{Jacob:1959at,  Trueman:1964zzb, Hara:1970gc, Hara:1971kj}, see the older review \cite{Martin:102663} for a more comprehensive list. See also \cite{Bellazzini:2016xrt, deRham:2017zjm} for more recent discussions.}  
We will use their language in this work. Each photon in \eqref{eq:scattering_under_consideration} has two helicities, as a result there are 16 scalar amplitudes which fully describe the process \eqref{eq:scattering_under_consideration}. For simplicity we assume parity invariance in this work.\footnote{Due to CPT symmetry in the case of neutral identical  particles (as in this paper) parity invariance implies time-reversal invariance.}
Taking into account the fact that the particles under consideration are also identical we are left with only  5 different amplitudes:

\begin{equation}
	\label{eq:helicities}
	++\to++,\qquad ++\to--,\qquad +-\to+-,\qquad +-\to-+,\qquad ++\to+-,
\end{equation}
where $\pm$ correspond to the helicities of the particles  in \eqref{eq:scattering_under_consideration}.
We denote the corresponding amplitudes in the center of mass frame by
\begin{equation}
	\label{eq:list_amplitudes}
	\Phi_1(s,t,u),\qquad \Phi_2(s,t,u),\qquad \Phi_3(s,t,u),\qquad \Phi_4(s,t,u),\qquad \Phi_5(s,t,u),
\end{equation}
where $s$, $t$ and $u$ are the Mandelstam variables describing the scattering process and obeying the standard relation
\begin{equation}
	\label{eq:condition_stu}
	s+t+u=0.
\end{equation}
Due to crossing equations only the $\Phi_1$, $\Phi_2$, and $\Phi_5$ amplitudes are independent, the rest can be related to these as
\begin{equation}
	\label{eq:crossing_intro}
	\Phi_3(s,t,u) = \Phi_1(u,t,s),\qquad
	\Phi_4(s,t,u) = \Phi_1(t,s,u).
\end{equation}
Moreover, crossing also implies that $\Phi_1(s,t,u)$ is symmetric under $t-u$ permutation and the amplitudes $\Phi_2(s,t,u)$ and $\Phi_5(s,t,u)$ are fully  symmetric under permutations of their arguments.
We give the precise definition of the amplitudes \eqref{eq:list_amplitudes} in appendix \ref{sec:S-matrix_setup}. There we also derive their crossing equations and unitarity constraints. For completeness we also explain how to define the amplitudes via tensor structures in appendix \ref{app:tensor_structures_main} both in vector and spinor formalisms.

\paragraph{Non-perturbative observables}
At low energy, the amplitudes $\Phi_1$, $\Phi_2$ and $\Phi_5$ have the following expansion:\footnote{Let us emphasize  that even though we write $\cO(s^n)$,  the expansion is  at small $s\sim t \sim u$ and $n$ is the total power of $s,t,u$. }
\begin{equation}
	\label{eq:amplitudes_EFT}
	\begin{aligned}
		\Phi_1(s,t,u) &= g_2 s^2 + g_3 s^3 + g_4 s^4+ g_4^\prime  s^2 t u + L_1( s| t, u) +\cO( s^5) , \\
		\Phi_2(s,t,u)  &=f_2 (s^2+ t^2 + u^2)+f_3 s t u + f_4(s^2+t^2+u^2)^2 \\
		&+ L_2(s,t,u)+\cO(s^5), \\
		\Phi_5(s,t,u)  &= h_3 \, s t u+\cO(s^5)~.
	\end{aligned}
\end{equation}
The two functions $L_1(s,t,u)$ and $L_2(s,t,u)$ are fixed by unitarity in terms of the polynomial terms in \eqref{eq:amplitudes_EFT}, see appendix \ref{sec:oneloopUnitarity}:
\begin{align}
	\nn
	L_1(s|t,u) &\equiv  s^2\left(\beta_{1,1}s^2+\beta_{1,2} t u\right)\log\left(- s \sqrt{g_2}\right) +\beta_{1,3}s^2\left( t^2\log\left(- t \sqrt{g_2}\right)+ u^2\log\left(-u\sqrt{g_2}\right)\right),\\
	L_2(s,t,u) &\equiv \beta_2\left(s^4 \log\left(-s \sqrt{g_2}\right) + t^4\log\left(-t \sqrt{g_2}\right) + u^4\log\left(-u \sqrt{g_2}\right) \right),
	\label{eq:loops}
\end{align}
where the coefficients $\beta$ read as
\begin{equation}
\label{betaEFT}
	\begin{aligned}
		\beta_{1,1}&= - \frac{14f_2^2+5g_2^2}{160\pi^2},\qquad
		&&\beta_{1,2}= \frac{f_2^2}{240\pi^2}\\
		\beta_{1,3}&= - \frac{g_2^2}{80\pi^2},\qquad
		&&\beta_{2}= - \frac{5f_2 g_2}{48 \pi^2}.
	\end{aligned}
\end{equation}
Equations \eqref{eq:amplitudes_EFT} - \eqref{betaEFT} follow from a few simple assumptions, which are compactly encoded by an effective field theory (EFT), to be discussed below.\footnote{The amplitudes \eqref{eq:amplitudes_EFT} can also be derived from softness, unitarity, crossing symmetry and kinematical contraints.
Softness is the assumption that the 2-to-2 amplitude scales like $(\textup{energy})^4$, and that the 2-to-$(n\ge 3)$ amplitude scales at least like $(\textup{energy})^6$ at low energy. 
Then unitarity fixes the coefficient of the $\log$ terms as in \eqref{betaEFT}, see appendix \ref{sec:oneloopUnitarity}.
Crossing symmetry restricts the polynomials of $s$, $t$ and $u$ that can appear in \eqref{eq:amplitudes_EFT}. Finally, the   kinematical contraints discussed in \cite{Hebbar:2020ukp} (see eq. (2.139)), imply that $\Phi_3=u^2 (...)$ and $\Phi_5=st u (...)$. }
For now, it suffices to know that the main ingredient is the absence of other massless degrees of freedom beyond the spin one particle $\gamma$.

We refer to the real parameters $g_n$, $f_n$ and $h_n$ in \eqref{eq:EFT_amplitudes} as non-perturbative observables. Our notation is almost identical to the one of \cite{Henriksson:2021ymi,Henriksson:2022oeu}\footnote{The difference between our observables (in black) and the ones of \cite{Henriksson:2021ymi,Henriksson:2022oeu} (in blue) is the coefficients $g_4$ and $g_4^\prime$ which are related to their $\textcolor{blue}{g_{4,1}}$ and $\textcolor{blue}{g_{4,2}}$ as follows
	\begin{align}
		g_4 = \textcolor{blue}{g_{4,1} + 2 g_{4,2}}, \qquad g_4^\prime= \textcolor{blue}{-2g_{4,2}}. \nn
\end{align} 
 } apart from the fact that we take into account the logarithmic branch cuts associated to intermediate massless particles.
As it will be shown, the parameter $g_2$ is always non-negative, thus the object $\sqrt{g_2}$ entering inside the log terms is unambiguous.

For complete clarity, let us emphasize that the observables $g_n$, $f_n$ and $h_n$ are well defined and measurable in terms of derivatives of the non-perturbative scattering amplitude. At the $n\leq 3$ level we have
\begin{equation}
g_2 = \frac{1}{2}\partial_s^2\Phi_1(0,0,0),\qquad
g_3= \frac{1}{3!}\partial_s^3\Phi_1(0,0,0),
\end{equation}
together with
\begin{equation}
2f_2 = \frac{1}{2}\partial_s^2\Phi_2(0,0,0),\qquad
-f_3 = \frac{1}{2}\partial_s^2\partial_t\Phi_2(0,0,0),\qquad
-h_3 = \frac{1}{2}\partial_s^2\partial_t\Phi_5(0,0,0).
\end{equation}
At the $n=4$ level we have instead
\begin{equation}
\begin{aligned}
	g_4    &=\lim_{t\rightarrow 0} \lim_{s\rightarrow 0} \frac{1}{4!}\partial_s^4\left(\Phi_1(s,t,-s-t)-L_1(s|t,-s-t)\right),\\
	-g'_4 &=\lim_{t\rightarrow 0} \lim_{s\rightarrow 0} \frac{1}{3!}\partial_s^3\partial_t\left(\Phi_1(s,t,-s-t)-L_1(s|t,-s-t)\right),\\
	4f_4  &=\lim_{t\rightarrow 0} \lim_{s\rightarrow 0} \frac{1}{4!}\partial_s^4\left(\Phi_2(s,t,-s-t)-L_2(s,t,-s-t)\right).
\end{aligned}
\end{equation}

We choose $g_2$ to define an energy scale. It is then convenient to define the following dimensionless observables
\begin{equation}
	\label{eq:coefficients}
	\bar f_2 \equiv \frac{f_2}{g_2},\quad
	\bar f_3\equiv \frac{f_3}{g_2^{3/2}},\quad
	\bar g_3\equiv \frac{g_3}{g_2^{3/2}},\quad
	\bar h_3\equiv \frac{h_3}{g_2^{3/2}},\quad
	\overline f_4\equiv \frac{f_4}{g_2^2},\quad
	\overline g_4\equiv \frac{g_4}{g_2^2},\quad
	\overline g_4^\prime \equiv \frac{g_4^\prime}{g_2^2}.
\end{equation}

\paragraph{Effective field theory}
From the QFT perspective there is only one consistent way to describe a massless spin one particle, namely as a $U(1)$ gauge theory. We use this fact to construct an effective field theory (EFT) Lagrangian density that can be used to describe the process \eqref{eq:scattering_under_consideration} at low energies. This is given by summing all possible linearly independent Lorentz invariants built out of the electromagnetic tensor $F_{\mu\nu}$ with some generic coefficients. The most general form of such a Lagrangian density reads as
\begin{align}
	\label{eq:EFT_lagrangian}
	\mathcal{L}_\text{EFT} = -\frac{1}{4}F_{\mu\nu} F^{\mu\nu} +\mathcal{L}_{6}+\mathcal{L}_{8}+\mathcal{L}_{10}+\mathcal{L}_{12} +\ldots,
\end{align}
where $\mathcal{L}_n$ denotes   terms   with mass dimension $n$. Explicitly they read as\footnote{We do not include terms with more than four factors of $F_{\mu\nu}$ because these do not contribute to the process \eqref{eq:scattering_under_consideration}  to the order in $s,\,t,\,u$ we are considering. }
	\begin{align}
		\nn
		\mathcal{L}_{6} &=0,\\
		\label{eq:lagrangian_coefficients}
		\mathcal{L}_8 &=  c_1 (F_{\mu\nu} F^{\nu\mu})(F_{\alpha\beta} F^{\beta\alpha})+ c_2 (F_{\mu\nu}F^{\nu\rho}F_{\rho\sigma}F^{\sigma\mu}),\\
		\nn
		\mathcal{L}_{10} &=  c_3 (F_{\alpha \beta}\partial^\beta F_{\mu\nu} \partial^\alpha F^{\nu\rho}{F_{\rho}}^{\mu})+ c_4 (\partial_\alpha F_{\mu\nu} \partial^\alpha F^{\nu\mu})(F_{\rho\sigma} F^{\sigma\rho}) + c_5 (\partial_\alpha F_{\mu\nu}F^{\nu\rho} \partial^\alpha F_{\rho\sigma}F^{\sigma\mu}),\\
		\nn
		\mathcal{L}_{12} &= c_6 ( \partial_\nu F_{\alpha \rho} \partial^\alpha F^{\nu \sigma} \partial^\delta F^{\beta \rho} \partial_\beta F_{\delta \sigma} )+ c_7 ( F_{\rho \beta} \partial_\alpha F_{\sigma \gamma} \partial^{\beta} F^{\sigma \delta} \partial_{\delta} \partial^{\gamma} F^{\rho \alpha}) + c_8 ( F_{\alpha \beta} F_{\sigma \gamma} \delta^\alpha \delta_{\rho} F^{\gamma \delta} \partial^\beta \partial_{\delta} F^{\rho \sigma}).
	\end{align}
The EFT description is valid up to some cut-off scale which we denote by $M$.
The real dimensionful coefficients $ c_i$ are called  Wilson coefficients. They have the following mass dimensions
\begin{equation}
	[c_1] =[c_2] = -4,\qquad
	[c_3] = [c_4] = [c_5] = -6,\qquad
	[c_6]= [c_7]= [c_8]= -8.
\end{equation}
Some recent experimental bounds on some of these Wilson coefficients can be found in \cite{Aaboud:2017bwk,Ellis:2022uxv}.

\begin{figure}
	\centering 
	\begin{tikzpicture}
	
	\draw[color={rgb:black,1;white,6}, fill={rgb:black,1;white,6}](0,0) circle (1.2);
		\draw[->] (-5.2,0)-- (5.2,0);
		\draw[->] (0,-1)--(0,2.8);

		\draw (4,2.5)--(4.5,2.5);
		\draw (4,2.5)--(4,3);
		\draw (4.3,2.8) node{$s$};
		
		\draw[red, decoration = {zigzag,segment length = 2mm, amplitude = 1mm}, decorate] (0,0) -- (5,0);
		 \draw[red] (0,0) node{$\bullet$};
		 \draw (-.15, -0.2) node{$0$};
		\draw[red, decoration = {zigzag,segment length = 2mm, amplitude = 1mm}, decorate] (-5,0) -- (.5,0);
		\draw[red] (.5,0) node{$\bullet$};
		\draw (.4, -0.3) node{$-t$};
		\draw[color=black!60, thick, dashed](0,0) circle (1.2);
		
		\draw[->] (0,0)-- (0.83,0.83);
		\draw (1.2, 1) node{$M^2$};
		
	\end{tikzpicture}
	\label{fig:EFT_ampNEW}
	\caption{Analytic structure of the full non-perturbative amplitudes $\Phi_i(s,t,u)$ in the complex plane $s$ at a fixed value of $t<0$. There are two branch cuts along the real axis, namely, for $s>0$ and for $u=-s-t>0$. The EFT amplitudes are good approximations of the full non-perturbative amplitude inside the grey region corresponding to $|s|\lesssim M^2$. 
	The full non-perturbative amplitude may have other branch points on the real axis. For example, in QED there are branch points at $s=4m_e^2$ and $u=4m_e^2$ associated with the threshold for electron-positron production.
	 }
\end{figure}
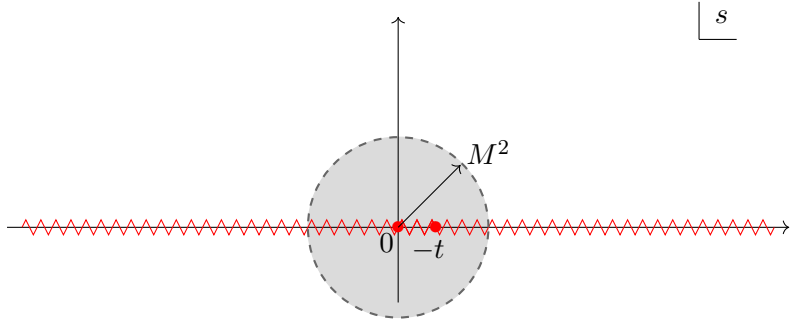

Using the EFT Lagrangian density we can compute scattering amplitudes of massless particles $\Phi_i(s,t,u)$. We denote them by $\Phi^\text{EFT}_i(s,t,u)$. The details of this computation, to 1-loop order, are provided in appendix \ref{app:computation_EFTs}. The amplitudes $\Phi^\text{EFT}_i(s,t,u)$ are good approximations of the full non-perturbative amplitudes $\Phi_i(s,t,u)$ in the regime $|s|\lesssim M^2$, namely
\begin{equation}
	\label{eq:EFT_amplitudes}
	 i=1,2,5:\qquad
	\Phi_i(s,t,u)\approx \Phi^\text{EFT}_i(s,t,u) ,\qquad |s|\lesssim M^2.
\end{equation}
This is depicted in figure \ref{fig:EFT_ampNEW}.
The amplitudes $\Phi^\text{EFT}_i(s,t,u)$ will have precisely the same form as in \eqref{eq:amplitudes_EFT} given the relations
\begin{equation}
	\label{eq:amp_coeffs_in_EFT}
	\begin{aligned}
		&g_2 = 2 (4c_1 + 3c_2), \qquad	f_2  = 2(4c_1 + c_2),\\
		&g_3 = 4 c_4, \qquad f_3 = 6(c_3+2c_4-c_5), \qquad h_3 = \frac{3}{2}c_3,\\
		&g_4 = \frac{1}{4}(-3 c_6 + 2 c_7),\qquad
		g_4^\prime = \frac{1}{2}(3 c_6 - 2 c_7 - c_8),\qquad
		f_4 = -\frac{c_6}{8}.
	\end{aligned}
\end{equation}
In the computation above we used dimensional regularization and we have chosen our renormalization scale $\mu$ to be $\mu^2 = \frac{1}{\sqrt{g_2}}$.

\paragraph{Partial amplitudes and unitarity}
Given the scattering amplitudes \eqref{eq:list_amplitudes} one defines partial amplitudes as follows
\begin{equation}
	\label{eq:partial_amplitudes_intro}
	{\Phi_i^{\ell}}(s) \equiv \frac{1}{32\pi}\int_0^\pi d\theta \sin \theta \,d^{\, \ell}_{\lambda_{12}, \lambda_{34}}(\theta)\; \Phi_i\left(s,t(s,\theta),u(s,\theta)\right)
\end{equation}
where the scattering angle $\theta$ is related to the Mandelstam variables as
\begin{equation}
	\label{eq:scattering_angle}
	t = - \frac{s}{2}(1-\cos\theta),\quad
	u = - \frac{s}{2}(1+\cos\theta).
\end{equation}
The small Wigner $d$-matrix is defined in \eqref{eq:wignerD}. The differences of helicities are defined as
\begin{equation}
	\lambda_{12} \equiv \lambda_1 -\lambda_2,\qquad
	\lambda_{34} \equiv \lambda_3 -\lambda_4.
\end{equation}
The values of helicities for the amplitudes $\Phi_i(s,t,u)$ are given in \eqref{eq:helicities}.

Using partial amplitudes we can write full non-linear unitarity constraints in simple positive semi-definite form as
\begin{equation}
	\label{eq:non-linear_unitarity}
\begin{pmatrix}
	\mathbb I & \mathbb S^{ \dagger}(s) \\
	\mathbb S (s) & \mathbb I
\end{pmatrix} \succeq 0,
\end{equation}
where $\mathbb I$ is the (appropriate) identity matrix and 
\begin{equation}
	\label{eq:entriesa_unitarity}
	\mathbb S (s) = \begin{cases}
		1+i(\Phi_1^0(s) + \Phi_2^0(s)),\qquad \ell=0,\\
		1+i(\Phi_1^\ell(s) - \Phi_2^\ell(s)),\qquad\ell \geq 0 \text{ (even)},\\
		1+2i\Phi_3^\ell(s) ,\qquad\qquad\quad\;\;\;\ell \geq 3 \text{ (odd)},\\
		\begin{pmatrix}
			1 & 0\\
			0 & 1
		\end{pmatrix}+
		i\begin{pmatrix}
			\Phi_1^\ell(s) + \Phi_2^\ell(s)  &  \qquad 2 \Phi_5^{\ell}(s) \\
			2 \Phi_5^\ell(s) &    \qquad 2\Phi_3^\ell(s)
		\end{pmatrix}
	,\qquad\ell \geq 2 \text{ (even)}.
	\end{cases}	
\end{equation}
These conditions hold in the physical regime $s\geq 0$.

There is a simpler subset of the above constraints (called positivity) given by
\begin{equation}
	\Im \mathbb F (s) \succeq 0,
\end{equation}
where  $\mathbb F $ is defined via $\mathbb S $ as follows
\begin{equation}
	\mathbb S  = \mathbb I +i \mathbb F \,.
\end{equation}
A subset (the linear part) of these positivity constraints  read as
\begin{equation}\label{eq:positivity}
	\begin{split}
		&\ell\geq 0 \text{ (even)}:\qquad\Im (\Phi_1^\ell(s)+\Phi^\ell_2(s))\geq 0,\\
		&\ell\geq 0 \text{ (even)}:\qquad\Im (\Phi^\ell_1(s)-\Phi^\ell_2(s))\geq0,\\
		&\ell\geq 2 :\qquad\qquad\quad \Im \Phi^\ell_3(s) \geq0.
	\end{split}
\end{equation}

\paragraph{Goal of the paper}
In this paper we study bounds on the dimensionless observables \eqref{eq:coefficients} coming from full non-linear unitarity \eqref{eq:non-linear_unitarity}.

\paragraph{Bounds from positivity}
Using dispersion relations it is relatively easy to incorporate the positivity constraints \eqref{eq:positivity} and obtain analytic bounds on the observables. This is addressed in section \ref{sec:positivity}. Here we briefly summarize the results from that section. 

For the observables $g_2$ and $f_2$ we obtain the following rigorous bounds 
\begin{equation}
	\label{eq:analytic_bounds}
 -1 \leq	\frac{f_2 }{g_2}\leq 1,\qquad
	g_2 \geq 0.
\end{equation}
We could not derive any bound on  $g_3$, $f_3$ and $h_3$ using dispersion relations and positivity.
There are simple dispersion relations for $g_4$ and $f_4$, however due to the presence of log terms in \eqref{eq:amplitudes_EFT} we cannot derive bounds on these observables.
More precisely, we find
\begin{equation}
		\label{eq:results_positivity}
		\begin{aligned}
			\bar{g}_4 \pm 2 \bar f_4 &\geq 0+\underbrace{ \frac{42 \bar f_2^2\pm50\bar f_2+21}{480 \pi ^2}}_{>0} \log(\hat{s} \sqrt{g_2}) + \cO(\hat{s} \sqrt{g_2})\,,
		\end{aligned}
\end{equation}
where $0<\hat{s} \lesssim M^2$.
The first (zero) term in the right-hand side of   \eqref{eq:results_positivity}  was obtained in \cite{Henriksson:2021ymi,Henriksson:2022oeu} neglecting the branch cuts from photon loops. The second term in these expressions is a novel result obtained by taking into account the $log$ terms. 
We would like to consider $\hat{s}\rightarrow 0$ to drop the error term.
However, in this limit, the bound \eqref{eq:results_positivity} is useless because the $log$ terms diverge.

More bounds similar to these can be derived---see for example \cite{Guerrieri:2020bto,Bellazzini:2020cot,Bellazzini:2021oaj} for  bounds including IR \emph{logs}. For bounds on EFTs from positivity derived in various other contexts, see  \cite{Caron-Huot:2020cmc,Caron-Huot:2021rmr,Davighi:2021osh,deRham:2021fpu,Henriksson:2021ymi,Henriksson:2022oeu,Caron-Huot:2022ugt,Tolley:2020gtv,deRham:2021bll,deRham:2022hpx,Chiang:2022ltp,Caron-Huot:2022jli}.

\paragraph{Bounds from full non-linear unitarity}
We use the primal numerical approach of \cite{Paulos:2017fhb,Homrich:2019cbt} to bound the observables \eqref{eq:coefficients} using full non-linear unitarity \eqref{eq:non-linear_unitarity}. This is done in section \ref{sec:numerics}. We briefly summarize our results here. 

\begin{figure}[h!]
	\centering
	\includegraphics[scale=1]{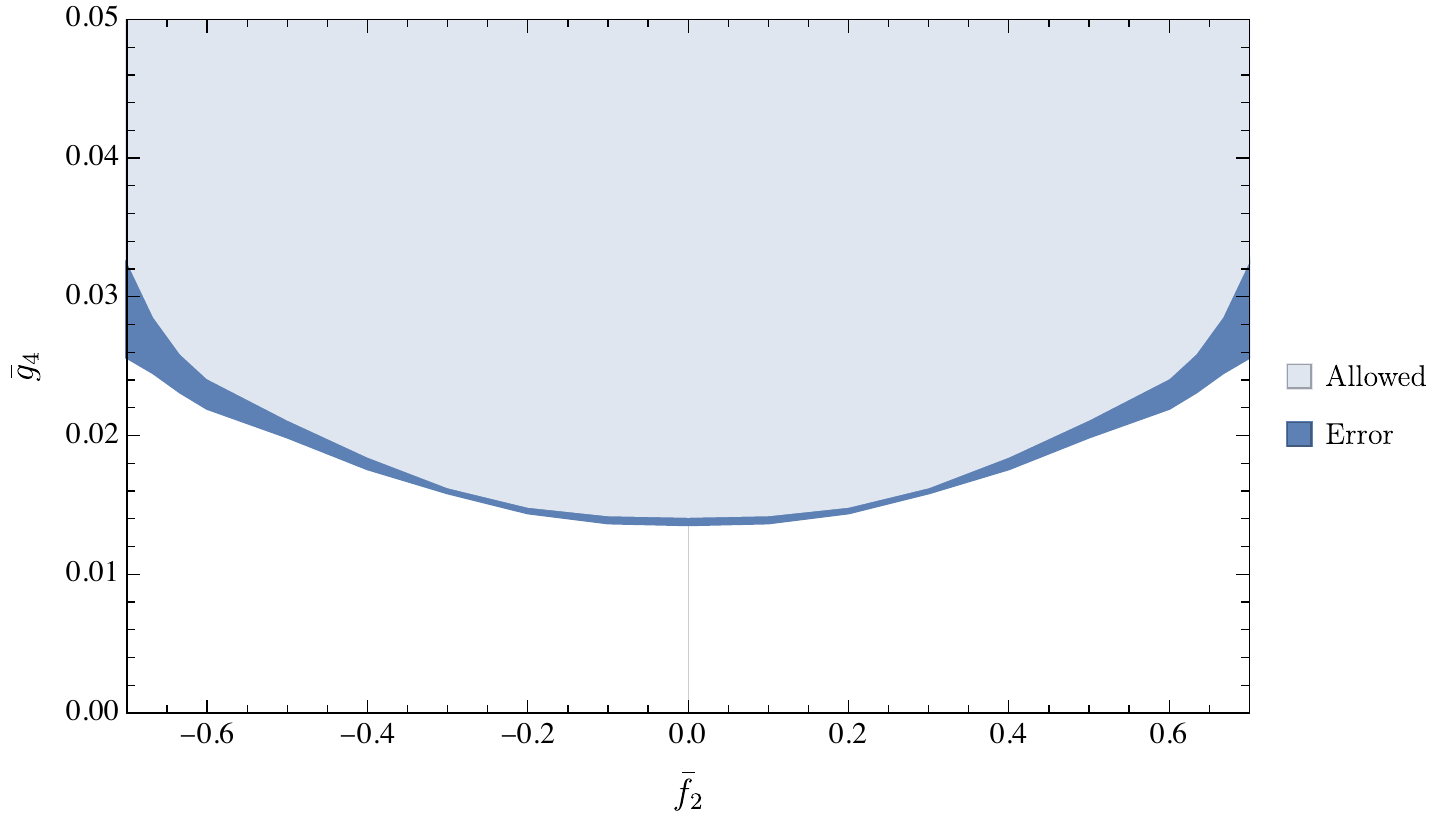}
	\caption{Lower bound on the observable $\bar g_4$ as a function of $\bar f_2$. The allowed region is shaded in light blue. The estimated error is indicated by dark blue.}
	\label{fig:lowerg1vsc1Extrapolated}
\end{figure}

\begin{figure}[h!]
	\centering
	\includegraphics[scale=1]{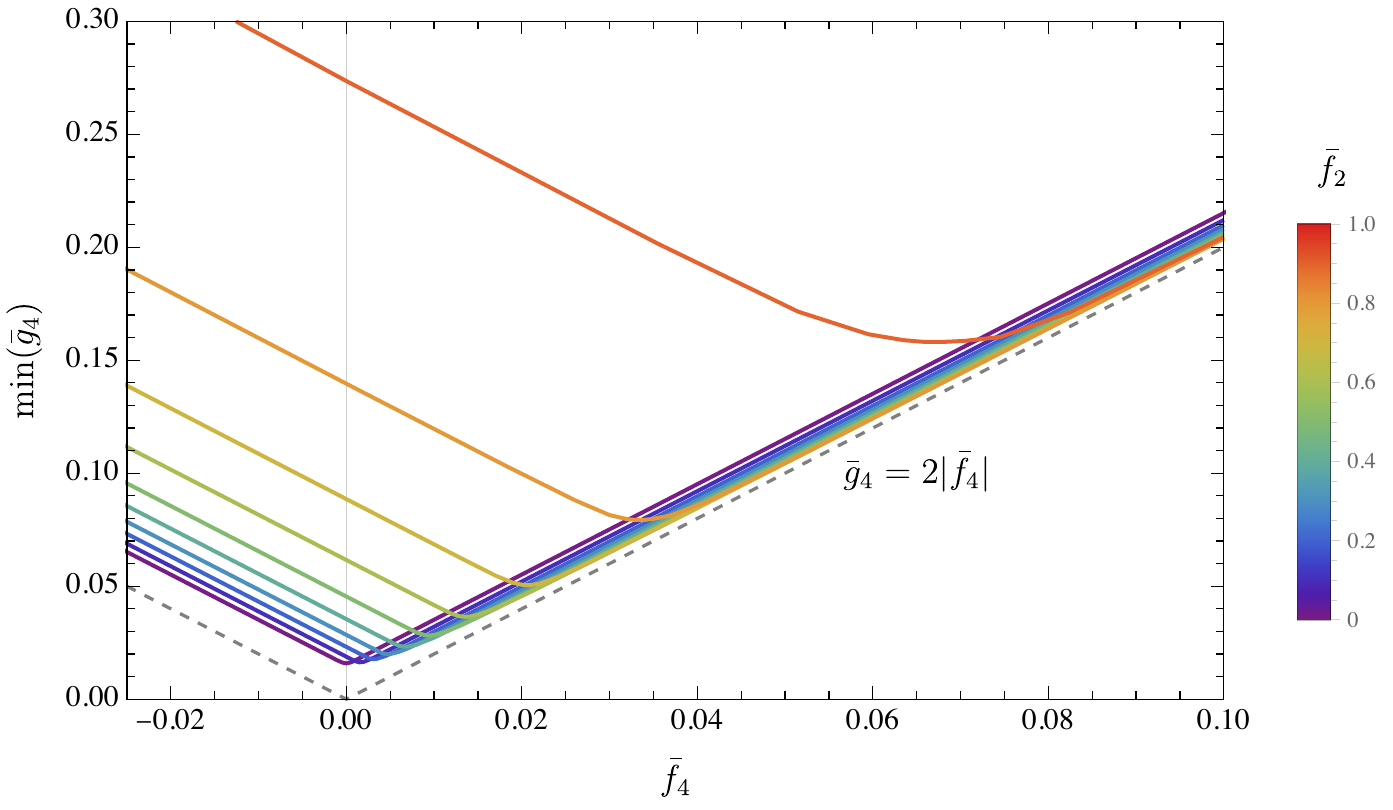}
	\caption{Lower bound on the $\bar g_4$ observable as a function $\bar{f}_4$ for a given value of $\bar{f}_2$. The allowed region lies above the colored lines. The dashed line described by $\bar g_4=2|\bar f_4|$ is placed for   reference.}
	\label{fig:g4vsf4vsf2INTRO}
	
	\vspace{10mm}
	
	\includegraphics[scale=1]{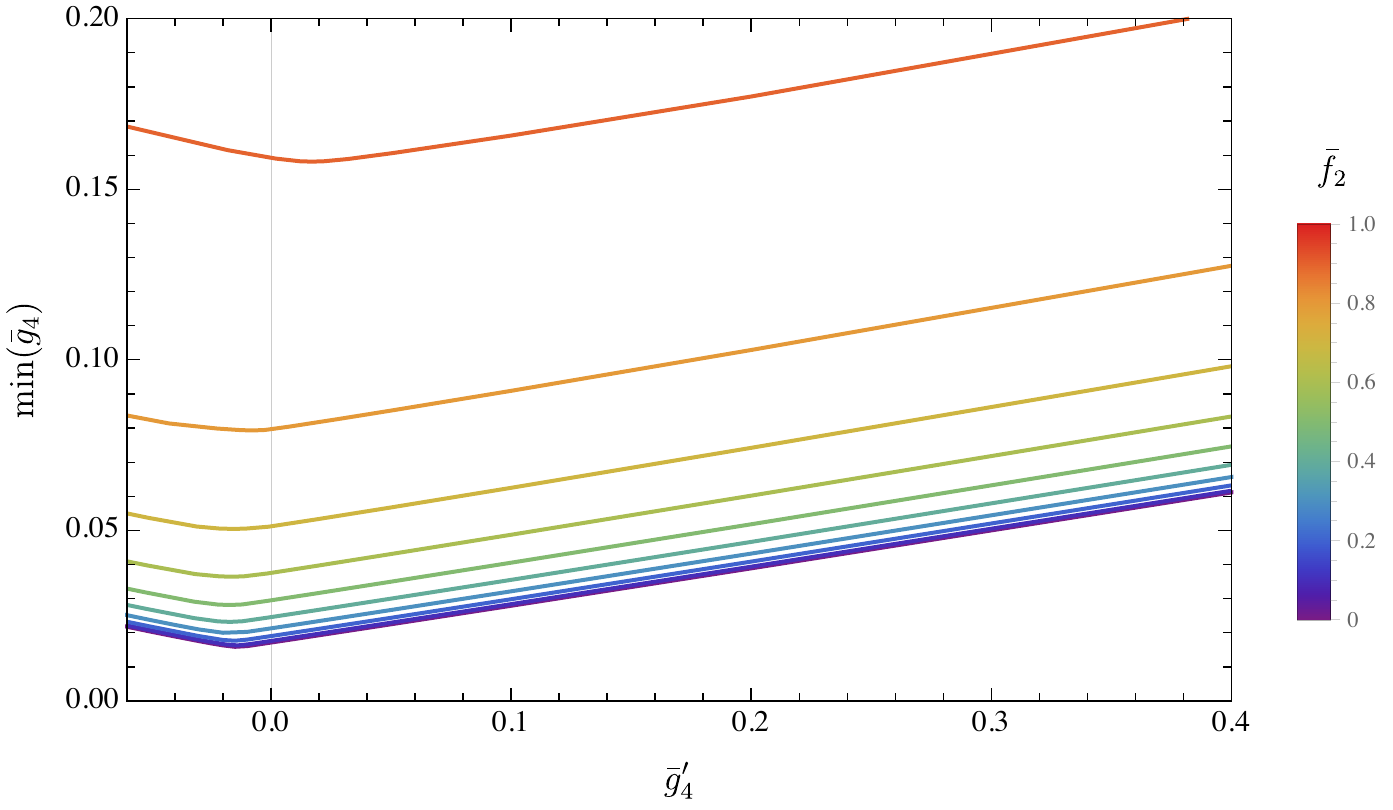}
	\caption{Lower bound on the $\bar g_4$ observable as a function $\bar{g}_4'$ for a given value of $\bar{f}_2$. The allowed region lies above the colored lines.}
	\label{fig:g4vsg4pvsf2INTRO}
\end{figure}

First, we found numerically that the bound on $\bar f_2$ is identical to \eqref{eq:analytic_bounds} which was found by using positivity only. Second, we found that neither upper nor lower bounds exist on the obervables $\bar g_3$, $\bar f_3$ and $\bar h_3$. Finally, we discovered that there exists a lower bound on $\bar g_4$. The bound is presented in figures \ref{fig:lowerg1vsc1Extrapolated} - \ref{fig:g4vsg4pvsf2INTRO}. There the lower bound on $\bar g_4$ is constructed as a function of $\bar f_2$, $\bar f_4$ and $\bar g_4'$ respectively.

All our numerical data can be downloaded from \href{https://doi.org/10.5281/zenodo.7308006}{https://doi.org/10.5281/zenodo.7308006}.

\paragraph{The absence of bounds} As stated in the previous paragraph, we often see that there is no bound on a given observable. We explain this fact in section \ref{sec:weak_coupling_models} by explicitly constructing weakly coupled theories which satisfy all our assumptions and have unbounded observables. For instance we show analytically that no bounds exist on $\bar g_3$, $\bar f_3$ and $\bar h_3$.

\newpage
\section{Bounds from positivity and constraints on EFTs}
\label{sec:positivity}

In this section we will write relations which allow to express the observables \eqref{eq:coefficients} or \eqref{eq:observables_EFT} as integrals of the amplitudes $\Phi_i(s,t,u)$. These relations are called dispersion relations. For simplicity we will work in the forward limit $t=0$. Combining dispersion relations with positivity \eqref{eq:positivity} allows us to bound our observables. In what follows we will focus only  on $g_2$, $f_2$, $g_4$ and $f_4$.

\paragraph{Positivity constraints}
Positivity constraints were given in \eqref{eq:positivity} in terms of partial amplitudes. These constraints can be equivalently translated into amplitudes in the forward limit using the inverse of \eqref{eq:partial_amplitudes_intro}. We get then
\begin{equation}\label{eq:positivity_forward}
	\Im (\Phi_1(s)\pm\Phi_2(s))\geq 0,\qquad
	\Im \Phi_3(s) \geq0.
\end{equation} 
We independently derive these constraints also in appendix \ref{app:ForwardLimit}.
Here and in the rest of this section we use the following short-hand notation for the forward amplitudes
\begin{equation}
	\label{eq:forward_amp}
	\Phi_i(s) \equiv \Phi_i(s, \, t=0,\, u=-s).
\end{equation}

\paragraph{Dispersion relations}
Let us start by defining the following functions
\begin{equation}
	\label{eq:Vn}
	V_n^\pm(s) \equiv\frac{\Phi_1(s)\pm\Phi_2(s)+ \Phi_3(s)}{s^{n+1}}
\end{equation}
in the $s$ complex plane. 
The functions $V_n^\pm(s)$ have an analytic structure inherited from the functions $\Phi_i(s,t,u)$ as depicted in figure \ref{fig:EFT_ampNEW} and an additional pole at $s=0$. The functions $V_n^\pm(s)$ obey $V_n^\pm(-s)=-V_n^\pm(s)$ for even $n$ due to  $s-u$ crossing symmetry in equation \eqref{eq:crossing_intro} and the discussion below it.  Finally, the imaginary part of the functions $V_n^\pm(s)$ is non-negative due to positivity constraints \eqref{eq:positivity_forward} for $s>0$.

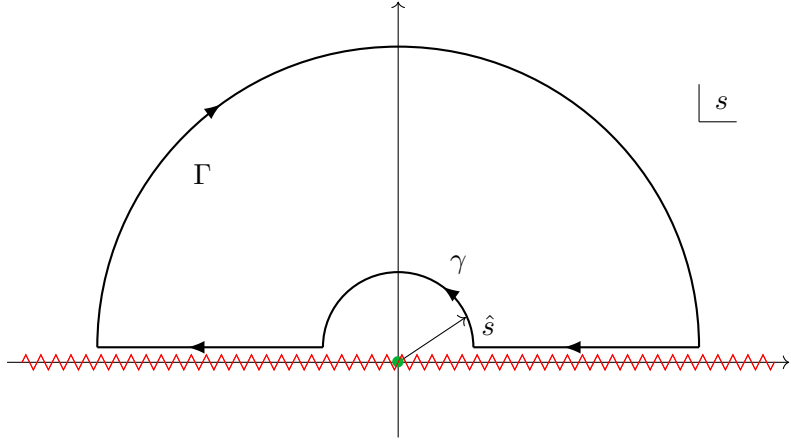
\begin{figure}
	\centering 
	\begin{tikzpicture}
		\draw[->] (-5.2,0)-- (5.2,0);
		\draw[->] (0,-1)--(0,4.8);
		
		\draw[thick,-, decoration={ markings,mark = at position 0.3 with {\arrow[line width=1.2pt]{latex}}}, postaction={decorate}]  		(-4,0.2) arc (-180:-360:4cm);
		\draw[thick,-, decoration={ markings,mark = at position 0.3 with {\arrow[line width=1.2pt]{latex}}}, postaction={decorate}] (1,0.2) arc (0:180:1cm);
		\draw[thick,-, decoration={ markings,mark = at position 0.6 with {\arrow[line width=1.2pt]{latex}}}, postaction={decorate}]  (4,0.2) --  (1,0.2);
		\draw[thick,-, decoration={ markings,mark = at position 0.6 with {\arrow[line width=1.2pt]{latex}}}, postaction={decorate}]  (-1,0.2) -- (-4,0.2);
		
		\draw (0.8,1.3) node{$\gamma$};
		\draw (-2.6,2.5) node{$\Gamma$};
		
		\draw (4,3.2)--(4.5,3.2);
		\draw (4,3.2)--(4,3.7);
		\draw (4.3,3.45) node{$s$};
		
		\draw[red, decoration = {zigzag, segment length = 2mm, amplitude = 1mm}, decorate] (0,0) -- (5,0) ;
		\draw[red, decoration = {zigzag,segment length = 2mm, amplitude = 1mm}, decorate] (-5,0) -- (0,0);
		
		\draw[->] (0,0)-- (0.9,0.6);
		\draw (1.2, 0.5) node{$\hat{s}$};
		
		\draw[verde] (0,0) node{$\bullet$};
		
		\draw[red, decoration = {zigzag,segment length = 2mm, amplitude = 1mm}, decorate] (-5,0) -- (5,0);
		
	\end{tikzpicture}
	\caption{Integration contour of the analytic functions $V_n^\pm$ defined in \eqref{eq:Vn} in the $s$ complex plane. The functions $V_n^\pm$ have the same analytic structures as in figure \ref{fig:EFT_ampNEW}. Compared to figure \ref{fig:EFT_ampNEW} they have however an additional pole at $s=0$ indicated by a green dot.}
	\label{fig:contour_arcs}
\end{figure}

Integrating $V_n^\pm(s)$ over a closed contour as depicted in figure \ref{fig:contour_arcs} we get
\begin{equation}
	\label{eq:integral_closed}
	\oint ds V_n^\pm(s)=0
\end{equation}
since $V_n^\pm(s)$ are analytic inside this contour of integration.
The contour has several pieces: the small arc with radius $\hat{s}$ denoted by $\gamma$, the big arc with infinitely large radius denoted by $\Gamma$ and two horizontal stretches.
 Splitting the integral in \eqref{eq:integral_closed} into these pieces we get
\begin{equation}
	\label{eq:sum_integrals}
	\int_\infty^{\hat{s}} ds V_n^\pm(s) +\int_{-\hat{s}}^{-\infty} ds V_n^\pm(s) +\int_\gamma ds V_n^\pm(s)  +\int_\Gamma ds V_n^\pm(s) =0.
\end{equation}
Assuming the analogue of the Martin-Froissart bound for spin one massless particles\footnote{For a recent discussion of the Froissart bound in the case of massless spin two particles see \cite{Haring:2022cyf}.} we get
\begin{equation}
	\lim_{|s|\rightarrow \infty} s\, V_n^\pm(s) =0 \, , \quad \text{for } n\geq 2
	\label{Froissart_n}
\end{equation}
As a result the integral over the large arc $\Gamma$ vanishes
\begin{equation}
	\label{eq:term_1}
	\int_\Gamma ds V_n^\pm(s) = 0.
\end{equation}
Using $s-u$ crossing symmetry we get\footnote{The first integral is evaluated slightly above the right branch cute. By crossing symmetry, the left integral is related to the integral slightly below the right branch cut. The sum of the two terms gives a discontinuity which in turn is related to the imaginary part of the amplitude.}
\begin{equation}
		\label{eq:term_2}
-\left(	\int_\infty^{\hat{s}} ds V_n^\pm(s) +\int_{-\hat{s}}^{-\infty} ds V_n^\pm(s)\right) =2i\, \int_{\hat{s}}^\infty ds \Im V_n^\pm(s).
\end{equation}
In order to evaluate the third integral in \eqref{eq:sum_integrals} we use the representation \eqref{eq:amplitudes_EFT} of the amplitudes which is valid for $|s|\lesssim M^2$. Performing the change of variables $s=\hat{s} \cos\phi$, where $\phi\in [0;\pi]$, we get  
\begin{equation}
		\label{eq:term_3}
	\begin{aligned}
		\int_\gamma ds V_2^\pm(s) &= \pi i\,2(g_2\pm f_2) +\mathcal{O}(\hat{s}^2/ M^8),\\
		\int_\gamma ds V_4^\pm(s) &= \pi i\,2\left(g_4 \pm 2f_4 + \left(\beta _{1,1}+\beta _{1,3}\ \pm \beta_{2}\right)\log\left(\hat{s}\sqrt{g_2}\right)\right)+\mathcal{O}(\hat{s}/M^{10})\,,
	\end{aligned}
\end{equation}
where we estimated the error using the expected EFT scaling of Wilson coefficients with the cutoff scale $M$.

Plugging equations \eqref{eq:term_1} - \eqref{eq:term_3} into \eqref{eq:sum_integrals} we finally obtain the following dispersion relations
\begin{equation}
	\label{eq:sum_rule_A}
	g_2 \pm f_2 +\mathcal{O}(\hat{s}^2/M^8) = \frac{1}{\pi} \int_{\hat{s}}^{\infty}ds\frac{\Im [\Phi_1(s)\pm\Phi_2(s)+ \Phi_3(s)]}{s^3},
\end{equation}
together with
\begin{multline}
\label{eq:sum_rule_B}
g_4 \pm 2f_4 + \left(\beta _{1,1}+\beta _{1,3}\ \pm \beta_{2}\right)\log\left(\hat{s}\sqrt{g_2}\right)  +\mathcal{O}(\hat{s}/M^{10}) =\\ \frac{1}{\pi} \int_{\hat{s}}^{\infty}ds\frac{\Im [\Phi_1(s)\pm\Phi_2(s)+ \Phi_3(s)]}{s^5}.
\end{multline}

\paragraph{Bounds on $g_2$ and $f_2$} Consider the dispersion relation \eqref{eq:sum_rule_A} and
take the limit $\hat{s}\rightarrow 0$. In this limit the combination $g_2\pm f_2$ is related to a certain integral of the amplitudes. Due to positivity in the form \eqref{eq:positivity_forward} this integral is non-negative and we obtain the following rigorous inequalities
\begin{equation}
	\label{eq:positivity_bound_A}
		g_2\geq0,\qquad -1\leq \frac{f_2}{g_2} \leq 1 .
\end{equation}

\paragraph{Bounds on $g_4$ and $f_4$} Consider now the dispersion relation \eqref{eq:sum_rule_B}. Naively it is impossible to take the limit $\hat{s}\rightarrow 0$ because there is a log term in the left-hand side which diverges in this limit. Nevertheless, using positivity conditions \eqref{eq:positivity_forward} we arrive at 
\begin{equation}
\label{boundg4f4}
	g_4 \pm 2f_4 + \underbrace{\left(\beta _{1,1}+\beta _{1,3}\ \pm \beta_{2}\right)}_{<0} \log\left(\hat{s}\sqrt{g_2}\right)  +\mathcal{O}(\hat{s}/M^{10}) \geq 0.
\end{equation}
This relation holds for any $\hat{s}$ in the range $\hat{s}\lesssim M^2$. 
Unfortunately this relation is not very useful because either the error term is large (for $\hat{s}$ large but still below the cutoff scale $M^2$) or the $\log$ term is large (for $\hat{s}$ small).
Dividing  by $g_2^2$, we obtain a dimensionless version of \eqref{boundg4f4}:
\begin{equation}
		\begin{aligned}
			\bar{g}_4 \pm 2 \bar f_4 &\geq 0+\underbrace{ \frac{42 \bar f_2^2\pm50\bar f_2+21}{480 \pi ^2}}_{>0} \log(\hat{s} \sqrt{g_2}) + \cO(\hat{s} \sqrt{g_2})\,,
		\end{aligned}
\end{equation}
 which we quoted in the introduction. Here we further assumed that $g_2 \sim M^{-4}$ to simplify the error term.

It is actually possible to take the limit $\hat{s}\rightarrow 0$ inside \eqref{eq:sum_rule_B}. To this end, consider the following representation of the logarithm
\begin{equation}
	\log(x) = \int_{x}^{\infty}dy\frac{-1}{y(1+y)} + \cO(x).
\end{equation}
It allows to bring the result \eqref{eq:sum_rule_B} into the following form
\begin{equation}
	\label{eq:sumrule_g4f4}
g_4 \pm 2 f_4= \frac{1}{\pi} \int_{\hat{s}}^{\infty}ds\left(\frac{\Im [\Phi_1(s)\pm \Phi_2(s) +\Phi_3(s)]}{s^5} + \frac{\pi (\beta_{1,1}+\beta_{1,3}\pm\beta_{2})}{s(1+s\sqrt{g_2})}\right) + \cO(\hat{s}/M^{10}).
\end{equation}
The integrand here is well-defined at $\hat{s}\rightarrow 0$ since the divergence of the first term governed by \eqref{eq:amplitudes_EFT} cancels the divergence of the second term. We can thus write  an explicit integral form of $g_4\pm 2f_4$ as
\begin{equation}
	\label{eq:sumrule_g4f4_new}
	g_4 \pm 2 f_4= \frac{1}{\pi} \int_{0}^{\infty}ds\left(\frac{\Im [\Phi_1(s)\pm \Phi_2(s) +\Phi_3(s)]}{s^5} + \frac{\pi (\beta_{1,1}+\beta_{1,3}\pm\beta_{2})}{s(1+s\sqrt{g_2})}\right).
\end{equation}

Let us now notice that
\begin{equation}
	\beta_{1,1}+\beta_{1,3}\pm\beta_{2} <0.
\end{equation}
This means that the second term in the integrand in \eqref{eq:sumrule_g4f4_new} is negative whereas the first term in the integrand is non-negative due to positivity. As a result the integrand in \eqref{eq:sumrule_g4f4_new} does not have definite sign and no positivity bound can be deduced on the simple combination $g_4 \pm 2 f_4$.

\section{Bounds from full non-linear unitarity}
\label{sec:numerics}

In this section we present our numerical bounds. We start in subsection \ref{sec:setup} by explaining our numerical setup. We will then use it to bound the coefficient $\bar f_2$ in subsection \ref{sec:bounds_f2}. We will explore bounds on $\bar g_3$, $\bar f_3$ and $\bar h_3$ in subsection \ref{sec:bounds_g3f3h3}. We will study bounds on $\bar g_4$, $\bar g_4'$ and $\bar f_4$ in subsection  \ref{sec:bounds_g4g4pf4}. In subsection \ref{subsec:linear_nogo} we show in all generality that parameters appearing linearly in a scattering amplitude are subject at most to one-sided bounds in the non-perturbative S-matrix bootstrap. 
Finally, in subsection \ref{sec:low_spin_dominance} we test the low spin dominance conjecture using our data.

We make our numerical data public. It can be downloaded from\\ \href{https://doi.org/10.5281/zenodo.7308006}{https://doi.org/10.5281/zenodo.7308006}.

\subsection{Numerical setup}
\label{sec:setup}

We consider the following non-perturbative ansatz for the three independent amplitudes\footnote{The prefactors in front of $\Phi_1\, , \Phi_5$  use the variable $\chi$ defined below \eqref{eq:chi_def}. They are included so that at low energy, the prefactors become those of the spinor-helicity amplitudes \eqref{eq:Phi_h_relation}.}
\begin{equation}
	\label{eq:ansatz}
	\begin{aligned}
		\Phi_1(s,t,u)  &=  \chi_s^2\sum_{a,b,c=0 }^{N_{\text{max}}} \alpha^1_{abc} \rho_s^a\rho_t^b\rho_u^c+ \mathbb{L}_1(s|t,u),\\
		\Phi_2(s,t,u)  &= \sum_{a,b,c=0 }^{N_{\text{max}}} \alpha^2_{abc} \rho_s^a\rho_t^b\rho_u^c+  \mathbb{L}_2(s,t,u),\\
		\Phi_5(s,t,u)  &=(-\chi_s\chi_t\chi_u)  \sum_{a,b,c=0 }^{N_{\text{max}}} \alpha^5_{abc} \rho_s^a\rho_t^b\rho_u^c.
	\end{aligned}
\end{equation}
Here $\alpha^i_{abc}$ are real dimensionless parameters. Due to crossing symmetry $\alpha^1_{abc}= \alpha^1_{acb}$ and $\alpha^2_{abc}$, $\alpha^5_{abc}$ are fully symmetric in their indices. The $\rho$-variable is defined as
\begin{equation}
	\rho_s \equiv \frac{\sqrt{-s_0}-\sqrt{-s}}{\sqrt{-s_0}+\sqrt{-s}}. 
\end{equation}
where $s_0<0$ is a free real parameter. The $\chi$-variable is defined as
\begin{equation}\label{eq:chi_def}
	\chi_s \equiv \frac{1}{4}(\rho_s-1)^2 - \frac{1}{4}(\rho_s-1)^3 = \frac{s}{s_0}-3\left(\frac{s}{s_0}\right)^2+ \cO(s^{5/2}).
\end{equation}
This variable was introduced in \cite{Guerrieri:2020bto}, and it is built in such a way that it becomes a constant at high energy.
It is also convenient to define the following object
\begin{equation}
	P \equiv \frac{1}{8}(1+\rho_s)(1+\rho_t)(1+\rho_u).
\end{equation}
At fixed scattering angle $\theta$ (defined using the Mandelstam variables in \eqref{eq:t_u_to_scattering_angle}) and small value of $s$ it behaves as $1$ and at fixed angle and large value of $s$ it decays as $\cO(s^{-3/2})$.
Finally the functions $\mathbb{L}_1$ and $\mathbb{L}_2$ are defined as\footnote{The prefactor $P$ is introduced here in order to make the functions $\mathbb{L}_1(s|t,u)$ and $\mathbb{L}_2(s,t,u)$ decay fast enough at large values of $s$. This allows to use the large energy constraints derived in appendix \ref{app:large_energy}.}
\begin{equation}
	\label{eq:loops_ansatz}
	\begin{aligned}
		\mathbb{L}_1(s|t,u) &\equiv s_0^{4}\,P\,\chi_s^2\left[\beta_{1,1} \chi_s^2 \log\chi_s+\beta_{1,2} \chi_t\chi_u \log\chi_s + \beta_{1,3} (\chi_t^2 \log\chi_t+\chi_u^2 \log\chi_u)\right],\\
		\mathbb{L}_2(s,t,u) &\equiv s_0^{4}\,P\,\beta_2(\chi_s^4\log\chi_s+\chi_t^4\log\chi_t+\chi_u^4\log\chi_u).
	\end{aligned}
\end{equation}
These functions %are constructed in such a way that at low energy they have the following form
have the following low energy expansion:
\begin{align*}
	\mathbb{L}_1(s|t,u)&= s^2\left[\beta_{1,1} s^2 \log(s/s_0)+\beta_{1,2} tu \log(s/s_0) + \beta_{1,3} (t^2 \log(t/s_0)+u^2 \log(u/s_0))\right] + \cO(s^5),\\
	\mathbb{L}_2(s,t,u)&=\beta_2(s^4\log(s/s_0)+t^4\log(t/s_0)+u^4\log(u/s_0))+ \cO(s^5).
\end{align*}
Comparing them with \eqref{eq:loops} we conclude that
\begin{subequations}
\begin{align}
	%\label{eq:expansion_Ls}
	\mathbb{L}_1(s|t,u) &= L_1(s|t,u)-
	s^2\left[\beta_{1,1} s^2+\beta_{1,2} tu+ \beta_{1,3} (t^2 +u^2 )\right]\log(-s_0\sqrt{g_2}) + \cO(s^5),\\
	\mathbb{L}_2(s,t,u) &= L_2(s,t,u)-\beta_2(s^4+t^4+u^4)\log(-s_0\sqrt{g_2})+ \cO(s^5).
\end{align}
\label{eq:expansion_Ls}
\end{subequations}

As was discussed in the introduction, scattering amplitudes of any massless spin one particles in the vicinity of $s=0$ have the representation \eqref{eq:amplitudes_EFT}. We, thus, need to expand \eqref{eq:ansatz} around $s=0$ taking into account \eqref{eq:expansion_Ls} and match the result with  \eqref{eq:amplitudes_EFT}. This procedure will generate a set of linear constraints on the parameters of the ansatz $\alpha$. Solving these constraints and plugging the solution back into \eqref{eq:ansatz} we obtain the final form of the ansatz which depends on the following parameters 
\begin{equation}
	\label{eq:coefficients_numerics}
	\{s_0; g_2, g_3, g_4, g_4', f_2, f_3,f_4,h_3,\, \text{rest of }\alpha s\}.
\end{equation}
All the coefficients in this list enter the ansatz linearly except for $s_0$, $g_2$ and $f_2$. Squares of $g_2$ and $f_2$ multiply the log terms. The former also enters inside the log terms.

The ansatz \eqref{eq:ansatz} has a finite number of terms controlled by the parameter $N_\text{max}$. All the numerical results depend on this parameter. The true bound is obtained by extrapolating the numerical results to $N_\text{max}=\infty$.

In order to impose non-linear unitarity we first compute partial amplitudes by plugging \eqref{eq:ansatz} into the definition \eqref{eq:partial_amplitudes_intro}. The integrals are evaluated numerically in Mathematica. They will depend on the set of parameters \eqref{eq:coefficients_numerics}. Plugging them into \eqref{eq:non-linear_unitarity} we obtain a set of unitarity constraints. We impose these constraints for a finite number of spins $\ell=0,\ldots, L_\text{max}$. All the numerical results also depend on the parameter $L_\text{max}$ and thus require an extrapolation to $L_\text{max}=\infty$. We carefully discuss the extrapolation procedure for both  $N_\text{max}$ and $L_\text{max}$  in subsection \ref{sec:g4vsf2}. In order to improve the convergence with $L_\text{max}$, we add the unitarity constraints  in the limit $L_\text{max}=\infty$. This is discussed in appendix \ref{app:large_spin}.  The convergence is also improved by adding the amplitude positivity constraint in the forward limit
\eqref{eq:positivity_forward_general}. As $\Lmax\to\infty$, these constraints are included in the full unitarity constraints imposed numerically \eqref{eq:non-linear_unitarity} via the positivity \eqref{eq:positivity}. However, for any finite $\Lmax$, \eqref{eq:positivity_forward_general} contains more information.

The unitarity constraints \eqref{eq:non-linear_unitarity} should be imposed for all $s\geq 0$. In practice we pick a finite grid of $s$ values where we impose unitarity. The points are chosen using the Chebyshev distribution in the $\rho_s$ variable (see \emph{e.g.} footnote 34 in \cite{Hebbar:2020ukp}). The number of points in this grid is denoted by $N_\text{grid}$. In all the calculations we use $N_\text{grid}=200$ for unitarity constraints with spins $\ell\leq 50$ and $N_\text{grid}=50$ for unitarity constraints with spins $\ell> 50$. We also include analytic constraints at $s=\infty$, for details see appendix \ref{app:large_energy}. 
On top of the unitarity constraints \eqref{eq:non-linear_unitarity} we also impose the positivity constraints %written in the form 
\eqref{eq:positivity_forward} at $N_\text{grid}=200$ values of $s$.

The amplitudes in $d=4$ are dimensionless. We can thus rewrite the ansatz \eqref{eq:ansatz} in terms of dimensionless quantities only. This is done by using $g_2$ as a scale and defining
\begin{equation}
	\label{eq:dimensionless_mandelstam}
	\bar s \equiv \sqrt{g_2}\, s,\qquad
	\bar t \equiv \sqrt{g_2}\, t,\qquad
	\bar u \equiv \sqrt{g_2}\, u.
\end{equation}
Identical definitions are understood to hold for $s_0$. 
  With these definitions the explicit dependence on $g_2$ completely disappears in \eqref{eq:ansatz} and the parameters we are left with are precisely the ones given in \eqref{eq:coefficients}. In other words the ansatz depends only on the following coefficients 
\begin{equation}
	\label{eq:coefficients_numerics_2}
	\{\bar s_0; \bar g_3, \bar g_4, \bar g_4', \bar f_2, \bar f_3, \bar f_4, \bar h_3,\, \text{rest of }\alpha s\}
\end{equation}
and $g_2$ is not a free variable.
We can chose the parameter  $\bar s_0$ %, $\bar t_0$ and $\bar u_0$ 
at our will. A particular choice does not play any role for significantly large values of $N_\text{max}$. It can happen however that for some values of these parameters the numerics converge better and it is desirable to search for their optimal values. In practice, this search is unfeasible since we do not want to numerically evaluate  the partial amplitudes for different choices of $\bar s_0$.

There is however a trick which allows to keep a tunable parameter in the numerical setup. To achieve that, we keep the ansatz in the form where $g_2$ enters explicitly. We then set  
\begin{equation}
	\label{eq:s0_choice}
	 s_0 = -1.
\end{equation}
This sets the scale of the problem which makes all the other variables and parameters (like $s$ or $g_2$) effectively dimensionless. The parameter $g_2$ is now free and can be set to any value. The bounds on \eqref{eq:coefficients_numerics_2} do not depend on this choice in the limit $N_\text{max}=\infty$. However, for a particular value of $g_2$ the numerics will converge faster. Faster convergence means in practice that at a fixed $N_\text{max}$ there is some value of $g_2$ which leads to a better result (\emph{e.g.} if we look for a minimal value of some parameter, better result means lower minimum of this parameter). For each optimization problem, we always perform a scan in $g_2$ first. This strategy was already used in \cite{Guerrieri:2020bto} and \cite{Guerrieri:2021ivu}.\footnote{We thank Andrea Guerrieri for emphasizing this idea.
} The explicit details of this scan are explained below.
Scanning over $\bar s_0$ and scanning over $g_2$ given \eqref{eq:s0_choice} is equivalent. This can be seen by using the definition \eqref{eq:dimensionless_mandelstam} and \eqref{eq:s0_choice} which lead to
\begin{equation}
	\bar s_0   = -\sqrt{g_2}.
\end{equation}

We solve the following optimization problem: find the values of parameters \eqref{eq:coefficients_numerics_2} such that the unitarity conditions \eqref{eq:non-linear_unitarity} are satisfied and one of the parameters in \eqref{eq:coefficients_numerics_2} has a minimal or maximal value. We construct optimization problems in Mathematica. For solving the optimization problems we use SDPB \cite{Simmons-Duffin:2015qma,Landry:2019qug}. The summary of the parameters used in our numerics can be found in table \ref{tab:params}.

\begin{table}
	\centering
	\begin{tabular}{c|c}
		\toprule
		SDPB precision &	$832$ (binary)		\\
		\midrule
		dualityGapThreshold	&	$10^{-8}$	\\
		\midrule
		integral precision  & $35$ (decimal)\\
		\midrule
		mathematica internal precision  & $200$ (decimal)\\
		\midrule
		$N_{\text{grid}}$	&	$\left\{\begin{matrix}200 &\ell\leq 50\\ 50  &\ell>50\end{matrix}\right.$	\\
		\bottomrule
	\end{tabular}
	\caption{Parameters in the numerical setup.}
	\label{tab:params}
\end{table}

\subsection{Bounds on $\bar f_2$}
\label{sec:bounds_f2}
The very first optimization problem we would like to address is: {\it what are the maximal and minimal allowed values of the observable $\bar f_2$}. For solving this optimization problem, we set the $\beta$'s as free linear parameters of our ansatz \eqref{eq:loops_ansatz}.\footnote{One may think that including the constraints \eqref{betaEFT} can lead to stronger bounds on $\bar{f}_2$. However, the extremal amplitudes that give $\bar{f}_2\to \pm 1$ are very weakly coupled and therefore the 1-loop $\log$ terms are irrelevant. }   The numerical solution of this optimization problem at a fixed value $N_\text{max}=20$ and several values of $g_2$ is presented in table \ref{tab:f2vsg2}.  Here we choose $L_\text{max}=50$. The numerics converges extremely fast for this run and no extrapolation to $\Nmax\rightarrow\infty$ and $\Lmax\rightarrow\infty$ is needed. From table \ref{tab:f2vsg2} we conclude that
\begin{equation}
	-1 \lessapprox\bar f_2\lessapprox +1,
\end{equation}
which is in perfect agreement with the positivity bound \eqref{eq:positivity_bound_A}. 
The solution is very stable and depends weakly on $g_2$. Nevertheless, we see that when $g_2 \rightarrow 0$ we get slightly better results (smaller minimum and greater maximum). This suggests that the extremal values of the bound $\bar f_2=\pm 1$ are saturated by free theories, which have $g_2=0$.\footnote{Strictly speaking, the amplitude vanishes in free theory, and there is no meaning for the observables \eqref{eq:coefficients}. What we mean by ``free theories'' in this section is, more precisely, the free limit of weakly coupled amplitudes, for which ratios of Wilson coefficients can have well defined limits.} 
	
Notice that the solution presented in table \ref{tab:f2vsg2} is symmetric under the $\bar f_2 \leftrightarrow -\bar f_2$ exchange. This comes from the fact that  the unitarity constraints \eqref{eq:non-linear_unitarity} are symmetric under $\Phi_2 \leftrightarrow -\Phi_2$ when $\Phi_5=0$. Investigating the solution leading to table \ref{tab:f2vsg2} we indeed see that $\Phi_5\approx0$ numerically. For further discussions of the symmetries of the unitarity constraints, see the last paragraph of appendix \ref{app:unitarity}. 

The result obtained here can be seen as a non-trivial check of our numerical setup. Interestingly, using full unitarity we do not get stronger bounds on $\bar f_2$ than equation \eqref{eq:positivity_bound_A}, which follows from positivity only. This is easy to understand in light of the observation above: the bounds are saturated by the free limit of weakly coupled theories, for which the constraints coming from full non-linear unitarity are irrelevant. We review the free theories saturating this bound in subsection \ref{subsec:rule_in_8}.

\begin{table}[h!]
	\centering
	\begin{tabular}{c|cc}
		\toprule
		$g_2$ & $\min(\bar f_2)$ &$\max(\bar f_2)$\\
		\midrule
		$10^{-2}$& $-0.998710$ & $0.998710$\\
		$1$& $-0.998619$& $0.998619$\\
		$10^2$&$-0.991237$& $0.991237$\\
		\bottomrule
	\end{tabular}
	\caption{Minimal and maximal allowed values of the $\bar{f}_2$ observable for different choices of $g_2$. These results are obtained with $\Nmax=20$ and $\Lmax=50$.}
\label{tab:f2vsg2}
\end{table}

\subsection{Bounding the $(\bar{g}_3,\bar{f}_3,\bar{h}_3)$ space}
\label{sec:bounds_g3f3h3}
Let us now study the upper and lower bounds on the $\bar g_3$, $\bar f_3$ and $\bar h_3$ observables.

Recall that our ansatz has log terms encoded into the objects $\mathbb{L}_1(s|t,u)$ and $	\mathbb{L}_2(s,t,u)$ defined in \eqref{eq:loops_ansatz}. The coefficients $\beta$ in these expressions depend on $g_2$ and $f_2$ quadratically according to \eqref{betaEFT}. Notice however that we can use our numerical procedure to determine only observables entering linearly in our ansatz. The terms with $g_2^2$ are not a problem since we always fix the value of $g_2$. However, we need to take some care of terms containing $f_2^2$. One option is to simply set $\mathbb{L}_1(s|t,u)=0$ and $	\mathbb{L}_2(s,t,u)=0$ in the ansatz since these terms only become important for bounding $\bar g_4$, $\bar g_4'$ and $\bar f_4$. Another option is to also fix the value of $f_2$ or equivalently the value of $\bar f_2$.\footnote{Yet another option is to keep $\beta$ as free linear parameters of the ansatz \eqref{eq:loops_ansatz}.} 
We have realized both options in practice, in this section we present only the results of the second one. It is important to notice, though, that these options are inequivalent in principle. We will comment on this fact in subsection \ref{subsec:linear_nogo}.

Let us now look for the minimal and maximal values of the $\bar g_3$ observable at some fixed value of $\bar f_2$. Concretely, we will use two values $\bar f_2=0$ and $\bar f_2=3/10$. We work at $\Nmax=20$ and $\Lmax=50$, and do not perform any extrapolation in these parameters. The solution of this optimization problem is presented in table \ref{tab:g3vsg2} for different values of $g_2$. From this table we see that the optimal value of $g_2$ is at $g_2=0$ (when $g_2\rightarrow0$ we get the lowest minimum and the highest maximum). The results of table \ref{tab:g3vsg2} suggest that the optimal solution is a free theory (because $g_2=0$) with
\begin{equation}
	-\infty \le \bar g_3 \le +\infty.
\end{equation}
Exactly the same conclusion holds for $\bar f_3$ and $\bar h_3$. We will confirm this finding by analytically constructing free theories with $g_2=0$, $\bar g_3=\pm \infty$, $\bar f_3=\pm \infty$ and $\bar h_3=\pm \infty$ in subsection \ref{subsec:rule_in_10}.

\begin{table}[h!]
	\centering
	\begin{tabular}{c|cc|cc}
		\toprule
			&\multicolumn{2}{c|}{$\bar{f}_2=0$}&\multicolumn{2}{c}{$\bar{f}_2=3/10$}	\\
		$g_2$	&	$\min(\bar{g}_3)$	&	$\max(\bar{g}_3)$	&	$\min(\bar{g}_3)$	&	$\max(\bar{g}_3)$	\\ \midrule
		$10^{-2}$	&	 $-546.9$ 	&	 $470.5$ 	&	 $-390.2$ 	&	 $466.4$ 	\\
		$1$	&	 $-54.67$ 	&	 $47.00$ 	&	 $-39.00$ 	&	 $46.59$ 	\\
		$10^2$	&	 $-5.397$ 	&	 $4.628$ 	&	 $-3.840$ 	&	 $4.585$ 	\\
		\bottomrule
	\end{tabular}
\caption{Minimal and maximal allowed values of $\bar{g}_3$ at a fixed value of $\bar f_2$ for different choices of $g_2$. These results are obtained with $\Nmax=20$ and $\Lmax=50$.}
\label{tab:g3vsg2}
\end{table}

\subsection{Bounding the $(\bar{g}_4,\bar{g}'_4,\bar{f}_4)$ space}
\label{sec:bounds_g4g4pf4}

Let us finally study bounds on the observables $\bar g_4$, $\bar g_4'$ and $\bar f_4$. As explained in the previous subsection, all the bounds are obtained at some fixed values of $\bar f_2$. In subsection \ref{sec:g4vsf2} we present our numerical results for the lower bound on $\bar g_4$ as a function of $\bar f_2$. In subsection \ref{sec:g4vsf2vsf4} we present our lower bound on $\bar g_4$ as a function of $\bar f_2$ and $\bar f_4$. Finally, in subsection \ref{sec:g4vsf2vsg4p} we present our lower bound on $\bar g_4$ as a function of $\bar f_2$ and $\bar g_4'$. No upper bound exists on $\bar g_4$ and $\bar g_4'$, and from the results below one can conclude that neither upper nor lower bounds exist on $\bar f_4$, as in the case of the $\bar g_3$, $\bar f_3$ and $\bar h_3$ observables studied in the previous subsection. As for the lower bound on $\bar g_4'$, the numerics in that region converge poorly, so no conclusion can be drawn.
We will partly explain the presence of unbounded directions in subsection \ref{subsec:rule_in_12}, by explicitly constructing free theories with infinitely large values for the Wilson coefficients. In particular, we will also show that $\bar g_4'$ cannot be bounded from below.

%%%%%%%%%%%%%%%%%
\subsubsection{$\min{\bar g_4}$ vs. $\bar f_2$}\label{sec:g4vsf2}
We start by minimizing $\bar g_4$ at several fixed values of $\bar f_2$. The results of the numerical optimization are given in table \ref{tab:g4vsg2}. This data is computed using $\Nmax=20$ and $\Lmax=50$. It is clear from the table that the best convergence is achieved at $g_2\sim 100$, since this value of $g_2$ gives the lowest minimum. In the reminder of subsection \ref{sec:bounds_g4g4pf4} we will always use $g_2=100$.

\begin{table}[h!]
	\centering
	\begin{tabular}{c|cccc}
		\toprule
		&	\multicolumn{4}{c}{$\min(\bar{g}_4)$}							\\ \midrule
		$g_2$ 	&	$\bar{f}_2=0$	&	$\bar{f}_2=3/10$	&	$\bar{f}_2=6/10$	&	$\bar{f}_2=9/10$	\\
		\midrule
		$10^{-2}$	&	$1.068$	&	$1.176$	&	$1.839$	&	$11.41$	\\
		$1$	&	$0.02478$	&	$0.03205$	&	$0.05527$	&	$0.2837$	\\
		$10$	&	$0.01711$	&	$0.02142$	&	$0.03974$	&	$0.1706$	\\
		$100$	&	$0.01588$	&	$0.01993$	&	$0.03646$	&	$0.1581$	\\
		$200$	&	$0.01583$	&	$0.01995$	&	$0.03645$	&	$0.1631$	\\
		$1000$	&	$0.01589$	&	$0.02025$	&	$0.03947$	&	$0.3270$	\\
		\bottomrule
	\end{tabular}
	\caption{Lower bound on $\bar{g}_4$ at different values of $g_2$ and $\bar f_2$. The optimal lowest minimum is achieved at $g_2\sim100$.  The results are obtained with $\Nmax=20$ and $\Lmax=50$.}
	\label{tab:g4vsg2}
\end{table}

In practice our bounds depend on the parameters $\Nmax$ and $\Lmax$. The correct bound is obtained only in the limit $\Nmax\to\infty$ and $\Lmax\to\infty$. Let us carefully discuss how one can estimate the correct lower bound on $\bar g_4$ in this limit. 

\begin{figure}[tb]
	\centering
	\includegraphics[width=1\linewidth]{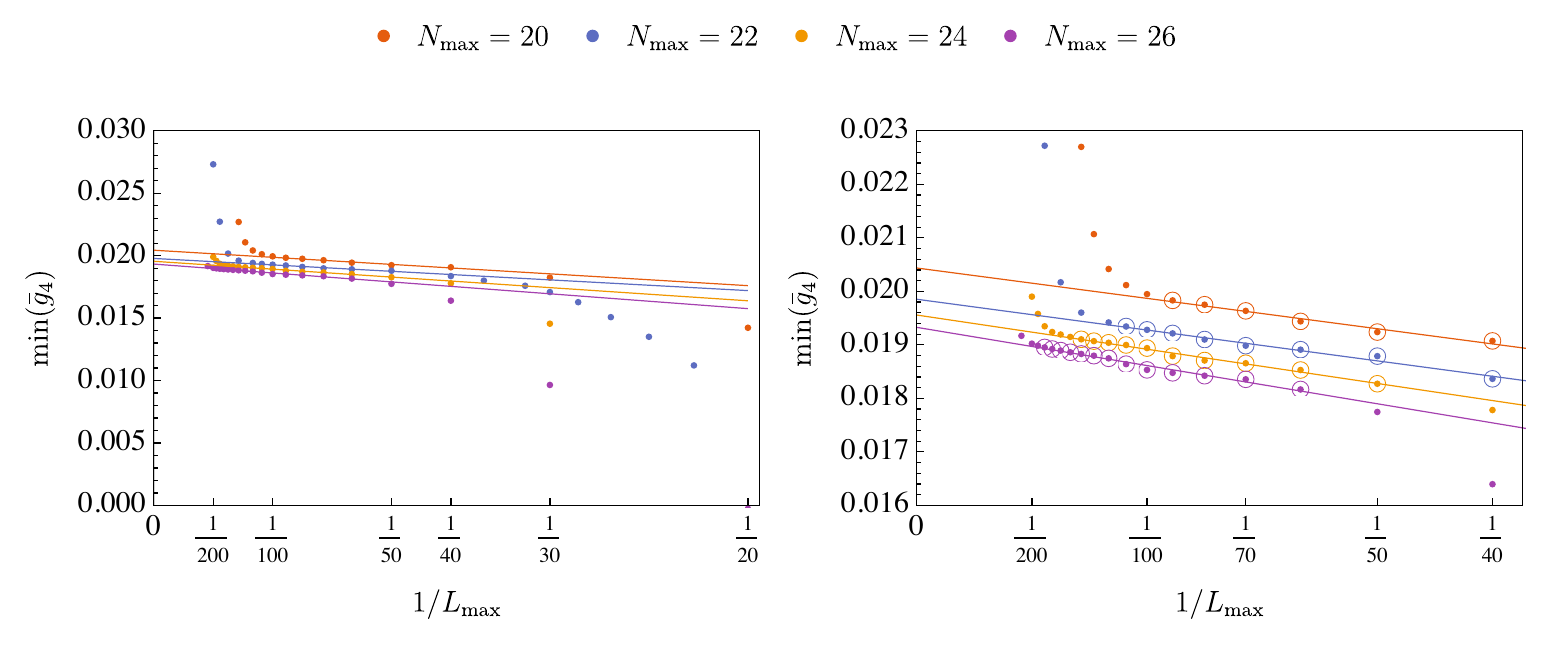}
	\caption{Lower bound on $\bar{g}_4$ at fixed $\bar{f}_2=-3/11$ (QED point).  Colored dots indicate the numerical data. Solid lines indicate linear extrapolations in $1/\Lmax$. The right plot is a zoomed version of the left one. The circles around the numerical data indicate where the data is linear in $1/\Lmax$. The linear regions are called the plateaus. The linear extrapolation is performed only within each plateau.}
	\label{fig:lowerg4_NmaxVSspin}
	
	\vspace{25mm}
	
	\includegraphics[scale=1]{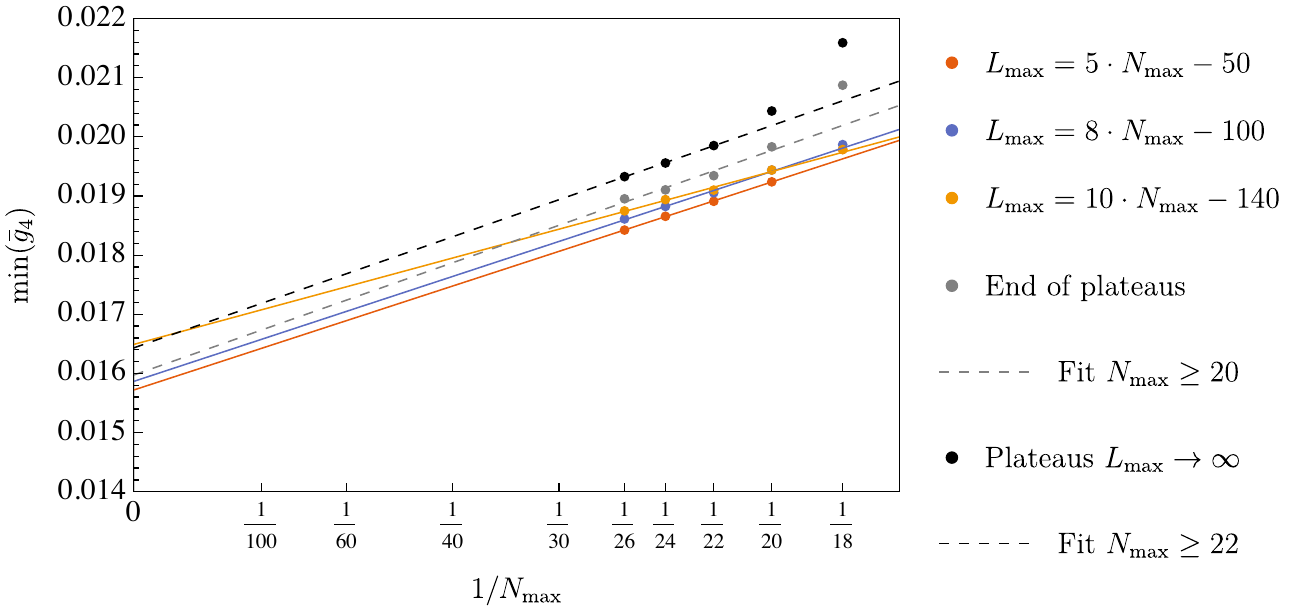}
	\caption{Lower bound on $\bar{g}_4$ at fixed $\bar{f}_2=-3/11$ (QED point).  Colored dots denote the numerical data obtained at various fixed values of $\Lmax$. Black dots are obtained instead by performing $\Lmax\rightarrow\infty$ extrapolation in figure \ref{fig:lowerg4_NmaxVSspin}.  Dashed lines indicate linear extrapolations in $1/\Nmax$ to $\Nmax\rightarrow\infty$.}
	\label{fig:lowerg4_FormulaVSNmax}
\end{figure}

Let us begin by using the strategy employed in \cite{Guerrieri:2021ivu}. In figure \ref{fig:lowerg4_NmaxVSspin} we study the dependence of the $\bar g_4$ bound on $\Lmax$ at fixed values of $\Nmax$. For concreteness we take $\bar f_2=-3/11$ which is the value in QED  at leading order in the coupling. The numerical data is indicated by colored dots. In the left plot of figure \ref{fig:lowerg4_NmaxVSspin} we see that the bound gets stronger when we increase $\Lmax$. For $\Nmax=20$ the bound stabilizes around $\Lmax=30$ and diverges around $\Lmax=100$. In the interval $\Lmax\in(30,100)$ the bound is linear. We refer to this interval as the plateau. For larger values of $\Nmax$ the divergence of the bounds begins at larger values of $\Lmax$. This can be explained as follows - for a fixed finite $\Nmax$, as we increase $\Lmax$, the number of unitarity constraints that the ansatz has to satisfy increases. At some point, the ansatz is not big enough to satisfy all of these constraints and therefore the bound that we get diverges. But if we then increase $\Nmax$, the ansatz is now bigger and therefore it can satisfy unitarity constraints for larger values of $\Lmax$. Nevertheless, even with this bigger $\Nmax$ and hence a bigger ansatz, the bound will still diverge at some (larger) $\Lmax$ and we must disregard the data after this point.

We therefore restrict our attention to the plateau\footnote{
	We find the   plateau using the discrete derivative $\texttt{dDer}(1/\Lmax)$. Then, we select  points such that  $\texttt{dDer}(1/\Lmax) \leq X\cdot \min(\texttt{dDer})$, where $X$ is a factor that can be chosen. In practice we chose $X=3$.}
and use linear extrapolation in $1/\Lmax$ to obtain the lower bound of $\bar g_4$ at $\Lmax\rightarrow \infty$ for fixed values of $\Nmax$. The extrapolations are indicated by solid lines in figure \ref{fig:lowerg4_NmaxVSspin}. The right plot in figure \ref{fig:lowerg4_NmaxVSspin} is  a zoomed in version of the left plot.  The circles around the numerical data there indicate the points which were included in the plateau. 

Once we obtain the extrapolated bounds at $\Lmax\rightarrow \infty$ we also do a linear extrapolation in $1/\Nmax$. The result is depicted in figure \ref{fig:lowerg4_FormulaVSNmax}  in black. Black dots indicate extrapolated $\Lmax\rightarrow\infty$ values. Black lines indicate linear extrapolation in $1/\Nmax$. From previous experience we expect the numerical data to be linear in $1/\Nmax$ starting from $\Nmax \sim 20$. This is compatible with the results depicted in figure \ref{fig:lowerg4_FormulaVSNmax}.

\begin{figure}[tb]
	\centering
	\includegraphics[scale=1]{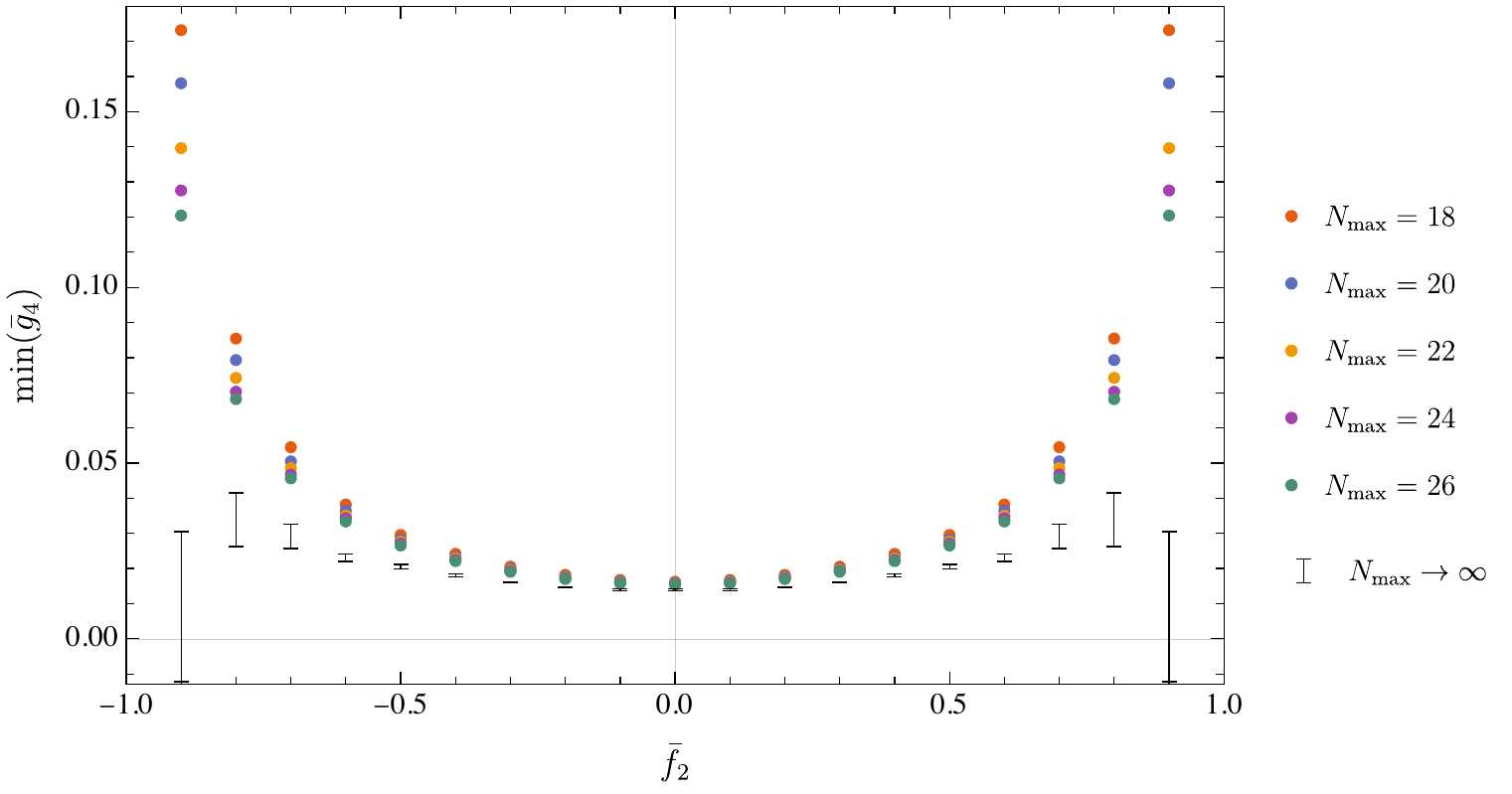}
	\caption{Lower bound on $\bar g_4$ as a function of $\bar{f}_2$. Colored dots represent the numericsl data obtained with $\Lmax$ defined in \eqref{eq:Lmax_NmaxFormula} and various values of $\Nmax$. Black bars represent the extrapolation of the numerical results to $\Nmax\rightarrow \infty$ and indicate the estimated errors.}
	\label{fig:lowerg4vsf2andNmax}
\end{figure}

Performing the above extrapolation in $\Lmax$ and $\Nmax$ is computationally very expensive since it requires obtaining a lot of numerical data. Let us explore a cheaper alternative. One could compute bounds at various values of $\Nmax$ with
\begin{equation}
	\label{eq:formula_Lmax}
	\Lmax = a \Nmax - b,
\end{equation}
with $a$ and $b$ some integer numbers. Once the data with this $\Lmax$ is obtained, we can perform a linear extrapolation in $1/\Nmax$ to $\Nmax\rightarrow\infty$. In figure \ref{fig:lowerg4_FormulaVSNmax} we make several choices of $a$ and $b$. All the resulting extrapolations lie very close to the one obtained by using the previous approach. Notice, however, that the choice of $a$ and $b$ cannot be completely arbitrary, the values of $\Lmax$ should always be inside the plateau. In figure \ref{fig:lowerg4_FormulaVSNmax} we also indicate the extrapolation using the points on the plateau right before the divergence. We denote this choice by ``End of plateaus''.
In the rest of this section, we shall use 
\begin{equation}\label{eq:Lmax_NmaxFormula}
	\Lmax=5\Nmax-50\,,
\end{equation}
to perform extrapolations to $\Nmax\rightarrow\infty$. From now on, we assume that \eqref{eq:Lmax_NmaxFormula} is a good choice for other values of $\bar f_2$ and all the other bounds computed below.

We are finally in a position to present our lower bound on $\bar g_4$ as a function of $\bar f_2$. It is given in figure \ref{fig:lowerg4vsf2andNmax}. The numerical data for different $\Nmax$ is depicted by colored dots. The extrapolated values are given by black bars which also reflect a rough error of the extrapolation.\footnote{To estimate the error on the extrapolated bound, we perform a linear interpolation using $\Nmax\geq N_*$ where $18\leq N_*\leq 24$ and compute the maximal difference between interpolated values. Therefore, this error indicates if the linear interpolation is a reasonable extrapolation of the bound to $\Nmax\to\infty$. }  Figure \ref{fig:lowerg1vsc1Extrapolated} presented in the introduction is obtained from \ref{fig:lowerg4vsf2andNmax} by using only the extrapolated results and connecting the points.

The lower bound on $\bar g_4$ is symmetric under $\bar f_2\to  -\bar f_2$ as a consequence of the $\Phi_2 \leftrightarrow -\Phi_2$ symmetry in the unitarity constraints \eqref{eq:non-linear_unitarity} when $\Phi_5=0$. Indeed, we observe that the optimal solution leads to $\Phi_5\approx0$ and thus the numerical solution is $\Phi_2 \leftrightarrow -\Phi_2$ symmetric. The same remains true in subsections \ref{sec:g4vsf2vsf4} and \ref{sec:g4vsf2vsg4p}.  The absolute minimum of the bound in figure  \ref{fig:lowerg1vsc1Extrapolated}  is achieved at $f_2=0$ and is given by
\begin{equation}\label{eq:absming4}
	\bar{g}_4\geq 0.0138\pm 0.0003.
\end{equation}

For larger values of $\bar f_2$ the errors quickly increase. In order to understand this, recall from subsection \ref{sec:bounds_f2} that the boundary values $\bar f_2 = \pm 1$ are saturated by free theories with $g_2=0$. Moreover, we shall prove in subsection \ref{subsec:rule_in_8} that free theories are in fact the only theories with $\bar f_2 = \pm 1$. In order to efficiently obtain bounds in this region we would need to keep $g_2\approx 0$, while in the present section we fix $g_2=100$. This leads to bad convergence when we approach $\bar f_2 = \pm 1$.

\subsubsection{$\min{\bar g_4}$ vs. $\bar f_2$ and $\bar f_4$}\label{sec:g4vsf2vsf4}

\begin{figure}[h!]
	\centering
	\includegraphics[scale=0.95]{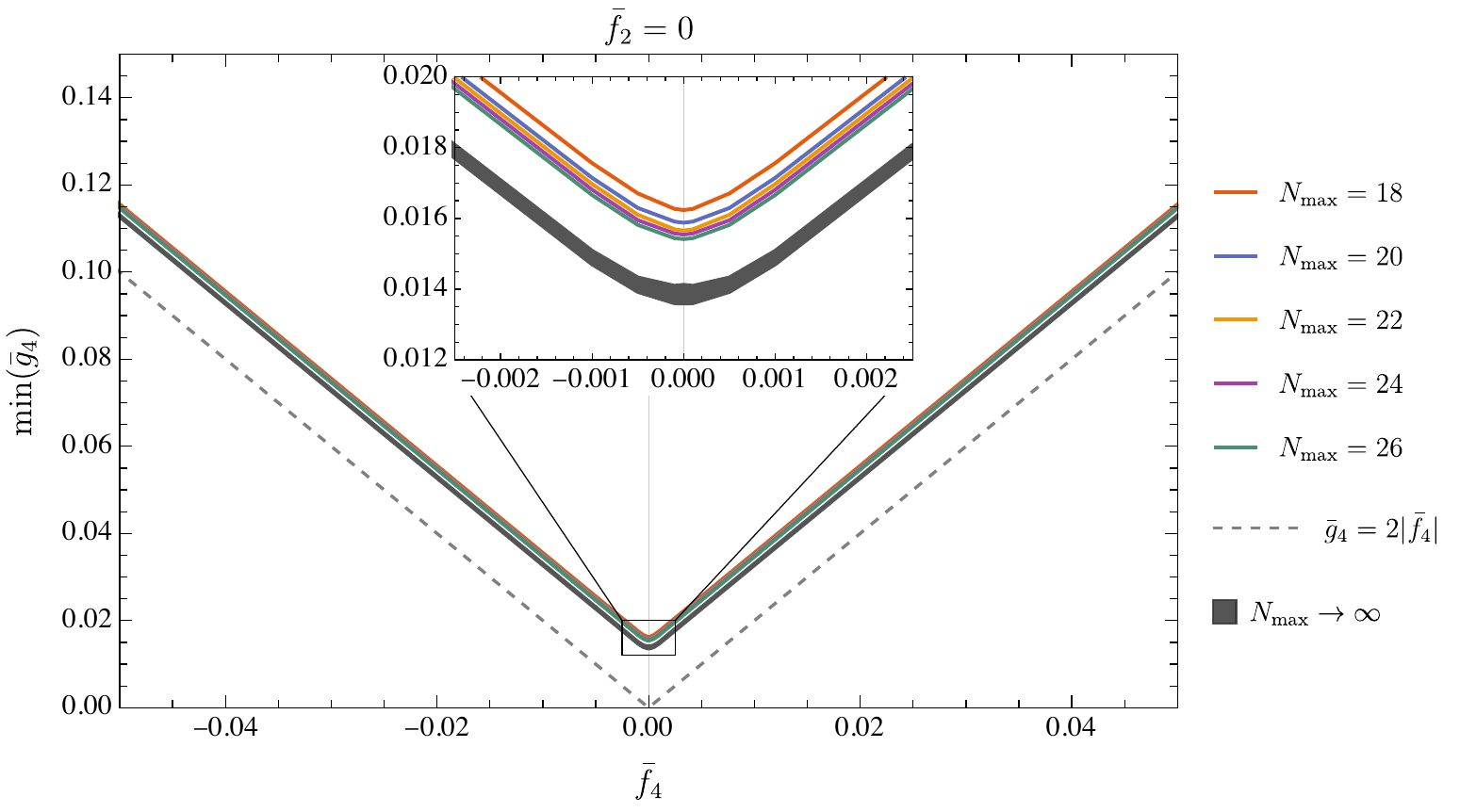}
	\caption{Lower bound on $\bar g_4$ as a function of $\bar{f}_4$ at the fixed value $\bar{f}_2=0$. Colored lines represent the numerical data obtained with $\Lmax$ defined in \eqref{eq:Lmax_NmaxFormula} and various $\Nmax$. The gray line represents the extrapolated bound to $\Nmax\rightarrow \infty$. The width of the gray line represents the estimated error. The extrapolated bound has the form $\bar{g_4}\geq (0.0129\pm 0.0001)+ (1.999\pm 0.001) |\bar{f}_4|$. The dashed line is drawn for reference. }
	\label{fig:g4vsf4atf20}
	
	\includegraphics[scale=0.95]{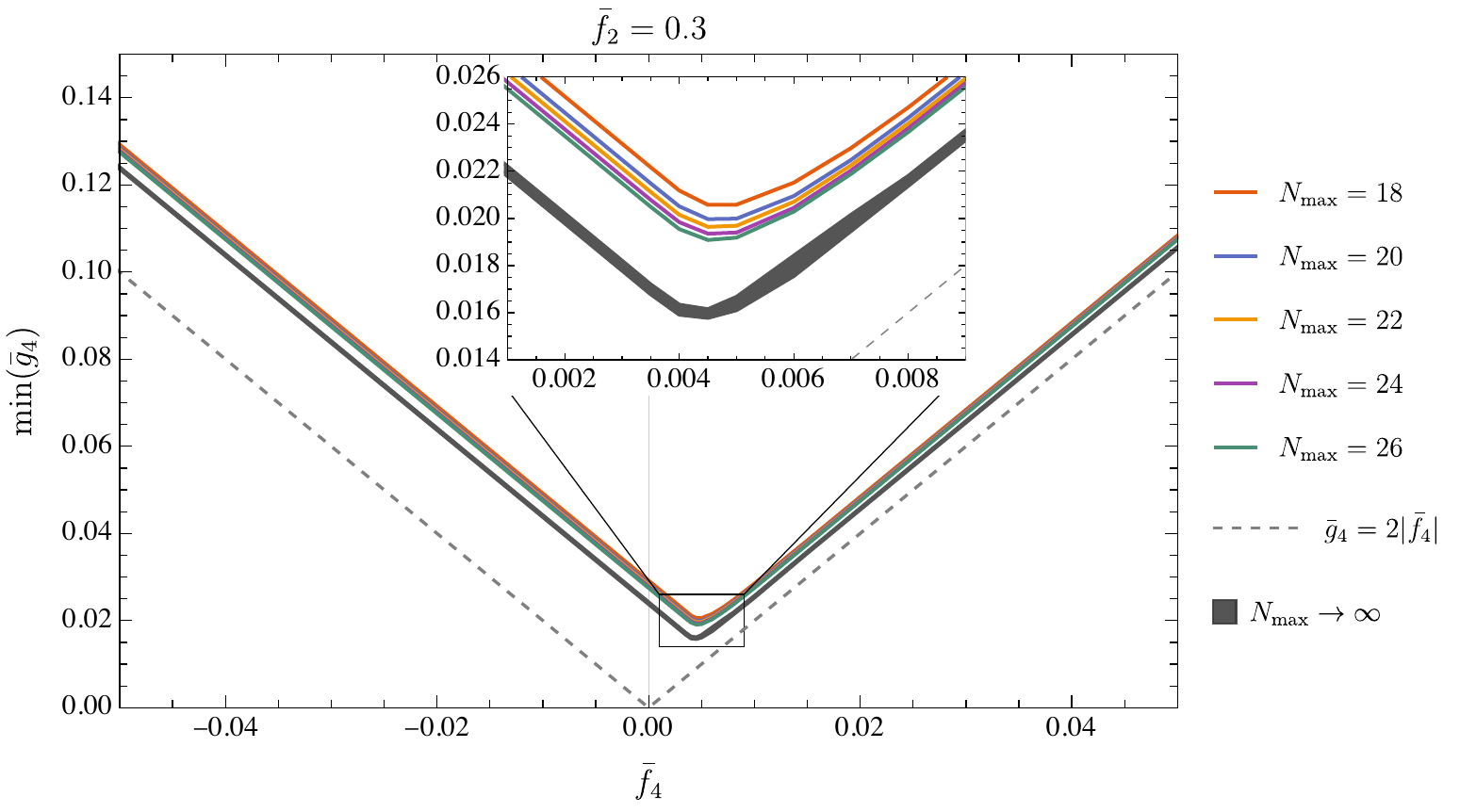}
	\caption{Lower bound on $\bar g_4$ as a function of $\bar{f}_4$ at the fixed value $\bar{f}_2=0.3$. The extrapolated bound to the  right of the minimum has the form $\bar g_4\geq (0.0055\pm 0.0001) + (2.000\pm 0.001) \bar f_4$. The extrapolated bound to the  left of the minimum has the form $\bar g_4\geq(0.0240\pm 0.0001)+(-1.997\pm 0.002)\bar f_4$ .
}
	\label{fig:g4vsf4atf203}
\end{figure}

The lower bound on $\bar g_4$ as a function of $\bar f_4$ at two fixed values $\bar{f}_2=0$ and $\bar{f}_2=0.3$ are shown in figures \ref{fig:g4vsf4atf20} and \ref{fig:g4vsf4atf203}. The colored lines indicate the numerical data for various values of $\Nmax$. The gray line indicates the extrapolated bound. The width of this line represents the extrapolation error. 

\begin{figure}[h!]
	\centering
	\includegraphics[scale=1]{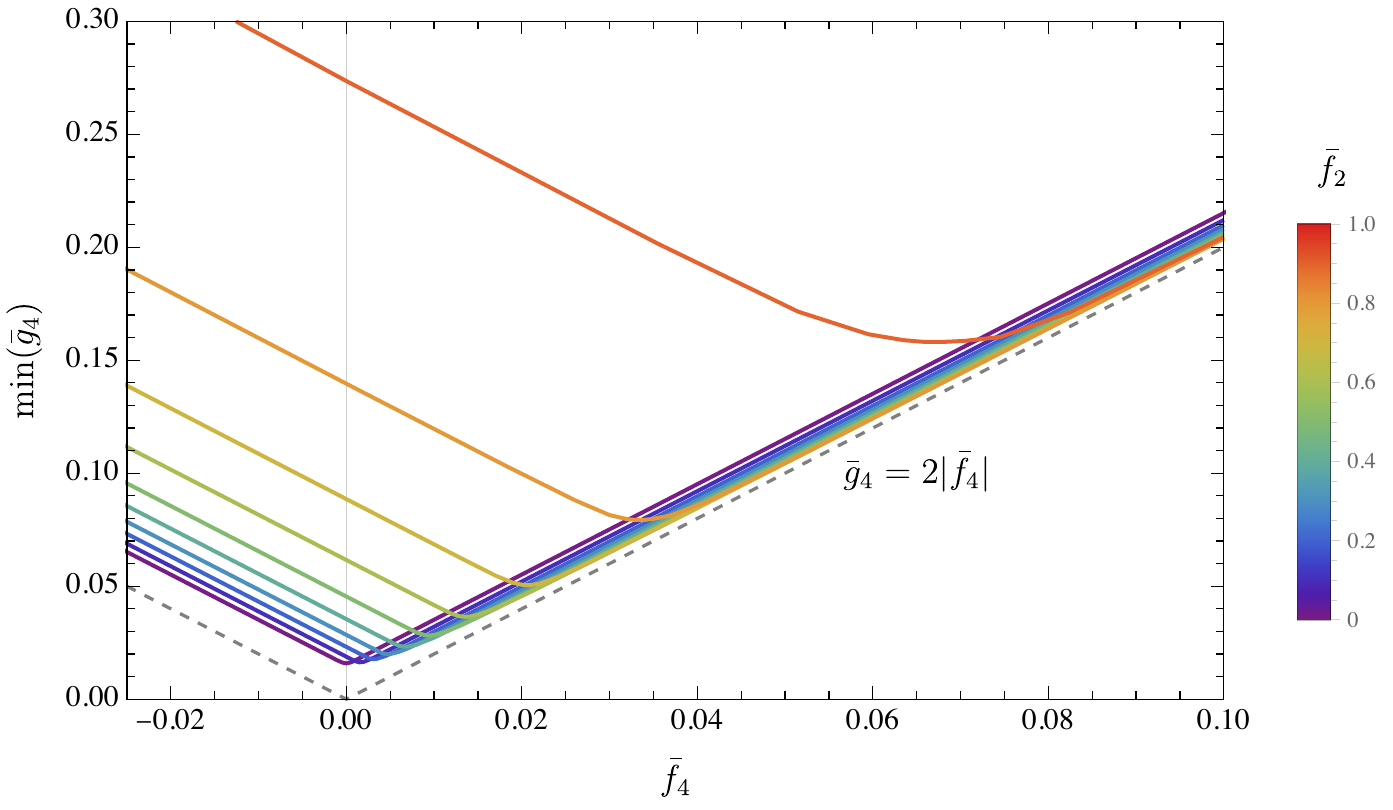}
	\caption{Lower bound on $\bar g_4$ as a function of $\bar{f}_4$ at fixed values of $\bar{f}_2$. This bound is constructed at $\Nmax=20$ and $\Lmax=50$.}
	\label{fig:g4vsf4vsf2}
\end{figure}

\begin{figure}[h!]
	\centering
	\includegraphics[scale=1]{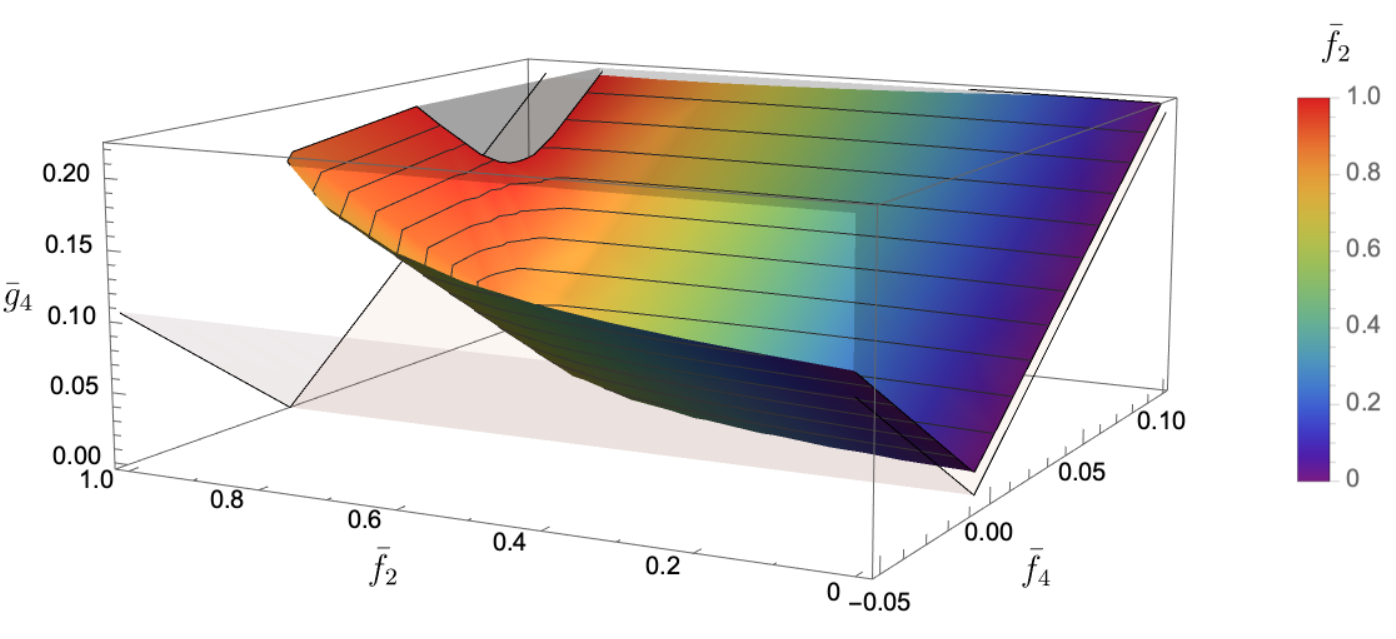}
	\caption{Lower bound on $\bar g_4$ as a function of $\bar{f}_4$ and $\bar{f}_2$. The transparent gray V-shape represents $\bar g_4 = 2\bar f_4$, it is reported for reference. The color gradient indicates places with the same values of $\bar{f}_2$ and is equivalent to the one used in figure \ref{fig:g4vsf4vsf2}. The solid lines indicate places with the same values of $\bar{g}_4$. The bound is constructed at $\Nmax=20$ and $\Lmax=50$.
	}
	\label{fig:3d_g4vsf4vsf2}
\end{figure}

As we can see from figures \ref{fig:g4vsf4atf20} and \ref{fig:g4vsf4atf203}, the extrapolated bound has exactly the same behavior as the finite $\Nmax$ one and lies very close to it. In what  follows we will focus on $\Nmax=20$ and $\Lmax=50$ in order to reduce the cost of our numerical computation. This allows us to scan over more values of $\bar{f}_2$. In figures \ref{fig:g4vsf4vsf2} and \ref{fig:3d_g4vsf4vsf2} we present the lower bound on $\bar g_4$ at $\bar{f}_2=n/10, \ n=0,1,... 9$. Using the $ \Phi_2 \leftrightarrow - \Phi_2$ symmetry discussed above we automatically obtain the bound also for $n=-9,... -1$.

It is interesting to notice that all our bounds in figures \ref{fig:g4vsf4atf20} - \ref{fig:g4vsf4atf203} satisfy the naive condition 
\begin{equation}
	\bar{g}_4 \pm 2 \bar f_4 \geq 0
\end{equation}
which is obtained from \eqref{eq:results_positivity} by dropping the log terms. We indicate this condition by a gray dashed line.


\subsubsection{$\min{\bar g_4}$ vs. $\bar f_2$ and $\bar g_4'$}\label{sec:g4vsf2vsg4p}
\begin{figure}[h!]
	\centering
	\includegraphics[scale=0.95]{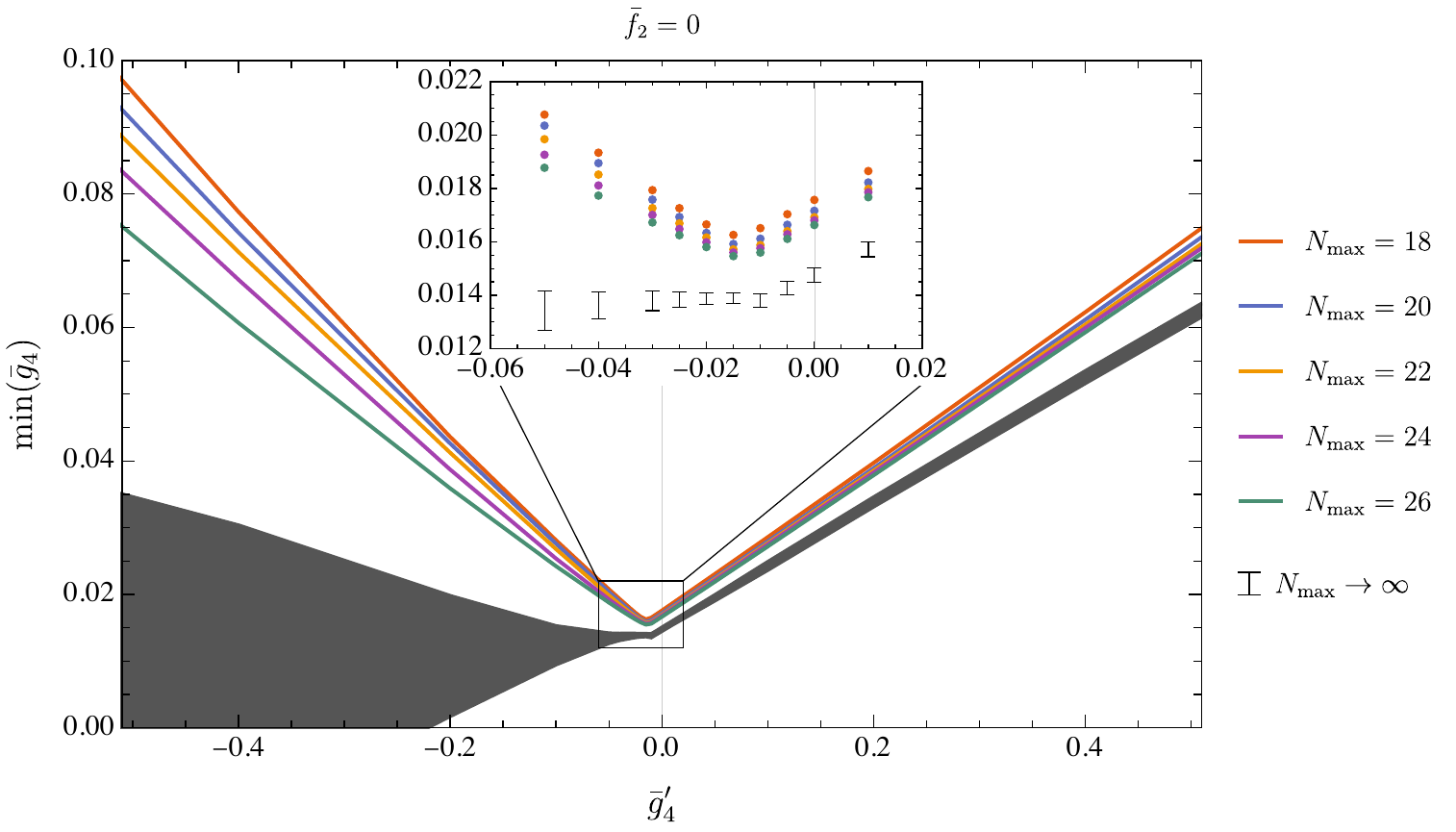}
	\caption{Lower bound on $\bar g_4$ as a function of $\bar{g}'_4$ at the fixed value $\bar{f}_2=0$. Colored lines (and dots in the zoomed part) represent the numerical data obtained with $\Lmax$ defined in \eqref{eq:Lmax_NmaxFormula} and various $\Nmax$. The gray line (and black bars in the zoomed part) represent the extrapolated bound to $\Nmax\rightarrow \infty$. The width of the gray line represents the estimated error. The extrapolation to the left of the minimum has an enormous estimated error and, thus, cannot be trusted.}
	\label{fig:g4vsg4p_f20}
	
	\includegraphics[scale=0.95]{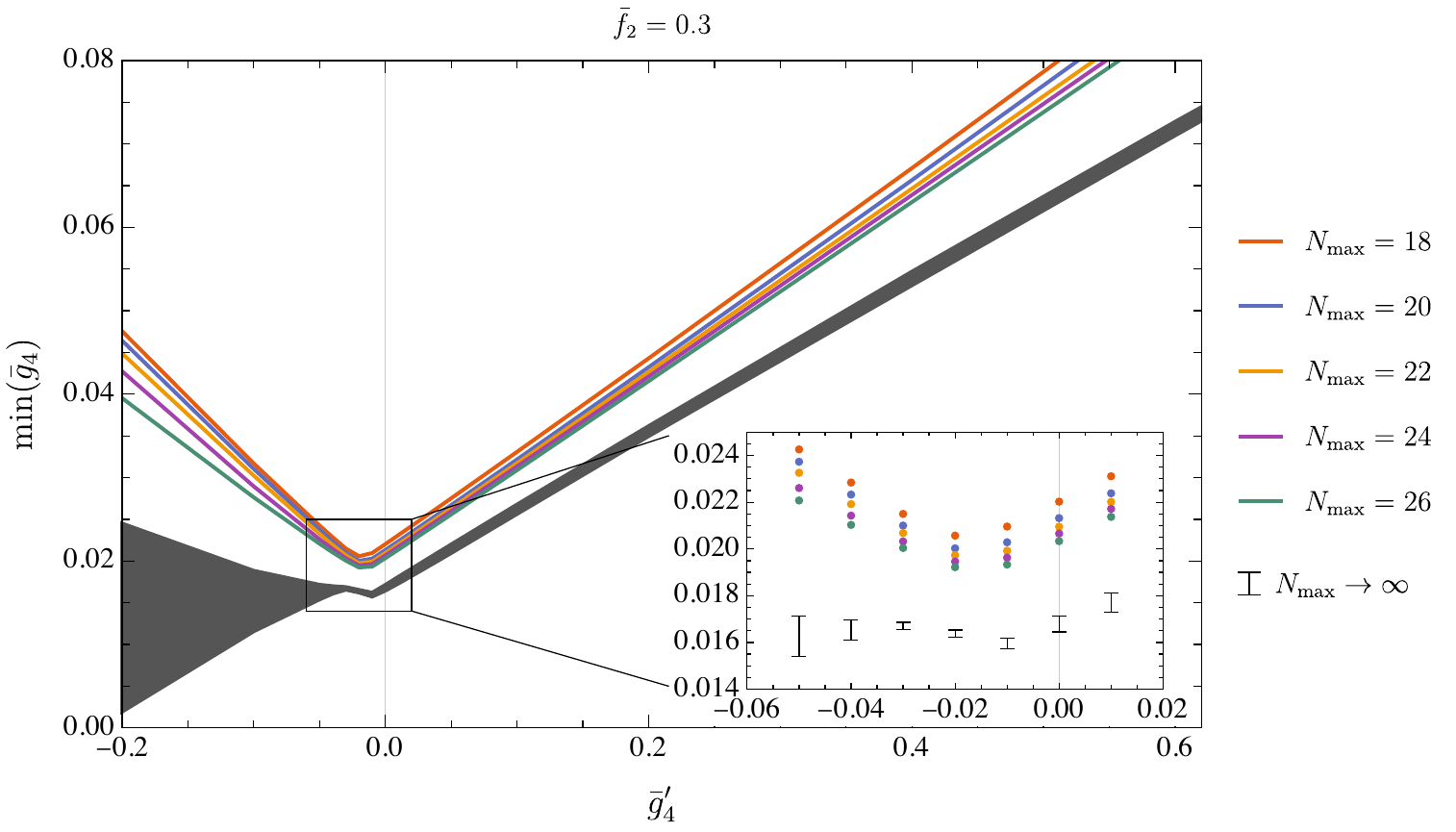}
	\caption{Lower bound on $\bar g_4$ as a function of $\bar{g}'_4$ at the fixed value $\bar{f}_2=0.3$. The gray line represents the extrapolation to $\Nmax\rightarrow \infty$.}
	\label{fig:g4vsg4p_f203}
\end{figure}

The lower bound on $\bar g_4$ as a function of $\bar g_4'$ at two fixed values $\bar{f}_2=0$ and $\bar{f}_2=0.3$ are shown in figures \ref{fig:g4vsg4p_f20} and \ref{fig:g4vsg4p_f203}. The colored lines indicate the numerical data for various values of $\Nmax$. The gray line indicates the extrapolated bound. The width of this line represents the extrapolation error. 
On these plots, we observed that for negative values of $\bar g_4'$ the extrapolation error grows very quickly. As a result, the extrapolation cannot be trusted in this region. 
A quick investigation shows that in this case there is no plateau in $\Lmax$ analogous to figure \ref{fig:lowerg4_NmaxVSspin}. Further work is needed in order to understand the issue.

\begin{figure}[tbh]
	\centering
	\includegraphics[scale=1]{g4vsg4pvsf2}
	\caption{Lower bound on $\bar g_4$ as a function of $\bar{g}_4'$ at fixed values of $\bar{f}_2$. This bound is constructed at $\Nmax=20$ and $\Lmax=50$.}
	\label{fig:g4vsg4pvsf2}
	
	\includegraphics[scale=0.8]{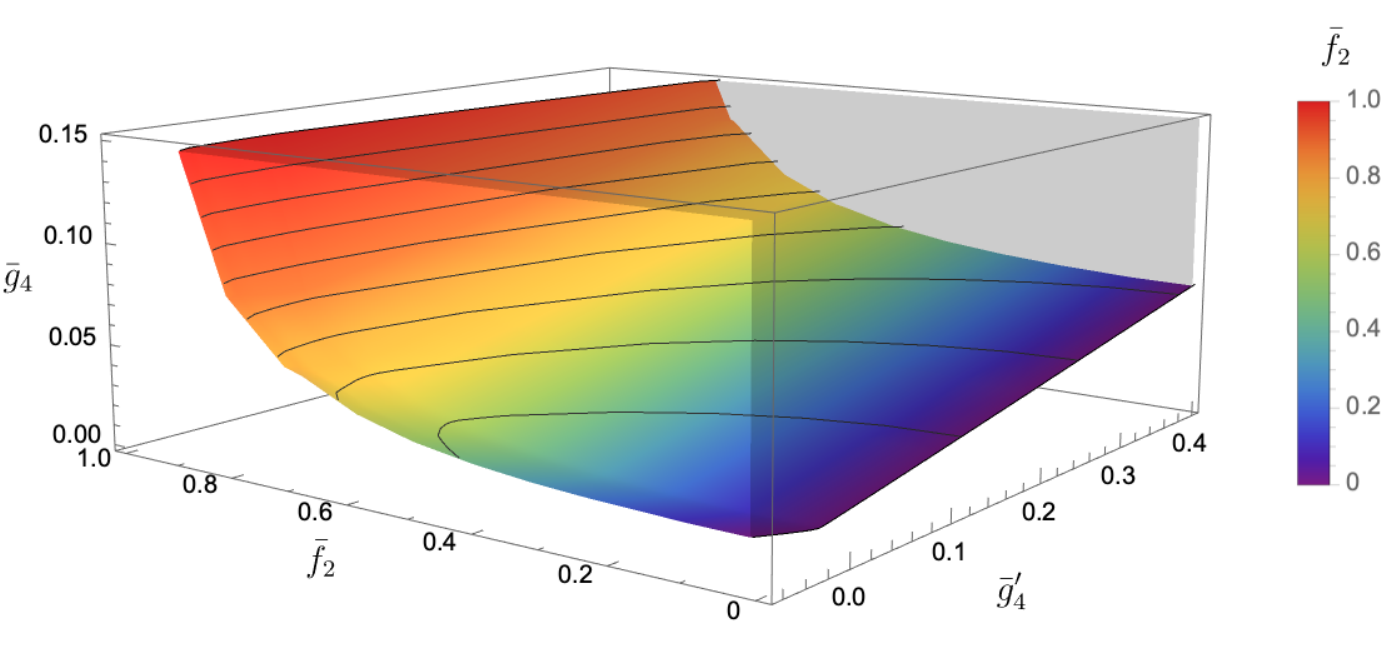}
	\caption{Lower bound on $\bar g_4$ as a function of $\bar{g}_4'$ and $\bar{f}_2$. The color gradient indicates places with the same values of $\bar{f}_2$ and is equivalent to the one used in figure \ref{fig:g4vsg4pvsf2}. The solid lines indicate places with the same values of $\bar{g}_4$. The bound is constructed at $\Nmax=20$ and $\Lmax=50$. }
	\label{fig:3d_g4vsg4pvsf2}
\end{figure}

Fixing $\Nmax=20$ and $\Lmax=50$ allows us to scan over more values of $\bar f_2$. As in the previous subsection, exactly the same bounds hold when $\bar f_2\to -\bar f_2$. We present our results in figures \ref{fig:g4vsg4pvsf2} and \ref{fig:3d_g4vsg4pvsf2}. There, we focus on the region where one could have performed a controlled extrapolation.

\subsection{Linear constraints and the numerical bounds}
\label{subsec:linear_nogo}

Let us now make a general remark on a limitation specific to convex optimization, when it comes to bounding observables like the ones in \eqref{eq:coefficients}. For simplicity, consider the simplified setting of a single amplitude $\mathcal{T}(s)$, function of a single Mandelstam invariant---for instance, a scalar 2-to-2 amplitude in two dimensions. Given an amplitude which satisfies unitarity and crossing, it is easy to see that the one parameter family 
\begin{equation}
\mathcal{T}_\lambda(s)=\lambda\mathcal{T}(s)~, \qquad \lambda\in [0,1]~,
\label{Tlambda}
\end{equation} 
does as well. This simple fact implies that \emph{no two-sided bound is possible for dimensionless ratios of coefficients of different mass dimension, if they appear linearly in the amplitude.} 

Concretely, suppose that the amplitude has a low energy expansion of the form
\begin{equation}
\mathcal{T}(s)=g_2s^2 +g_3 s^3 + \dots~.
\end{equation}
Then, equation \eqref{Tlambda} generates a one-parameter family of allowed values for the dimensionless ratio:
\begin{equation}
\bar{g}_3(\lambda) = \frac{g_3(\lambda)}{{[g_2(\lambda)]}^{3/2}} = 
\frac{1}{\lambda^{1/2}}\frac{g_3}{g_2^{3/2}}~.
\end{equation}
If $g_3$ is positive (negative) in the original solution to unitarity and crossing, $\bar{g}_3$ is unbounded from above (below).

It is worth emphasizing that the argument above is not sufficient to forbid two-sided bounds in physical EFTs: elastic unitarity imposes that Wilson coefficients appear nonlinearly in the low energy expansion---see \eqref{eq:amplitudes_EFT} - \eqref{betaEFT}. Hence, if particle production happens at some higher order in $s$, the reasoning must be modified. As mentioned in subsection \ref{sec:bounds_g3f3h3}, we did add nonlinear terms in the ansatz, in particular involving $g_2$, with the exception of the runs to bound $\bar{f}_2$. However, as a matter of fact, in our numerical runs we only found two-sided bounds precisely for $\bar{f}_2$. The latter is a ratio of coefficients with the same mass dimension, and does not scale with the parameter $\lambda$. 

This raises the question whether the absence of two-sided bounds is indeed a hallmark of Wilson coefficients of any EFT. In a certain region of parameter space, the answer is affirmative, as we shall see in detail in the next section. Indeed, weakly coupled theories precisely come equipped with a small parameter $\lambda$, and the low energy Wilson coefficients are linear in it at leading order---see table \ref{tab:wilson}. Elastic unitarity is perturbatively satified order by order in $\lambda$, hence one can consider these theories---if they can be UV completed---as a non-linear completion of the example in \eqref{Tlambda}.
We will indeed use weakly coupled theories to explain some features of the plots presented in this section.

\subsection{Numerical tests of low spin dominance} 
\label{sec:low_spin_dominance}
In this subsection, we test low spin dominance which was conjectured in \cite{Bern:2021ppb}.\footnote{See also \cite{Chowdhury:2021ynh, Ghosh:2022net} where low spin dominance was shown to follow from locality in tree level EFTs.} Consider the observables 
\begin{equation}
	\begin{aligned}
	 \langle	\rho_\ell^{+} \rangle_k &\equiv 16 (\ell +1) \int_0^\infty  \frac{ \text{Im} \left( \Phi_1^\ell(s) + \Phi_2^\ell(s)\right)  }{s^{k+1}} ds \,, \\
	 \langle	\rho_\ell^{-} \rangle_k &\equiv 16 (\ell +1) \int_0^\infty  \frac{ \text{Im} \left( \Phi_1^\ell(s) - \Phi_2^\ell(s)\right)  }{s^{k+1}} ds \,,\\
	 \langle	\rho_\ell^{3} \rangle_k &\equiv 16 (\ell +1) \int_0^\infty  \frac{ \text{Im} \left( \Phi_3^\ell(s) \right)  }{s^{k+1}} ds\,.
	\end{aligned} 
\end{equation}
In our case, only the $k=2$ integrals converge because for higher values of $k$, the logarithms in \eqref{eq:amplitudes_EFT} spoil  convergence. Note that in terms of these positive observables, the sum rules \eqref{eq:sum_rule_A} can be written (after taking the limit $\hat s \rightarrow $ 0 there) as 
\begin{equation}
	\label{eq:sum_rule_pwave_moments}
	\begin{aligned}
	g_2 + f_2 &= \sum_{l = 0, 2 \ldots}  \langle	\rho_\ell^{+} \rangle_2 + \sum_{l =  2, 3 \ldots} \langle	\rho_\ell^{3} \rangle_2 \,,\\
	g_2 - f_2 &=\sum_{l = 0, 2 \ldots}  \langle	\rho_\ell^{-} \rangle_2 + \sum_{l =  2, 3 \ldots} \langle	\rho_\ell^{3} \rangle_2\,.
	\end{aligned}
\end{equation}
The weak version of low spin dominance states that in each channel, the contribution of the lowest spin dominates over all the higher spins i.e.
\begin{equation}
	 \frac{ \langle	\rho_0^{+} \rangle_2 }{\langle	\rho_{\ell >0}^{+} \rangle_2 } \geq 1 ,  \quad  \frac{ \langle	\rho_0^{-} \rangle_2 }{\langle	\rho_{\ell >0}^{-} \rangle_2 } \geq 1 , \quad  \frac{ \langle	\rho_0^{3} \rangle_2 }{\langle	\rho_{\ell >0}^{3} \rangle_2 } \geq 1\,.
\end{equation}
The strong version of low spin dominance instead states that in fact the contribution from the lowest spin is much larger than the other spins
\begin{equation}
	\frac{ \langle	\rho_0^{+} \rangle_2 }{\langle	\rho_{\ell >0}^{+} \rangle_2 } \geq \alpha ,  \quad  \frac{ \langle	\rho_0^{-} \rangle_2 }{\langle	\rho_{\ell >0}^{-} \rangle_2 } \geq \alpha, \quad  \frac{ \langle	\rho_0^{3} \rangle_2 }{\langle	\rho_{\ell >0}^{3} \rangle_2 } \geq \alpha\,,
\end{equation}
with $\alpha \sim 100$ in \cite{Bern:2021ppb}. In our work, since we construct scattering amplitudes, we test this conjecture and find that the weak version of low spin dominance seems to always be true, however the strong version need not be. More precisely we find amplitudes where $\alpha < 10$. We illustrate this by considering two points from   figure  \ref{fig:g4vsf4atf20} in tables \ref{tab:sum_rule_1} and \ref{tab:sum_rule_2}. Note that in this subsection we analyzed the data at $\Nmax=26$ and $\Lmax=80$ and checked that the behavior is stable as we change $\Lmax$ and $\Nmax$. In table \ref{tab:sum_rule_1} which is the absolute minimum value of $\bar g_4$, we see that\footnote{Note that when $\bar f_2 = 0$ and $\bar f_4 =0$, the amplitude $\Phi_2 =0$ and therefore $\Phi^-_\ell = \Phi^+_\ell$. } 
\begin{equation}
	\bar f_2 = 0 \, , \bar f_4 = 0 \, : \quad
	\frac{ \langle	\rho_0^{+} \rangle_2 }{\langle	\rho_{\ell >0}^{+} \rangle_2 }  = 5.7 ,  \quad   \quad  \frac{ \langle	\rho_0^{3} \rangle_2 }{\langle	\rho_{\ell >0}^{3} \rangle_2 } = 6.3 \,,
\end{equation}
which indicates some low spin dominance but not the strong form. We contrast this in table \ref{tab:sum_rule_2} with the point  $\bar f_2 = 0$ and $\bar f_4 = 0.05$  which lies to right hand side of the minimum where the bound between $\bar f_4$ and $\bar g_4$ appears to be linear. Here we observe much stronger low spin dominance, albeit only in the $\rho_\ell^+$ channel: 
\begin{equation}
	\bar f_2 = 0 \, , \bar f_4 = 0.05 \, : \quad
	\frac{ \langle	\rho_0^{+} \rangle_2 }{\langle	\rho_{\ell >0}^{+} \rangle_2 }  = 21.9 ,  \quad \frac{ \langle	\rho_0^{-} \rangle_2 }{\langle	\rho_{\ell >0}^{-} \rangle_2 } = 3.3,   \quad  \frac{ \langle	\rho_0^{3} \rangle_2 }{\langle	\rho_{\ell >0}^{3} \rangle_2 } = 8.8 \,.
\end{equation}

\vspace{10mm}
\begin{table}[h!]
	\centering
	\begin{tabular}{cccccccc}
		\toprule
		&	\multicolumn{6}{c}{
			$\bar{f}_2 = 0$, \, $\bar{f}_4 = 0$, \, $\bar{g}_4 \approx 0.015$ 
		}							\\ \midrule
		&	$\ell=0$	&	$\ell=1$	&	$\ell=2$	&	$\ell=3$	&$\ell=4$	&$\ell=20$   & $\sum_\ell$ 	\\
		\midrule
		$\langle	\rho_{\ell}^{+} \rangle_2$	&	$27.9$	&  &	$4.90$	& &	$1.62$	& $0.03	$ & $36.1 $\\
		$\langle	\rho_{\ell}^{-} \rangle_2$	&	$27.9$	&  &	$4.90$	& &	$1.62$	& 	$0.03$ & $36.1$	 \\
		$\langle	\rho_{\ell}^{3} \rangle_2$	 &  &	&	$50.4$	&	$8.04$	&	$2.86$	&	$0.01$  & $63.9	$\\
		\bottomrule
	\end{tabular}
	\caption{Sum rules for the minimum point on the boundary of allowed region in figure \ref{fig:g4vsf4atf20}. Note that at this point, the amplitude $\Phi_2 =0$ and therefore $\Phi^+ = \Phi^-$. The last column is the sum of  partial wave contributions up to spin $\ell = 20$ and we see that \eqref{eq:sum_rule_pwave_moments} is saturated already, keeping in mind that $g_2 = 100 $ and $f_2 =0$.}
	\label{tab:sum_rule_1}
\end{table}
\vspace{10mm}
\begin{table}[h!]
	\centering
	\begin{tabular}{cccccccc}
		\toprule
		&	\multicolumn{6}{c}{ $\bar{f}_2 = 0$, \, $\bar{f}_4 = 0.05$, \, $\bar{g}_4  \approx 0.11$}							\\ \midrule
		&	$\ell=0$	&	$\ell=1$	&	$\ell=2$	&	$\ell=3$	&$\ell=4$	&$\ell=20$   & $\sum_\ell$ 	\\
		\midrule
		$\langle	\rho_{\ell}^{+} \rangle_2$	&	$40.6$	&  &	$1.85$	& &	$0.46$	& $0.01$ 	 & $43.4$\\
		$\langle	\rho_{\ell}^{-} \rangle_2$	&	$25.3$	&  &	$7.69$	& &	$5.12$	& 	$0.02$ &	$43.4$ \\
		$\langle	\rho_{\ell}^{3} \rangle_2$	 &  &	&	$47.8$	&	$5.42$	&	$1.51$	&	$0.02$  & $56.6$	\\
		\bottomrule
	\end{tabular}
	\caption{Sum rules for a  point on the boundary of the allowed region in figure \ref{fig:g4vsf4atf20}. The last column is the sum of  partial wave contributions up to spin $\ell = 20$ and we once again see that \eqref{eq:sum_rule_pwave_moments} is saturated already, since $g_2 = 100 $ and $f_2 =0$.}
	\label{tab:sum_rule_2}
\end{table}

\newpage

We also display two points from   figure \ref{fig:g4vsf4atf203} in tables \ref{tab:sum_rule_3} and \ref{tab:sum_rule_4} where we observe similar phenomena. 
In particular notice in table \ref{tab:sum_rule_4} that at the point $\bar f_2 =0.3$ and $\bar f_4 = 0.05$  which lies in the linear region, the $\rho^+_\ell$ channel has strong low spin dominance while the $\rho^-_\ell$ channel is opposite. 
\begin{equation}
	\begin{aligned}
		\bar f_2 = 0.3 \, , \bar f_4 = 0.0045 \, : \quad
		\frac{ \langle	\rho_0^{+} \rangle_2 }{\langle	\rho_{\ell >0}^{+} \rangle_2 }  &= 4.7,  \quad \frac{ \langle	\rho_0^{-} \rangle_2 }{\langle	\rho_{\ell >0}^{-} \rangle_2 } = 12.9,   \quad  \frac{ \langle	\rho_0^{3} \rangle_2 }{\langle	\rho_{\ell >0}^{3} \rangle_2 } = 8.7 \,, \\
		\bar f_2 = 0.3 \, , \bar f_4 = 0.05 \, : \quad
		\frac{ \langle	\rho_0^{+} \rangle_2 }{\langle	\rho_{\ell >0}^{+} \rangle_2 }  &= 53.1,  \quad \frac{ \langle	\rho_0^{-} \rangle_2 }{\langle	\rho_{\ell >0}^{-} \rangle_2 } = 2.2,   \quad  \frac{ \langle	\rho_0^{3} \rangle_2 }{\langle	\rho_{\ell >0}^{3} \rangle_2 } = 9.4 \,.
	\end{aligned}
\end{equation}
We also remark that if we instead consider points with negative values of $\bar f_4$, say $\bar f_4 = -0.05$, we find that it is now the $\rho^-_\ell$ channel which has strong low spin dominance while the $\rho^+_\ell$ channel has weaker low spin dominance.

\vspace{5mm}
\begin{table}[h!]
	\centering
	\begin{tabular}{cccccccc}
		\toprule
		&	\multicolumn{6}{c}{$\bar{f}_2 = 0.3 $, \, $\bar{f}_4 = 0.0045$, \, $\bar{g}_4 \approx 0.02 $ }							\\ \midrule
		&	$\ell=0$	&	$\ell=1$	&	$\ell=2$	&	$\ell=3$	&$\ell=4$	&$\ell=20$   & $\sum_\ell$ 	\\
		\midrule
		$\langle	\rho_{\ell}^{+} \rangle_2$	&	$50.5$	&  &	$10.8$	& &	$5.05$	& $0.03$ 	 & $70.7$ \\
		$\langle	\rho_{\ell}^{-} \rangle_2$	&	$9.33$	&  &	$0.72$	& &	$0.27$	& 	$0.02$ & $10.7$	 \\
		$\langle	\rho_{\ell}^{3} \rangle_2$	 &  &	&	$49.8$	&	$5.72$	&	$2.01$	&	$0.02$  & $59.3$	\\
		\bottomrule
	\end{tabular}
\caption{Sum rules for the minimum point on the boundary of figure \ref{fig:g4vsf4atf203}. The last column is the sum of  partial wave contributions up to spin $\ell = 20$ and we see that \eqref{eq:sum_rule_pwave_moments} is saturated already, keeping in mind that $g_2 = 100 $ and $f_2 =30$.}
\label{tab:sum_rule_3}
\end{table}
\vspace{5mm}
\begin{table}[h!]
	\centering
	\begin{tabular}{cccccccc}
		\toprule
		&	\multicolumn{6}{c}{$\bar{f}_2 = 0.3$, \, $\bar{f}_4 = 0.05$, \,  $\bar{g}_4 \approx 0.11 $}							\\ \midrule
		&	$\ell=0$	&	$\ell=1$	&	$\ell=2$	&	$\ell=3$	&$\ell=4$	&$\ell=20$   & $\sum_\ell$ 	\\
		\midrule
		$\langle	\rho_{\ell}^{+} \rangle_2$	&	$86.5$	&  &	$1.63$	& &	$0.37$	&  $0.01$ 	 & $88.9$ \\
		$\langle	\rho_{\ell}^{-} \rangle_2$	&	$14.6$	&  &	$6.66$	& &	$4.06$	& 	$0.02$ & 	$28.8$ \\
		$\langle	\rho_{\ell}^{3} \rangle_2$	 &  &	&	$34.9$	&	$3.70$	&	$1.00$	&	$0.02$  &	$41.1$\\
		\bottomrule
	\end{tabular}
	\caption{Sum rules for a  point on the boundary of the allowed region in figure \ref{fig:g4vsf4atf203}. The last column is the sum of  partial wave contributions up to spin $\ell = 20$ and we once again see that \eqref{eq:sum_rule_pwave_moments} is saturated already, keeping in mind that $g_2 = 100 $ and $f_2 =30$.}
	\label{tab:sum_rule_4}
\end{table}

\section{Allowed amplitudes from perturbation theory}\label{sec:weak_coupling_models}

In the first part of this work, we saw how to explore the space of Wilson coefficients using the numerical S-matrix bootstrap. This section is dedicated to answering the following question: how much of this space can be understood analytically? Our main tool will be a set of effective field theories which are under perturbative control, and which will be used to populate the space of Wilson coefficients. 
Most of the models we consider were shown in \cite{Henriksson:2021ymi} to be compatible with analytic positivity bounds, however we also discuss two cases which are in tension with the same bounds.  These theories are obtained by weakly coupling the photon to a heavy particle at tree level (Yukawa-like theories) or at one loop (QED-like theories). The resulting theories have the following feature: they all have a dimensionless parameter $\lambda$, such that at the point $\lambda=0$ the heavy particle decouples and the photons are free. These theories are not necessarily UV complete on their own when $\lambda\neq0$, but for some of them one can exhibit explicit UV completions.\footnote{More precisely, what one can concoct is a renormalizable action with marginal couplings. This still leaves open the possibility of a Landau pole, see appendix \ref{app:landauPoleQED}.} Furthermore, they all obey the classical Regge growth conjecture \cite{Chowdhury:2019kaq}, which states that the Regge intercept should be bounded by two for tree-level scattering amplitudes. The non-perturbative version of the bound is not known in the case of massless spin one particles, but the same statement was recently proven for graviton scattering \cite{Haring:2022cyf}, and we shall assume that it holds for photons as well. Even when we do not exhibit an explicit UV completion, we take the Regge boundedness as a reason to trust in its existence.

 The rest of the section is organized as follows. In subsection \ref{subsec:wilson}, we gather the Wilson coefficients of interest, while a detailed analysis of the theories that produce them is relegated to appendix \ref{app:Wilson}. The following subsection is dedicated to explaining which linear combinations of the basis of amplitudes are compatible with unitarity. In the three subsections \ref{subsec:rule_in_8}, \ref{subsec:rule_in_10}, \ref{subsec:rule_in_12}, we rule in portions of the space of physical observables defined in \eqref{eq:coefficients}. 

\subsection{Wilson coefficients of models with a small parameter}
\label{subsec:wilson}

As mentioned above, the class of theories we consider are obtained by integrating out a particle at tree level or at one-loop. At tree level, we consider resonances of spin bounded by two. The criterion for this choice is the requirement of Regge boundedness discussed above. In appendix \ref{app:Wilson} we give more details on the construction of these amplitudes, and in subsections \ref{subsec:rule_in_8}, \ref{subsec:rule_in_10} we will further comment on the case of a spin two resonance, which leads to some subtleties. The second class of theories comprises the coupling of a photon to an electrically charged particle: the particle is only created in pairs, and therefore contributes to the photon EFT at one loop. We collect the Wilson coefficients in table \ref{tab:wilson}, where $\lambda$ is a dimensionless parameter. In all cases, the coefficients are only correct to leading order in $\lambda$, which should therefore be taken to be small. In the following, we shall often denote the couplings by the dimension of the corresponding operators in the Lagrangian \eqref{eq:EFT_lagrangian}:
$(g_2,f_2)$ are the dimension~8 Wilson coefficients, $(g_3,f_3,h_3)$ are the dimension 10 and $(g_4,g_4',f_4)$ are the dimension 12 coefficients.

\begin{table}[t]
\noindent \makebox[\textwidth]{
\newcolumntype{C}{>{$\displaystyle}c<{$}}
\begin{tabularx}{1.09\textwidth}{l*{8}{C}}
\toprule
  & g_2 & f_2 & g_3 & f_3 & h_3 & g_4 & g'_4 & f_4   \\
  \midrule
  \bfseries Spin $0$  & & & & & & & & \\
  scalar & \frac{\lambda^2}{m^4} & \frac{\lambda^2}{m^4} & \frac{\lambda^2}{m^6} & \frac{3\lambda^2}{m^6}& 0 & \frac{\lambda^2}{m^8}& 0& \frac{\lambda^2}{2m^8}\\ [10pt]
  axion & \frac{\lambda^2}{m^4} & -\frac{\lambda^2}{m^4} & \frac{\lambda^2}{m^6} & -\frac{3\lambda^2}{m^6}& 0 & \frac{\lambda^2}{m^8}& 0& -\frac{\lambda^2}{2m^8} \\ [10pt]
  \midrule
  \bfseries Spin $2$   & & & & & & & & \\ [10pt]
  %parity even I   & \frac{\lambda^2}{m^4} & \frac{\lambda^2}{m^4} & \frac{\lambda^2}{m^6} & -\frac{15\lambda^2}{m^6}& 0 & \frac{\lambda^2}{m^8}  & -\frac{6\lambda^2}{m^8} & \frac{\lambda^2}{2 m^8}  \\ [10pt]
  parity even I   & \frac{\lambda^2}{m^4} & \frac{\lambda^2}{m^4} & \frac{\lambda^2}{m^6} & 0 & 0 & \frac{\lambda^2}{m^8}  & -\frac{6\lambda^2}{m^8} & \frac{\lambda^2}{2 m^8}  \\ [10pt]
  parity even II  & \frac{\lambda^2}{m^4} & 0 & -\frac{\lambda^2}{m^6} & 0 & 0 & \frac{\lambda^2}{m^8}  & -\frac{2\lambda^2}{m^8} & 0 \\ [10pt]
  %parity odd & \frac{\lambda^2}{m^4} & -\frac{\lambda^2}{m^4} & \frac{\lambda^2}{m^6} & \frac{15\lambda^2}{m^6}& 0 & \frac{\lambda^2}{m^8}  & -\frac{6\lambda^2}{m^8} & -\frac{\lambda^2}{2 m^8}  \\ [10pt]
  parity odd & \frac{\lambda^2}{m^4} & -\frac{\lambda^2}{m^4} & \frac{\lambda^2}{m^6} & 0 & 0 & \frac{\lambda^2}{m^8}  & -\frac{6\lambda^2}{m^8} & -\frac{\lambda^2}{2 m^8}  \\ [10pt]
  \midrule
  \bfseries One loop & & & & & & & & \\ [10pt]
  scalar QED & \frac{2\lambda^2}{45 m^4} & \frac{\lambda^2}{30 m^4} & \frac{\lambda^2}{210 m^6} & \frac{\lambda^2}{63 m^6}& \frac{\lambda^2}{630 m^6} & \frac{11\lambda^2}{9450m^8}  & -\frac{\lambda^2}{3780m^8} & \frac{\lambda^2}{1890 m^8 }  \\ [10pt]
  spinor QED & \frac{11\lambda^2}{45 m^4} & -\frac{\lambda^2}{15 m^4} & \frac{4\lambda^2}{315 m^6} & -\frac{2\lambda^2}{63 m^6}& -\frac{\lambda^2}{315 m^6} & \frac{13\lambda^2}{2700m^8}  & -\frac{\lambda^2}{378m^8} & -\frac{\lambda^2}{945 m^8 }  \\ [10pt]
  vector QED & \frac{14\lambda^2}{5 m^4} & \frac{\lambda^2}{10 m^4} & -\frac{47\lambda^2}{630 m^6} & \frac{\lambda^2}{21 m^6}& \frac{\lambda^2}{210 m^6} & \frac{131\lambda^2}{3150m^8}  & -\frac{23\lambda^2}{420m^8} & \frac{\lambda^2}{630 m^8 }  \\ [10pt]
 \bottomrule
\end{tabularx} }
\caption{Wilson coefficients of photons EFTs obtained by integrating out particles at tree level or one loop.}
\label{tab:wilson}
\end{table}

\subsection{The rules of the game}
\label{subsec:rules}

If the amplitudes in table \ref{tab:wilson} are to be treated as UV complete, their convex hull should lie within the numerical bounds found in section \ref{sec:numerics}. However,  violations of unitarity only disappear in the $\lambda \to 0$ limit, in which case all the dimensionless observables defined in \eqref{eq:coefficients} are pushed to infinity, with the exception of $\bar{f_2}$. For instance, the non-vanishing observables associated with the first row of table \ref{tab:wilson} are
\begin{equation}
\bar{f}_2=1~, \quad \bar{g}_3=\frac{1}{\lambda}~, \quad 
\bar{f}_3=\frac{3}{\lambda}~, \quad \bar{g}_4=\frac{1}{\lambda^2}~,
\quad \bar{f}_4=\frac{1}{2\lambda^2}~.
\end{equation}
 We conclude that the amplitudes considered in this section mark \emph{unbounded directions} in the space of photon EFTs at dimension larger than 8.

In the rest of this section, we shall find such directions. A simplifying feature of the problem is that the convex hull of these amplitudes degenerates into a cone in the weak coupling limit. Indeed, let us define the vectors $v_i(\lambda_i)$, where the index $i$ labels the rows of table \ref{tab:wilson} and whose components are the Wilson coefficients:
\begin{equation}
v=(g_2,f_2\, |\, g_3,f_3,h_3 \, |\, g_4,g'_4,f_4)
= (v^{(2)}\, |\, v^{(3)} \, |\, v^{(4)})~.
\label{WilsonVector}
\end{equation}
The vectors spanning subspaces at each mass dimension are distinguished by the upper index.
The following obvious equality,
\begin{equation}
\sum_i \alpha_i v_i(\lambda_i) = \sum_i \beta_i v_i(\lambda'_i)~, \qquad \beta_i=\frac{\alpha_i}{(\kappa_i)^2}~,\quad\lambda'_i=\lambda_i \kappa_i~, \quad \kappa_i>0~,
\label{alphaRescale}
\end{equation}
makes the condition $\sum_i \alpha_i =1$ irrelevant. In the $\lambda_i \to 0$ limit, all linear combinations of the amplitudes with coefficients $\beta_i>0$ are allowed. In other words, perturbative amplitudes are only constrained by positivity, rather than by full nonlinear unitarity.

A second simplification comes from the fact that we are allowed to take \emph{independent} linear combinations at dimension 8 and 10, or at dimension 8 and 12. This follows from the freedom of choosing the mass of the exchanged particle, which introduces new dimensionless parameters $m_i/m_j$. For instance, the following change of variables,
\begin{equation}
x=\frac{\lambda^2}{m^4}~, \quad y=\frac{\lambda^2}{m^6}~,
\end{equation}
allows to write the Wilson coefficients at dimension 8 and 10 as linear functions of different parameters. Referring to the definition \eqref{WilsonVector},
\begin{equation}
v^{(2)}=x\, \hat{v}^{(2)}~, \quad v^{(3)}=y\, \hat{v}^{(3)}~, \quad v^{(4)}=\frac{y^2}{x} \hat{v}^{(4)}~,
\end{equation}
with hatted vectors being purely numerical. Then we can extend the observation \eqref{alphaRescale} to
\begin{multline}
\sum_i \alpha_i v_i(x_i,y_i) =\sum_i  \left(\beta_i^{(2)} v_i^{(2)}(x'_i)\,|\,\beta_i^{(3)} v_i^{(3)}(y'_i)\,|\,\beta^{(4)}_i v_i^{(4)}(y'^2_i/x'_i)\right)~, \\
\beta_i^{(2)}=\frac{\alpha_i}{\kappa^{(2)}_i}~, \ \beta_i^{(3)}=\frac{\alpha_i}{\kappa^{(3)}_i}~,
\ \beta_i^{(4)}=\alpha_i\frac{\kappa^{(2)}_i}{\big(\kappa^{(3)}_i\big)^2}~,
\qquad x'_i=x_i \kappa^{(2)}_i~, \ y'_i=y_i \kappa^{(3)}_i~, \quad \kappa_i^{(2,3)}>0~.
\end{multline}
This means that, for any choice of positive $\beta_i^{(2)},\,\beta_i^{(3)}$, one can find an amplitude in the convex hull of the perturbative theories defined by table \ref{tab:wilson} as $\lambda \to 0.$\footnote{Notice that instead $\beta_i^{(4)}=\big(\beta_i^{(3)}\big)^2/\beta_i^{(2)}.$} An analogous statement holds for the subspace of dimension 8 and dimension 12 coefficients.

A simple consequence of this observation is that \emph{the asymptotic allowed values of the dimension 10 or of the dimension 12 coefficients do not depend on $\bar{f_2}$.}

More generally, we are lead to a simple recipe to find which directions are necessarily unbounded in the space of couplings of each mass dimension:
\vspace{10pt}

\emph{Consider the vectors of Wilson coefficients in table \ref{tab:wilson}, restricted to a given mass dimension larger than 8. Set $\lambda=m=1$. Compute the one-sided cone generated by positive linear combinations of the numerical vectors thus obtained. The intersection of this cone with the sphere at infinity marks the unbounded directions.}
\vspace{10pt}

The basic technology to solve this kind of problems is reviewed in \cite{Arkani-Hamed:2020blm}. We shall only report the results in the rest of the section.

\subsection{The $\bar{f}_2$ space}
\label{subsec:rule_in_8}

The spin 0 exchanges in table \ref{tab:wilson} are sufficient to cover the space of couplings at dimension 8, which is parametrized by $|\bar{f_2}|\leq 1$. Indeed, the scalar amplitude has $\bar{f_2}=1$ and the axion amplitude has $\bar{f_2}=-1$. Their convex hull generates the whole allowed space.

The other amplitudes in table \ref{tab:wilson} all lie inside this interval. In the case of the spin 2 resonance, this fact deserves a comment. The exchange of a resonance determines the amplitude uniquely only at the location of the pole. In particular, the amplitudes constructed in appendix \ref{app:Wilson} are ambiguous up to the addition of a polynomial in the Mandelstam invariants, of the kind showed in \eqref{eq:amplitudes_EFT}. Such polynomials correspond to contact interactions for the photons, as explained in the introduction. A polynomial of degree two can be tuned so that the Regge limit of the spin two resonance obeys \eqref{Froissart_n}, and, as a consequence, the dispersion relation \eqref{eq:sum_rule_A}. This implies $|\bar{f}_2|\leq 1$. Let us emphasize that \eqref{Froissart_n} is strictly weaker than the Regge bound mentioned in the previous subsection, since it only holds at $t=0$, for a specific linear combination of amplitudes. In fact, it is not possible to obtain a Regge intercept strictly less than 2, for all values of $t$, by adding a finite degree polynomial.

Let us finally comment on the constraints imposed by dispersion relations and crossing onto the amplitudes living at the boundary of the allowed region. We start from the boundary $\bar{f_2}=1$. Equations  \eqref{eq:sum_rule_A} and \eqref{eq:positivity_forward} imply
\begin{equation}
\Im\left(\Phi_1(s)-\Phi_2(s)\right)=0~, \qquad \Im\Phi_3(s)=0~,
\label{ImConstraintf21}
\end{equation} 
where we recall that the notation means the amplitudes are evaluated at $t=0$. Let us now decompose \eqref{ImConstraintf21} in partial waves, using \eqref{eq:PartialWave_Expansion}. The small Wigner $d-$matrix are non negative at $\theta=0$, as it is easy to verify from the definition \eqref{eq:wignerD}. Together with the positivity constraint \eqref{eq:positivity}, this means that \eqref{ImConstraintf21} is valid at the level of partial waves:
\begin{equation}
\Im\left(\Phi_1^\ell(s)-\Phi_2^\ell(s)\right)=0~, \qquad \Im\Phi_3^\ell(s)=0~.
\label{PartialConstraintf21}
\end{equation}
Furthermore, the unitarity constraints \eqref{eq:unitarity_1} - \eqref{eq:unitarity_2} force these combinations of partial waves to vanish:
\begin{equation}
\left(\Phi_1^\ell(s)-\Phi_2^\ell(s)\right)=0~, \qquad \Phi_3^\ell(s)=0~.
\label{ZeroConstraintf21}
\end{equation}
This is easily seen for the $\Phi_1^\ell(s)-\Phi_2^\ell(s)$ combination, which obeys the usual spinless unitarity constraint $|1+i \mathcal{T}|<1$.\footnote{Recall that $t-u$ symmetry of $\Phi_1$ and $\Phi_2$ and the property $d^\ell_{0,0}(\pi-\theta)=(-1)^\ell d^\ell_{0,0}(\theta)$ make $\Phi_1^\ell(s)-\Phi_2^\ell(s)$ vanish for $\ell$ odd.} With some more work, one can extract from \eqref{eq:unitarity_2} the following necessary conditions valid for even spin:
\begin{equation}
\left|1+i (\Phi_1^\ell(s)+\Phi_2^\ell(s)) \right|<1~, \qquad
\left|1+2i \Phi_3^\ell(s) \right|<1~, \qquad
\left|\Phi_5^\ell(s) \right|<1~.
\end{equation}
The second of these conditions, together with \eqref{eq:unitarity_3}, implies the third equality in \eqref{ZeroConstraintf21}.\footnote{Recall that $\Phi_3^\ell=0$ for $\ell=0,\,1$.} 

If we now plug \eqref{ZeroConstraintf21} back into \eqref{eq:unitarity_2}, we discover that also
\begin{equation}
\Phi_5^\ell(s)=0~.
\end{equation}
All in all, we found that the following amplitudes vanish if $\bar{f_2}=1$:
\begin{equation}
\Phi_1(s,t,u)-\Phi_2(s,t,u)=0~, \qquad
\Phi_3(s,t,u)=0~, \qquad
\Phi_5(s,t,u)=0~.
\end{equation}
Using crossing---equation \eqref{eq:crossing_intro}---we conclude that in fact all the amplitudes must vanish. A similar reasoning also leads to the same conclusion for the other boundary of the space allowed by unitarity, $\bar{f_2}=-1$.

Notice that this conclusion does not contradict the statement that scalar and pseudoscalar resonances saturate these bounds. Indeed, as emphasized, the theories considered in this section are only compatible with unitarity in the limit $\lambda \to 0$, \emph{i.e.} precisely when they become free. In other words, in any interacting UV completion of the EFT obtained integrating out a scalar or an axion, the value of $\bar{f_2}$ is corrected at higher orders in $\lambda$.

\subsection{The $(\bar{g}_3,\bar{f}_3,\bar{h}_3)$ space}
\label{subsec:rule_in_10}

In subsection \ref{sec:bounds_g3f3h3}, we saw that the numerics indicates the absence of bounds for the dimension 10 observables. Here, we show that the full space is in fact covered by (linear combinations of) perturbative completions of the photon effective lagrangian.  Following the recipe given in subsection \ref{subsec:rules}, we consider the cone generated by the following amplitudes: scalar, axion, scalar QED, spinor QED and vector QED. One can easily see that the cone coincides with the whole three-dimensional space spanned by $g_3$, $f_3$ and $h_3$, thus proving that the sphere at infinity in the normalized space $(\bar{g}_3,\bar{f}_3,\bar{h}_3)$ is populated by weakly coupled theories. 

Notice that, since the unitarity constraints are convex, no lower bound on $\bar{g}_3$, $\bar{f}_3$, $\bar{h}_3$ is possible either. Let us emphasize that it was not necessary to include spin two resonances in order to achieve this result. In other words, bounds on dimension 10 Wilson coefficients are impossible already within amplitudes with Regge intercept smaller than two.

\subsection{The $(\bar{g}_4,\bar{g}'_4,\bar{f}_4)$ space}
\label{subsec:rule_in_12}

The space of dimension 12 Wilson coefficients is the first one which is not fully covered by (linear combinations of) weakly coupled amplitudes. This is in accordance with the results of section \ref{sec:numerics}, where we presented evidence of non-trivial bounds in the $(\bar{g}_4,\bar{g}'_4,\bar{f}_4)$ space. 

\begin{figure}[t]
\centering
\includegraphics[scale=0.8]{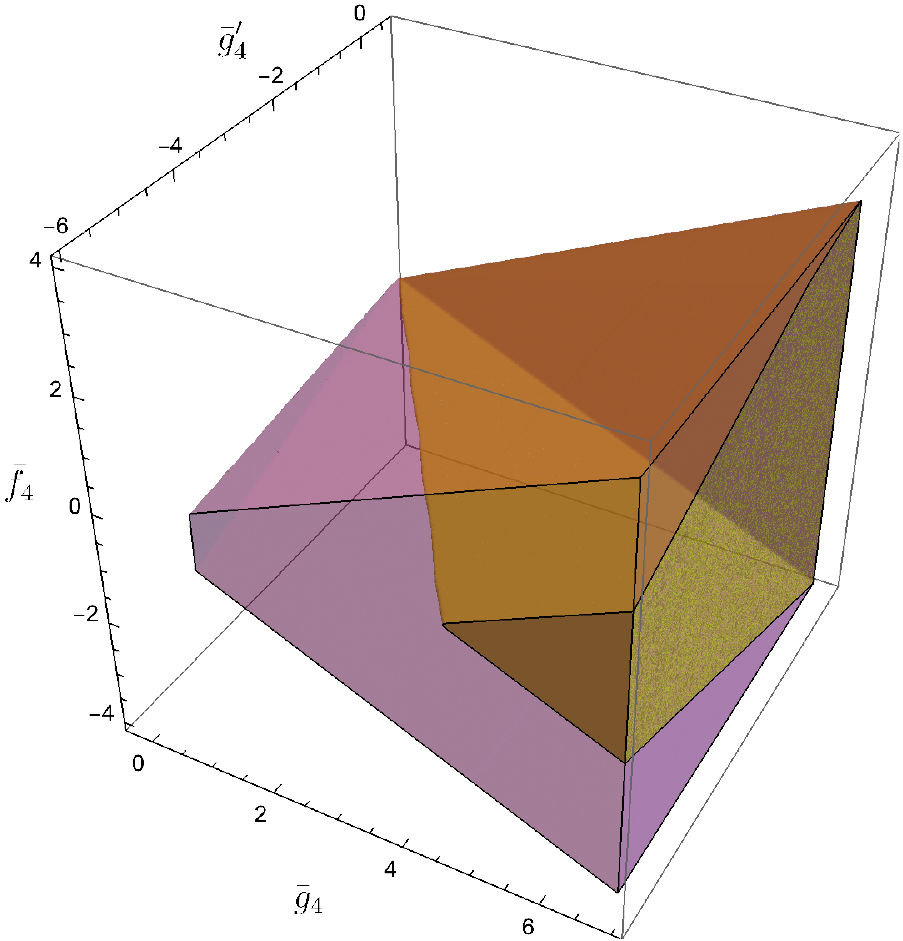}
\caption{A portion of the cone generated by positive linear combinations of the perturbative amplitudes of table \ref{tab:wilson}, restricted to the space of dimension 12 Wilson coefficients. The smaller cone (yellow) only includes amplitudes which grow slower than $s^2$ in the Regge limit. The wider cone (purple) includes all the rows of table \ref{tab:wilson}.}
\label{fig:3dcones}
\end{figure}

The boundaries of the cone generated by the perturbative amplitudes depend on which portion of table \ref{tab:wilson} is included in the analysis. Let us begin by including all the rows of the table. These vectors generate a cone bounded by the following four faces:
\begin{equation}
g_4 -2 f_4 \geq 0~, \qquad g_4+2f_4 \geq 0~, \qquad g_4' \leq 0~, \qquad
g_4'+6g_4 \geq 0~. 
\label{dim12coneAll}
\end{equation}
This implies in particular $g_4\geq0$. If instead we restrict to theories whose Regge intercept is strictly less than two, \emph{i.e.} we exclude the spin two resonances, we get the following inequalities:
\begin{equation}
-10 f_4 + 5 g_4 + 2 g_4'\geq 0~, \qquad
-157 f_4 + 85 g_4 + 60 g_4'\geq 0~, \qquad
g_4' \leq 0~,\qquad
230 f_4 + 115 g_4 + 94 g_4'\geq 0~.
\label{dim12coneSlow}
\end{equation}
Again, only the half-space $g_4\geq0$ is populated. The two cones are shown in figure \ref{fig:3dcones}.

It is interesting to compare these findings with the amplitudes constructed numerically in section \ref{sec:numerics}, and with the analytic results obtained in \cite{Arkani-Hamed:2020blm,Henriksson:2021ymi,Henriksson:2022oeu}. The numerical results are showcased in figures \ref{fig:g4vsf4atf20} - \ref{fig:3d_g4vsf4vsf2} and \ref{fig:g4vsg4p_f20} - \ref{fig:3d_g4vsg4pvsf2}, for the $(\bar{g}_4,\bar{f}_4)$ and for the $(\bar{g}_4,\bar{g}'_4)$ planes, respectively. In particular, it is clear that the $(\bar{g}_4,\bar{f}_4)$ space is bounded at infinity by the two directions $\bar{g}_4/\bar{f}_4= \pm 2$, for all values of $\bar{f}_2$. This matches both the wider and the narrower cones, equations  \eqref{dim12coneAll} and \eqref{dim12coneSlow}. The two boundaries are generated by the scalar and axion amplitudes respectively (and also by the parity even I and by the parity odd amplitudes) in table \ref{tab:wilson}. 

The status of the $(\bar{g}_4,\bar{g}'_4)$ plane appears less clear. The bounds in figures \ref{fig:g4vsg4p_f20} - \ref{fig:3d_g4vsg4pvsf2} are well converged only when $\bar{g}'_4 \gtrsim 0$, while all the perturbative amplitudes we consider have non-positive values for the same Wilson coefficient. The two plots are therefore hardly comparable. 
It would be interesting to find a perturbative amplitude with positive $g_4'$ that could explain our numerical results.
For reference, the sections of the two cones in \eqref{dim12coneAll} and \eqref{dim12coneSlow} are bounded, respectively, as
\begin{equation}
-6\leq \frac{g_4'}{g_4}\leq 0~, \qquad g_4\geq0~,
\label{gBoundAll}
\end{equation}
and
\begin{equation}
-\frac{345}{262}\leq \frac{g_4'}{g_4}\leq 0~, \qquad g_4\geq0~.
\label{gBoundSlow}
\end{equation}
The lower bound in the first equation is reached by the parity even I and parity odd spin 2 resonances. If those are excluded, smallest ratio belongs to vector QED. One can also combine the numerical results to the weakly coupled theories to obtain a larger convex hull. In particular, one can consider figure \ref{fig:g4vsg4p_f20}, and draw a line with slope $-1/6$ or $-262/345$ (depending on the reader's opinion about the spin 2 resonances) which intersects the $\bar{g}_4'=0$ axis in correspondence of the numerical lower bound. The region to the right of this line and above the bound is populated by linear combinations of numerical and perturbative amplitudes.

Finally, in the $(\bar{g}_4',\bar{f}_4)$ plane, the perturbative amplitudes cover the semi-circle at infinity with $\bar{g}_4'<0$. Therefore, their convex hull covers the whole half-plane $(\bar{g}_4'<0,\bar{f}_4)$. Again, one can combine this knowledge with the numerical results. From figures \ref{fig:g4vsg4p_f20} and \ref{fig:g4vsg4p_f203} it is clear that there is no upper bound on $\bar{g}'_4$. 
Consider then an allowed point with $\bar{g}_4'=x>0$ in the $(\bar{g}_4',\bar{f}_4)$ plane. The convex hull of this point and the half plane $\bar{g}_4'<0$ is the region to the left of the line $\bar{g}_4'=x$. Now, since $x$ can go to infinity, we conclude that there are no bounds in the $(\bar{g}_4',\bar{f}_4)$ plane.

In \cite{Henriksson:2021ymi,Henriksson:2022oeu}, the space of Wilson coefficients was bounded analytically, by means of linear positivity---see section \ref{sec:positivity}---and tree level crossing, \emph{i.e.} null constraints. Depending on how many null constraints were imposed, the following bounds were found:
\begin{align}
&-\frac{18}{7}\leq\frac{\bar{g}_4'}{\bar{g}_4}\leq \frac{370}{29}~,\quad g_4\geq0~,
&\textup{\cite{Henriksson:2021ymi}}~, \label{gPisaI} \\
&-\frac{12}{5}\leq\frac{\bar{g}_4'}{\bar{g}_4}\leq 0~,\quad g_4\geq0~,
&\textup{\cite{Henriksson:2022oeu}}~. \label{gPisaII}
\end{align}
On the other hand, the paper \cite{Arkani-Hamed:2020blm} found the following bounds:
\begin{equation}
-6\leq\frac{\bar{g}_4'}{\bar{g}_4}\leq \frac{30}{7}~,\quad g_4\geq0~.
\label{gEFTHedron}
\end{equation}
\begin{figure}[t]
\centering
\includegraphics[scale=1]{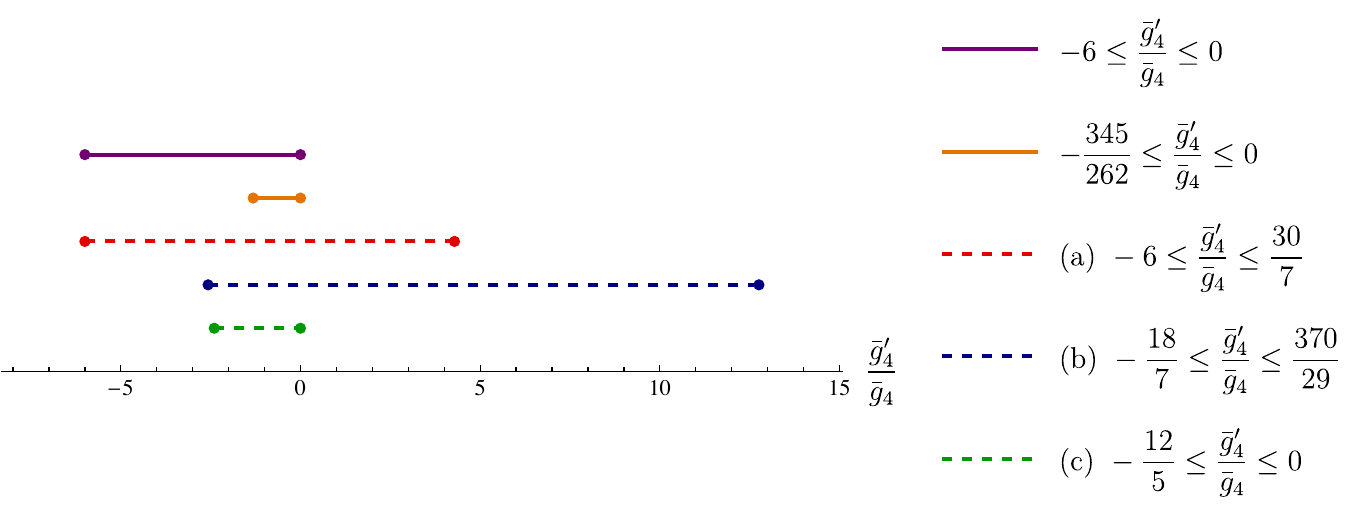}
\caption{Comparison between \eqref{gBoundAll}, \eqref{gBoundSlow}---purple and orange solid lines respectively---and the analytic bounds imposed by linear positivity and tree level crossing. The allowed region (a) was derived in \cite{Arkani-Hamed:2020blm}, (b)  in \cite{Henriksson:2021ymi} and updated by the same authors in \cite{Henriksson:2022oeu} (c). Notice that part of the purple region lies outside the bounds which rely on null constraints.}
\label{fig:g4g4pAnVsRuleIn}
\end{figure}
\begin{figure}[t]
\centering
\includegraphics[scale=1]{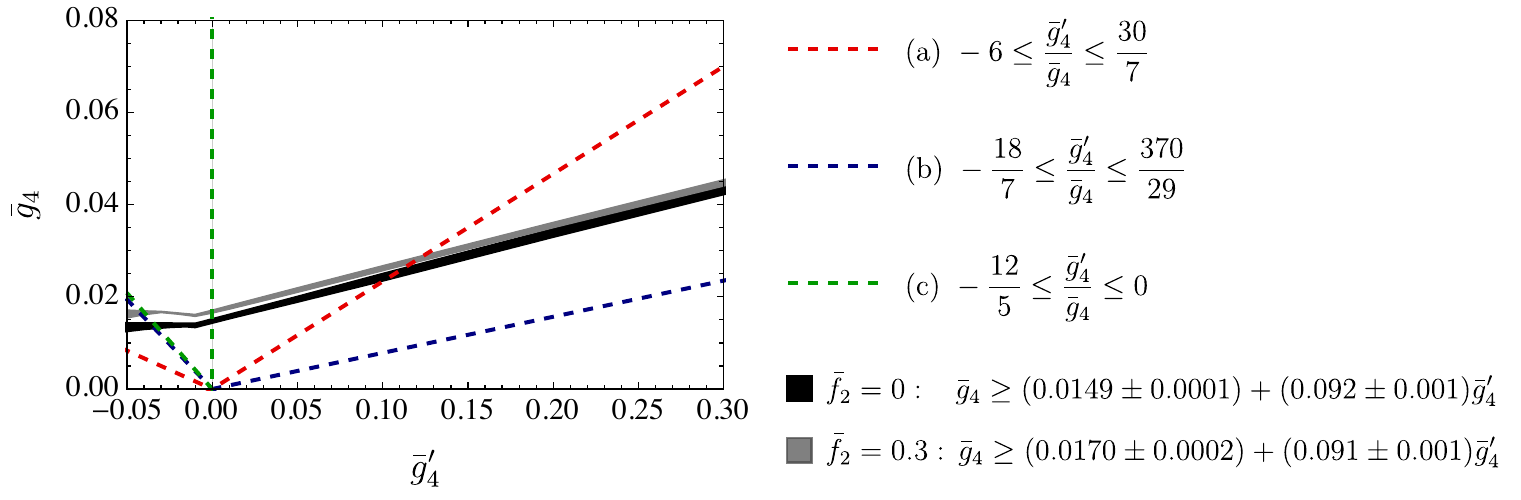}
\caption{Comparison between the numerical bounds obtained in section \ref{sec:numerics} and the analytic bounds (a) \cite{Arkani-Hamed:2020blm}, (b) \cite{Henriksson:2021ymi}  and (c) \cite{Henriksson:2022oeu}. The allowed regions lie in the $\bar{g}_4>0$ half-plane, and are delimited by the dashed lines (notice that the blue and green left edges almost coincide in the picture).}
\label{fig:g4g4pAnVsNum}
\end{figure}
The bounds \eqref{gPisaII} - \eqref{gEFTHedron} are compared to the weakly coupled theories (\ref{gBoundAll}) and (\ref{gBoundSlow}) in figure \ref{fig:g4g4pAnVsRuleIn}, and to the numerical results in figure \ref{fig:g4g4pAnVsNum}. It is interesting to notice that the cone generated by our perturbative amplitudes extends beyond the bounds found in \cite{Henriksson:2021ymi,Henriksson:2022oeu}. This is not a contradiction, since the use of null constraints requires the Regge limit to be strictly softer than $s^2$ beyond the forward limit. Indeed, excluding the spin 2 exchanges we get a convex hull squarely contained within their bounds. However, \cite{Henriksson:2021ymi} did consider the exchange of a massive graviton, finding it compatible with the analytic bounds. As we further discuss in appendix \ref{app:Wilson}, the amplitude proposed in \cite{Henriksson:2021ymi} coincides with our parity even II up to operators of dimension 8 in the EFT. One can easily check that the Wilson coefficients of dimension 10 and 12 in table \ref{tab:wilson} coincide with the ones reported in table 2 of \cite{Henriksson:2021ymi}. At the level of our analysis, which is based on the classical Regge bound, there is no reason to prefer this combination to the other two spin 2 resonances: it would be interesting to understand if, in some way, the other amplitudes are exonerated from the null constraints of  \cite{Henriksson:2021ymi,Henriksson:2022oeu}, while this one must obey them. In appendix \ref{app:Wilson}, we offer a few additional comments on the status of the parity even I and parity odd EFTs.

\newpage
\section{Discussion}

Photons are the particles of light. To the best of our knowledge, they are massless spin one particles. Assuming four dimensional Lorentz invariance, quantum mechanics and the absence of other massless particles, the low energy dynamics of photons must be described by the EFT Lagrangian \eqref{eq:EFT_lagrangian}. Of course, we expect photons to also interact with gravitons, but this is a excindingly small effect at reasonable energies.\footnote{For example, the  exclusive cross section of two photons into two photons from 1-loop effects in QED is of order $\sigma_{QED} \sim \alpha^4 s^3 /m_e^8$ and from tree-level graviton exchange is of order $\sigma_{grav} \sim G_N^2 s \sim s/m_{P}^4$. Therefore, the ratio $\sigma_{grav} / \sigma_{QED} \sim
   m_e^8/(m_P^4 \alpha^4  s^{2})  \sim  10^{-58} \left( 1\, eV / \sqrt{s} \right)^4$   is tiny for visible (or higher frequency) photons.
}
The Wilson coefficients $c_i$ in the EFT Lagrangian \eqref{eq:EFT_lagrangian} 
parametrize our ignorance about the high energy behavior of the theory.
Of course, in the real world we expect these to be dominated by QED effects. Nevertheless, in this paper we asked ourselves what values can these numbers take compatible with the S-matrix bootstrap principles of Lorentz invariance, unitarity and analyticity.
Using a numerical algorithm, we estimated several non-perturbative bounds as described in section \ref{sec:numerics}. QED seems to live well inside the allowed region. 
 Some bounds are saturated by weakly coupled amplitudes as we discuss in section \ref{sec:weak_coupling_models} but others are not. For example, the amplitude with minimal $\bar g_4$ is strongly coupled. Is there a physical theory that realizes such a small value of 
 $\bar g_4$? This is an open question for the future.

\begin{figure}
	\centering 
	\begin{tikzpicture}
		\draw[->] (-5.2,0)-- (5.2,0);
		\draw[->] (0,-1)--(0,2.8);

		\draw (4,2.5)--(4.5,2.5);
		\draw (4,2.5)--(4,3);
		\draw (4.3,2.8) node{$s$};
		
		\draw (2.2, -0.39) node{$M^2$};
        \draw[blue] (2,0) node{$\bullet$};
        
        \draw (-2, -0.39) node{$-M^2-t$};
        \draw[blue] (-1.5,0) node{$\bullet$};
		
		\draw[blue, decoration = {zigzag, segment length = 2mm, amplitude = 1mm}, decorate] (2,0) -- (5,0) ;
		\draw[blue, decoration = {zigzag,segment length = 2mm, amplitude = 1mm}, decorate] (-5,0) -- (-1.5,0);
		
	\end{tikzpicture}
	\caption{Analytic structure of the amplitudes $\Phi_i(s,t,u)$ assumed in the positivity study \cite{Henriksson:2021ymi}.
	The branch cuts associated to photon loops are neglected due to the assumption of weak coupling.  The  beginning of the cut at $s=M^2$ defines an energy scale $M$ that  is interpreted as the UV cutoff of the photon  EFT.
	\label{fig:s-plane-nocuts}}
\end{figure}
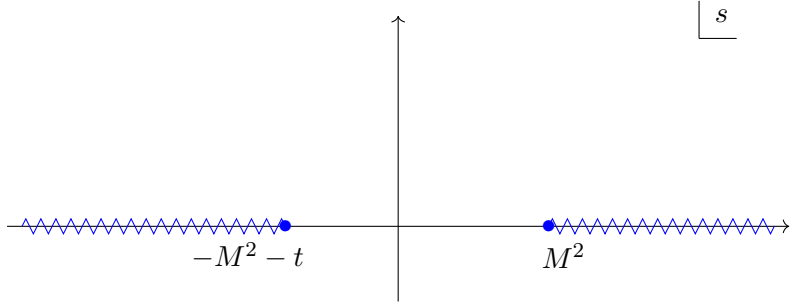

Let us comment on the bounds derived in \cite{Henriksson:2021ymi} using positivity.
In that work, the authors assumed weak coupling below the scale $M$ and neglected the branch cuts of the amplitude from photon loops.  More precisely, they assumed the analytic structure depicted in figure \ref{fig:s-plane-nocuts} and defined the scale $M$ from the position of the branch point at $s=M^2$. This assumption makes it possible to derive bounds on new dimensionless quantities like
\begin{equation}
	\label{eq:observables_EFT}
	\frac{g_3 M^2}{g_2},\qquad
	\frac{g_4 M^4}{g_2},\qquad
	\frac{g_4' M^4}{g_2},\qquad\ldots
\end{equation}
It would be interesting to derive similar bounds without completely ignoring the $\log$ terms in \eqref{eq:amplitudes_EFT}.
This requires a different definition of the scale $M$. For example, we may impose that the discontinuity of the amplitude is bounded by \eqref{eq:amplitudes_EFT} up to $s=M^2$ and let it free for $s>M^2$ (still imposing unitarity).
Such scenario can be easily studied with our primal numerical methods (see \cite{Miro:2022cbk} for a concrete implementation of a similar idea). 
We leave this exploration for the future.

The primal numerical S-matrix bootstrap becomes very expensive in the presence of massless particles. This has been noticed before in \cite{Guerrieri:2020bto, Guerrieri:2021ivu} but it was more extreme in this work. For this reason, we are in great need of more efficient numerical methods. Hopefully, the dual methods of \cite{Guerrieri:2021tak, He:2021eqn} can be generalized to massless particles.

It would be very interesting to consider the joint system of scattering amplitudes between photons and charged particles. However, this idea collides with the well known problem of IR divergences in 4 space-time dimensions. The same applies to scattering amplitudes involving photons and gravitons. These are also not well defined non-perturbatively in 4 dimensions. 
Hopefully,   better (inclusive?) observables can be constructed with properties amenable to a bootstrap approach.
 In the meantime, one can extend the present S-matrix bootstrap study in two main directions. Firstly, we can go to higher dimensions to avoid the IR divergences. It would be very interesting to study photon-graviton scattering in higher dimensions in the context of the weak gravity conjecture \cite{Henriksson:2022oeu}.
 Secondly, we can consider photons as probes of any QFT with a continuous global symmetry.
 In particular, we can study scattering of pions and photons in QCD in order to probe the chiral anomaly, analogously to the use of dilatons to probe  the $a$-anomaly \cite{Karateev:2022jdb}.
 Furthermore, this system can be further simplified by working in the planar limit of QCD as in \cite{Albert:2022oes}. In this case, non-linear unitarity is reduced to positivity.

\section*{Acknowledgements}
We thank Kausik Ghosh, Andrea Guerrieri, Sébastien Reymond, Lorenzo Ricci, Francesco Riva, Biswajit Sahoo, Balt van Rees, Pedro Vieira and Alexander Zhiboedov for very useful conversations. We also thank Andrea Guerrieri for reading the draft and giving valuable comments.

DK is supported by the SNSF Ambizione grant PZ00P2\_193411. 
The work
of KH and JP is supported by the Simons Foundation grant 488649 (Simons Collaboration on the Nonperturbative Bootstrap) and by the Swiss National Science Foundation
through the project 200020\_197160 and through the National Centre of Competence in Research SwissMAP.
MM is supported by the SNSF Ambizione grant PZ00P2\_193472. AH is supported by the Simons Foundation grant 488659 (Simons Collaboration on the Nonperturbative Bootstrap). AH thanks the participants of the  `S-matrix Bootstrap IV' workshop in Crete, Greece and the `Non-perturbative Methods in Quantum Field Theory' workshop at CERN, Switzerland, where this work was presented, for interesting comments. 
\appendix

\newpage
\section{Amplitudes, crossing and unitarity}
\label{sec:S-matrix_setup}
In this appendix we define scattering amplitudes of spin one massless particles. Focusing on the center of mass frame we will derive crossing equations and unitarity conditions they must obey. For completeness, in appendix \ref{app:tensor_structures_main} we will study these amplitudes in a generic frame and derive crossing equations also there. We will find that the results of the two appendices are in perfect agreement.

In this paper we work in the mostly plus metric
\begin{equation}\label{eq:metric}
	\eta^{\mu\nu} = \{-,+,\ldots,+,+\}\,.
\end{equation}
Throughout our work we always focus on the case of $d=4$ space-time dimensions except for appendices \ref{app:basis_d>4} and \ref{app:crossing_d>4} where we stay in general number of space-time dimensions $d$.

\subsection{Amplitudes for spin one massless particles}
\label{app:amplitudes_COM}

We start by defining scattering amplitudes in a general frame. We then focus on the center of mass frame.

\paragraph{Scattering amplitudes in a general frame}
Consider the 2-to-2 (in-out) scattering process of massless spin one particles, schematically $12\rightarrow 34$. It is described by the scattering amplitude $	\mathcal{S}{}_{\lambda_1, \lambda_2}^{\lambda_3,\lambda_4}(p_1,p_2,p_3,p_4)$ defined via the following matrix element
\begin{equation}
	\label{eq:scattering_amplitude}
	(2\pi)^4\delta^{(4)}(p_1^\mu + p_2^\mu - p_3^\mu - p_4^\mu) \times
	\mathcal{S}{}_{\lambda_1, \lambda_2}^{\lambda_3,\lambda_4}(p_1,p_2,p_3,p_4)\equiv
	{}_{id}\langle\kappa_3,\kappa_4|S|\kappa_1,\kappa_2\rangle_{id}.
\end{equation}
Here $S$ is the scattering operator.  The particles with 4-momenta $p_1$ and $p_2$ are incoming, and the particles with momenta $p_3$ and $p_4$ are outgoing. We adopt the convention that helicities of incoming particles are always placed downstairs, instead helicities of outgoing particles are always placed upstairs. The two-particle state describing the system of two identical massless particles is defined as
\begin{equation}
	\label{eq:2PS}
	|\kappa_1,\kappa_2\rangle_{id} \equiv \frac{1}{\sqrt 2}\big(
	|\vec p_1; \lambda_1\>\otimes |\vec p_2; \lambda_2\>+
	|\vec p_2; \lambda_2\>\otimes |\vec p_1; \lambda_1\>\big),
\end{equation}
where the symbol $\otimes$ stands for the ordered tensor product.  By construction it obeys the condition $|\kappa_1,\kappa_2\rangle_{id}=|\kappa_2,\kappa_1\rangle_{id}$. The $\sqrt{2}$ factor in the definition is part of our conventions. The one-particle states entering \eqref{eq:2PS} are denoted by
\begin{equation}
	\label{eq:1PS}
	|\vec p; \lambda\>\equiv |m=0,\vec p; j=1,\lambda\>.
\end{equation}
The right-hand side of \eqref{eq:1PS} is standard notation for one-particle states, where $m$ is the mass of the particle, $j$ is its spin, $\vec p$ is the spatial momentum and $\lambda$ is the helicity. For massless spin $j=1$ particles helicity can only take two values $\lambda=\pm 1$.
The normalization of one-particle states is given by
\begin{equation}
	\label{eq:normalization_1PS}
	\<\vec p_1; \lambda_1|\vec p_2; \lambda_2\> = 2\myP_1 \delta_{\lambda_1}^{\lambda_2}
	\times(2\pi)^3\delta^{(3)}(\vec p_1-\vec p_2)\,,
\end{equation}
where we defined $\myP_i = |\vec{p}_i|$.

Let us now define the interacting part of the scattering operator $T$ as follows
\begin{equation}
	\label{eq:operator_T}
	S=1+i T.
\end{equation}
This leads to the definition of the interacting scattering amplitude $\mathcal{T}{}_{\lambda_1, \lambda_2}^{\lambda_3,\lambda_4}(p_1,p_2,p_3,p_4)$, namely
\begin{equation}
	\label{eq:scattering_amplitude_interacting}
	(2\pi)^4\delta^{(4)}(p_1^\mu + p_2^\mu - p_3^\mu - p_4^\mu) \times
	\mathcal{T}{}_{\lambda_1, \lambda_2}^{\lambda_3,\lambda_4}(p_1,p_2,p_3,p_4)\equiv
	{}_{id}\langle\kappa_3,\kappa_4|T|\kappa_1,\kappa_2\rangle_{id},
\end{equation}
Using \eqref{eq:operator_T} we can write the relation between the scattering amplitude and its interacting part. It reads
\begin{equation}
	\label{eq:relation_ST}
	\mathcal{S}{}_{\lambda_1, \lambda_2}^{\lambda_3,\lambda_4}(p_1,p_2,p_3,p_4) =
	\frac{{}_{id}\langle\kappa_3,\kappa_4|\kappa_1,\kappa_2\rangle_{id}}{(2\pi)^4\delta^{(4)}(p_1^\mu + p_2^\mu - p_3^\mu - p_4^\mu)}+i\mathcal{T}{}_{\lambda_1, \lambda_2}^{\lambda_3,\lambda_4}(p_1,p_2,p_3,p_4).
\end{equation}
The first term in the right-hand side of \eqref{eq:relation_ST} is a formal expression. It can be straightforwardly evaluated for example in the center of mass frame in spherical coordinates (e.g. see footnote 15 in \cite{Hebbar:2020ukp}). The normalization of two-particle states follows from \eqref{eq:2PS} and \eqref{eq:normalization_1PS}. It reads
\begin{equation}
	\label{eq:normalization_2PS}
	{}_{id}\langle\kappa_3,\kappa_4|\kappa_1,\kappa_2\rangle_{id}=
	4\myP_1\myP_2(2\pi)^6	\left(\delta^{(3)}(\vec{p_1}-\vec{p_3})\delta^{(3)}(\vec{p_2}-\vec{p_4})\delta_{\lambda_1}^{\lambda_3}\delta_{\lambda_2}^{\lambda_4} + (3 \leftrightarrow 4) \right).
\end{equation}

Using the 4-momenta $p_i^\mu$  one can form three scalar quantities called the Mandelstam variables. For the in-out amplitudes their standard form reads as
\begin{equation}
	\label{eq:mandelstam_variables}
	s\equiv -(p_1+p_2)^2,\quad
	t\equiv -(p_1-p_3)^2,\quad
	u\equiv -(p_1-p_4)^2,\quad
	s+t+u=0.
\end{equation}
Their physical range is
\begin{equation}
	\label{eq:physical_range}
	s\geq 0,\qquad
	t\in[-s,\,0],\qquad
	u\in[-s,\,0].
\end{equation}

\paragraph{Scattering amplitudes in the center of mass frame}
The center of mass (COM) frame is defined by the following configuration of the 4-momenta
\begin{equation}
	\label{eq:COM_frame}
	\begin{aligned}
		&p_1^{\text{com}}\equiv\frac{\sqrt{s}}{2}\times(1,0,0,+1),\\
		&p_2^{\text{com}}\equiv\frac{\sqrt{s}}{2}\times(1,0,0,-1),\\
		&p_3^{\text{com}}\equiv\frac{\sqrt{s}}{2}\times(1,+\sin \theta ,0,+\cos \theta),\\
		&p_4^{\text{com}}\equiv\frac{\sqrt{s}}{2}\times(1,-\sin \theta ,0,-\cos \theta),
	\end{aligned}
\end{equation}
where $\theta \in [0,\pi]$ is the scattering angle. Plugging these into the definition of the Mandelstam variables \eqref{eq:mandelstam_variables} we find that
\begin{equation}
	\label{eq:t_u_to_scattering_angle}
	\sin\theta =\frac{2\sqrt{tu}}{s},\quad
	\cos\theta =\frac{t-u}{s}
	\quad\Leftrightarrow\quad
	t = - \frac{s}{2}(1-\cos\theta),\quad
	u = - \frac{s}{2}(1+\cos\theta).
\end{equation}
The center of mass amplitudes are defined as
\begin{equation}
	\label{eq:com_amplitudes}
	\mathcal{T}{}_{\lambda_1, \lambda_2}^{\lambda_3,\lambda_4}(s,t,u) \equiv
	\mathcal{T}{}_{\lambda_1, \lambda_2}^{\lambda_3,\lambda_4}(p_1^\text{com},p_2^\text{com},p_3^\text{com},p_4^\text{com}).
\end{equation}

Due to the presence of identical particles and parity symmetry, there are only 5 distinct center of mass amplitudes, our choice here is
\begin{equation}
	\label{eq:helicity_amplitude_def}
	\begin{aligned}
		\Phi_1 (s,t,u) &\equiv \mathcal{T}^{++}_{++}(s,t,u),\\
		\Phi_2(s,t,u)  &\equiv \mathcal{T}^{--}_{++}(s,t,u),\\
		\Phi_3(s,t,u)  &\equiv \mathcal{T}^{+-}_{+-}(s,t,u),\\
		\Phi_4(s,t,u)  &\equiv \mathcal{T}^{-+}_{+-}(s,t,u),\\
		\Phi_5(s,t,u)  &\equiv \mathcal{T}^{+-}_{++}(s,t,u).
	\end{aligned}
\end{equation}
The rest of the center of mass amplitudes are related to the above 5 ones via the 11 relations. Due to the presence of identical particles we get the following 9 constraints
\begin{equation}
	\label{eq:relations_com_1}
	\begin{aligned}
		\mathcal{T}_{--}^{--}&=\Phi_1,\quad\;\;\;
		\mathcal{T}_{--}^{+-}=\mathcal{T}_{--}^{-+},\quad
		\mathcal{T}_{-+}^{--}=\Phi_5,\quad\;\;\;
		\mathcal{T}_{-+}^{-+}=\Phi_3,\quad
		\mathcal{T}_{-+}^{+-}=\Phi_4,\\
		\mathcal{T}_{-+}^{++}&=\mathcal{T}_{--}^{-+},\quad
		\mathcal{T}_{+-}^{--}=\Phi_5,\quad\;\;\;
		\mathcal{T}_{+-}^{++}=\mathcal{T}_{--}^{-+},\quad
		\mathcal{T}_{++}^{-+}=\Phi_5.
	\end{aligned}
\end{equation}
Due to the requirement of parity invariance we get in addition another two relations which read
\begin{equation}
	\label{eq:relations_com_2}
	\mathcal{T}_{--}^{-+}=\Phi_5,\qquad
	\mathcal{T}_{--}^{++}=\Phi_2.
\end{equation} 
Both \eqref{eq:relations_com_1} and \eqref{eq:relations_com_2} follow straightforwardly from equations (2.64), (2.86), (2.89) and (2.90) in \cite{Hebbar:2020ukp}.

\subsection{Crossing equations in the center of mass frame}

The $s-t$ and $s-u$ crossing equations for the center of mass amplitudes in the case of massless $j=1$ particles can be obtained up to an overall phase using the arguments of Trueman and Wick \cite{Trueman:1964zzb}. They were derived in detail for example in \cite{Hebbar:2020ukp}, see equations (2.81) and (2.82). Specializing to the case of photon scattering, their result reads

\begin{equation}
	\label{eq:crossing_COM}
	\begin{aligned}
		\mathcal{T}_{\lambda_1, \lambda_2}^{\lambda_3,\lambda_4}(s,t,u) &=
		\chi_{st}\,
		\mathcal{T}_{-\lambda_1, +\lambda_3}^{+\lambda_2,-\lambda_4}(t,s,u),\\
		\mathcal{T}_{\lambda_1, \lambda_2}^{\lambda_3,\lambda_4}(s,t,u) &=
		\chi_{su}\,
		\mathcal{T}_{-\lambda_1, +\lambda_4}^{-\lambda_3,+\lambda_2}(u,t,s),
	\end{aligned}
\end{equation}
where the overall phases $\chi_{st}$ and $\chi_{su}$ remain undetermined. Using \eqref{eq:helicity_amplitude_def}, \eqref{eq:relations_com_1} and \eqref{eq:relations_com_2} we can rewrite the crossing equations \eqref{eq:crossing_COM} in terms of the 5 center of mass amplitudes $\Phi_i(s,t,u)$ only. They read
\begin{equation}
	\label{eq:crossing_com}
	\Phi_I(s,t,u) = \chi_{st}\sum_{J=1}^5 C^{st}_{IJ} \Phi_J(t,s,u),\qquad
	\Phi_I(s,t,u) = \chi_{su}\sum_{J=1}^5 C^{su}_{IJ} \Phi_J(u,t,s),
\end{equation}
where the crossing matrices read
\begin{equation}
	\label{eq:matrices_C}
	C^{st} \equiv \begin{pmatrix}
		0 & 0 & 0 & 1 & 0 \\
		0 & 1 & 0 & 0 & 0 \\
		0 & 0 & 1 & 0 & 0 \\
		1 & 0 & 0 & 0 & 0 \\
		0 & 0 & 0 & 0 & 1
	\end{pmatrix},\qquad
	C^{su} \equiv \begin{pmatrix}
		0 & 0 & 1 & 0 & 0 \\
		0 & 1 & 0 & 0 & 0 \\
		1 & 0 & 0 & 0 & 0 \\
		0 & 0 & 0 & 1 & 0 \\
		0 & 0 & 0 & 0 & 1
	\end{pmatrix}.
\end{equation}
Both matrices have the following eigenvalues $\{-1,1,1,1,1\}$. They cannot however be simultaneously diagonalized.

The easiest way to fix the unknown phases $\chi_{st}$ and $\chi_{su}$ in \eqref{eq:crossing_com} is to plug the explicit expressions of $\Phi_i(s,t,u)$ given in \eqref{eq:amplitudes_EFT} at the lowest order in $s$ (i.e. using only $s^2$ terms)  into the crossing equations \eqref{eq:crossing_com}. One then immediately concludes that
\begin{equation}
	\label{eq:phases}
	\chi_{st}=\chi_{su}=+1.
\end{equation}

As an alternative approach, in appendix \ref{app:LSZ}, we will derive the crossing equation \eqref{eq:crossing_COM} explicitly using LSZ. There, we also show that the phases are given by \eqref{eq:phases}.

\subsection{Unitarity}\label{app:unitarity}
We now discuss the constraints on the amplitudes due to unitarity. This subsection is an application of the general construction presented in \cite{Hebbar:2020ukp}, which the reader is referred to for more details. We begin by defining a short-hand notation for the two-particle state \eqref{eq:2PS} evaluated in the center of mass frame, namely
\begin{equation}
	\label{eq:2PS_com}
	|(\myP, \theta,\phi) ; \lambda_1, \lambda_2\rangle_{id} \equiv  \frac{1}{\sqrt{2}} (|\vec p; \lambda_1 \rangle \otimes |-\vec p; \lambda_2  \rangle + |-\vec p; \lambda_2 \rangle \otimes |\vec p; \lambda_1  \rangle).
\end{equation}
Here $(\theta, \phi)$ are the angular coordinates of $\vec p$ and $\myP \equiv |\vec p\,|$. The state \eqref{eq:2PS_com} transforms in the reducible representation of the Poincar\'e group. Let us now define a two particle state which transforms in the irreducible representation of the Poincar\'e group instead. It reads
\begin{multline}
	\label{eq:irrep}
	|c, \vec 0, \ell, \lambda; \lambda_1 , \lambda_2 \rangle_{id} \equiv\\ \frac{2 \ell +1}{4 \pi \sqrt{2} C_\ell }\int_{0}^{2 \pi} d \phi \int_{0}^{\pi} d \theta \sin \theta e^{-i \phi (\lambda_1 + \lambda_2-\lambda)} d^{\,\ell}_{\lambda, \lambda_{12}} (\theta) |(\myP, \theta,\phi) ; \lambda_1, \lambda_2\rangle_{id},
\end{multline}
where $\lambda_{12} \equiv \lambda_1 - \lambda_2$, $\ell=0,1,2,\ldots$ and $\lambda$ are the  total spin and helicity, $c \equiv  2 \myP =\sqrt{s} $ is the center of mass energy and
\begin{equation}
	C_\ell  \equiv \sqrt{8 \pi (2\ell +1)}\,.
\end{equation}
Since the states \eqref{eq:irrep} transform in irreducible representations of the Poincar\'e group  we conclude that their inner product with the scattering operator $T$ have the following most general form
\begin{equation}
	\label{eq:irrep_matrix_element}
	{}_{id}\langle c', \vec p\,', \ell', \lambda' ; \lambda_3 , \lambda_4 |T|c, \vec p, \ell, \lambda ; \lambda_1 , \lambda_2 \rangle_{id} = (2 \pi)^4 \delta^{4}(p^\mu - p'^\mu)\delta_{\ell \ell'} \delta_{\lambda \lambda'} {\mathcal T_{\ell}}^{\, \lambda_3, \lambda_4}_{\, \lambda_1, \lambda_2}(s).
\end{equation}
Here the functions ${\mathcal T_{\ell}}^{\, \lambda_3, \lambda_4}_{\, \lambda_1, \lambda_2}(s)$ are called (interacting part of) partial amplitudes. They are related to the interacting part of  scattering amplitudes via the following integral transform
\begin{equation}
	\label{eq:partial_amplitudes}
	{\mathcal T_{\ell}}^{\, \lambda_3, \lambda_4}_{\, \lambda_1, \lambda_2}(s) = \frac{1}{32\pi}\int_0^\pi d\theta \sin \theta \,d^{\, \ell}_{\lambda_{12}, \lambda_{34}}(\theta)  \mathcal{T}_{\lambda_1, \lambda_2}^{\lambda_3,\lambda_4}\Big(s,t(s,\theta),u(s,\theta)\Big),
\end{equation}
with $t(s,\theta)$ and $u(s,\theta)$ given in \eqref{eq:t_u_to_scattering_angle}. Here $d^{\, \ell}_{\lambda_{12}, \lambda_{34}}(\theta)$ stand for small Wigner d-matrices, they are defined by 
\begin{align}
	d^\ell_{\lambda\lambda'}(\beta) &= \sqrt{(\ell+\lambda)!(\ell-\lambda)!(\ell+\lambda')!(\ell-\lambda')!}\nn\\
	&\qquad \times \sum_{\nu=0}^{2j}(-1)^{3\nu +\lambda -\lambda'} \frac{(\cos(\beta/2))^{2\ell +\lambda'-\lambda -2\nu} (\sin(\beta/2))^{\lambda-\lambda'+2\nu}}{\nu!(\ell -\lambda -\nu)! (\ell+\lambda'-\nu)!(\nu +\lambda-\lambda')!} \label{eq:wignerD}\\
	&=\left(\cos\left(\frac{\beta}{2}\right)\right)^{\lambda'+\lambda} \left(\sin\left(\frac{\beta}{2}\right)\right)^{\lambda'-\lambda}\sqrt{\frac{(l-\lambda )! (\lambda'+l)!}{(\lambda +l)! (l-\lambda')!} } \nn \\
	&\qquad \times \frac{\, _2F_1\left(\lambda'-\ell,\ell+\lambda'+1;-\lambda +\lambda'+1;\sin ^2\left(\frac{\beta }{2}\right)\right)}{\Gamma (-\lambda +\lambda'+1)}\,,
\end{align}
 where for completeness, we wrote two equivalent definitions.\footnote{Note that $\texttt{Mathematica}$ implements the small Wigner d-matrices with a different sign convention  $\texttt{WignerD}[\{j,\lambda, \lambda'\}, \beta] = (-1)^{\lambda-\lambda'}d^j_{\lambda\lambda'}(\beta)$.}
Using properties of the Wigner $d$-matrices, see for example appendix A.1 of \cite{Hebbar:2020ukp}, the integral transform \eqref{eq:partial_amplitudes} can be inverted and we obtain the usual partial wave expansion 
\begin{equation}\label{eq:PartialWave_Expansion}
	 \mathcal{T}_{\lambda_1, \lambda_2}^{\lambda_3,\lambda_4}\Big(s,t(s,\theta),u(s,\theta)\Big) = \sum_{\ell} 16\pi(2\ell+1) \, {T_{\ell}}^{\, \lambda_3, \lambda_4}_{\, \lambda_1, \lambda_2}(s) \, d^{\, \ell}_{\lambda_{12}, \lambda_{34}}(\theta).
\end{equation}
In the case of two identical spin one massless particles there are three possible two particle Poincar\'e irreps.  We list them using the notation $+$ for helicity $+1$ and $-$ for helicity $-1$\footnote{Note that $|c,\vec p, \ell, \lambda; - , + \rangle_{id} = (-1)^{\ell} |c,\vec p, \ell, \lambda; + , - \rangle_{id}$ by Bose symmetry.} 
\begin{equation}
	|c,\vec p, \ell, \lambda; + , + \rangle_{id}, \quad |c,\vec p, \ell, \lambda; - , - \rangle_{id}, \quad |c,\vec p, \ell, \lambda; + , - \rangle_{id}
\end{equation}
Bose symmetry of identical particles also implies the selection rule that the first two states in the list above only exist for even $\ell$. The third state exists for all spin $\ell \geq 2$.\footnote{The total spin $\ell$ must always be greater than the difference in helicity $\lambda_1 - \lambda_2$ of the two particles.} Under parity transformation, the three states transform as follows 
\begin{equation}
	\begin{aligned}
		\mathcal P|c,\vec 0, \ell, \lambda; + , + \rangle_{id} &= (-1)^{\ell} |c,\vec 0, \ell, \lambda; - , - \rangle_{id},\\
		\mathcal P|c,\vec 0, \ell, \lambda; - , - \rangle_{id} &= (-1)^{\ell} |c,\vec 0, \ell, \lambda; + , + \rangle_{id},\\
		\mathcal P|c,\vec 0, \ell, \lambda; + , - \rangle_{id} &=  |c,\vec 0, \ell, \lambda; + , - \rangle_{id}.
	\end{aligned}
	\label{parityOnPartialWaves}
\end{equation}
Since we consider parity invariant theories, it is convenient to define new linear combinations which are parity eigenstates:
\begin{equation}
	\label{eq:parity_even_two_photon_irreps}
	\text{\bf parity even:}\quad
	\begin{aligned}
		|1\rangle &\equiv \frac{1}{\sqrt{2}}\Big(|c,\vec p, \ell, \lambda; + , + \rangle_{id} +  |c,\vec p, \ell, \lambda; - , - \rangle_{id}\Big), \quad \ell \geq 0\;(\text{even}),\\
		|2 \rangle &\equiv  \sqrt{2}\,|c,\vec p, \ell, \lambda; + , - \rangle_{id}, \quad \ell \geq 2,
	\end{aligned}
\end{equation}
\begin{equation}
	\label{eq:parity_odd_two_photon_irrep}
		\text{\bf parity odd:}\quad
	|3\rangle \equiv \frac{1}{\sqrt{2}}\Big(|c,\vec p, \ell, \lambda; + , + \rangle_{id} -  |c,\vec p, \ell, \lambda; - , - \rangle_{id}\Big), \quad \ell \geq 0\; (\text{even}).\\
\end{equation}
In a unitary quantum theory, the norm of any state in the theory must be non-negative. Consider the following set of six states:
\begin{equation}
		|1\rangle_{in},\qquad |2\rangle_{in}, \qquad |3\rangle_{in},  \qquad
		|1\rangle_{out}, \qquad |2\rangle_{out},  \qquad |3\rangle_{out}.
\end{equation}
Any linear combination of these states must have non-negative norm. This statement is equivalent to the statement that the $6 \times 6$ Hermitian matrix formed by the inner products between the six states is positive semi-definite. Factoring out the overall delta functions we write
\begin{equation}
	\begin{pmatrix}
		{}_{in}\langle a' | b \rangle_{in} & {}_{in}\langle a' | b \rangle_{out} \\
		{}_{out}\langle a' | b \rangle_{in} & {}_{out}\langle a' | b \rangle_{out}
	\end{pmatrix} =\mathcal H_{\ell}(s) \times (2 \pi)^4 \delta^4(p^\mu - p'^\mu) \delta_{\ell \ell'} \delta_{\lambda \lambda'}\,.
\end{equation} 
Unitarity as stated above then implies that 
\begin{equation}
	\label{eq:positivity_matrix_generic}
	\mathcal H_{\ell}(s) \succeq 0 \,,\quad \forall \  \ell \quad \text{and} \quad s \geq 0\,.
\end{equation}
The inner products between two incoming states or two outgoing states are fixed by the normalization of these states, namely\footnote{The pre-factors in \eqref{eq:parity_even_two_photon_irreps} and \eqref{eq:parity_odd_two_photon_irrep} ensure that all three states have the same normalization}
\begin{equation}
	\label{eq:inner_product_inin_outout}
	{}_{in}\langle a' | b \rangle_{in} = {}_{out}\langle a' | b \rangle_{out} = \delta_{a'b} \times \delta_{\ell \ell'} \delta_{\lambda \lambda'} (2\pi)^4 \delta^4(p^\mu - p'^\mu)\,.
\end{equation}
The inner products between incoming and outgoing states are, by definition, the matrix elements of the scattering operator $S = 1 + i T$ and therefore due to \eqref{eq:irrep_matrix_element} we have  
\begin{equation}
	\label{eq:inner_product_out_in}
	\begin{aligned}
		{}_{out}\langle 1' | 1 \rangle_{in} &= \delta_{\ell \ell'} \delta_{\lambda \lambda'} (2\pi)^4 \delta^4(p^\mu - p'^\mu) \left(1 + i\left({\mathcal T_{\ell}}^{\, +, +}_{\, +, +}(s)+{\mathcal T_{\ell}}^{\, -, -}_{\, +, +}(s)\right)\right) \\
		{}_{out}\langle 2' | 2 \rangle_{in} &= \delta_{\ell \ell'} \delta_{\lambda \lambda'} (2\pi)^4 \delta^4(p^\mu - p'^\mu) \left(1 + 2i\,{\mathcal T_{\ell}}^{\, +, -}_{\, +, -}(s)\right) \\
		{}_{out}\langle 2' | 1 \rangle_{in} &= \delta_{\ell \ell'} \delta_{\lambda \lambda'} (2\pi)^4 \delta^4(p^\mu - p'^\mu) \left( 2i\,{\mathcal T_{\ell}}^{\, +, -}_{\, +, +}(s)\right)\\
		{}_{out}\langle 1' | 2 \rangle_{in} &= \delta_{\ell \ell'} \delta_{\lambda \lambda'} (2\pi)^4 \delta^4(p^\mu - p'^\mu) \left( 2i\,{\mathcal T_{\ell}}^{\, +, -}_{\, +, +}(s)\right)\\
		{}_{out}\langle 3' | 3 \rangle_{in} &= \delta_{\ell \ell'} \delta_{\lambda \lambda'} (2\pi)^4 \delta^4(p^\mu - p'^\mu) \left(1 + i\left({\mathcal T_{\ell}}^{\, +, +}_{\, +, +}(s)-{\mathcal T_{\ell}}^{\, -, -}_{\, +, +}(s)\right)\right)\,. \\
	\end{aligned}
\end{equation}

Because of the invariance under parity there is no scattering between states with different parity eigenvalues. Hence the inner products ${}_{out}\langle 3' | 1 \rangle_{in}$ and ${}_{out}\langle 3' | 2 \rangle_{in}$ are all zero and the positive semi-definite condition \eqref{eq:positivity_matrix_generic} simplifies into smaller matrices. Taking into account the parity selection rules, we arrive at two separate sectors, namely the parity even and parity odd sectors.
\paragraph{Parity even sector} 
We begin by considering parity even eigenstates \eqref{eq:parity_even_two_photon_irreps}. For even spin $\ell \geq 2$, we have
\begin{align}
	\label{eq:matrix_1}
	\mathcal H^{+}_{\ell}(s)
	\times\delta_{\ell \ell'}\delta_{\lambda\lambda'}(2\pi)^4\delta^{(4)}(p-p')
	&\equiv \begin{pmatrix}
		{}_{in}\langle 1' | 1 \rangle_{in} & {}_{in}\langle 1' | 2 \rangle_{in} & {}_{in}\langle 1' | 1 \rangle_{out} & {}_{in}\langle 1' | 2 \rangle_{out} \\
		{}_{in}\langle 2' | 1 \rangle_{in} & {}_{in}\langle 2' | 2 \rangle_{in} & {}_{in}\langle 2' | 1 \rangle_{out} & {}_{in}\langle 2' | 2 \rangle_{out} \\
		{}_{out}\langle 1' | 1 \rangle_{in} & {}_{out}\langle 1' | 2 \rangle_{in} & {}_{out}\langle 1' | 1 \rangle_{out} & {}_{out}\langle 1' | 2 \rangle_{out} \\
		{}_{out}\langle 2' | 1 \rangle_{in} & {}_{out}\langle 2' | 2 \rangle_{in} & {}_{out}\langle 2' | 1 \rangle_{out} & {}_{out}\langle 2' | 2 \rangle_{out} 
	\end{pmatrix}   
\end{align}
The case of spin $\ell = 0$ is special because the state $|2\rangle$  does not exist and therefore we get a smaller matrix:
\begin{equation}
	\label{eq:matrix_2}
	\mathcal H^+_0(s)\times \delta_{\lambda\lambda'}(2\pi)^4\delta^{(4)}(p-p')  \equiv \begin{pmatrix}
		{}_{in}\langle 1' | 1 \rangle_{in} & {}_{in}\langle 1' | 1 \rangle_{out} \\
		{}_{out}\langle 1' | 1 \rangle_{in} & {}_{out}\langle 1' | 1 \rangle_{out}
	\end{pmatrix},                  
\end{equation}                        
We now consider odd $\ell \geq 3$, in which case the only state that exists is state $|2\rangle$ and therefore we have
\begin{align}                  
	\label{eq:matrix_3}
	\mathcal H^{+}_{\ell}(s)
	\times\delta_{\ell \ell'}\delta_{\lambda\lambda'}(2\pi)^4\delta^{(4)}(p-p')  &\equiv \begin{pmatrix}
		{}_{in}\langle 2' | 2 \rangle_{in} & {}_{in}\langle 2' | 2 \rangle_{out} \\
		{}_{out}\langle 2' | 2 \rangle_{in} & {}_{out}\langle 2' | 2 \rangle_{out}
	\end{pmatrix}
\end{align}
\paragraph{Parity odd sector} We now turn to the parity odd eigenstate \eqref{eq:parity_odd_two_photon_irrep} which exists for even spin $\ell \geq 0$. We have
\begin{equation}
	\label{eq:matrix_4}
	\mathcal H^-_\ell(s)\times\delta_{\ell \ell'}\delta_{\lambda\lambda'}(2\pi)^4\delta^{(4)}(p-p')  \equiv \begin{pmatrix}
		{}_{in}\langle 3' | 3 \rangle_{in} & {}_{in}\langle 3' | 3 \rangle_{out} \\
		{}_{out}\langle 3' | 3 \rangle_{in} & {}_{out}\langle 3' | 3 \rangle_{out}
	\end{pmatrix}.   
\end{equation}

\paragraph{Final summary}
We can now plug equations \eqref{eq:inner_product_inin_outout}, \eqref{eq:inner_product_out_in} into \eqref{eq:matrix_1} - \eqref{eq:matrix_4} to obtain the final form of the unitarity constraints. Below we will use the following notation for the partial amplitudes in the center of mass frame
\begin{equation}
	\begin{aligned}
		\Phi^\ell_1 (s) \equiv \mathcal T_\ell{}_{++}^{++}(s), \\
		\Phi^\ell_2 (s) \equiv \mathcal T_\ell{}_{++}^{--}(s), \\
		\Phi^\ell_3 (s) \equiv \mathcal T_\ell{}_{+-}^{+-}(s), \\
		\Phi^\ell_4 (s) \equiv \mathcal T_\ell{}_{+-}^{-+}(s), \\
		\Phi^\ell_5 (s) \equiv \mathcal T_\ell{}_{++}^{+-}(s).
	\end{aligned}
	\label{eq:photon_independent_partial}
\end{equation}
These are consistent with the definitions given in \eqref{eq:helicity_amplitude_def}. We finally get
\begin{align}
	\label{eq:unitarity_1}
\ell \geq 0 \text{ (even)}:\quad&	\begin{pmatrix}
	1 & 1\\
	1 & 1
\end{pmatrix}+
i\begin{pmatrix}
	0 & -\Phi_1^{\ell*}(s) + \Phi_2^{\ell*}(s) \\
	\Phi_1^\ell(s) - \Phi_2^\ell(s) & 0 
\end{pmatrix} \succeq 0\,,\\[4pt]
	\label{eq:unitarity_3}
\ell \geq 3 \text{ (odd)}:\quad	&	\begin{pmatrix}
	1 & 1\\
	1 & 1
\end{pmatrix}+
2i\begin{pmatrix}
	0 & -\Phi_3^{\ell*}(s)  \\
	\Phi_3^\ell(s)  & 0 
\end{pmatrix} \succeq 0\,,\\[4pt]
	\label{eq:unitarity_4}
\ell = 0 :\quad	&	\begin{pmatrix}
	1 & 1\\
	1 & 1
\end{pmatrix}+
i\begin{pmatrix}
	0 & -\Phi_1^{0*}(s) - \Phi_2^{0*}(s) \\
	\Phi_1^0(s) + \Phi_2^0(s) & 0
\end{pmatrix} \succeq 0\,,\\[4pt]
	\label{eq:unitarity_2}
\ell \geq 2 \text{ (even)}:\quad&		\begin{pmatrix}
	\mathbb I_{2\times 2} & \mathbb S^{\ell\dagger}_{2\times 2}(s) \\
	\mathbb S^\ell_{2\times 2}(s) & \mathbb I_{2\times 2} 
\end{pmatrix} \succeq 0,
\end{align}
where in \eqref{eq:unitarity_2} we have defined 
\begin{equation}\label{eq:unitarity_2_details}
	\mathbb I_{2\times 2} \equiv
	\begin{pmatrix}
		1 & 0\\
		0 & 1
	\end{pmatrix},\qquad
	\mathbb S^\ell_{2\times 2}(s) \equiv
	\begin{pmatrix}
		1 & 0\\
		0 & 1
	\end{pmatrix}+
	i\begin{pmatrix}
		\Phi_1^\ell(s) + \Phi_2^\ell(s)  &  \qquad 2 \Phi_5^{\ell}(s) \\
		2 \Phi_5^\ell(s) &    \qquad 2\Phi_3^\ell(s)
	\end{pmatrix}\,.
\end{equation}
The unitarity constraints presented in \eqref{eq:non-linear_unitarity} and \eqref{eq:entriesa_unitarity} are a compact rewriting of the above conditions.

We conclude this subsection by commenting on a curious symmetry of the above unitarity constraints.
The partial amplitude $\Phi_5^\ell$ appears only in the $4$ by $4$ matrix \eqref{eq:unitarity_2} via  \eqref{eq:unitarity_2_details}. It is straightforward to see that semidefinite positivity of \eqref{eq:unitarity_2} is  invariant under $\Phi_5\leftrightarrow-\Phi_5$. If in some particular theory $\Phi_5=0$ (and as a consequence $\Phi_5^\ell=0$), the matrix \eqref{eq:unitarity_2} can be brought into a block diagonal form and as a result the semidefinite positivity then simply reduces to a $2$ by $2$ condition 
\begin{equation}
		\label{eq:unitarity_5}
	\ell \geq 2 \text{ (even)}:\quad	\begin{pmatrix}
		1 & 1\\
		1 & 1
	\end{pmatrix}+
	i\begin{pmatrix}
		0 & +\Phi_1^{\ell*}(s) + \Phi_2^{\ell*}(s) \\
		\Phi_1^\ell(s) + \Phi_2^\ell(s) & 0 
	\end{pmatrix} \succeq 0,
\end{equation}
and a $2$ by $2$ condition on $\Phi_3^\ell$, same as \eqref{eq:unitarity_3} but now also for even spin
\begin{equation}
	\label{eq:unitarity_6}
\ell \geq 2 \text{ (even)}:\quad		\begin{pmatrix}
	1 & 1\\
	1 & 1
\end{pmatrix}+
2i\begin{pmatrix}
	0 & -\Phi_3^{\ell*}(s)  \\
	\Phi_3^\ell(s)  & 0 
\end{pmatrix} \succeq 0\,.\\[4pt]
\end{equation}
The unitarity conditions \eqref{eq:unitarity_1} -  \eqref{eq:unitarity_4} together with \eqref{eq:unitarity_5} and \eqref{eq:unitarity_6} are now $\Phi_2\leftrightarrow-\Phi_2$ symmetric.

\subsection{Forward limit}\label{app:ForwardLimit}
Consider the scattering operator $S$. It is unitary, namely
\begin{equation}
	S^\dagger S =1.
\end{equation}
Splitting $S$ into its trivial and interacting part according to \eqref{eq:operator_T}, the above constraint can be rewritten in the following form
\begin{equation}
	T-T^\dagger = i\, T^\dagger T.
\end{equation}
Taking the expectation value in some state $|\text{state}\>$ we get 
\begin{equation}
	\label{eq:optical_theorem}
	\Im\<\text{state}|T|\text{state}\>= \frac{1}{2}\<\text{state}|TT^\dagger|\text{state}\> \geq 0.
\end{equation}
The last inequality holds because any norm in a unitary theory should be non-negative. The result \eqref{eq:optical_theorem} is known as the optical theorem.

We have defined two-particle states of identical spin one massless particles in \eqref{eq:2PS}. Let us consider the situation when these states are in the center of mass (COM) frame given by \eqref{eq:COM_frame}. In what follows we will use the following short-hand notation
\begin{equation}
	|\lambda_1,\lambda_2\> \equiv |\kappa_1,\kappa_2\rangle_{id}^{\text{COM}},\qquad
	\<\lambda_3,\lambda_4| \equiv {}_{id}^{\text{COM}}\<\kappa_3,\kappa_4|,
\end{equation}
where $\kappa_i$ are the particles participating in the scattering process $12 \rightarrow 34$ with the 4-momenta $p_i^\mu$.
We would also like to define the following state
\begin{equation}\label{eq:general_state}
	|\text{state}\>=\alpha_1 |+,+\>+\alpha_2 |-,-\> + \alpha_3 |+,-\> +\alpha_4 |-,+\>,
\end{equation}
with an analogous definition for $\<\text{state}|$.
Here $\vec \alpha \equiv \{\alpha_1, \alpha_2, \alpha_3, \alpha_4\}$ is a vector of complex numbers. Plugging \eqref{eq:general_state} into \eqref{eq:optical_theorem}, using the definitions \eqref{eq:helicity_amplitude_def} and evaluating the expression in the forward limit given by $\theta=0$ (or equivalently $t=0$) we obtain the following constraint\footnote{In writing this equation we have dropped the overall delta-function $(2\pi)^4\delta^{(4)}(0)$.}
\begin{align}
	\label{eq:forward_constraint}
 \vec{\alpha}^\dagger  \Im\begin{pmatrix}
		\Phi_1(s)&\Phi_2(s) & \Phi_5(s) &  \Phi_5(s) \\
		\Phi_2(s)& \Phi_1(s) &  \Phi_5(s) &  \Phi_5(s) \\
		 \Phi_5(s) &  \Phi_5(s)& \Phi_3(s)&  \Phi_4(s) \\
		 \Phi_5(s) &  \Phi_5(s) &  \Phi_4(s) &\Phi_3(s)
	\end{pmatrix} \vec{\alpha} \geq 0.
\end{align}
Here we have also defined the short-hand notation of the amplitudes in the forward limit
\begin{equation}
	\label{eq:forward_amp_app}
	\Phi_i(s) \equiv \Phi_i(s, \, t=0,\, u=-s).
\end{equation}
In the forward limit we have a special situation because
\begin{equation}
	\label{eq:special_situation}
	\Phi_4(s) = \Phi_5(s) = 0.
\end{equation}
This follows from  the representation of amplitudes in terms of partial amplitudes \eqref{eq:PartialWave_Expansion} and  the fact that
\begin{equation}
d^\ell_{2,-2}(\theta=0) =d^\ell_{2,0}(\theta=0)=0.
\end{equation}
Using \eqref{eq:special_situation} the constraint \eqref{eq:forward_constraint} can be rewritten in its final form
\begin{align}
	\label{eq:forward_constraint_2}
\Im \begin{pmatrix}
		\Phi_1(s)&\Phi_2(s) & 0& 0 \\
		\Phi_2(s)& \Phi_1(s) & 0&0\\
		0 & 0& \Phi_3(s)&0 \\
		0 & 0& 0&\Phi_3(s)
	\end{pmatrix}\succeq 0.
\end{align}
Due to the Sylvester's criterion the positive semi-definite condition \eqref{eq:forward_constraint_2} is equivalent to
\begin{equation}\label{eq:positivity_forward_general}
	\left\{ \begin{split}
		&\Im (\Phi_1(s)+\Phi_2(s))\geq 0\\
		&\Im (\Phi_1(s)-\Phi_2(s))\geq0\\
		&\Im \Phi_3(s) \geq0
	\end{split}\right.  \qquad \qquad \forall s\geq0 \ .
\end{equation} 
The result \eqref{eq:positivity_forward_general} can be alternatively obtained from the unitarity constraints \eqref{eq:unitarity_1} - \eqref{eq:unitarity_2} and  \eqref{eq:PartialWave_Expansion} by taking the forward limit $\theta=0$.

\section{Tensor structures}
\label{app:tensor_structures_main}

Consider the in-out interacting amplitude $\mathcal{T}{}_{\lambda_1, \lambda_2}^{\lambda_3,\lambda_4}(p_1,p_2,p_3,p_4)$ describing the scattering process $12\rightarrow 34$. Recall that $\lambda_1$ and $\lambda_2$ are helicities of the incoming particles with 4-momenta $p_1$ and $p_2$, and $\lambda_3$ and $\lambda_4$ are helicities of the outgoing particles with 4-momenta $p_3$ and $p_4$.
It is convenient to factorize the interacting scattering amplitude of spinning particles in the following way
\begin{equation}
	\label{eq:decomposition_general}
	\mathcal{T}{}_{\lambda_1, \lambda_2}^{\lambda_3,\lambda_4}(p_1,p_2,p_3,p_4) =
	\sum_{I=1}^{N}\mathcal{A}_I(s,t,u)
	\mathbf{T}_I{}_{\lambda_1, \lambda_2}^{\lambda_3,\lambda_4}(p_1,p_2,p_3,p_4).
\end{equation}
Here the objects $\mathbf{T}_I$ take care of the correct Lorentz transformation properties. They are called tensor structures. Their form is completely fixed by  kinematics. There are $N$ linearly independent tensor structures. The objects $\mathcal{A}_I(s,t,u)$ are referred to as the components of the interacting scattering amplitude and are invariant under Lorentz transformations. As a result they depend only on the Mandelstam variables defined in \eqref{eq:mandelstam_variables}.
Contrary to tensor structures, the components of interacting amplitudes $\mathcal{A}_I(s,t,u)$ carry dynamical information of a particular theory.

The goal of this appendix is to explicitly construct a linearly independent basis of tensor structures $\mathbf{T}_I$ in the case of identical spin one massless particles. We will do it using two different formalisms: the vector formalism presented in appendix \ref{sec:tensor_struct_vector_main} and the spinor formalism presented in appendix \ref{sec:tensor_struct_spinHel_main}. While the former works for any space-time dimension, the latter is somewhat more efficient but also dimension dependent: here we present its $d=4$ incarnation. (We will discuss group theory behind the construction of tensor structures in appendix \ref{subsec:gaugeToSH}). We will find that in general space-time dimensions $d>4$ there are $N=7$ linearly independent tensor structures, instead in $d=4$ space-time dimensions there are $N=5$ tensor structures \cite{Chowdhury:2019kaq}.

In equation \eqref{eq:com_amplitudes}, we have defined scattering amplitudes in the center of mass frame in $d=4$. Evaluating \eqref{eq:decomposition_general} in the center of mass frame and plugging the result into \eqref{eq:com_amplitudes} we conclude that
\begin{equation}
	\label{eq:tensor_structures_COM}
		\mathcal{T}{}_{\lambda_1, \lambda_2}^{\lambda_3,\lambda_4}(s,t,u) \equiv
\sum_{I=1}^{5}\mathcal{A}_I(s,t,u)
\mathbf{T}_I{}_{\lambda_1, \lambda_2}^{\lambda_3,\lambda_4}(p_1^\text{com},p_2^\text{com},p_3^\text{com},p_4^\text{com})
\end{equation}
Let us recall that in $d=4$ there are only 5 distinct center of mass amplitudes. They were chosen in \eqref{eq:helicity_amplitude_def} and denoted by $\Phi_I(s,t,u)$. Equation \eqref{eq:tensor_structures_COM} gives an explicit relation between the two sets of objects $\Phi_I(s,t,u)$ and $\mathcal{A}_I(s,t,u)$ describing the scattering process $12\rightarrow 34$, namely
\begin{equation}
	\label{eq:relation_amplitudes}
	\Phi_I (s,t,u) = \sum_{J=1}^5M_{IJ}(s,t,u) \mathcal{A}_J(s,t,u).
\end{equation}
The matrix $M$ depends on the explicit form of the tensor structures $\mathbf{T}_I$.

When constructing tensor structures it is often convenient to work with all-in amplitudes describing the process $1234\rightarrow \text{vacuum}$. We denote them by
\begin{equation}
	\mathcal{T}{}_{\lambda_1, \lambda_2,\lambda_3,\lambda_4}(p_1,p_2,p_3,p_4).
\end{equation}
We remind the reader that helicities placed downstairs represent incoming particles, instead helicities placed upstairs represent outgoing particles. This convention allows to quickly distinguish between the in-out amplitudes and the all-in amplitudes in formulas.
We will define the all-in amplitudes in appendix \ref{app:all-in_amplitudes} and explain how they are related to the in-out amplitudes $\mathcal{T}{}_{\lambda_1, \lambda_2}^{\lambda_3,\lambda_4}(p_1,p_2,p_3,p_4)$.

\subsection{All-in vs. in-out amplitudes}
\label{app:all-in_amplitudes}
We focus here on $d=4$ space-time dimensions where helicities $\lambda_i$ are simply numbers. All the conclusions of this subsection however hold in generic $d\geq4$.

We defined the in-out interacting part of the scattering amplitudes describing the scattering process $12\rightarrow 34$ in \eqref{eq:scattering_amplitude_interacting}. Analogously we can define all-in scattering amplitudes by crossing both of the outgoing particles i.e. by analytic continuation of the in-out amplitude
\begin{equation}
	\label{eq:connection}
	\mathcal{T}{}_{\lambda_1, \lambda_2,\lambda_3,\lambda_4}(p_1,p_2,p_3,p_4)\equiv
	\mathcal{T}{}_{+\lambda_1, +\lambda_2}^{-\lambda_3,-\lambda_4}(p_1,p_2,-p_3,-p_4).
\end{equation}
The all-in amplitudes are non-physical because the process $1234\rightarrow \text{vacuum}$ does not obey the energy-momentum conservation for physical particles with $p_i^0\geq 0$, $i=1,2,3,4$. 
Crossing is a very non-trivial operation, see for example appendix E in \cite{Hebbar:2020ukp} for a detailed discussion. The relation \eqref{eq:connection} can be seen as the definition of the all-in amplitude in terms of the in-out amplitude via the following analytic continuation\footnote{This procedure is ambiguous. We make a particular choice of the analytic continuation, which in spherical coordinates reads as
	\begin{equation*}
		\label{eq:ac}
		p^0 \rightarrow -p^0,\quad
		\myP \rightarrow -\myP,\quad
		\theta \rightarrow \theta,\quad
		\phi \rightarrow \phi.
\end{equation*}}
\begin{equation}
	\label{eq:pminusp}
	i=3,4:\qquad
	p_i^\mu \rightarrow \text{complex values} \rightarrow-p_i^\mu.
\end{equation}
The Mandelstam variables describing the process $1234\rightarrow \text{vacuum}$ are defined as
\begin{equation}
	\label{eq:mandelstam_variables_all_in}
	s\equiv -(p_1+p_2)^2,\quad
	t\equiv -(p_1+p_3)^2,\quad
	u\equiv -(p_1+p_4)^2,\quad
	s+t+u=0.
\end{equation}
Notice the difference between \eqref{eq:mandelstam_variables_all_in} and  \eqref{eq:mandelstam_variables}. The two are equivalent however if one takes into account \eqref{eq:pminusp}.
Analogously to \eqref{eq:decomposition_general} we can decompose the all-in scattering amplitude into tensor structures
\begin{align}
	\mathcal{T}_{\lambda_1, \lambda_2,\lambda_3,\lambda_4}(p_1,p_2,p_3,p_4)
	= \sum_{I=1}^{N} \mathcal{A}_I(s,t,u) 	\mathbf{T}_I{\,}_{\lambda_1, \lambda_2,\lambda_3,\lambda_4}(p_1,p_2,p_3,p_4),
	\label{eq:decomposition_crossing_def}
\end{align}
where  we recall that $\mathcal{A}_I$ are the unknown functions containing the dynamics of the theory and $\mathbf{T}_I$ are the basis of tensor structures whose form is completely fixed by kinematics.

The benefit of working with all-in amplitudes is that they are $S_4$ permutation  symmetric in the case of identical particles. Concretely, the all-in amplitudes obey the following constraints
\begin{equation}
	\label{eq:crossing_in}
	\begin{aligned}
		\mathcal{T}_{\lambda_1, \lambda_2,\lambda_3,\lambda_4}(p_1,p_2,p_3,p_4) &=
		\mathcal{T}_{\lambda_1, \lambda_3,\lambda_2,\lambda_4}(p_1,p_3,p_2,p_4),\\
		\mathcal{T}_{\lambda_1, \lambda_2,\lambda_3,\lambda_4}(p_1,p_2,p_3,p_4) &=
		\mathcal{T}_{\lambda_1, \lambda_4,\lambda_3,\lambda_2}(p_1,p_4,p_3,p_2),\\
		&\ldots
	\end{aligned}
\end{equation}
These are nothing but the crossing equations. Crossing equations for in-out amplitudes are slightly more complicated. We will derive them in appendix  \ref{app:LSZ}, they read
\begin{equation}
	\label{eq:crossing_general_frame}
	\begin{aligned}
		\mathcal{T}_{\lambda_1, \lambda_2}^{\lambda_3, \lambda_4}(p_1, p_2, p_3, p_4) = \mathcal{T}_{+\lambda_1, -\lambda_3}^{-\lambda_2, +\lambda_4}(p_1, -p_3, -p_2, p_4),\\
		\mathcal{T}_{\lambda_1, \lambda_2}^{\lambda_3, \lambda_4}(p_1, p_2, p_3, p_4) = 
		\mathcal{T}_{+\lambda_1, -\lambda_4}^{+\lambda_3, -\lambda_2}(p_1, -p_4, p_3, -p_2).
	\end{aligned}
\end{equation}
The crossing equations for the all-in amplitudes \eqref{eq:crossing_in} and the in-out amplitudes \eqref{eq:crossing_general_frame} are equivalent if we take into account the relation \eqref{eq:connection} and its variations.

Let us conclude this subsection by being more precise about the symmetry of \eqref{eq:decomposition_crossing_def}. As we already explained, due to the fact that the particles are identical and neutral, \eqref{eq:decomposition_crossing_def} must be invariant under $S_4$ permutation symmetry. There is a special normal subgroup of $S_4$ which is $\mathbb{Z}_2\times\mathbb{Z}_2$. This subgroup is generated by the $\{(2, 1, 4, 3), (3, 4, 1, 2)\}$ permutations and leaves the Mandelstam variables \eqref{eq:mandelstam_variables_all_in} invariant. As a result the functions $\mathcal{A}_I(s,t,u)$ are $\mathbb{Z}_2\times\mathbb{Z}_2$ invariant. When constructing the basis of tensor structures  $	\mathbf{T}_I$ we will require them to be $\mathbb{Z}_2\times\mathbb{Z}_2$ symmetric. The remaining symmetry of  \eqref{eq:decomposition_crossing_def} is
\begin{equation}
	\label{eq:remaining_crossing_symmetry}
	S_4/(\mathbb{Z}_2\times\mathbb{Z}_2)= S_3\,
\end{equation}  
which is precisely the $s-t-u$ crossing symmetry. The basis of tensor structures $\mathbf{T}_I$ will transform in some generally non-trivial representation of $S_3$. In order to make \eqref{eq:decomposition_crossing_def} invariant under $S_3$ we will demand that the amplitudes $\mathcal{A}_I(s,t,u)$ transform such that they compensate for the non-trivial transformation of  $\mathbf{T}_I$. This solves crossing, the details are in subsections \ref{app:crossing_d>4} and \ref{sec:parityAndPerm_SpinHel} for  the vector and the spinor formalism respectively.

\subsection{Group theory of tensor structures}
\label{subsec:gaugeToSH}

As is standard, we require that our quantum system is invariant under the restricted Poincar\'e group, which consists of the group of translations and the proper orthochronous Lorentz group $SO^+(1,3)$. (Au contraire, the full Poincar\'e group also contains two additional discrete symmetries: parity and time-reversal). In fact, quantum mechanics requires the symmetry group to be centrally extended to 
$SL(2, \mathbb C)$. The two groups have the same algebra. In this work we are often imprecise and refer to both $SO^+(1,3)$ and $SL(2, \mathbb C)$ as the Lorentz group.

In this section we work in $d=4$ space-time dimensions. There are 6 generators of the $so^+(1,3)$ algebra, namely 3 generators of rotations $J_i$ and 3 generators of boosts $K_i$. At the (complexified) algebra level 
\begin{equation}
	\label{eq:Lorentz_algebra}
	sl(2,\mathbb{C}) = su_L(2) \oplus su_R(2).
\end{equation}
The generators of the latter algebra are denoted by $M_i^L$ and $M_i^R$. The relation to the generators $J_i$ and $K_i$ is given by
\begin{equation}
	M_i^L = \frac{1}{2}(J_i+iK_i),\qquad
	M_i^R = \frac{1}{2}(J_i-iK_i).
\end{equation}
For later it is also convenient to define
\begin{equation}
	M^L_\pm \equiv M_1^L\pm M_2^L,\qquad
	M^R_\pm \equiv M_1^R\pm M_2^R.
\end{equation}

Particles are defined as a special set of irreducible representations of the Poincaré group. They are classified using representations of the so called Little group, a subgroup of the Lorentz group which leaves invariant the momentum of the particle in the standard center of mass frame. For massless particles, such standard momentum is 
\begin{equation}
k^\mu\equiv \{E,0,0,E\}~, \qquad E \geq 0~,
\end{equation}
and the little group is $ISO(2)$.
Its algebra has 3 generators commonly denoted by $A$, $B$ and $J$. They are related to the Lorentz generators as
\begin{equation}
	A = -J_1-K_2 = - M_+^L - M_-^R, \quad B =J_2-K_1 = i( - M_+^L + M_-^R),
	\label{ABtoM}
\end{equation}
together with
\begin{equation}
	\label{eq:Little_group_spin}
	J=J_3=M_3^L+M_3^R.
\end{equation}
Massless particles are assumed to transform trivially under transformations generated by $A$ and $B$ (a requirement coming from experiments). Thus, massless particles 
in the standard frame are labeled only by their helicity $\lambda$. In other words, 
\begin{equation}
	A| \vec{k}, \lambda \rangle = 0,\qquad B| \vec{k}, \lambda \rangle=0, \qquad
	J| \vec{k}, \lambda \rangle = \lambda | \vec{k}, \lambda \rangle. \label{PhotLGirrep}
\end{equation}
The general one particle state is then defined as the result of applying a boost to the state of the particle in the standard frame. A Lorentz transformation $\Lambda$ acts on a 4-vector as
\begin{equation}
p^\mu \rightarrow p'{}^{\mu} = \Lambda^\mu{}_\nu p^\nu.
\end{equation}
For each light-like momentum $p$, one can define a standard Lorentz transformation such that
\begin{equation}
p^\mu=L(p)^\mu{}_\nu k^\nu~.
\label{standardBoost}
\end{equation}
Then the general one-particle state is
\begin{equation}
| \vec{p}, \lambda \rangle = U(L(p)) | \vec{k}, \lambda \rangle~. 
%\qquad p^\mu=L(p)^\mu{}_\nu k^\nu~.
\label{InducedStates}
\end{equation} 
$U(\Lambda)$ is a (infinite dimensional) unitary representation of the Lorentz group. \eqref{InducedStates}, together with \eqref{PhotLGirrep}, uniquely identifies the action of such unitary operator on any single particle state:
\begin{equation}
	| \vec{p}, \lambda \rangle \rightarrow
	U(\Lambda)| \vec{p}, \lambda \rangle = e^{-i \lambda\,\theta} 
	| \vec{\Lambda p}, \lambda \rangle,
	\label{StatesLorentz}
\end{equation}
where $\theta$ depends on $\Lambda$ and $p$ and is the angle of rotation around the third axis defined by the following transformation:
\begin{equation}
L^{-1}(\Lambda p) \Lambda L(p)~,
\end{equation}
which belongs to the little group.
As a result, from the definition of the scattering amplitude \eqref{eq:scattering_amplitude_interacting}  and \eqref{eq:connection}, we deduce the following transformation property
\begin{equation}
	\mathcal{T}_{\lambda_1,\lambda_2,\lambda_3,\lambda_4}(\Lambda p_1, \Lambda p_2,\Lambda p_3, \Lambda p_4) = 
	t_1^{2\lambda_1}t_2^{2\lambda_2}t_3^{2\lambda_3}t_4^{2\lambda_4}
	\mathcal{T}_{\lambda_1,\lambda_2,\lambda_3,\lambda_4}(p_1, p_2, p_3, p_4),
	\label{TCovariance}
\end{equation}
where we have defined the short-hand notation  
\begin{equation}
	t_i \equiv e^{i \theta_i/2}.
\end{equation}

The tensor structures $\mathbf T_i$ introduced in \eqref{eq:decomposition_general} take care of the correct transformation properties of \eqref{eq:decomposition_general} given by \eqref{TCovariance}. In order to find them, it is useful to consider building blocks which have an index in a representation of the Lorentz group and another in the representation \eqref{PhotLGirrep} of the little group:
\begin{equation}
\Lambda_A{}^B \Xi_{\lambda,B}(p)=t^{-2\lambda}\,\Xi_{\lambda,A}(\Lambda p)~.
\label{GeneralInterpolator}
\end{equation}
These objects can be called  interpolators between the full Lorentz group and its Little group sub-group. Their usefulness stems from the fact that, contracting away the Lorentz index $A$ in \eqref{GeneralInterpolator} with other objects transforming in the same representation of the Lorentz group, one automatically gets the correct transformation law for the amplitude, equation \eqref{TCovariance}. The interpolators might be contracted with each other, or with the momenta. Hence, the most natural choice is to pick $A$ to be an index in the vector representation of the Lorentz group. A different choice is given by the spin $1/2$ representation, \emph{i.e.} the pair of fundamental representations of the $su(2)$ factors in \eqref{eq:Lorentz_algebra}. We call the former and the latter the vector and spinor formalisms respectively, and in the following we explain how to construct both classes of interpolators.

If we specify \eqref{GeneralInterpolator} to a transformation $W$ of the little group, we get
\begin{equation}
W_A{}^B \Xi_{\lambda,B}(k) = t^{-2\lambda}\,\Xi_{\lambda,A}(k)~.
\end{equation}
In other words, $\Xi(k)$ is a finite dimensional representation of the little group. Hence, the first step in constructing the interpolators is to check if the vector and the spinor representations contain the irreducible representation \eqref{PhotLGirrep} in their branching rules. Once this is done, the general interpolator is obtained via the same rule \eqref{InducedStates} which defined the state:
\begin{equation}
 \Xi_{\lambda,A}(p) = L_A{}^B(p)\, \Xi_{\lambda,B}(k)~,
 \label{XiToGenFrame}
\end{equation}
where $L_A{}^B(p)$ is the standard boost \eqref{standardBoost} in the vector or spinor representations.
This guarantees that the phase $\theta$ in \eqref{GeneralInterpolator} and \eqref{StatesLorentz} coincide. 

Starting with the spinors, equations \eqref{ABtoM} - \eqref{PhotLGirrep} tell us how the spin one-half irreps of $sl(2,\mathbb{C})$ transform under $iso(2)$. In particular, the spin up spinor of the $(1/2,0)$ has $A=B=0$ and $J=1/2$, while the spin down spinor of the $(0,1/2)$ has $A=B=0$ and $J=-1/2$. These are the interpolators we are looking for, and they are usually called spinor helicities. On the other hand, if we turn to the vector formalism, we face a problem. The $(1/2,1/2)$ representation of the Lorentz group decomposes into three representations of $iso(2)$, but none is of the form \eqref{PhotLGirrep}. The product of the spinor helicities is a singlet, hence proportional to $k^\mu$. The other two representations are two dimensional, and have a top component with $J=\pm 1$ respectively. Acting with $A$ and $B$ on it, we do not get zero, instead we get the singlet, $k^\mu$.\footnote{Recall that $iso(2)$ is not semi-simple.} The top components are the polarization vectors, and the action of $A$ and $B$ has the effect of a gauge transformation on them. The request that the tensor structures are invariant by shifting the polarization vector by the momentum will ensure that \eqref{TCovariance} is satisfied.

In the following two paragraphs, we explicitly define the interpolators in the vector and spinor formalisms in an arbitrary reference frame: they are easily obtained by applying \eqref{XiToGenFrame} to the representations we just discussed.

\paragraph{Vector formalism}
The interpolator which has the Lorentz index in the vector representation is denoted by
\begin{equation}
	\label{eq:polarizations_interpolator}
	\Xi_{\lambda,A}(p) \to \epsilon_{\lambda,\mu}(p).
\end{equation}
It is usually called polarization in the literature. Here $\lambda$ is the helicity and $\mu=0,1,2,3$ is the Lorentz index. As explained, representation theory allows for $\lambda=\pm 1$, and the tensor structures must be identified under the following equivalence relation:
\begin{equation}
	\label{eq:gauge_transformation}
	\epsilon_{\lambda,\mu}(p)\sim \epsilon_{\lambda,\mu}(p) + c_\lambda(p)\, p_\mu\,,
\end{equation}
with $c_\lambda(p)$ an arbitrary function.
The requirement \eqref{eq:gauge_transformation} can be solved by using the following building blocks
\begin{equation}
	\label{eq:H_object}
	H_{\lambda,\alpha\beta} \equiv p_{\alpha}\, \epsilon_{\lambda,\beta}(p) -  p_{\beta}\, \epsilon_{\lambda,\alpha}(p)
\end{equation}
instead of the polarizations \eqref{eq:polarizations_interpolator}. If we want an explicit form for $\epsilon_{\lambda,\mu}$, we can start in the standard frame, where the group theory explained above dictates
\begin{equation}
\epsilon_{\lambda,\mu}(k)=\frac{1}{\sqrt{2}} 
\begin{pmatrix}
		0\\
		1\\
		i \lambda\\
       0
	\end{pmatrix}~, \qquad \lambda=\pm 1~.
\end{equation}
The polarization vector in a general frame, which is obtained via \eqref{XiToGenFrame}, is easily written in terms of the components of the 4-momentum $p^\mu$ in spherical coordinates:
\begin{equation}\label{eq:momenta_sphericalCoord}
	p^\mu = \{p^0, \vec p\,\},\qquad
	\vec p = \{\myP\cos\phi\sin\theta,\, \myP\sin \phi\sin \theta,\, \myP\cos \theta\},\qquad
	\myP\equiv |\vec p\,|~.
\end{equation}
We obtain
\begin{equation}
	\label{eq:polarizations}
	\epsilon_{\lambda,\mu}(p) =
	\frac{e^{i \lambda \phi } }{\sqrt{2}}
	\begin{pmatrix}
		0\\
		\cos \theta \cos \phi -i \lambda \sin \phi\\
		\cos \theta \sin\phi  +i \lambda \cos\phi\\
		-\sin \theta 
	\end{pmatrix}~.
\end{equation}
Notice that \eqref{eq:polarizations} obeys
%This requirement can be obeyed by imposing 
\begin{equation}
	\label{eq:transversality}
	p^\mu \epsilon_{\lambda,\mu}(p) =0~,
\end{equation}
which is compatible with the equivalence relation \eqref{eq:gauge_transformation}.

\paragraph{Spinor formalism}
Now, consider an interpolator with its Lorentz index in the spinor representation of \eqref{eq:Lorentz_algebra}. There are two such representations, $(1/2,0)$ and $(0,1/2)$. As a result we will have two different interpolators, we denote them by\footnote{Notice that in the literature, the spinor-helicity variables are usually called $\lambda_\al$ and $\tilde \lambda_{\aldot}$.}
\begin{equation}
\Xi_{\lambda,A} \to \qquad	\xi_{\lambda,\alpha}\quad\text{and}\quad \tilde \xi_{\lambda,\dot\alpha}.
\end{equation}
The indices $\alpha$ and $\dot{\alpha}$ take the values 0 or 1. As described above, the two representations allow for $\lambda=\pm 1/2$ and specifically we shall see that $\lambda=+1/2$ for $\xi_{\lambda,\alpha}$ and $\lambda=-1/2$ for $\tilde\xi_{\lambda,\dot\alpha}$.
We will thus simply write
\begin{equation}
	\label{eq:spinors_xi}
	\xi_{\alpha}\qquad\text{and}\qquad \tilde \xi_{\dot\alpha}.
\end{equation}
In the standard frame, if we use the conventions of appendix A of \cite{Wess:1992cp},\footnote{For generic spinors $\psi_\alpha$ and $\tilde{\psi}_{\dot{\alpha}}$, the generators are
\begin{align}
&(1/2,0) &&M^L_i:\ \delta\psi_\alpha = \frac{i}{2} \sigma^i_{\alpha\beta}\psi_\beta~, &&M^R_i:\ \delta\psi_\alpha=0~, \\
& (0,1/2) && M^L_i:\ \delta\tilde{\psi}_{\dot{\alpha}}=0~,  && M^R_i:\ \delta\tilde{\psi}_{\dot{\alpha}} = -\frac{i}{2} \sigma^i_{\dot{\beta}\dot{\alpha}}\tilde{\psi}_{\dot{\beta}}~. \label{rightSpinorRule}
\end{align}
Here, $\sigma^i$ are the Pauli matrices and repeated indices are summed. Comparing this transformation law with \eqref{ABtoM} - \eqref{eq:Little_group_spin}, one finds \eqref{StandSpinorHelicity}. Notice in particular that $\tilde{\xi}(k)$ has negative eigenvalue under $M_3^R$.} we get
\begin{equation}
\xi_\al(k)=\sqrt{2E}
\begin{pmatrix}
1 \\ 0
\end{pmatrix}~, \qquad 
\tilde{\xi}_{\aldot}(k)= \sqrt{2E}
\begin{pmatrix}
1 \\ 0
\end{pmatrix}~.
\label{StandSpinorHelicity}
\end{equation} 
 The normalization is chosen so that their product obeys the identity
\begin{equation}
k^\mu = -\frac{1}{2} \bar{\sigma}^{\mu\, \aldot \al} \xi_\al(k)\,\tilde{\xi}_{\aldot}(k)  ~, \qquad
\bar{\sigma}^{\mu} = (-1_{2\times 2},-\sigma^i)~.
\label{pFromLambda}
\end{equation}
This equation explicitly confirms that the tensor product of the spinor helicities gives the singlet under the little group.
The usual procedure leads to the following expression for a generic frame:
\begin{equation}\label{eq:spinHel_GeneralMomentum}
	\xi_{\alpha}(p)= \sqrt{2E} \begin{pmatrix}\cos{\frac{\theta}{2}}\\ \sin{\frac{\theta}{2}}e^{i\phi} \end{pmatrix}, \quad
	\tilde \xi_{\dot\alpha}(p)= \sqrt{2E}\begin{pmatrix}\cos{\frac{\theta}{2}}\\ \sin{\frac{\theta}{2}}e^{-i\phi} \end{pmatrix}\,.
\end{equation}
Equation \eqref{pFromLambda} is still obeyed:
\begin{equation}
	p^\mu = -\frac{1}{2} \bar{\sigma}^{\mu\, \aldot \al} \sh_\al(p)\tsh_{\aldot}(p)~.
\end{equation}
As wished, under Lorentz transformations we have
\begin{equation}\label{eq:spinor_transform}
	\xi_{\alpha} \rightarrow t \,\sh_{\alpha} \qquad\text{and}\qquad \tsh_{\dot\alpha} \rightarrow t^{-1} \tilde \xi_{\dot\alpha}.
\end{equation}

Let us conclude by relating the spinor connectors \eqref{eq:spinors_xi} and vector ones \eqref{eq:polarizations}. Let us first rewrite the latter in the following way 
\begin{equation}
	\epsilon_{\lambda, \alpha\dot\alpha} (p) = \epsilon_{\lambda,\mu} (p) \sigma^\mu_{\alpha\dot\alpha}~, \qquad \sigma^{\mu} = (-1_{2\times 2},\sigma^i)~.
\end{equation}
In the standard frame, the relation is fixed by choosing spinors whose tensor product has $\lambda = \pm 1$:
\begin{equation}
	\epsilon_{+, \al \aldot}(k) = \sh_\al(k)\, \tilde{\eta}_{\aldot}, \qquad
	\epsilon_{-, \al \aldot}(k) = \eta_\al\, \tsh_{\aldot}(k)~,
	\label{StandpolSpinor}
\end{equation}
where the auxiliary spinors $\eta$ and $\tilde{\eta}$ are both proportional to $(0,1)$. If we instead pick generic auxiliary spinors, the polarization vectors will differ from \eqref{StandpolSpinor} by a gauge transformation \eqref{eq:gauge_transformation}, in accordance with the features of the representation they belong to. Hence, one can promote \eqref{StandpolSpinor} to a generic frame, and the choice of $\eta$ and $\tilde{\eta}$ is irrelevant in any gauge invariant expression.

\subsection{Tensor structures basis in vector formalism}\label{sec:tensor_struct_vector_main}

In this section we construct a linearly independent basis of tensor structures $\mathbf{T}_I$ in vector formalism for scattering amplitudes of identical spin one massless particles. Recall that tensor structures were defined in \eqref{eq:decomposition_general} for the in-out amplitudes and in \eqref{eq:decomposition_crossing_def} for the all-in amplitudes. In practice it will be easier to work in this appendix with all-in amplitudes. Once the basis of tensor structures for the all-in amplitudes is constructed, the basis for in-out amplitudes simply follows due to the relation \eqref{eq:connection}. The discussion of this section closely follows the logic of \cite{Chowdhury:2019kaq}.

This appendix is organized as follows. In appendix \ref{app:basis_d>4} we will construct a basis of tensor structures in $d>4$ space-time dimensions. In appendix \ref{app:basis_d=4} we will construct a basis of tensor structures in the special case of $d=4$ dimensions. Given the latter basis, we will also compute the matrix $M$ defined in \eqref{eq:relation_amplitudes} in $d=4$. In appendix \ref{app:crossing_d>4}, we will solve the crossing equations explicitly in $d>4$. In appendix \ref{app:crossing_d=4}, we will do the same for the special case of $d=4$.

\subsubsection{Basis of tensor structures in $d>4$}
\label{app:basis_d>4}
The only ingredients we have at our disposal to construct the tensor structures $\mathbf{T}_I$ defined in \eqref{eq:decomposition_crossing_def} are the 4-momenta $p^\alpha$ and the invariant blocks $H_{\lambda,\alpha\beta}$ defined in \eqref{eq:H_object}.\footnote{We remind the reader that we denote helicities using $\lambda$, whereas we denote Lorentz indices with $\alpha, \beta$. }

Parity even tensor structures are then constructed as all possible Lorentz invariant products between the 4-momenta of particles $p^\alpha$ and the basic building blocks $H_{\lambda}^{\alpha\beta}$, schematically
\begin{equation}
\label{eq:structures_schematic}
(p_1)^{n_1} (p_2)^{n_2}(p_3)^{n_3} (p_4)^{n_4} H_{\lambda_1} H_{\lambda_2} H_{\lambda_3} H_{\lambda_4}\biggr\rvert_{\mathbb{Z}_2\times\mathbb{Z}_2 \text{ symmetrized}}^{\text{Lorentz contracted}}.
\end{equation}
We also require ${\mathbb{Z}_2\times\mathbb{Z}_2}$ symmetrization according to the discussion at the end of appendix \ref{app:all-in_amplitudes}. Parity odd tensor structures would also contain a single Levi-Cevita symbol $\varepsilon^{\alpha_1\dots\alpha_d}$. We do not discuss such structures in this paper and instead focus only on parity even ones.
Let us denote the total power of the 4-momenta in \eqref{eq:structures_schematic} by
\begin{equation}
	n\equiv n_1+n_2+n_3+n_4.
\end{equation}
In order to be able to fully contract all the Lorentz indices in \eqref{eq:structures_schematic} $n$ must be even. We perform all possible Lorentz contractions in \eqref{eq:structures_schematic} using Mathematica. At the $n=0$ level there are 6 linearly independent tensor structures. At the $n=2$ level there is only one additional linearly independent tensor structure.\footnote{Notice that when checking linear dependences one is allowed to multiply tensor structures with any product of the Mandelstam variables. Such relation were called linear up to \textit{descendants} in \cite{Chowdhury:2019kaq}.}
%from the 6 ones obtained at $n=0$ level. 
At the $n\geq 4$ there are no more linearly independent tensor structures apart from the ones already obtained at the $n=0$ and $n=2$ levels. Thus, we arrive at the conclusion that in $d>4$ there are 7 linearly independent tensor structures. We will see in appendix \ref{app:basis_d=4} that in $d=4$, only 5 structures are independent.

Our choice for the basis of 7 tensor structures reads as
\begin{align}
	\nn
	\mathbf{T}_1{\,}_{\lambda_1, \lambda_2,\lambda_3,\lambda_4}(p_1,p_2,p_3,p_4) &\equiv E_1^{(1)} \equiv \tr(H_{\lambda_1} H_{\lambda_2}) \tr(H_{\lambda_3} H_{\lambda_4}),\\
	\nn
	\mathbf{T}_2{\,}_{\lambda_1, \lambda_2,\lambda_3,\lambda_4}(p_1,p_2,p_3,p_4)&\equiv  E_1^{(2)}\equiv \tr(H_{\lambda_1} H_{\lambda_3}) \tr(H_{\lambda_2} H_{\lambda_4}),\\
	\nn
	\mathbf{T}_3{\,}_{\lambda_1, \lambda_2,\lambda_3,\lambda_4}(p_1,p_2,p_3,p_4) &\equiv  E_1^{(3)}\equiv  \tr(H_{\lambda_1} H_{\lambda_4}) \tr(H_{\lambda_2} H_{\lambda_3}),\\
	\label{eq:basis_general_d}
	\mathbf{T}_4{\,}_{\lambda_1, \lambda_2,\lambda_3,\lambda_4}(p_1,p_2,p_3,p_4) &\equiv  E_2^{(1)}\equiv \tr(H_{\lambda_1} H_{\lambda_3} H_{\lambda_2} H_{\lambda_4}),\\
	\nn
\mathbf{T}_5{\,}_{\lambda_1, \lambda_2,\lambda_3,\lambda_4}(p_1,p_2,p_3,p_4) &\equiv  E_2^{(2)}\equiv \tr(H_{\lambda_1} H_{\lambda_2} H_{\lambda_3} H_{\lambda_4}),\\
	\nn
	\mathbf{T}_6{\,}_{\lambda_1, \lambda_2,\lambda_3,\lambda_4}(p_1,p_2,p_3,p_4) &\equiv  E_2^{(3)}\equiv  \tr(H_{\lambda_1} H_{\lambda_2} H_{\lambda_4} H_{\lambda_3}),\\
	\nn
	\mathbf{T}_7{\,}_{\lambda_1, \lambda_2,\lambda_3,\lambda_4}(p_1,p_2,p_3,p_4) &\equiv  E_3\;\;\equiv p_2  H_{\lambda_1}  p_3\text{tr}\left(H_{\lambda_2} H_{\lambda_3} H_{\lambda_4}\right) -p_1  H_{\lambda_2} p_4\text{tr}\left(H_{\lambda_1} H_{\lambda_3} H_{\lambda_4}\right)\\
	\nn
	&  +\, p_2  H_{\lambda_4}  p_3\text{tr}\left(H_{\lambda_1} H_{\lambda_2} H_{\lambda_3}\right)-p_1 H_{\lambda_3}  p_4\text{tr}\left(H_{\lambda_1} H_{\lambda_2} H_{\lambda_4}\right).
\end{align}
We emphasize that these structures are manifestly $\mathbb{Z}_2\times\mathbb{Z}_2$ invariant. For later convenience we have also denoted our 7 tensor structures by $E_i$.
In \eqref{eq:basis_general_d} we used the following short-hand notation for contractions of Lorentz indices:
\begin{equation}
	\begin{aligned}
		\tr(H_{\lambda_1} H_{\lambda_2}) &\equiv H_{\lambda_1,\alpha\beta}H_{\lambda_2}^{\beta\alpha},\\
		\tr(H_{\lambda_1} H_{\lambda_2} H_{\lambda_3} H_{\lambda_4}) &\equiv H_{\lambda_1,\alpha\beta}H_{\lambda_2}^{\beta\gamma}
		H_{\lambda_3,\gamma\rho}H_{\lambda_4}^{\rho\alpha},\\
		\left(p_2 H_{\lambda_1} p_3\right) &\equiv p_{2,\alpha} H_{\lambda_1}^{\alpha\beta} p_{3,\beta}.
	\end{aligned}
\end{equation}

In \cite{Costa:2011mg} it was noticed that the number of independent scattering amplitudes (or equivalently the number of linearly independent tensor structures) in $d$ dimensions is equal to the number of tensor structures describing the four-point functions of local operators with the same spin in conformal field theory in $d-1$ dimensions. This was proven in \cite{Kravchuk:2016qvl} by arguing that even though the details of both constructions are different, the Little group analysis in both situations is the same and thus the counting should match. We confirm this equivalence for our particular example of scattering of spin one massless particles, see table 1 of \cite{Dymarsky:2013wla} with the summary of the conformal field theory results.\footnote{In the particular case of $d=5$ scattering amplitudes,  the associated CFT counting is also summarized in table 1 of \cite{CastedoEcheverri:2015mkz}. There it was also shown that no parity odd structures are allowed in this particular dimension.} We notice the special feature that in $d\geq 5$ independently of the number of dimensions there are always 7 structures. Looking at the CFT results, we foresee that in $d=4$ there are two more linear relations among the structures \eqref{eq:basis_general_d}, such that we get only 5 linearly independent tensor structures. We address the special case of $d=4$ in appendix \ref{app:basis_d=4}. For more examples of this correspondence see subsection 2.4 in \cite{Hebbar:2020ukp}.

\subsubsection{Basis of tensor structures in $d=4$}
\label{app:basis_d=4}
In the special case of $d=4$, one can verify the existence of two relations among the 7 tensor structures in \eqref{eq:basis_general_d}. They read
\begin{align}
	0&=	s^2 E_2^{(2)} - t^2 E_2^{(1)}-\;  \frac{s^2-t^2}{4}\left(E_1^{(1)}+E_1^{(2)}+E_1^{(3)}\right), \label{eq:relstructst} \\
	0&=	s^2 E_2^{(3)} - u^2 E_2^{(1)}- \frac{s^2-u^2}{4}\left(E_1^{(1)}+ E_1^{(2)}+E_1^{(3)}\right).\label{eq:relstructsu}
\end{align}
These relations should be used to eliminate two structures from the basis \eqref{eq:basis_general_d}. There are several ways to do this. For example, we can decide to eliminate $E_2^{(2)}$ and $E_2^{(3)}$. As a result, the basis in $d=4$ consists of the following 5 structures
\begin{equation}
	\{E_1^{(1)} ,E_1^{(2)} ,E_1^{(3)} , E_2^{(1)}, E_3 \}.
\end{equation}
Thus, our choice for the basis of tensor structures in $d=4$ reads as
\begin{align}
	\nn
	\mathbf{T}_1{\,}_{\lambda_1, \lambda_2,\lambda_3,\lambda_4}(p_1,p_2,p_3,p_4) &\equiv \tr(H_{\lambda_1} H_{\lambda_2}) \tr(H_{\lambda_3} H_{\lambda_4})\\
	\nn
	\mathbf{T}_2{\,}_{\lambda_1, \lambda_2,\lambda_3,\lambda_4}(p_1,p_2,p_3,p_4)&\equiv  \tr(H_{\lambda_1} H_{\lambda_3}) \tr(H_{\lambda_2} H_{\lambda_4})\\
	\nn
	\mathbf{T}_3{\,}_{\lambda_1, \lambda_2,\lambda_3,\lambda_4}(p_1,p_2,p_3,p_4) &\equiv  \tr(H_{\lambda_1} H_{\lambda_4}) \tr(H_{\lambda_2} H_{\lambda_3})\\
	\label{eq:basis_d=4_all-in}
	\mathbf{T}_4{\,}_{\lambda_1, \lambda_2,\lambda_3,\lambda_4}(p_1,p_2,p_3,p_4) &\equiv  \tr(H_{\lambda_1} H_{\lambda_3} H_{\lambda_2} H_{\lambda_4})\\
	\nn
	\mathbf{T}_5{\,}_{\lambda_1, \lambda_2,\lambda_3,\lambda_4}(p_1,p_2,p_3,p_4) &\equiv  p_2  H_{\lambda_1}  p_3\text{tr}\left(H_{\lambda_2} H_{\lambda_3} H_{\lambda_4}\right) -p_1  H_{\lambda_2} p_4\text{tr}\left(H_{\lambda_1} H_{\lambda_3} H_{\lambda_4}\right)\\
	\nn
	&  +\, p_2  H_{\lambda_4}  p_3\text{tr}\left(H_{\lambda_1} H_{\lambda_2} H_{\lambda_3}\right)-p_1 H_{\lambda_3}  p_4\text{tr}\left(H_{\lambda_1} H_{\lambda_2} H_{\lambda_4}\right).
\end{align}

\paragraph{Basis for in-out amplitudes}
So far we have only discussed tensor structures for the all-in amplitudes. Let us now explain how to obtained the basis for in-out amplitudes in $d=4$ from \eqref{eq:basis_d=4_all-in}. 

We start by noticing that the following relations hold
\begin{equation}
	\label{eq:property_H_ac}
	\epsilon_{\lambda}^{\alpha}(-p) = \epsilon_{\lambda}^{\alpha}(p),\qquad
	H_{\lambda_i}^{\alpha\beta}(-p) = 
	-H_{\lambda_i}^{\alpha\beta}(p).
\end{equation}
They follow from the definitions \eqref{eq:polarizations} and \eqref{eq:H_object}. It is also convenient to define
\begin{equation}\label{eq:def_Hstar}
	H^{\lambda_i,\alpha\beta}(p) \equiv \left(H_{\lambda_i}^{\alpha\beta}(p)\right)^* = H_{-\lambda_i}^{\alpha\beta}(p).
\end{equation}
Here the second equality holds due to \eqref{eq:polarizations}. Using the relation between the all-in and in-out amplitudes \eqref{eq:connection}, and taking into account \eqref{eq:property_H_ac} and \eqref{eq:def_Hstar}, we obtain the basis of structures for the in-out amplitudes from \eqref{eq:basis_d=4_all-in}. It reads
\begin{equation}
	\label{eq:basis}
	\begin{aligned}
		\mathbf{T}_1{}_{\lambda_1, \lambda_2}^{\lambda_3,\lambda_4} &= \tr(H_{\lambda_1} H_{\lambda_2}) \tr(H^{\lambda_3} H^{\lambda_4}),\\
		\mathbf{T}_2{}_{\lambda_1, \lambda_2}^{\lambda_3,\lambda_4} &= \tr(H_{\lambda_1} H^{\lambda_3}) \tr(H_{\lambda_2} H^{\lambda_4}),\\
		\mathbf{T}_3{}_{\lambda_1, \lambda_2}^{\lambda_3,\lambda_4} &= \tr(H_{\lambda_1} H^{\lambda_4}) \tr(H_{\lambda_2} H^{\lambda_3}),\\
		\mathbf{T}_4{}_{\lambda_1, \lambda_2}^{\lambda_3,\lambda_4} &= \tr(H_{\lambda_1} H^{\lambda_3} H_{\lambda_2} H^{\lambda_4}),\\
		\mathbf{T}_5{}_{\lambda_1, \lambda_2}^{\lambda_3,\lambda_4} &=
		- \left(p_2 H_{\lambda_1} p_3\right)\;\,\text{tr}\left(H_{\lambda_2} H^{\lambda_3} H^{\lambda_4}\right) + \left(p_1 H_{\lambda_2} p_4\right)\;\,\text{tr}\left(H_{\lambda_1} H^{\lambda_3} H^{\lambda_4}\right)\\
		&\;\;\;\;-\left(p_2 H^{\lambda_4} p_3\right)\text{tr}\left(H_{\lambda_1} H_{\lambda_2} H^{\lambda_3}\right)+\left(p_1 H^{\lambda_3} p_4\right)\text{tr}\left(H_{\lambda_1} H_{\lambda_2} H^{\lambda_4}\right).
	\end{aligned}
\end{equation}

\paragraph{Center of mass frame}
The scattering of identical spin one massless particles can be either described by the five center of mass amplitudes $\Phi_I$ or the five amplitude components $\mathcal{A}_I$. The relation between the two was given in \eqref{eq:tensor_structures_COM} and \eqref{eq:relation_amplitudes}. Plugging the explicit basis \eqref{eq:basis} into \eqref{eq:tensor_structures_COM} we conclude that the matrix $M$ defined in \eqref{eq:relation_amplitudes} has the following explicit form
\begin{equation}
	\label{eq:matrix_M}
	M(s,t,u) =
	\begin{pmatrix}
		s^2 & 0   & 0   &\frac{s^2}{4} & 0\\
		s^2 & t^2 & u^2 &-\frac{tu}{2} & stu\\
		0   & 0   & u^2 &\frac{u^2}{4} & 0\\
		0   & t^2 & 0   &\frac{t^2}{4} & 0\\
		0   & 0   & 0   &0             & \frac{stu}{4}
	\end{pmatrix}.
\end{equation}

\paragraph{Crossing equations for in-out amplitudes}
Let us now inspect the basis of tensor structures \eqref{eq:basis}. Using the properties \eqref{eq:property_H_ac} and \eqref{eq:def_Hstar} we can straightforwardly write the following relations
\begin{equation}
	\label{eq:relations_structures}
	\begin{aligned}
			\mathbf{T}_I{}_{+\lambda_1, -\lambda_3}^{-\lambda_2, +\lambda_4}(p_1, -p_3, -p_2, p_4) &= \sum_{J=1}^5 \widetilde C^{st}_{IJ}(s,t,u)\, \mathbf{T}_J{}_{\lambda_1, \lambda_2}^{\lambda_3,\lambda_4}(p_1, p_2, p_3, p_4),\\
			\mathbf{T}_I{}_{+\lambda_1, -\lambda_4}^{+\lambda_3, -\lambda_2}(p_1, -p_4, p_3, -p_2) &= \sum_{J=1}^5 \widetilde C^{su}_{IJ}(s,t,u)\, \mathbf{T}_J{}_{\lambda_1, \lambda_2}^{\lambda_3,\lambda_4}(p_1, p_2, p_3, p_4),
		\end{aligned}
\end{equation}
where the matrices $\widetilde C^{st}$ and $\widetilde C^{su}$ read as
\begin{equation}
	\label{eq:crossing_matrices}
	\widetilde C^{st}_{IJ}(s,t,u) =
	\begin{pmatrix}
			0 & 1 & 0 & 0 & 0 \\
			1 & 0 & 0 & 0 & 0\\
			0 & 0 & 1 & 0 & 0 \\
			\frac{s^2-t^2}{4 s^2} & \frac{s^2-t^2}{4 s^2} & \frac{s^2-t^2}{4 s^2} & \frac{t^2}{s^2} & 0 \\
			0 & 0 & 0 & 0 & 1
		\end{pmatrix},\,
	\widetilde C^{su}_{IJ}(s,t,u) =
	\begin{pmatrix}
			0 & 0 & 1 & 0 & 0 \\
			0 & 1 & 0 & 0 & 0\\
			1 & 0 & 0 & 0 & 0 \\
			\frac{s^2-u^2}{4 s^2} & \frac{s^2-u^2}{4 s^2} & \frac{s^2-u^2}{4 s^2} & \frac{u^2}{s^2} & 0 \\
			0 & 0 & 0 & 0 & 1
		\end{pmatrix}.
\end{equation} 
Plugging the decomposition \eqref{eq:decomposition_general} into the crossing equations for the in-out amplitudes \eqref{eq:crossing_general_frame} and using the relations \eqref{eq:relations_structures} we obtain
\begin{equation}
	\label{eq:crossing_T}
	\mathcal{A}_I(s,t,u) = \sum_{J=1}^5 \mathcal{A}_J(t,s,u)\, \widetilde C^{st}_{JI}(s,t,u),\qquad
	\mathcal{A}_I(s,t,u) = \sum_{J=1}^5 \mathcal{A}_J(u,t,s)\, \widetilde C^{su}_{JI}(s,t,u).
\end{equation}
Using \eqref{eq:relation_amplitudes} we can rewrite these equations in the form \eqref{eq:crossing_com}. The following consistency relations must hold then
\begin{equation}
	\label{eq:consistency_relations}
	\begin{aligned}
			\chi_{st}C^{st}  &= M(s,t,u) (\widetilde C^{st}(s,t,u))^T(M(t,s,u))^{-1},\\
			\chi_{su}C^{su} &= M(s,t,u) (\widetilde C^{su}(s,t,u))^T(M(u,t,s))^{-1}.
		\end{aligned}
\end{equation}
Plugging here \eqref{eq:matrix_M}, \eqref{eq:crossing_matrices} together with \eqref{eq:matrices_C}, \eqref{eq:phases} we explicitly verify the validity of the relations \eqref{eq:consistency_relations}. This can be seen as an excellent consistency check.

\subsubsection{Solving crossing equations in $d>4$}
\label{app:crossing_d>4}
Let us focus on $d>4$ space-time dimensions in this subsection and address the remaining crossing $S_3$ permutation symmetry, see \eqref{eq:remaining_crossing_symmetry} and the discussion around it. Inspecting the basis of tensor structures \eqref{eq:basis_general_d} we quickly see that the following triplets of tensor structures
\begin{equation}
	\left(E_1^{(1)},E_1^{(2)},E_1^{(3)}\right)
	\quad\text{and}\quad
	\left(E_2^{(1)},E_2^{(2)},E_2^{(3)}\right)
\end{equation}
independently form three dimensional irreducible representations of $S_3$. Moreover the tensor structure $E_3$
transforms as the singlet of $S_3$ (trivial representation). 

In order to enforce the $S_3$ symmetry on \eqref{eq:decomposition_crossing_def} we need to require that the amplitude components $\mathcal{A}_I$ defined in \eqref{eq:decomposition_crossing_def} transform in the same representation as the associated tensor structures. In other words
\begin{equation}
	\left(\mathcal{A}_1, \mathcal{A}_2, \mathcal{A}_3\right)
	\quad\text{and}\quad
	\left(\mathcal{A}_4, \mathcal{A}_5, \mathcal{A}_6\right)
\end{equation}
must independently form triplet representations of $S_3$ and $\mathcal{A}_7$ must be an $S_3$ singlet. Let us introduce the three functions
\begin{equation}
	f_1(s|t,u),\quad
	f_2(s|t,u),\quad
	f_3(s,t,u),
\end{equation}
which obey the following properties
\begin{equation}
	\begin{aligned}
		f_1(s|t,u)&=f_1(s|u,t),\\
		f_2(s|t,u)&=f_2(s|u,t),\\
		f_3(s,t,u)&=f_3(t,s,u)=f_3(u,t,s).\\
	\end{aligned}
\end{equation}
The required symmetry properties of the amplitude components $\mathcal{A}_I$  explained above are achieved by
\begin{equation}
	\begin{aligned}
		\mathcal{A}_1(s,t,u) &= f_1(s|t,u), \quad
		\mathcal{A}_2(s,t,u) = f_1(t|u,s), \quad
		\mathcal{A}_3(s,t,u) = f_1(u|s,t), \\
		\mathcal{A}_4(s,t,u) &= f_2(s|t,u), \quad
		\mathcal{A}_5(s,t,u) = f_2(t|u,s), \quad
		\mathcal{A}_6(s,t,u) = f_2(u|s,t)
	\end{aligned}
\end{equation}
together with 
\begin{equation}
	\mathcal{A}_7(s,t,u) = f_3(s,t,u).
\end{equation}
As a result we obtain the following decomposition of the all-in scattering amplitude into a basis of tensor structures which is automatically fully crossing invariant
\begin{equation}
	\label{eq:amplitude_ansatzE}
	\begin{split}
		\mathcal{T}_{\lambda_1, \lambda_2,\lambda_3,\lambda_4}(p_1,p_2,p_3,p_4) &= f_1(s|t,u) E_1^{(1)}+ f_1(t|u,s) E_1^{(2)} + f_1(u|s,t)E_1^{(3)}\\
		&+ f_2(s|t,u)E_2^{(1)}+ f_2(t|u,s)E_2^{(2)}+f_2(u|s,t)E_2^{(3)}\\
		&+f_3(s,t,u)E_3.
	\end{split} 
	\raisetag{\baselineskip}
\end{equation}
The helicity indices in the right-hand side are implicit.

\subsubsection{Solving crossing equations in $d=4$}
\label{app:crossing_d=4}
The expression \eqref{eq:amplitude_ansatzE} is automatically crossing invariant in $d\geq4$. If we focus on the specific case of $d=4$ there are two additional relations between the 7 tensor structures given by \eqref{eq:relstructst} and \eqref{eq:relstructsu}. We use them to eliminate the structures $E_2^{(2)}$ and $E_2^{(3)}$ from \eqref{eq:amplitude_ansatzE}. As a result we obtain the following expression
\begin{align}
	\nn
	\mathcal{T}_{\lambda_1, \lambda_2,\lambda_3,\lambda_4}(p_1,p_2,p_3,p_4)&= \left(f_1(s|t,u)+\frac{f_2(u|s,t)+f_2(t|s,u)}{4}-\frac{u^2f_2(u|s,t)+t^2f_2(t|s,u)}{4s^2}\right)E_1^{(1)}\nn \\
	\nn
	&+\left(f_1(t|s,u)+\frac{f_2(u|s,t)+f_2(t|s,u)}{4}-\frac{u^2f_2(u|s,t)+t^2f_2(t|s,u)}{4s^2}\right)E_1^{(2)}\nn \\
	\nn
	&+\left(f_1(u|s,t)+\frac{f_2(u|s,t)+f_2(t|s,u)}{4}-\frac{u^2f_2(u|s,t)+t^2f_2(t|s,u)}{4s^2}\right)E_1^{(3)}\nn \\
	\nn
	&+\left(f_2(s|t,u)+\frac{u^2f_2(u|s,t)+t^2f_2(t|s,u)}{s^2}\right) E_2^{(1)}\nn \\
	\label{eq:decomposition_4d}
	&+f_3(s,t,u)E_3.
\end{align}
The helicity indices in the right-hand side are implicit.
Comparing with \eqref{eq:decomposition_crossing_def} and taking into account \eqref{eq:basis_d=4_all-in} we conclude that
\begin{equation}
	\label{eq:solution_crossing}
	\begin{aligned}
		\mathcal{A}_1(s,t,u) &=
		f_1(s|t,u)+\frac{f_2(u|s,t)+f_2(t|s,u)}{4}-\frac{u^2f_2(u|s,t)+t^2f_2(t|s,u)}{4s^2},\\
		\mathcal{A}_2(s,t,u) &= f_1(t|s,u)+\frac{f_2(u|s,t)+f_2(t|s,u)}{4}-\frac{u^2f_2(u|s,t)+t^2f_2(t|s,u)}{4s^2},\\
		\mathcal{A}_3(s,t,u) &= f_1(u|s,t)+\frac{f_2(u|s,t)+f_2(t|s,u)}{4}-\frac{u^2f_2(u|s,t)+t^2f_2(t|s,u)}{4s^2},\\
		\mathcal{A}_4(s,t,u) &= f_2(s|t,u)+\frac{u^2f_2(u|s,t)+t^2f_2(t|s,u)}{s^2},\\
		\mathcal{A}_5(s,t,u) &= f_3(s,t,u).
	\end{aligned}
\end{equation}
The solution for the amplitude components \eqref{eq:solution_crossing} was derived in the case of the all-in amplitudes. It holds however also in the case of in-out amplitudes if we change the meaning of the Mandelstam variables from \eqref{eq:mandelstam_variables_all_in} to \eqref{eq:mandelstam_variables}. Then plugging \eqref{eq:solution_crossing} and \eqref{eq:basis} into \eqref{eq:decomposition_general} we can check that the in-out crossing equations \eqref{eq:crossing_general_frame} are automatically satisfied. This is a non-trivial consistency check.

If we insist on keeping the manifestly crossing invariant decomposition of the all-in amplitude \eqref{eq:decomposition_4d} in $d=4$ there is another price to pay. The functions $f_i$ contain some redundancies. This implies the existence of an equivalence class of functions $f_i$ that generates the same amplitude components \eqref{eq:solution_crossing}. In what follows we show that this equivalence class has the following most generic form:\footnote{In \cite{Chowdhury:2019kaq}, this equivalence relation was already investigated. We notice that, probably due to a typo, the result reported in the paper is incorrect.}
\begin{equation}\label{eq:equivalence_f12}
	\begin{split}
		f_1(s|t,u) &\sim f_1(s|t,u) -\frac{1}{4}\left\{ g'(s|t,u)+ g'(t|s,u)+ g'(u|s,t)\right\},\\
		f_2(s|t,u) &\sim f_2(s|t,u) + g'(s|t,u),
	\end{split}
\end{equation}
where 
\begin{equation}\label{eq:gprimewithga}
		 g'(s|t,u) \equiv t^2 g_A(u;s,t) + u^2g_A(t;s,u),
\end{equation}
and $g_A(s;t,u)$ is some function antisymmetric in the last two variables.

To begin with, notice, that the relations \eqref{eq:relstructst} and \eqref{eq:relstructsu} already imply a redundancy of the function describing the same amplitude
\begin{equation}\label{eq:equivalent_class_fct_def}
	\begin{split}
		f_1(s|t,u)&\sim f_1(s|t,u) + \hat f_1(s|t,u),\\
		f_2(s|t,u)&\sim f_2(s|t,u) + \hat f_2(s|t,u),
	\end{split}
\end{equation}
if $\hat f_1$ and $\hat f_2$ obey the following relation
\begin{equation}
	\label{eq:vanishingamplitude}
	\begin{aligned}
		0&= \hat f_1(s|t,u) E_1^{(1)} + \hat f_1(t|u,s) E_1^{(2)} + \hat f_1(u|s,t)E_1^{(3)}\\
		&+ \hat f_2(s|t,u)E_2^{(1)}+ \hat f_2(t|u,s)E_2^{(2)}+\hat f_2(u|s,t)E_2^{(3)}.
	\end{aligned}
\end{equation}
Let us now derive the most generic form of $\hat f_1$ and $\hat f_2$ which obey \eqref{eq:vanishingamplitude}. 

 In order to do it systematically we take a  $t-u$ symmetric combination of the relations  \eqref{eq:relstructst}  and \eqref{eq:relstructsu}, more precisely
\begin{equation}
	\begin{split}
		0={}& k(t,u) \left[s^2 E_2^{(2)} - t^2 E_2^{(1)}-  \frac{s^2-t^2}{4}\left(E_1^{(1)}+E_1^{(2)}+E_1^{(3)}\right)\right]\\
		+{\ }& k(u,t)\left[s^2 E_2^{(3)} - u^2 E_2^{(1)}- \frac{s^2-u^2}{4}\left(E_1^{(1)}+ E_1^{(2)}+E_1^{(3)}\right)\right],
	\end{split}
\end{equation}
where $k(x,y)$ is a function with no particular symmetry. This expression, together with those obtained by permutation, form a triplet of $S_3$. A crossing symmetric solution (i.e. singlet) can therefore be constructed by taking the ``inner product'' with symmetric function $h(x,y)=h(y,x)$ (as they also transform in a triplet of $S_3$) and we obtain:
	\begin{align}
		\nn
		0&=\frac{1}{4} \left(E_1^{(1)}+ E_1^{(2)}+ E_1^{(3)} \right)\Big\{s^2 \left[h(s,t) k(s,t)+h(s,u) k(s,u)-h(t,u) k(t,u)-h(t,u) k(u,t)\right]\\
		\nn
		&\qquad \qquad \qquad \qquad \qquad  \quad +t^2 \left[h(s,t)k(t,s)-h(s,u) k(s,u)-h(s,u) k(u,s)+h(t,u) k(t,u)\right]\\
		\nn
		&\qquad \qquad \qquad \qquad \qquad  \quad  +u^2 \left[-h(s,t) k(s,t)-h(s,t)
		k(t,s)+h(s,u) k(u,s)+h(t,u) k(u,t)\right]\Big\}\\
		\label{eq:amprelationgeneralhk}
		&+E_2^{(1)}\left\{t^2 [h(s,u) k(s,u)-h(t,u) k(t,u)]+u^2 [h(s,t) k(s,t)-h(t,u) k(u,t)]\right\}\\
		\nn
		&+E_2^{(2)}\left\{s^2 [h(t,u) k(t,u)-h(s,u) k(s,u)]+u^2 [h(s,t) k(t,s)-h(s,u) k(u,s)]\right\}\\
		\nn
		&+E_2^{(3)}\left\{s^2 [h(t,u) k(u,t)-h(s,t) k(s,t))+t^2 [h(s,u) k(u,s)-h(s,t) k(t,s)]\right\}.
	\end{align}

The expression \eqref{eq:amprelationgeneralhk} can be simplified by defining an antisymmetric function $g_A(s;t,u)$ in the last two variables as
\begin{equation}
	g_A(s;t,u)= h(t,s)k(t,s)-h(u,s)k(u,s)\,.
\end{equation} 
With the help of this function the relation \eqref{eq:amprelationgeneralhk} takes the form:
\begin{equation}\label{eq:amprelationgeneral}
	\begin{split}
		0&= \left(E_1^{(1)}+ E_1^{(2)}+ E_1^{(3)} \right) \frac{1}{4} \Big\{s^2 [g_A(t;s,u)+g_A(u;s,t)]+t^2 [g_A(s;t,u) +g_A(u;t,s)]\\ 
		&\qquad \qquad \qquad\qquad\qquad\qquad+u^2[g_A(s;u,t)+g_A(t;u,s)]\Big\}\\
		&+E_2^{(1)}\left\{t^2 g_A(u;s,t)+u^2g_A(t;s,u)\right\}
		+E_2^{(2)}\left\{s^2 g_A(u;t,s)+u^2 g_A(s;t,u)\right\}\\
		&+E_2^{(3)}\left\{s^2 g_A(t;u,s)+t^2 g_A(s;u,t)\right\}\,.
	\end{split}
\end{equation}
We can further simplify this expression by using another function $g'$ defined in \eqref{eq:gprimewithga}.

Plugging \eqref{eq:gprimewithga} into \eqref{eq:amprelationgeneral} we finally get
\begin{equation}\label{eq:amprelationgeneralgprime}
	\begin{split}
		0=  &-\frac{1}{4} \left\{g'(s|t,u)+g'(t|s,u)+g'(u|s,t) \right\}\left(E_1^{(1)}+ E_1^{(2)}+ E_1^{(3)} \right)\\
		&+g'(s|t,u)E_2^{(1)}+g'(t|u,s)E_2^{(2)}+g'(u|s,t)E_2^{(3)} \ .
	\end{split}
\end{equation}
Making the following identification
\begin{equation}\label{eq:f12hat}
	\begin{split}
		\hat f_1(s|t,u) &= -\frac{1}{4}\left\{ g'(s|t,u)+ g'(t|s,u)+ g'(u|s,t)\right\},\\
		\hat f_2(s|t,u) &=  g'(s|t,u),
	\end{split}
\end{equation}
we obtain the earlier announced answer \eqref{eq:equivalence_f12} with \eqref{eq:gprimewithga}.

\subsection{Tensor structures basis in spinor formalism in $d=4$}\label{sec:tensor_struct_spinHel_main}
In this section we will show how to write the basis of tensor structures $\mathbf{T}_I$ for the all-in amplitudes defined in \eqref{eq:decomposition_crossing_def} in the case of identical massless  particles using spinor formalism in $d=4$ space-time dimensions. This formalism is commonly referred to as the spinor-helicity formalism, for a review complementary to ours see for example \cite{Cheung:2017pzi}. The spinor formalism can also be used in the case of massive particles, see \cite{Arkani-Hamed:2017jhn} and appendix H in \cite{Hebbar:2020ukp}.

In order to proceed in a coherent way let us make a summary of what we have found using vector formalism first.
In the case of four different massless particles there are at most 16 independent functions which describe the scattering process. This means that in general we can write
\begin{align}
	\label{eq:decomposition_vec}
	\mathcal{T}_{\lambda_1, \lambda_2,\lambda_3,\lambda_4}(p_1,p_2,p_3,p_4)
	= \sum_{I=1}^{16} \mathcal{A}_I(s,t,u) 	\mathbf{T}_I{\,}_{\lambda_1, \lambda_2,\lambda_3,\lambda_4}(p_1,p_2,p_3,p_4).
\end{align}
Since in our case all the particles are identical there is an $S_4$ permutation symmetry. As discussed in the end of appendix \ref{app:all-in_amplitudes} we can impose its normal subgroup $\mathbb{Z}_2\times\mathbb{Z}_2$ which leads to 11 relations among the 16 amplitudes components $\mathcal{A}_I$ and we are left with only 5 distinct amplitude components $\mathcal{A}_I$. This was shown in a slightly different but equivalent way in appendix \ref{app:basis_d=4}, where we found that there are only 5 independent tensor structures which are $\mathbb{Z}_2\times\mathbb{Z}_2$ symmetric. Further imposing the rest of $S_4$ constraints we have concluded in appendix \ref{app:crossing_d=4} that there are only 3 independent functions describing the scattering process of identical massless spin one particles. We denoted these 3 functions by $f_i$. The 5 distinct amplitude components $\mathcal{A}_I$ were expressed in terms of the 3 functions $f_i$ in \eqref{eq:solution_crossing}.

In what follows we will repeat all these steps using the spinor formalism. As we will see shortly, in spinor formalism the decomposition \eqref{eq:decomposition_vec} diagonalizes and one can simply write
\begin{align}
	\label{eq:decomposition_spin}
	\mathcal{T}_{\lambda_1, \lambda_2,\lambda_3,\lambda_4}(p_1,p_2,p_3,p_4)
	= h_{(\lambda_1, \lambda_2,\lambda_3,\lambda_4)}(s,t,u) 	\mathbf{T}_{\lambda_1, \lambda_2,\lambda_3,\lambda_4}(p_1,p_2,p_3,p_4).
\end{align}
Here there is no summation over the repeated indices. This relation should be interpreted in the following way: to each helicity configuration $(\lambda_1, \lambda_2,\lambda_3,\lambda_4)$ corresponds a single amplitude component $h_{(\lambda_1, \lambda_2,\lambda_3,\lambda_4)}$. In what follows we will first construct the 16 tensor structures $\mathbf{T}_{\lambda_1, \lambda_2,\lambda_3,\lambda_4}$. Then using $S_4$ permutation symmetry we will write down all the relations between the amplitude components  $h_{(\lambda_1, \lambda_2,\lambda_3,\lambda_4)}$. We will find that there are only three independent amplitude components, we will denote them by $h_i$.

We will conclude this appendix by relating  the functions $h_i$, the functions $f_i$ of vector formalism and the center of mass frame amplitudes $\Phi_I$ of appendix \ref{sec:S-matrix_setup} to each other.

\subsubsection{Tensor structures}
\label{subsec:TensStructSH}
Using the spinors $\sh_\al$ and $\tsh_{\aldot}$ defined in \eqref{eq:spinors_xi} one can construct two Lorentz invariant building blocks, namely
\begin{equation}
	\plub{ij}\equiv (\sh_i)_\al(\sh_j)_\be \epsilon^{\al\be}, \qquad
\minb{ij}\equiv (\tsh_i)_{\aldot}(\tsh_j)_{\bedot} \epsilon^{\aldot\bedot},
	\label{SpinHalfPairings}
\end{equation}
where we used the short-hand notation $ \sh_i \equiv \sh(p_i)$ and $\tsh_i\equiv \tsh(p_i)$.

Let us now recall the Little group tranformation property of scattering amplitudes of massless particles. It reads 
\begin{equation}
	\mathcal{T}_{\la_1,\la_2,\la_3,\la_4}(p_1,p_2,p_3,p_4) \to t_1^{2\la_1}t_2^{2\la_2}t_3^{2\la_3}t_4^{2\la_4}
	\mathcal{T}_{\la_1,\la_2,\la_3,\la_4}(p_1,p_2,p_3,p_4),
	\label{Ttchange}
\end{equation}
where $t_i$ are some real scalar objects. 
Notice that imposing \eqref{Ttchange} assuming all $t_i$'s are independent parameters implies \eqref{TCovariance}. However, one could worry that there are solutions to \eqref{TCovariance} that do not respect  \eqref{Ttchange}.
In practice, this is not relevant in our case because as we will see below we will find independent tensor structures for every choice of helicities.
Recall that the spinors \eqref{eq:spinors_xi} transform as \eqref{eq:spinor_transform}
\begin{equation}
	\xi_i \rightarrow t_i \sh_i \qquad\text{and}\qquad \tsh_i \rightarrow t_i^{-1} \tilde \xi_i ,
	\label{lambdatchange}
\end{equation}
which means that the Lorentz invariant blocks \eqref{SpinHalfPairings} transform as 
\begin{equation}
	\label{eq:transformation_bl}
	\plub{ij} \to t_i t_j \plub{ij}, \qquad
	\minb{ij} \to t_i^{-1} t_j^{-1}\minb{ij}.
\end{equation}

We can construct tensor structures $\mathbf{T}_{\lambda_1, \lambda_2,\lambda_3,\lambda_4}$ in \eqref{eq:decomposition_spin} as products of Lorentz invariant blocks \eqref{SpinHalfPairings}. The exponents in these products are fixed by requiring correct transformation properties of the amplitude \eqref{Ttchange}. In our case of identical spin one massless particles we can write for instance
\begin{equation}
	\label{eq:structures_spinor}
	\begin{aligned}
		\mathbf{T}_{+,+,+,+}(p_1,p_2,p_3,p_4) &= (\plub{12}\plub{34})^2+(\plub{13}\plub{24})^2+(\plub{14}\plub{23})^2,\\
		\mathbf{T}_{+,+,-,-}(p_1,p_2,p_3,p_4) &=(\plub{12}\minb{34})^2,\\
		\mathbf{T}_{+,-,+,-}(p_1,p_2,p_3,p_4) &=(\plub{13}\minb{24})^2,\\
		\mathbf{T}_{+,-,-,+}(p_1,p_2,p_3,p_4) &=(\plub{14}\minb{23})^2,\\
		\mathbf{T}_{+,+,+,-}(p_1,p_2,p_3,p_4) &= (\minb{41}\plub{12}\plub{13})^2,\\
		\mathbf{T}_{+,+,-,+}(p_1,p_2,p_3,p_4) &= (\minb{31}\plub{12}\plub{14})^2,\\
		\mathbf{T}_{+,-,+,+}(p_1,p_2,p_3,p_4) &= (\minb{21}\plub{13}\plub{14})^2,\\
		\mathbf{T}_{-,+,+,+}(p_1,p_2,p_3,p_4) &= (\minb{12}\plub{23}\plub{24})^2.
	\end{aligned}
\end{equation}
Due to the transformation properties \eqref{eq:transformation_bl} square blocks take care of positive helicities and angular blocks take care of negative helicities.

In \eqref{eq:structures_spinor} we have defined tensor structures for 8 helicity configurations. The other 8 configurations which have opposite helicities to \eqref{eq:structures_spinor} are obtained from \eqref{eq:structures_spinor} by exchanging square and angular blocks, namely
\begin{equation}
	\plub{ij}  \leftrightarrow \minb{ij}.
\end{equation}

It is interesting to notice that the following identity exists:
\begin{equation}\label{eq:identity_spinor}
	(\plub{12}\plub{34})^2+(\plub{13}\plub{24})^2+(\plub{14}\plub{23})^2 = (s^2+t^2+u^2)\frac{\plub{12}\plub{34}}{\minb{12}\minb{34}}.
\end{equation}
The tensor structure $\frac{\plub{12}\plub{34}}{\minb{12}\minb{34}}$ was used for example in \cite{Arkani-Hamed:2020blm}. The latter choice might be more familiar to some readers.

\subsubsection{Parity and permutation symmetry $S_4$}\label{sec:parityAndPerm_SpinHel}
We have constructed the basis of 16 tensor structures given by 8 structures written explicitly in \eqref{eq:structures_spinor} together with the 8 related ones (where all the helicities have an opposite sign) which are obtained from \eqref{eq:structures_spinor} as explained at the end of the last subsection. Plugging these 16 structures into \eqref{eq:decomposition_spin} we obtain the final decomposition of the amplitude into tensor structures in spinor formalism. 
The goal of this subsection is to impose the constraints of parity and permutation symmetry $S_4$ on the decomposition  \eqref{eq:decomposition_spin} in order to get relations between the amplitude components $h_{(\lambda_1, \lambda_2,\lambda_3,\lambda_4)}$.

Let us start with parity. According to equation (2.64) in \cite{Hebbar:2020ukp} parity in our case requires
\begin{equation}
	\mathcal{T}{}_{\lambda_1, \lambda_2,\lambda_3,\lambda_4}(p_1,p_2,p_3,p_4) =	\mathcal{T}{}_{-\lambda_1, -\lambda_2,-\lambda_3,-\lambda_4}(p_1,p_2,p_3,p_4).
\end{equation}
Since under parity the following holds
\begin{equation}
	(\xi_i)_\al \xleftrightarrow{\mathcal{P}}(\tilde \xi_i)_{\aldot},\qquad
	\plub{ij}  \xleftrightarrow{\mathcal{P}} \minb{ij},
\end{equation}
we conclude from \eqref{eq:decomposition_spin} that
\begin{equation}\label{eq:parity_consequence_spinor}
	h_{(\lambda_1, \lambda_2,\lambda_3,\lambda_4)}(s,t,u) = h_{(-\lambda_1, -\lambda_2,-\lambda_3,-\lambda_4)}(s,t,u).
\end{equation}
We are thus left only with 8 independent functions $h_{(\lambda_1, \lambda_2,\lambda_3,\lambda_4)}$. 

Let us now impose the $S_4$ permutation symmetry on the 8 amplitudes not related by parity. The amplitude which has all plus helicities is
\begin{equation}
	\mathcal{T}_{+,+,+,+}.
\end{equation}
It is obviously invariant under permutation of any particles. In particular
\begin{align}
	\mathcal{T}_{+,+,+,+}(p_1,p_2,p_3,p_4)=\mathcal{T}_{+,+,+,+}(p_3,p_2,p_1,p_4),\\
	\mathcal{T}_{+,+,+,+}(p_1,p_2,p_3,p_4)=\mathcal{T}_{+,+,+,+}(p_4,p_2,p_3,p_1).
\end{align}
From the explicit form of the $\mathbf{T}_{+,+,+,+}$ tensor structure given in the first line of \eqref{eq:structures_spinor} we see that this tensor structure is also invariant under any permutations of particles, as a result we conclude that the function $h_{(+,+,+,+)}(s,t,u) $ in \eqref{eq:decomposition_spin} must be fully crossing symemtric in its arguments, namely
\begin{equation}\label{eq:constr_1}
	h_{(+,+,+,+)}(s,t,u) = h_{(+,+,+,+)}(t,s,u) =h_{(+,+,+,+)}(u,t,s) .
\end{equation}

Now consider amplitudes with two plus and two minus helicities, they are\footnote{Such amplitudes are often referred to in the literature as ``Maximally Helicity Violating'' (MHV) amplitudes.} 
\begin{equation}
	\label{eq:list_double_minus}
	\mathcal{T}_{+,+,-,-},\qquad
	\mathcal{T}_{+,-,+,-},\qquad
	\mathcal{T}_{+,-,-,+}.
\end{equation}
Permutation symmetry relates them as
\begin{equation}
	\mathcal{T}_{+,+,-,-}(p_1,p_2,p_3,p_4)=
	\mathcal{T}_{+,-,+,-}(p_1,p_3,p_2,p_4)=
	\mathcal{T}_{+,-,-,+}(p_1,p_3,p_4,p_2).
\end{equation}
Moreover, the first amplitude is also symmetric under both $1\leftrightarrow 2$ and $3\leftrightarrow 4$ permutations. (Similar statement can be made about the other two amplitudes in \eqref{eq:list_double_minus}). Investigating how the tensor structures in \eqref{eq:structures_spinor} transform under all these permutations we conclude that the following relations must be satisfied
\begin{equation}
	\label{eq:constr_2}
	\begin{aligned}
		h_{(++--)}(s,t,u) = h_{(++--)}(s,u,t) &= h_{(+-+-)}(t,s,u) =  h_{(+-+-)}(t,u,s)\\
		&=h_{(+--+)}(u,t,s) = h_{(+--+)}(u,s,t).
	\end{aligned}
\end{equation}

Finally, let us consider the amplitudes with only one minus helicity, namely
\begin{equation}
	\label{eq:list_one_minus}
	\mathcal{T}_{+,+,+,-},\qquad
	\mathcal{T}_{+,+,-,+},\qquad
	\mathcal{T}_{+,-,+,+},\qquad
	\mathcal{T}_{-,+,+,+}.
\end{equation}
They are all related by permutation symmetry
\begin{multline}
	\mathcal{T}_{+,+,+,-}(p_1,p_2,p_3,p_4)=
	\mathcal{T}_{+,+,-,+}(p_1,p_2,p_4,p_3)=\\
	\mathcal{T}_{+,-,+,+}(p_1,p_4,p_2,p_3)=
	\mathcal{T}_{-,+,+,+}(p_4,p_1,p_2,p_3).
\end{multline}
Moreover, the first amplitude in \eqref{eq:list_one_minus} is symmetric under any permutation of particles 1, 2 and $3$. (Similar statements hold for the rest of the amplitudes in \eqref{eq:list_one_minus}). Inspecting the permutation properties of the tensor structures in \eqref{eq:structures_spinor} we conclude that the following relations must be satisfied 
\begin{equation}	\label{eq:constr_3a}
h_{(+++-)}(s,t,u)=h_{(+++-)}(t,s,u)=h_{(+++-)}(u,t,s),
\end{equation}
together with
\begin{equation}\label{eq:constr_3b}
	h_{(+++-)}(s,t,u) = h_{(++-+)}(s,t,u) = h_{(+-++)}(s,t,u) = h_{(-+++)}(s,t,u).
\end{equation}
In order to find this result notice, that the following relation holds due to momentum conservation
\begin{equation}\label{key}
	(\minb{41}\plub{12}\plub{13})^2 = (\minb{42}\plub{12}\plub{23})^2 =	(\minb{43}\plub{32}\plub{13})^2.
\end{equation}

Summarizing, out of 16 amplitude components $h_{(\lambda_1, \lambda_2,\lambda_3,\lambda_4)}$ there are only three independent ones. We denote them as
\begin{equation}
	\label{eq:three_functions_h}
	\begin{aligned}
		h_1(s|t,u)  &\equiv h_{(++--)}(s,t,u),\\
		h_2(s,t,u) &\equiv (s^2+t^2+u^2)^{-1} h_{(++++)}(s,t,u),\\
		h_3(s,t,u) &\equiv h_{(+++-)}(s,t,u).
	\end{aligned}
\end{equation}
The rest of the functions $h_{(\lambda_1, \lambda_2,\lambda_3,\lambda_4)}$ can be obtained from these ones by using  \eqref{eq:parity_consequence_spinor} which holds due to parity and the relations \eqref{eq:constr_1}, \eqref{eq:constr_2}, \eqref{eq:constr_3a} and \eqref{eq:constr_3b} which hold due to permutation symmetry $S_4$. Notice, that the functions $h_2$ and $h_3$ are fully crossing symmetric in their arguments, whereas the function $h_1(s|t,u)$ is symmetric only in its last two arguments. The prefactor in the definition of $h_2$ is introduced for later convenience.

\subsubsection{Center of mass frame}
We have worked so far with all-in amplitudes. Let us now use \eqref{eq:connection} in order to obtain in-out amplitudes from the all-in ones. Explicitly, we have
\begin{equation}
	\label{eq:explict_relations_all-in-in-out}
	\begin{aligned}
\mathcal{T}^{++}_{++}(p_1,p_2,p_3,p_4)=\mathcal{T}_{++--}(p_1,p_2,-p_3,-p_4),\\
\mathcal{T}^{--}_{++}(p_1,p_2,p_3,p_4)=\mathcal{T}_{++++}(p_1,p_2,-p_3,-p_4),\\
\mathcal{T}^{+-}_{+-}(p_1,p_2,p_3,p_4)=\mathcal{T}_{+--+}(p_1,p_2,-p_3,-p_4),\\
\mathcal{T}^{-+}_{+-}(p_1,p_2,p_3,p_4)=\mathcal{T}_{+-+-}(p_1,p_2,-p_3,-p_4),\\
\mathcal{T}^{+-}_{++}(p_1,p_2,p_3,p_4)=\mathcal{T}_{++-+}(p_1,p_2,-p_3,-p_4).
	\end{aligned}
\end{equation}
The amplitudes in the right-hand side can still be decomposed into tensor structures \eqref{eq:decomposition_spin} with the basis given by \eqref{eq:structures_spinor}. Evaluating \eqref{eq:explict_relations_all-in-in-out} in the center of mass frame defined in \eqref{eq:COM_frame} and using \eqref{eq:helicity_amplitude_def} together with \eqref{eq:decomposition_spin}, \eqref{eq:structures_spinor} and \eqref{eq:three_functions_h} we conclude that
\begin{equation}
	\label{eq:relation_com_spin}
	\begin{aligned}
		\Phi_1(s,t,u)&= (\plub{12}\minb{34})^{2}h_1(s|t,u)\Big|_\text{COM}, \\
		\Phi_2(s,t,u)&= \left((\plub{12}\plub{34})^2+(\plub{13}\plub{24})^2+(\plub{14}\plub{23})^2\right)  (s^2+t^2+u^2)^{-1} h_2(s,t,u)\Big|_\text{COM}, \\
		\Phi_3(s,t,u)&= (\plub{14}\minb{23})^{2}h_1(u|t,s)\Big|_\text{COM},  \\
		\Phi_4(s,t,u)&= (\plub{13}\minb{24})^{2}h_1(t|s,u)\Big|_\text{COM},  \\
		\Phi_5(s,t,u)&= (\minb{31}\plub{12}\plub{14})^2 h_3(s,t,u)\Big|_\text{COM}.
	\end{aligned}
\end{equation}

In order to evaluate the tensor structures in \eqref{eq:relation_com_spin} in the center of mass frame we first apply the analytic continuation \eqref{eq:pminusp} to \eqref{eq:spinHel_GeneralMomentum}. We then plug in the center of mass frame values for the 4-momenta given in \eqref{eq:COM_frame}.  This leads us to
\begin{align}
	\xi_1^\text{COM} &= s^{1/4} \begin{pmatrix}1\\0 \end{pmatrix},
	\qquad\;\; 
	\xi_2^\text{COM} = s^{1/4} \begin{pmatrix}0\\1 \end{pmatrix}, \\
	\quad \xi_3^\text{COM} &= i s^{1/4} \begin{pmatrix}\cos\frac{\theta}{2}\\ \sin\frac{\theta}{2} \end{pmatrix},
	\quad \xi_4^\text{COM} = i s^{1/4} \begin{pmatrix}\sin\frac{\theta}{2}\\ -\cos\frac{\theta}{2} \end{pmatrix},
	\label{SpinHelCMin}
\end{align}
together with
\begin{equation}
	\tilde{\xi}_{1} ^\text{COM}= \xi_{1}^\text{COM},\qquad
	\tilde{\xi}_{2}^\text{COM} = \xi_{2}^\text{COM},\qquad
	\tilde{\xi}_{3}^\text{COM} = \xi_{3}^\text{COM},\qquad
	\tilde{\xi}_{4}^\text{COM} = \xi_{4}^\text{COM}.
\end{equation}
Where the $i$ in $\sh^{\text{COM}}_{3,4}$ and $\tsh^{\text{COM}}_{3,4}$ comes from the analytic continuation from incoming to outgoing particle.

Plugging these into the definitions \eqref{SpinHalfPairings} we conclude
\begin{equation}
	\label{eq:Phi_h_relation}
	\begin{aligned}
			\Phi_1(s,t,u)&= s^2 h_1(s|t,u), \\
			\Phi_2(s,t,u)&= h_2(s,t,u),\\
			\Phi_3(s,t,u)&= u^2 h_1(u|t,s),\\
			\Phi_4(s,t,u)&= t^2 h_1(t|s,u),\\
			\Phi_5(s,t,u)&= stu\, h_3(s,t,u).
	\end{aligned}
\end{equation}

The center of mass amplitudes $\Phi_I$ are related to the amplitude components $\mathcal{A}_I$ via \eqref{eq:relation_amplitudes} and \eqref{eq:matrix_M}. The amplitude components  $\mathcal{A}_I$ in turn are expressed in terms of the 3 functions $f_i$ according to \eqref{eq:solution_crossing}. Using \eqref{eq:Phi_h_relation} we can then relate $h_i$ and $f_i$ functions. We get
\begin{equation}
	\label{eq:f_h_relation}
	\begin{aligned}
		h_1(s|t,u)&= f_1(s|t,u) + \frac{1}{4}(f_2(s|t,u)+ f_2(t|s,u)+ f_2(u|s,t)),\\
		h_2(s,t,u)&= s^2 f_1(s|t,u) + t^2 f_1(t|s,u)+ u^2f_1(u|s,t)\\
		&- \frac{1}{2}(tuf_2(s|t,u) + su f_2(t|s,u)+ st f_2(u|s,t)) + stu f_3(s,t,u),\\
		h_3(s,t,u)&= \frac{1}{4} f_3(s,t,u).
	\end{aligned}
\end{equation}
Notice that the functions $h_i$ written in terms of functions $f_i$ here are invariant under the equivalence class transformation \eqref{eq:equivalence_f12}. This is a powerful consistency check.

\section{One-loop from elastic unitarity}\label{sec:oneloopUnitarity} 
In this appendix, we will compute the one-loop amplitude at threshold using elastic unitarity.  This computation follows the idea of \cite{EliasMiro:2019kyf,Guerrieri:2020bto,Guerrieri:2021ivu}, the novelty compared to those references is the presence of spin and we will see that the general result can be easily extended.

The (tree-level) amplitudes are schematically given at threshold by \eqref{eq:amplitudes_EFT}
\begin{equation}
	\mathcal{T} = a_2 \bar{s}^2 + a_3 \bar{s}^3 + \cO(\bar{s}^4)\,,
\end{equation}
where we wrote generic dimensionless coefficients $a_i\in\mathbb{R}$ and used the dimensionless Mandelstam variables \eqref{eq:dimensionless_mandelstam}.  From unitarity ($\Im \cT \sim \cT^2 $), we expect the amplitude to acquire an imaginary part at $\cO(\bar{s}^4)$. This discontinuity corresponds to EFT loops and, as we are considering massless particles, it is given at threshold by \emph{logs}. This leads to the ansatz \eqref{eq:amplitudes_EFT}, with the one-loop discontinuity given by \eqref{eq:loops}.

We will now show how to fix the coefficients $\beta_i$ to \eqref{betaEFT} by imposing unitarity at threshold.  To begin, we decompose the ansatz \eqref{eq:amplitudes_EFT} in the partial waves amplitudes $\mathbb{S}^\ell(s)$, defined in \eqref{eq:entriesa_unitarity}. 

Elastic unitarity is broken due to particle production starting at $\mathcal{T}(2\to 4) = \cO(\bar s^3)$.  Therefore at threshold, unitarity implies
\begin{equation}\label{eq:unitarityAtThreshold}
	\text{Eigenvalues}\left[(\mathbb{S}^\ell(s))^\dagger\mathbb{S}^\ell(s)\right] = 1 + \cO(\bar s^6)\,.
\end{equation}
We can expand the partial wave amplitude as follows:
\begin{equation}
	\mathbb{S}^\ell(s) =  \mathbb I +i (\mathbb F_2^\ell \bar s^2 + \mathbb F_3^\ell \bar s^3 +\mathbb F_4^\ell \bar s^4 ) + \cO(\bar s^4 \log \bar{s})\,,
\end{equation}
and \eqref{eq:unitarityAtThreshold} implies
\begin{equation}
	2 \Im \mathbb F_4^\ell = (\mathbb F_2^\ell)^2\,,  \quad \forall\ \ell \,.
\end{equation}
We recall here that $\mathbb F^\ell$ are one-by-one or two-by-two matrices \eqref{eq:entriesa_unitarity}.  Imposing this equation allows us to fix the coefficients $\beta_i$  using for the logarithm the prescription  $\log(-s) = \log(|s|) - i\pi \theta(s)$.  The result matches the explicit loop computation described in appendix \ref{app:computation_EFTs}.  Note that this result could be extended to $\cO(\bar s^5)$, as elastic unitarity is only broken at $\cO(\bar s^6)$ \eqref{eq:unitarityAtThreshold}.

\section{Computation of scattering amplitudes in EFTs}
\label{app:computation_EFTs}
In this appendix, we will compute scattering amplitudes from the general EFT written in \eqref{eq:EFT_lagrangian} to eighth order in derivatives i.e. fourth order in $s$. This means that we must go to one-loop order in the dim-8 vertices $\mathcal{L}_8$ given in \eqref{eq:lagrangian_coefficients} but only to linear order in the dim-10 $\mathcal{L}_{10}$ and dim-12  $\mathcal{L}_{12}$ vertices.  The one-loop diagrams computed using dim-8 vertices will turn out to be UV-divergent and therefore we will need to regularize the integral and add counter terms to cancel these UV-divergences, using say the $\overline {MS} $ prescription. The calculations in this section were done with the help of software packages FeynCalc \cite{Mertig:1990an,Shtabovenko:2016sxi,Shtabovenko:2020gxv}, FeynArts \cite{Hahn:2000kx}, FeynHelpers \cite{Shtabovenko:2016whf}, FeynRules \cite{Alloul:2013bka} and PackageX \cite{Patel:2016fam}.
We  now proceed order by order in derivatives or equivalently in $s$. 

\subsubsection*{O($s^2$):} 
This corresponds to tree level in the dimension 8 lagragian $\mathcal{L}_8$. Conceptually it is straightforward to compute the Feynman amplitude from the Lagrangian. However, on a technical level, the process is a bit tricky due to the various Lorentz indices and different possible contractions between them. Using computer algebra to perform the contractions, we arrive at 
\begin{equation}
	M_{\mu \nu \rho \sigma} (p_1, p_2, p_3, p_4)= 8 c_1 \eta_{\mu \nu} \eta_{\rho \sigma} s^2 + 4 c_2 {p_4}_{\mu} {p_3}_{\nu} \eta_{\rho \sigma} s + \ldots 
\end{equation} 
where, for brevity, we only displayed a couple of terms. From this Feynman amplitude, one can derive the scattering amplitude by contracting the Lorentz indices with photon polarization vectors in the COM frame: 
\begin{equation}
	\cT_{\lambda_1 \lambda_2}^{\lambda_3 \lambda_4}(s,t,u) = \epsilon_{\lambda_1}^\mu (p_1^{\text{com}}) \epsilon_{\lambda_2}^\nu(p_2^{\text{com}}) \epsilon_{\lambda_3}^{\rho *}(p_3^{\text{com}})\epsilon_{\lambda_4}^{\sigma *} (p_4^{\text{com}})M_{\mu \nu \rho \sigma} (p_1^{\text{com}},p_2^{\text{com}},p_3^{\text{com}},p_4^{\text{com}})~,
\end{equation}
where the centre of mass frame momenta $p_i^{com}$ were defined in \eqref{eq:COM_frame} and the polarization vectors $\epsilon_{\lambda}^{\mu}(p)$ were defined in \eqref{eq:polarizations}. With this, we arrive at our result, which is as follows:
\begin{equation}
	\begin{aligned}
		\Phi_1(s,t,u)\big|_{s^2}& = 2 (4c_1 + 3 c_2) s^2~, \\
		\Phi_2(s,t,u)\big|_{s^2} &= 2(4c_1 + c_2) s^2~, \\
		\Phi_5(s,t,u)\big|_{s^2}&= 0~,
	\end{aligned}
\end{equation}
were we used the notation $	\Phi_i(s,t,u)\big|_{s^n}$ to emphasize that we only write the $\cO(s^n)$ terms. 
Comparing with \eqref{eq:amplitudes_EFT}, we deduce the following relations between  the Wilson  coefficients in the Lagrangian and the coefficients in the expansion of the amplitudes:
\begin{eqnarray}
	\begin{aligned}
		g_2 &= 2 (4c_1 + 3 c_2)~,\\
		f_2 &= 2(4c_1 +c_2)~.
	\end{aligned}
\end{eqnarray}

\subsubsection*{O($s^3$):} 
At this order, only the tree level amplitude of the dimension 10 Lagrangian $\mathcal{L}_{10}$ contributes. Once again using FeynCalc, we arrive at
\begin{equation}
	M_{\mu \nu \rho \sigma} (p_1, p_2, p_3, p_4)= 4 c_4 \, \eta_{\mu \nu} \eta_{\rho \sigma} s^3 + \frac{1}{2}  c_3 \, {p_4}_{\mu} {p_3}_{\nu} \eta_{\rho \sigma} s t + \ldots 
\end{equation} 
Repeating the same steps as before, \emph{i.e.} going to the COM frame and contracting with the polarization vectors, we compute the scattering amplitudes
\begin{equation}
	\begin{aligned}
		\Phi_1(s,t,u)\big|_{s^3}& = -4 c_4 s^3~,\\
		\Phi_2(s,t,u)\big|_{s^3} &= -6 (c_3 +2 c_4 - c_5)s t u~, \\
		\Phi_5(s,t,u)\big|_{s^3}&= -\frac{3}{2} c_3 s t u~,
	\end{aligned}
\end{equation}
and upon comparing with \eqref{eq:amplitudes_EFT}, we see that
\begin{eqnarray}
	\begin{aligned}
		g_3 &= 4 c_4~,\\
		f_3 &=  6(c_3 + 2 c_4 - c_5)~, \\
		h_3 &= \frac{3}{2} c_3~.
	\end{aligned}
\end{eqnarray}

\subsubsection*{O($s^4$):} 
We now have one-loop diagrams from the dimension 8 Lagrangian $\mathcal{L}_8$ as well as tree level diagrams from the dimension 12 Lagrangian $\mathcal{L}_{12}$. 
The one-loop diagrams are UV-divergent, using dimensional regularization ($d = 4- 2\epsilon$), they can be written in the following form:

\begin{equation}
	\begin{aligned}
		\Phi_1(s,t,u)\big|_{s^4}& = \left(\frac{\xi_1}{\epsilon} + \xi_0\right) s^4 + \left(\frac{\tilde \xi_1}{\epsilon} + \tilde \xi_0\right) s^2 t u + s^2\left(\beta_{1,1}s^2+\beta_{1,2} t u\right)\log\left(- \frac{s}{\mu^2} \right)\\
		& \qquad + \beta_{1,3}s^2\left( t^2\log\left(-\frac{t}{\mu^2}\right)+ u^2\log\left(-\frac{u}{\mu^2}\right)\right),\\
		\Phi_2(s,t,u)\big|_{s^4} &= \left(\frac{\kappa_1}{\epsilon}  + \kappa_0\right) (s^2 + t^2 +u^2)^2 + \beta_2\left(s^4 \log\left(-\frac{s}{\mu^2}\right) + t^4\log\left(-\frac{t}{\mu^2}\right) + u^4\log\left(-\frac{u}{\mu^2}\right) \right),\\
		\Phi_5(s,t,u)\big|_{s^4}&= 0~,
	\end{aligned}
\end{equation}
where $\mu^2$ is the dimensional regularization scale \footnote{To be precise, this is actually the redefined scale $\mu^2 \rightarrow  \frac{e^{\gamma_E}}{4 \pi}\mu^2$ which is used to get rid of factors of $4 \pi$ and the Euler-Mascheroni constant $\gamma_E$.} and the various coefficients in the equation are given by
\begin{equation}
	\begin{aligned}
		\xi_1 &= \frac{1}{120\pi^2}\left(1008 c_1^2 + 840 c_1 c_2 + 231 c_2^2\right)~, \\
		\xi_0 &= \frac{1}{1200 \pi^2}\left(16912 c_1^2 + 15000 c_1 c_2 + 4489 c_2^2\right)~,\\
		\tilde \xi_1 &= -\frac{1}{120 \pi^2} \left(224 c_1^2 + 304 c_1 c_2 + 110c_2^2 \right)~,\\
		\tilde \xi_0 &= -\frac{1}{1800 \pi^2} \left(7664 c_1^2 + 9304 c_1 c_2 + 2495 c_2^2 \right)~,\\
		\kappa_1 &= \frac{5}{24\pi^2}\left(16 c_1^2 + 16 c_1 c_2 + 3 c_2^2\right)~, 
		\\
		\kappa_0 &= \frac{1}{1440 \pi^2} \left(6992 c_1^2 + 8888 c_1 c_2 + 1773 c_2^2 \right)~, \\
		\beta_{1,1} &= \frac{1}{120\pi}\left(912 c_1^2 + 696 c_1 c_2 + 177 c_2^2\right)~,\\
		\beta_{1,2} &= -\frac{1}{60\pi} \left(16 c_1^2 +8 c_1 c_2 + c_2^2\right)~,\\
		\beta_{1,3} &= \frac{1}{20\pi}\left(4c_1+3c_2\right)^2~,\\
		\beta_2 &= \frac{5}{12 \pi^2}\left(16 c_1^2 + 16 c_1 c_2 + 3 c_2^2 \right)~.
	\end{aligned}
\end{equation}
The UV divergences, which are now neatly captured by the $\frac{1}{\epsilon}$ poles, can be cancelled by introducing dimension 12 counter-terms of the same form as in $\mathcal L_{12}$. The $ \overline{MS} $ regularization scheme corresponds to choosing the coefficients of the counter-terms such that they only cancel the divergences, without changing the finite parts.   Explicitly, in this scheme we choose the following the counter-terms:
\begin{equation}
	\begin{aligned}
		\delta c_6 &=  \frac{8\kappa_1}{\epsilon}~, \\
		\delta c_7 & = \frac{12 \kappa_1 + \xi_1} \epsilon~,\\
		\delta c_8 &= \frac{4 \xi_1 +2 \tilde \xi_1}{\epsilon}~.
	\end{aligned}
\end{equation}
Having done this, and also including the contribution from tree level $\mathcal{L}_{12}$ terms, we have the following result for the amplitude at order $s^4$ in dimensional regularization with $\overline{MS} $ scheme:

\begin{equation}
	\begin{aligned}
		\Phi_1(s,t,u)\big|_{s^4}& = \left(\frac{1}{4}(2c_7 - 3 c_6) + \xi_0\right) s^4 + \left(\frac{1}{2}(3c_6- 2c_7-c_8) + \tilde \xi_0\right) s^2 t u \\
		&\qquad +s^2\left(\beta_{1,1}s^2 +\beta_{1,2} t u\right)\log\left(- \frac{s}{\mu^2} \right)
		+ \beta_{1,3}s^2\left( t^2\log\left(-\frac{t}{\mu^2}\right)+ u^2\log\left(-\frac{u}{\mu^2}\right)\right),\\
		\Phi_2(s,t,u)\big|_{s^4} &= \left(-\frac{1}{8}c_6 + \kappa_0\right) (s^2 + t^2 +u^2)^2 + \beta_2\left(s^4 \log\left(-\frac{s}{\mu^2}\right) + t^4\log\left(-\frac{t}{\mu^2}\right) 
		+ u^4\log\left(-\frac{u}{\mu^2}\right) \right),\\
		\Phi_5(s,t,u)\big|_{s^4}&= 0~.
	\end{aligned}
\end{equation}
However, it is more convenient for our purposes  to choose a different subtraction scheme. Firstly, we choose the dimensional regularization scale $\mu^2$ to be related to the coefficient $g_2$:
\begin{equation}
	\mu^2 = \frac{1}{\sqrt{g_2} } = \frac{1}{\sqrt{2(4c_1 + 3c_2)}}.
\end{equation}
We then choose the following subtraction scheme---the counter-terms also cancel out the finite pieces $\xi_0$, $\tilde \xi_0$ and $\kappa_0$. Explicitly, 
\begin{equation}
	\begin{aligned}
		\delta c_6 &=  \frac{8\kappa_1}{\epsilon} +8\kappa_0~,\\
		\delta c_7 & = \frac{12 \kappa_1 + \xi_1} \epsilon + 12 \kappa_0 + \xi_0~,\\
		\delta c_8 &= \frac{4 \xi_1 +2 \tilde \xi_1}{\epsilon} + 4 \xi_0 + 2\tilde \xi_0~.
	\end{aligned}
\end{equation}

With this new scheme, we reach our final result:

\begin{equation}
	\begin{aligned}
		\Phi_1(s,t,u)\big|_{s^4}& = \frac{1}{4}(2c_7 - 3 c_6)  s^4 +\frac{1}{2}(3c_6- 2c_7-c_8)   s^2 t u + s^2\left(\beta_{1,1}s^2+\beta_{1,2} t u\right)\log\left(- s\sqrt{g_2}\right)\\
		& \qquad + \beta_{1,3}s^2\left( t^2\log\left(-t\sqrt{g_2}\right)+ u^2\log\left(-u\sqrt{g_2}\right)\right),\\
		\Phi_2(s,t,u)\big|_{s^4} &= -\frac{1}{8}c_6(s^2 + t^2 +u^2)^2 + \beta_2\left(s^4 \log\left(-s\sqrt{g_2}\right) + t^4\log\left(-t\sqrt{g_2}\right) + u^4\log\left(-u\sqrt{g_2}\right) \right),\\
		\Phi_5(s,t,u)\big|_{s^4}&= 0~,
	\end{aligned}
\end{equation}
and comparing with \eqref{eq:amplitudes_EFT} leads to the identification
\begin{equation}
	\begin{aligned}
		g_4 &= \frac{1}{4}(2c_7 - 3 c_6)~, \\
		g_4' &=\frac{1}{2}(3c_6- 2c_7-c_8)~, \\
		f_4 &=-\frac{1}{8}c_6~.
	\end{aligned}
\end{equation}
As it is clear from the computation, the above relations are only valid for the chosen subtraction scheme, and a different choice of subtraction scheme, say for example $\overline{MS}$, would lead to terms involving $c_1$ and $c_2$ in the above equation.

\section{Wilson coefficients in various models}
\label{app:Wilson}

In this appendix, we give some details on the amplitudes whose Wilson coefficients we use in section \ref{sec:weak_coupling_models}. The corresponding EFTs are all obtained by integrating out massive particles which are weakly coupled at their production threshold. We consider resonances which can be integrated out at tree level (Yukawa-like theories) and charged particles which can be integrated out at one-loop (QED-like theories). Notice that these theories are not necessarily UV complete on their own: new particles might be necessary to restore unitarity above a larger energy scale.

\subsection{Yukawa-like theories}

This subsection collects a few examples where a low spin resonance is integrated out at tree level. We consider resonances of spin 0 and 2, even or odd under parity. The resulting Wilson coefficients are summarized in the first five rows of table \ref{tab:wilson}. Notice that a massive spin 1 resonance cannot couple to two photons. This is known as the Landau-Yang theorem \cite{Landau:1948kw,Yang:1950rg}, and it simply follows from \eqref{eq:parity_even_two_photon_irreps},  \eqref{eq:parity_odd_two_photon_irrep}. Indeed, a resonance of spin $\ell$ only contributes to the partial wave of the same spin in the production channel, and there is no overlap of any two photon state with a spin 1 state.

There are multiple ways to derive the EFT. Since the field corresponding to the massive particle appears quadratically in the Lagrangian, one can evaluate the action on shell and expand at low energies, as done in \cite{Henriksson:2021ymi}. Alternatively, one can compute the 2-to-2 tree level amplitude starting from the Feynman rules, or directly, using factorization and crossing, as explained in \cite{Arkani-Hamed:2017jhn}. We will mostly follow the latter method, and we will comment on the relation to the Lagrangian construction only for the case of a spin 2 resonance.

We now illustrate the method, which will also serve as a lightning review. This subsection is not meant to be self-contained, see \cite{Arkani-Hamed:2017jhn} for details. At tree level the on-shell pole due to exchange of a massive particle of spin $S$ and mass $m$ can be written as
\begin{equation}
	\label{eq:local_gluing}
	\frac{M_{L,\; \lambda_1, \lambda_2}^{ \{I_1, I_2 .. I_{2S} \}} M_{ R,\;\lambda_3, \lambda_4, \; \{I_1, I_2 \ldots I_{2S} \}}}{P^2 + m^2}
\end{equation}
where $M_L$ and $M_R$ are three-point amplitudes, $\lambda_i$ are the photon helicities and ${I_1, I_2 \ldots I_{2S}} $ are the little group indices of the massive particle. Consider now the construction of the two photon to one massive particle amplitude, with particle 3 being the massive one. This amplitude must be constructed out of the spinor-helicity variables\footnote{See appendix \ref{subsec:gaugeToSH} for an introduction to spinor helicities for massless particles. In the massive case, the little group is $SO(3)$, and the index $I=1,\,2$ is in the spin $1/2$ representation, \emph{i.e.} the fundamental of $su(2)$. } 
\begin{equation}
	\xi_1^\alpha , \tilde{\xi}_1^{\dot \alpha}, \xi_2^\alpha , \tilde{\xi}_2^{\dot \alpha}, {\xi_3}^{I \, \alpha}, {\tilde \xi_3}^{I \, \dot \alpha}
\end{equation}
such that it has the right transformation property under the respective little groups of the three particles. For example the coupling of two positive helicity photons to spin 0 and spin 2 particles can be written as 
\begin{equation}
	\label{eq:massive_spinor_helicity_couplings}
	\begin{aligned}
\text{Spin 0}:\quad	M_{++}^{\text{spin 0}} &= \frac{g}{m} [12]^2, \\
\text{Spin 2}:\quad M_{++}^{\text{spin 2}} & = \frac{g}{m^5} \langle 1 {\bf 3} \rangle  \langle 2 {\bf 3} \rangle [ 1 {\bf 3}][2 {\bf 3}][12]^2,
	\end{aligned}
\end{equation}
where
 \begin{equation}
[i \, {\bfj}] \equiv \xi_i^{\alpha} \xi_{j \, \alpha}^I \qquad	\langle i \,{\bf j} \rangle \equiv \tilde \xi_{j\,\dot \alpha} \,\tilde \xi_{j }^{I \,\dot \alpha}\,.
 \end{equation}
We can now ``glue" three-point couplings in \eqref{eq:local_gluing} to deduce the four particle amplitude, keeping in mind that this prescription computes the all incoming amplitude. This procedure fixes the on shell residue of the amplitudes that we will present later in this section. Note that by choosing the coupling constants for $+$ and $-$ helicity photons appropriately, we can ensure that the pole lies in the parity even or odd channel. In principle one could have had other couplings, for example in the spin 2 case, 
\begin{equation}
	M_{++}^{\text{spin 2}}  = \frac{g}{m^7} \langle 1 {\bf 3} \rangle  \langle 2 {\bf 3} \rangle \langle 1 {\bf 3}\rangle \langle 2 {\bf 3}\rangle [12]^4 .
\end{equation}
However these are equivalent at the on shell mass pole i.e. $s = m^2$ because
\begin{equation}
			\langle 1 {\bf 3} \rangle \equiv \tilde \xi_{1 \, \dot \alpha} \, \tilde \xi_{3} ^{I \dot \alpha}
			= -\frac{\tilde \xi_{1 \, \dot \alpha} \, p_{3 }^{ \alpha \dot \alpha} \, \xi_{3 \, \alpha}^{I }}{m}
			= \frac{\tilde \xi_{1 \, \dot \alpha} \, p_{2}^{  \alpha \dot \alpha}\xi_{3 \, \alpha}^{I }}{m}
			= \frac{\langle 1 2\rangle [2 {\bf 3}]}{m},
\end{equation}
and 
\begin{equation}
	\langle 1 2 \rangle^2 = [12]^2 = s ,
\end{equation}
and as mentioned before, this procedure only fixes the on shell residue. Therefore we can trade powers of $s$ for powers of $m^2$. Another way to understand this is that these couplings are the same up to contact terms. We choose the former coupling in \eqref{eq:massive_spinor_helicity_couplings} since it is consistent with low energy $s \rightarrow 0$ and the high energy $s \rightarrow \infty $  behaviour that we assume in this work. This point will be explained in more detail later in the appendix. 

\subparagraph{Scalar} A parity even neutral scalar particle of mass $m$ can generically decay into two photons via the tree level coupling $\phi F_{\mu\nu}F^{\mu\nu}$, where $\phi$ is the field describing the particle. This interaction is not renormalizable, therefore the theory needs a UV completion. However, it is easy to imagine at least one: we can resolve the effective coupling by adding a charged particle (a scalar or a fermion) of mass $M>m$, with a Yukawa interaction with $\phi$. The resulting model is renormalizable. Here, we are interested in the EFT obtained by integrating out the charged particle. Its 2-to-2 photon amplitude at tree level can be easily constructed with the recipe given in \cite{Arkani-Hamed:2017jhn}:
\begin{subequations}
\begin{align}
\Phi_1(s,t,u) &= - \frac{\lambda^2}{m^2}\frac{s^2}{s-m^2}~, \\
\Phi_2(s,t,u) &= - \frac{\lambda^2}{m^2}\left(\frac{s^2}{s-m^2}+\frac{t^2}{t-m^2}+\frac{u^2}{u-m^2}\right)~, \\
\Phi_5(s,t,u) &= 0~,
\end{align}
\label{ampEFTscalar}
\end{subequations}
while $\Phi_3$ and $\Phi_4$ are obtained by crossing---see \eqref{eq:crossing_intro}.
If, for instance, we complete the theory by coupling the scalar to a spin 1/2 charged fermion of mass $M$, the dimensionless coupling $\lambda$ is proportional to   the ratio $m/M$.
Regardless, the amplitude \eqref{ampEFTscalar} shows that $m/\lambda$ is the cutoff of the EFT. Indeed, when the Mandelstam invariants are of order $(m/\lambda)^2$, the amplitudes become large, the theory is strongly coupled and something new must happen to unitarize. On the other hand, for fixed $s$ and $\lambda$ small, the violations of unitarity contained in \eqref{ampEFTscalar} can be cured by higher order terms in $\lambda$, as usual in effective field theory.

\subparagraph{Axion} The case of a parity odd scalar proceeds much as in the previous example. This time the interaction vertex is $\tilde{\phi} F_{\mu\nu} \tilde{F}^{\mu\nu}$, $\tilde{F}^{\mu\nu}$ being the Hodge dual of the field strength, and the amplitudes are
\begin{subequations}
\begin{align}
\Phi_1(s,t,u) &= - \frac{\lambda^2}{m^2}\frac{s^2}{s-m^2}~, \\
\Phi_2(s,t,u) &=  \frac{\lambda^2}{m^2}\left(\frac{s^2}{s-m^2}+\frac{t^2}{t-m^2}+\frac{u^2}{u-m^2}\right)~, \\
\Phi_5(s,t,u) &= 0~.
\end{align}
\label{ampEFTaxion}
\end{subequations}
The only difference with the scalar case is a overall minus sign in the $\Phi_2$ amplitude, which moves the production pole of the scalar resonance from the parity even to the parity odd partial wave.

\subparagraph{Parity even spin 2} 

If we construct an amplitude for the tree level exchange of a spin 2 resonance following \cite{Arkani-Hamed:2017jhn}, we find that there are multiple options. The ambiguity is parametrized by the three-point coupling between two photons and a massive spin 2 particle. A spin 2 massive particle can either decay into two photons with the same helicities, or into two photons with opposite helicities.\footnote{It is worth noticing that the spinor helicity structure for the decay into 2 photons with the same helicity is only unique once we ask for the absence of kinematic singularities, \emph{i.e.} we demand that the $\Phi_1$ amplitude has a low energy expansion of the kind \eqref{eq:amplitudes_EFT}.} The latter state has even parity---see \eqref{parityOnPartialWaves}---hence it only exists for a parity even spin 2 resonance. On the other hand, a parity even resonance couples to the $(++)$ and the $(--)$ states equally, while a parity odd one couples to them with opposite sign. All in all, we have a two-parameters family of couplings of photons to a parity even spin 2 resonance. For simplicity, we only consider the limiting cases where the resonance does not couple to both the $(+-)$ and the $++$ state, \emph{i.e.} we set the amplitude $\Phi_5$ to zero. This is not necessary, but it is enough for our purposes. The two amplitudes are  
\begin{subequations}
\begin{align}
\Phi^{\textup{I}}_1(s,t,u) &= - \lambda^2 \frac{s^2}{m^6}\frac{t^2-4 t u+u^2}{s-m^2}+\textup{polynomial}~, \\
\Phi^\textup{I}_2(s,t,u) &=  -\frac{\lambda^2}{m^6}\left(s^2\,\frac{t^2-4 t u+u^2}{s-m^2}+t^2\,\frac{s^2-4 s u+u^2}{t-m^2}+u^2\,\frac{s^2-4 s t +t^2}{u-m^2}\right)
+\textup{polynomial}~, \\
\Phi^\textup{I}_5(s,t,u) &= 0~,
\end{align}
\label{ampEFTParityEvenI}
\end{subequations}
for the coupling to equal helicity photons, and
\begin{subequations}
\begin{align}
\Phi^{\textup{II}}_1(s,t,u) &= - \frac{\lambda^2}{m^2}\left(\frac{s^2}{t-m^2}+\frac{s^2}{u-m^2}\right)+\textup{polynomial}~, \\
\Phi^{\textup{II}}_2(s,t,u) &=  0~, \\
\Phi^{\textup{II}}_5(s,t,u) &= 0~,
\end{align}
\label{ampEFTParityEvenII}
\end{subequations}
for the coupling to photons with opposite helicities. In \eqref{ampEFTParityEvenI} and \eqref{ampEFTParityEvenII}, we allowed for yet unspecified polynomials in the Mandelstam invariants. Clearly, a polynomial does not change the residues of the amplitude. Therefore, these are ambiguities in the EFT of the resonance and the photons. Such degrees of freedom are parametrized by the Wilson coefficients as in \eqref{eq:amplitudes_EFT}, because the latter provide the most general polynomial solution to crossing without kinematic singularities. In other words, the polynomials correspond to contact interactions among the photons. We use this ambiguity to improve the Regge limit of the amplitudes \eqref{ampEFTParityEvenI} and \eqref{ampEFTParityEvenII}. In particular, the type I amplitudes can be made compatible with the classical Regge growth conjecture \cite{Chowdhury:2019kaq}, \emph{i.e.} their growth can be limited to being $\mathcal{O}(s^2,u^2,t^2)$, depending on which variable is taken large. This requires adding a homogeneous polynomial of degree 3. Furthermore, as explained in subsection \ref{subsec:rule_in_8}, we can add a homogeneous polynomial of degree 2 to further improve the Regge limit in the forward kinematics, so that both type I and type II amplitudes obey the dispersion relation \eqref{eq:sum_rule_A}. Notice that none of the additions modify the Wilson coefficients at dimension 12 ($g_4,\,g_4',\,f_4$). The final results are
\begin{subequations}
\begin{align}
\Phi^{\textup{I}}_1(s,t,u) &= - \lambda^2 \frac{s^2}{m^6}\frac{t^2-4 t u+u^2}{s-m^2}+
~\frac{\lambda^2}{m^6} \left(m^2 s^2+s^3\right), \\
\Phi^{\textup{I}}_2(s,t,u) &=  -\frac{\lambda^2}{m^6}\left(s^2\,\frac{t^2-4 t u+u^2}{s-m^2}+t^2\,\frac{s^2-4 s u+u^2}{t-m^2}+u^2\,\frac{s^2-4 s t +t^2}{u-m^2}\right) \notag \\
&+\frac{\lambda^2}{m^4}\left(s^2+t^2+u^2\right)~,  \\
\Phi^{\textup{I}}_5(s,t,u) &= 0~,
\end{align}
\label{ampEFTParityEvenIimproved}
\end{subequations}
and
\begin{subequations}
\begin{align}
\Phi^{\textup{II}}_1(s,t,u) &= - \frac{\lambda^2}{m^2}\left(\frac{s^2}{t-m^2}+\frac{s^2}{u-m^2}\right)-\frac{\lambda^2}{m^4} s^2 ~, \\
\Phi^{\textup{II}}_2(s,t,u) &=  0~, \\
\Phi^{\textup{II}}_5(s,t,u) &= 0~.
\end{align}
\label{ampEFTParityEvenIIimproved}
\end{subequations}

Let us now make contact with the work \cite{Henriksson:2021ymi}. There, a single parity even coupling was considered, which in the present language is the type II coupling, without the addition of contact terms. The authors explicitly used a Lagrangian, and parametrized the coupling to the spin 2 resonance via
\begin{equation}
h^{\mu\nu} \left(\frac{u}{M} F_{\mu\rho}F_\nu{}^\rho+\frac{u'}{M} \eta_{\mu\nu}F_{\rho\sigma}F^{\rho\sigma}\right)~,
\label{hFsqCoupling}
\end{equation}
where $h$ is the field describing the resonance, $\eta$ is the metric and $u,\,u'$ are coupling constants. Since the trace of $h$ does not propagate, the $u'$ coupling does not produce poles, and, as pointed out in \cite{Henriksson:2021ymi}, it is in fact equivalent to contact interactions, which can be fine tuned to produce an amplitude which does not grow more than $s^2$ in the Regge limit. Since \eqref{hFsqCoupling} is the most general parity even cubic coupling at lowest derivative order, our additional type I amplitudes must correspond to a higher derivative interaction. This is indeed the case, as it can be reversed engineered from the amplitude, following \cite{Arkani-Hamed:2020blm}. The interaction producing equation \eqref{ampEFTParityEvenI} is
\begin{equation}
\frac{\lambda}{m^3} \left( \, h^{\mu \nu} \partial_\mu F_{\rho \sigma} \partial_\nu F^{\rho \sigma} - \frac{1}{4} \partial_\mu \partial_\nu h^{\mu \nu} F_{\rho \sigma} F^{\rho \sigma} \right)
\label{hHigherDer}
\end{equation}
As explained above, photon contact interactions can be added in order to obtain the Regge bounded amplitudes \eqref{ampEFTParityEvenIimproved}: if needed, their Lagrangian can be deduced combining \eqref{ampEFTParityEvenIimproved} with \eqref{eq:lagrangian_coefficients} and \eqref{eq:amp_coeffs_in_EFT}. 
It would be interesting to investigate the possible UV completions of the EFT of a spin 1 massless particle and a neutral spin 2 resonance, perhaps coupling them via a charged vector boson, and understand if the Lagrangian \eqref{hHigherDer} can arise at low energies.

\subparagraph{Parity odd spin 2}

As mentioned in the previous paragraph, the lowest derivative order cubic coupling with a parity odd spin 2 particle involves two photons with equal helicities (type I). The corresponding amplitude, equipped with contact terms as above, is
\begin{subequations}
\begin{align}
\Phi_1(s,t,u) &= - \lambda^2 \frac{s^2}{m^6}\frac{t^2-4 t u+u^2}{s-m^2}+
~\frac{\lambda^2}{m^6} \left(m^2 s^2+s^3\right), \\
\Phi_2(s,t,u) &=  \frac{\lambda^2}{m^6}\left(s^2\,\frac{t^2-4 t u+u^2}{s-m^2}+t^2\,\frac{s^2-4 s u+u^2}{t-m^2}+u^2\,\frac{s^2-4 s t +t^2}{u-m^2}\right) \notag \\
&-\frac{\lambda^2}{m^4}\left(s^2+t^2+u^2\right)~,  \\
\Phi_5(s,t,u) &= 0~.
\end{align}
\label{ampEFTParityOddIimproved}
\end{subequations}
All the same observations about the parity even type I amplitudes apply to this case as well. In particular, the Lagrangian which generates it is
\begin{equation}
\frac{\lambda}{m^3}\left( \, h^{\mu \nu} \partial_\mu F_{\rho \sigma} \partial_\nu \tilde F^{\rho \sigma} - \frac{1}{4} \partial_\mu \partial_\nu h^{\mu \nu} F_{\rho \sigma} \tilde F^{\rho \sigma} \right)
\end{equation}

\subsection{QED-like theories}

Let us briefly consider scalar, spinor and vector QED.
 By integrating out the massive charged particle (i.e. respectively the fermion, scalar and vector), one obtains the Wilson coefficients of the photon Lagrangian \eqref{eq:EFT_lagrangian}. The case of standard QED leads to the well known Euler-Heisenberg Lagrangian \cite{Heisenberg:1936nmg}. 
 
Here, we comment on the Regge limit of the one-loop amplitudes, from which the value in table \ref{tab:wilson} are easily obtained. 
In all three cases, the amplitudes can be explicitly written in a basis of three integrals \cite{berestetskii2012quantum,Yang:1994nu,PhysRevD.52.5018}. 
These integrals can be estimated when the Mandelstam invariants become large, and one can conclude that these one-loop amplitudes grow in the Regge limit at most as $\cO((\log s)^2)$.

Finally, note that these QED theories are not UV complete on their own (see appendix \ref{app:landauPoleQED}) and should be considered as partial UV completion of the photon EFT.

\section{UV incompleteness of QED}\label{app:landauPoleQED}
This appendix is a short summary of the results derived in \cite{Peskin:1995ev} for quantum electrodynamics (QED).

Let us consider the Lagrangian density of QED, it reads
\begin{equation}
	\label{eq:lagrangian_QED}
	\mathcal{L}_{QED} =\bar{\psi}(i\slashed D -m)\psi -\frac{1}{4}F_{\mu \nu}F^{\mu\nu} -\frac{1}{2\xi}(\pder{\mu}A^\mu)^2 +\text{counter terms}.
\end{equation}
Here $\psi(x)$ is the Dirac fermion field with the physical mass $m$ and the covariant derivative is
\begin{equation}
	D_\mu \equiv \pder{\mu} + ieA_\mu,
\end{equation}
where $e$ is the electric charge of $\psi$.
The excitations of $\psi(x)$ describe electrons and positrons with mass $m$.
The electromagnetic field strength tensor $F_{\mu\nu}(x)$ is given by
\begin{equation}\label{eq:electromagnetic_tensor}
	F_{\mu\nu}(x) \equiv \partial_\mu A_\nu(x) - \partial_\nu A_\mu(x),
\end{equation}
where $A_\mu(x)$ is the electromagnetic potential. The excitations of $A_\mu(x)$ describe photons. One usually also defines the electromagnetic constant $\alpha$ as
\begin{equation}
	\label{eq:alpha_constant}
	\alpha \equiv e^2/4\pi.
\end{equation}

Notice that the counter terms in \eqref{eq:lagrangian_QED} are adjusted in such a way that $m$ and $\alpha$ are physical finite constants. QED describes interaction between electrons, positrons and light in our world. The experimentally measured value of $\alpha$ is 
\begin{equation}
	\label{eq:results_alpha}
	\alpha \approx \frac{1}{137.035}.
\end{equation}
Let us discuss what observables can be computed in QED and how the result \eqref{eq:results_alpha} can be measured experimentally.

The electromagnetic form factor is defined as the following matrix element
\begin{equation}
	{}_\text{out}^{\lambda_1\lambda_2}\<m, \vec p_1; m, \vec p_2| J^\mu(x=0)|0\>,
\end{equation}
where $J^\mu(x)$ is the electromagnetic current and in the left-hand side we have a two-particle ``out'' asymptotic state built out of an electron and positron with helicities $\lambda_1$ and $\lambda_2$ and 3-momenta $\vec p_1$ and $\vec p_2$. 
Due to the Lorentz invariance this form factor can be decomposed into tensor structures as
\begin{equation}
	{}_\text{out}^{\lambda_1\lambda_2}\<m, \vec p_1; m, \vec p_2| J^\mu(x=0)|0\> =
	F_1(q^2) \times (\bar u_{\lambda_1}\gamma^\mu u_{\lambda_2})+
	F_2(q^2) \times \frac{i q_\nu}{2m}\,(\bar u_{\lambda_1}\sigma^{\mu\nu} u_{\lambda_2}),
\end{equation}
where $u_\lambda$ is the 4-component solution of the Dirac equation.
Here $F_1$ and $F_2$ are the scalar  components of the electromagnetic form factor. The total momentum $q$ of the two-particle state is 
\begin{equation}
	q^\mu \equiv p_1^\mu+p_2^\mu.
\end{equation}
For a generic discussion of form factors see \cite{Karateev:2019ymz,Karateev:2020axc}.

The functions $F_1$ and $F_2$  are computed to one-loop in section 6 in \cite{Peskin:1995ev}, see equations (6.56) and (6.57). Using the (semi-classical) Born approximation one can compute the expression for the magnetic moment of the electron in terms of the Landé g-factor which reads as
\begin{equation}
	g= 2 +2F_2(0), \qquad F_2(0) = \frac{\alpha}{2\pi}.
\end{equation}
By measuring $g$ one can determine $\alpha$ via the above relation.

The effective Coulomb potential was discussed in subsection 7.5 in \cite{Peskin:1995ev}. One finds that
\begin{equation}
	V(r) = - \frac{f(r)}{r},
\end{equation}
where the function $f(r)$ computed in QED to one loop has the form
\begin{equation}
	\label{eq:perturbative_result}
	f(r) = \alpha + \alpha^2\, h(r) + O(\alpha^3).
\end{equation}
The function $f(r)$ is often called the running coupling constant. The function $h(r)$ can be evaluated analytically at large and small distances. One gets
\begin{equation}
	\label{eq:perturbative_result_h}
	h(r)\underset{mr\gg 1}{=}\frac{1}{4\sqrt\pi} \frac{e^{-2mr}}{(mr)^{3/2}},\qquad
	h(r)\underset{mr\ll 1}{=}-\frac{1}{54\pi}(5/6+\gamma +\log(mr)),
\end{equation}
where $\gamma$ is the Euler gamma. Looking at \eqref{eq:perturbative_result} and \eqref{eq:perturbative_result_h} one finds the following asymptotic behavior
\begin{equation}
	f(\infty)=\alpha,\qquad
	f(0)=\infty.
\end{equation} 
At small distance or equivalently at high energies the function $f(r)$ blows up. This indicates that the theory is not well-defined at arbitrary high energies. 
In fact, one expects the coupling to diverge at a finite energy scale.
This problem is known as the Landau pole. The Landau pole issue was discovered perturbatively but is believed to hold non-perturbatively.

\section{LSZ derivation of crossing equations}
\label{app:LSZ}
We begin in subsection \ref{app:gauge_fixed_LSZ} with a derivation of crossing in general frame using the LSZ prescription, as defined in many textbooks, with the gauge fixed correlator of the $A_\mu$ fields. However, we find this prescription not as satisfactory as working with the gauge invariant correlator of field strengths $F_{\mu\nu}$. Therefore, in subsection \ref{app:gauge_inv_LSZ} we will use the latter to re-derive crossing in both general and COM frame.

\subsection{Gauge fixed LSZ prescription}
\label{app:gauge_fixed_LSZ}
The LSZ reduction formula for the scattering process $12 \rightarrow 34$ of four massless  spin-1 particles can be written in the following form \cite{Srednicki:2007qs}
\begin{eqnarray}
\label{eq:lsz_reduction}
\cT_{12\rightarrow 34}{}_{\lambda_1, \lambda_2}^{\lambda_3, \lambda_4}(p_1, p_2, p_3, p_4) &=& \int d^4 x_1 \, d^4 x_2\, d^4 x_3 \,d^4 x_4 \nonumber \\
&& \quad \times 
\quad e^{-i p_3 x_3}\, \epsilon^{\mu_3 *}_{\lambda_3} (p_3) (-\partial^2_3) \nonumber \\
&& \quad \times 
\quad e^{-i p_4 x_4}\,\epsilon^{\mu_4 *}_{\lambda_4} (p_4) (-\partial^2_4) \nonumber \\
&&  \quad \times \quad \langle \Omega | T\{A_{\mu_4}(x_4)A_{\mu_3}(x_3)A_{\mu_1}(x_1) A_{\mu_2}(x_2)\}| \Omega \rangle_{connected} \nonumber \\
&&  \quad \times \quad (-\overleftarrow{\partial_1^2}) \epsilon^{\mu_1 }_{\lambda_1}(p_1)\, e^{i p_1 x_1}\nonumber \\
&&\quad \times \quad (-\overleftarrow{\partial_2^2}) \epsilon^{\mu_2 }_{\lambda_2}(p_2)\, e^{i p_2 x_2},
\end{eqnarray}
where $|\Omega\>$ denotes the vacuum state and $A_{\mu_i}(x)$ are massless spin 1 fields in the Lorentz gauge. Similarly we can write the LSZ reduction formula for the process $1\bar 3 \rightarrow \bar 2 4$
\begin{eqnarray}
\label{eq:lsz_crossed}
\cT_{1\bar 3\rightarrow \bar 24}{}_{\lambda_1, \lambda_3}^{\lambda_2, \lambda_4}(p_1, p_3, p_2, p_4) &=& \int d^4 x_1 \, d^4 x_2\, d^4 x_3 \,d^4 x_4 \nonumber \\
&& \quad \times 
\quad e^{-i p_2 x_2}\, \epsilon^{\mu_2 *}_{\lambda_2} (p_2) (-\partial^2_2) \nonumber \\
&& \quad \times 
\quad e^{-i p_4 x_4}\,\epsilon^{\mu_4 *}_{\lambda_4} (p_4) (-\partial^2_4) \nonumber \\
&&  \quad \times \quad \langle \Omega | T\{A_{\mu_4}(x_4)A_{\mu_2}(x_2)A_{\mu_1}(x_1) A_{\mu_3}(x_3)\}| \Omega \rangle_{connected} \nonumber \\
&&  \quad \times \quad (-\overleftarrow{\partial^2_1}) \epsilon^{\mu_1 }_{\lambda_1}(p_1)\, e^{i p_1 x_1}\nonumber \\
&&\quad \times \quad (-\overleftarrow{\partial^2_3}) \epsilon^{\mu_3 }_{\lambda_3}(p_3)\, e^{i p_3 x_3},
\end{eqnarray}
Note that the amplitudes above are defined for positive energy momenta, namely $p_i^0>0$.  Crossing symmetry is the statement that the amplitudes for the two processes above are related by analytic continuation. Consider the $1\bar 3 \rightarrow 2 4$ process and analytically continue the expression \ref{eq:lsz_crossed} in $p_2$ and $p_3$ to allow for negative energies. 
\begin{eqnarray}
\label{eq:lsz_crossing_intermediate}
\cT_{1\bar 3\rightarrow \bar 24}{}_{\lambda_1, \lambda_3}^{\lambda_2, \lambda_4}(p_1, -p_3, -p_2, p_4) &=& \int d^4 x_1 \, d^4 x_2\, d^4 x_3 \,d^4 x_4 \nonumber \\
&& \quad \times 
\quad e^{i p_2 x_2}\, \epsilon^{\mu_2 *}_{\lambda_2} (-p_2) (-\partial^2_2) \nonumber \\
&& \quad \times 
\quad e^{-i p_4 x_4}\,\epsilon^{\mu_4 *}_{\lambda_4} (p_4) (-\partial^2_4) \nonumber \\
&&  \quad \times \quad \langle \Omega | T\{A_{\mu_4}(x_4)A_{\mu_2}(x_2)A_{\mu_1}(x_1) A_{\mu_3}(x_3)\}| \Omega \rangle_{connected} \nonumber \\
&&  \quad \times \quad (-\overleftarrow{\partial^2_1}) \epsilon^{\mu_1 }_{\lambda_1}(p_1)\, e^{i p_1 x_1}\nonumber \\
&&\quad \times \quad (-\overleftarrow{\partial^2_3}) \epsilon^{\mu_3 }_{\lambda_3}(-p_3)\, e^{-i p_3 x_3},
\end{eqnarray}
We see that the above expression looks very similar to \eqref{eq:lsz_reduction} except for the correlator and the polarization vectors which are evaluated at negative energies.
Using the fact that bosonic operators commute we see that the correlation functions are equal
\begin{multline}
\label{eq:permutations}
\langle \Omega | T\{A_{\mu_4}(x_4)A_{\mu_3}(x_3)A_{\mu_1}(x_1) A_{\mu_2}(x_2)\}| \Omega \rangle  =\\
\langle \Omega | T\{A_{\mu_4}(x_4)A_{\mu_2}(x_2)A_{\mu_1}(x_1) A_{\mu_3}(x_3)\}| \Omega \rangle
\end{multline}
We now consider the negative energy polarization vectors $\epsilon^\mu_{\lambda}(-p)$. They are the analytic continuations of the positive energy polarizations, which were defined in \ref{eq:polarizations}. Since we deal with massless particles in this work, we choose the following analytic continuation for the momenta 
\begin{equation}
p^0 \rightarrow -p^0 \quad, \myP \rightarrow -\myP, \quad \theta \rightarrow \theta, \quad \phi \rightarrow \phi
\end{equation}
which ensures that $p^\mu \rightarrow -p^\mu$. See \cite{Hebbar:2020ukp} for more details. Under this analytic continuation we see from the explicit form \ref{eq:polarizations} that
\begin{equation}
\label{eq:polarization_continuation_2}
\epsilon^\mu_{\lambda}(-p) = \epsilon^\mu_{\lambda}(p) = \epsilon^{\,\mu *}_{-\lambda}(p)
\end{equation}
Thus by using \ref{eq:polarization_continuation_2} and \ref{eq:permutations} in \ref{eq:lsz_crossing_intermediate} and then comparing with \ref{eq:lsz_reduction} we have 
\begin{equation}
\cT_{12\rightarrow 34}{}_{\lambda_1, \lambda_2}^{\lambda_3, \lambda_4}(p_1, p_2, p_3, p_4) = \cT_{1\bar 3\rightarrow \bar 24}{}_{+\lambda_1, -\lambda_3}^{-\lambda_2, +\lambda_4}(p_1, -p_3, -p_2, p_4).
\end{equation}
Analogously one derives the other three crossing equations. The complete summary of crossing equations reads
\begin{equation}
\label{eq:crossing_LSZ_ac2}
\begin{aligned}
%\label{eq:crossin_1-4_LSZ_2}
\cT_{12\rightarrow 34}{}_{\lambda_1, \lambda_2}^{\lambda_3, \lambda_4}(p_1, p_2, p_3, p_4) = 
\cT_{\bar 4 2\rightarrow 3\bar 1}{}_{-\lambda_4, +\lambda_2}^{+\lambda_3, -\lambda_1}(-p_4, p_2, p_3, -p_1),\\
%\label{eq:crossin_2-3_LSZ_2}
\cT_{12\rightarrow 34}{}_{\lambda_1, \lambda_2}^{\lambda_3, \lambda_4}(p_1, p_2, p_3, p_4) = \cT_{1\bar 3\rightarrow \bar 24}{}_{+\lambda_1, -\lambda_3}^{-\lambda_2, +\lambda_4}(p_1, -p_3, -p_2, p_4),\\
%\label{eq:crossin_1-3_LSZ_2}
\cT_{12\rightarrow 34}{}_{\lambda_1, \lambda_2}^{\lambda_3, \lambda_4}(p_1, p_2, p_3, p_4) = 
\cT_{\bar 3 2\rightarrow \bar 1 4}{}_{-\lambda_3, +\lambda_2}^{-\lambda_1, +\lambda_4}(-p_3, p_2, -p_1, p_4),\\
\cT_{12\rightarrow 34}{}_{\lambda_1, \lambda_2}^{\lambda_3, \lambda_4}(p_1, p_2, p_3, p_4) = 
\cT_{1\bar 4\rightarrow 3\bar 2}{}_{+\lambda_1, -\lambda_4}^{+\lambda_3, -\lambda_2}(p_1, -p_4, p_3, -p_2).
\end{aligned}
\end{equation}
Specializing to the case of scattering of identical neutral massless spin 1 particles, all the processes in the above 4 equations are the same and thus the crossing equations express a symmetry of the scattering amplitude.
\paragraph{All incoming amplitude}
Using the LSZ reduction formula \ref{eq:lsz_reduction} it is possible to define an unphysical 4 photons to nothing amplitude by analytic continuation: 
\begin{eqnarray}
\cT_{\lambda_1, \lambda_2,\lambda_3, \lambda_4}(p_1, p_2, p_3, p_4) &\equiv& \cT_{\,\lambda_1, \, \lambda_2}^{-\lambda_3, -\lambda_4}(p_1, p_2, -p_3, -p_4)\\
  &=& \int d^4 x_1 \, d^4 x_2\, d^4 x_3 \,d^4 x_4 \nonumber \\
&& \quad \times 
\quad e^{i p_3 x_3}\, \epsilon^{\mu_3}_{\lambda_3} (p_3) (-\partial^2_3) \nonumber \\
&& \quad \times 
\quad e^{i p_4 x_4}\,\epsilon^{\mu_4}_{\lambda_4} (p_4) (-\partial^2_4) \nonumber \\
&&  \quad \times \quad \langle \Omega | T\{A_{\mu_4}(x_4)A_{\mu_3}(x_3)A_{\mu_1}(x_1) A_{\mu_2}(x_2)\}| \Omega \rangle_{connected} \nonumber \\
&&  \quad \times \quad (-\overleftarrow{\partial_1^2}) \epsilon^{\mu_1}_{\lambda_1}(p_1)\, e^{i p_1 x_1}\nonumber \\
&&\quad \times \quad (-\overleftarrow{\partial_2^2}) \epsilon^{\mu_2}_{\lambda_2}(p_2)\, e^{i p_2 x_2},
\end{eqnarray}
where we used \ref{eq:polarization_continuation_2}. The benefit of defining this unphysical amplitude is that it is manifestly $S_4$ permutation invariant.

\subsection{Gauge invariant LSZ prescription}
\label{app:gauge_inv_LSZ}
We begin by considering the residue at the on-shell pole of a 4 point correlator of electromagnetic tensor $F_{\mu \nu}$, see for example Eq.10.3.2 of \cite{Weinberg:1995mt}:
\begin{equation}
	\begin{aligned}
	\sum_{\lambda_1,\lambda_2, \lambda_3, \lambda_4} H_{\mu_1 \nu_1 }^{\lambda_1 \,}(p_1) &H_{\mu_2 \nu_2 }^{\lambda_2 \, }(p_2)		H_{\lambda_3, \mu_3 \nu_3 }(p_3) H_{\lambda_4, \mu_4 \nu_4 }(p_4)	
	\; \cT_{12\rightarrow 34}{}_{\lambda_1, \lambda_2}^{\lambda_3, \lambda_4}(p_1, p_2, p_3, p_4)	\\
	 & =	\int d^4 x_1 \, d^4 x_2\, d^4 x_3 \,d^4 x_4
	\,e^{-i p_3 x_3}\, 	e^{-i p_4 x_4}\, (-\partial^2_3) \,
	(-\partial^2_4)  \\
	 & \quad  \quad \times \,\langle \Omega | T\{F_{\mu_4 \nu_4}(x_4)F_{\mu_3 \nu_3}(x_3)F_{\mu_1 \nu_1}(x_1) F_{\mu_2 \nu_2}(x_2)\}| \Omega \rangle_{connected}  \\
		&  \qquad \qquad (-\overleftarrow{\partial_1^2}) (-\overleftarrow{\partial_2^2})	\, e^{i p_1 x_1} \, e^{i p_2 x_2} 
	\end{aligned}
\end{equation}
where the object $H_{\lambda, \mu \nu}$  is the same as \eqref{eq:H_object} because in free theory we have
\begin{equation}
	H_{\lambda, \mu \nu}(p) = \langle 0 | F_{\mu \nu}(0) | p, \lambda \rangle \,.
\end{equation}
Unfortunately it is not easy to invert this object in a covariant way to extract the scattering amplitude. Instead we use the following relation which follows from the orthogonality and transversality of photon polarization vectors: 
\begin{equation}
\epsilon^{\mu \, *}_{\lambda'} (p) H_{\lambda, \mu \nu}(p) = p_\nu \delta_{\lambda' \lambda}
\end{equation}
to arrive at
\begin{equation}
	\begin{aligned}
		\label{eq:covariant_GI_LSZ}
	{p_1}_{\nu_1} {p_2}_{\nu_2} {p_3}_{\nu_3} {p_4}_{\nu_4}
		\; \cT_{12\rightarrow 34}{}_{\lambda_1, \lambda_2}^{\lambda_3, \lambda_4}&(p_1, p_2, p_3, p_4)	\\
		& =	\int d^4 x_1 \, d^4 x_2\, d^4 x_3 \,d^4 x_4 \\
  & \quad \times \quad	e^{-i p_3 x_3}\, \epsilon^{\mu_3 *}_{\lambda_3} (p_3) (-\partial^2_3)\\
		& \quad \times \quad e^{-i p_4 x_4}\,\epsilon^{\mu_4 *}_{\lambda_4} (p_4) (-\partial^2_4)  \\
		& \quad  \times \quad \,\langle \Omega | T\{F_{\mu_4 \nu_4}(x_4)F_{\mu_3 \nu_3}(x_3)F_{\mu_1 \nu_1}(x_1) F_{\mu_2 \nu_2}(x_2)\}| \Omega \rangle_{connected}  \\
		&  \quad \times \quad (-\overleftarrow{\partial_1^2}) \epsilon^{\mu_1 }_{\lambda_1}(p_1)\, e^{i p_1 x_1} \\
		&\quad \times \quad (-\overleftarrow{\partial_2^2}) \epsilon^{\mu_2 }_{\lambda_2}(p_2)\, e^{i p_2 x_2},
	\end{aligned}
\end{equation}
At this stage, we can evaluate the above equation in the COM frame \eqref{eq:COM_frame} and extract the amplitude as follows:
\begin{equation}
	\begin{aligned}
		\label{eq:COM_frame_GI_LSZ}
			\; \cT_{12\rightarrow 34}{}_{\lambda_1, \lambda_2}^{\lambda_3, \lambda_4}(s,t,u)	& =	\int d^4 x_1 \, d^4 x_2\, d^4 x_3 \,d^4 x_4 \\
		& \quad \times \quad	e^{-i p_3^{\text{com}} x_3}\, \epsilon^{\mu_3 *}_{\lambda_3} (p_3^{\text{com}}) \mathfrak{v}^{\nu_3}(-\partial^2_3)\\
		& \quad \times \quad e^{-i p_4^{\text{com}} x_4}\,\epsilon^{\mu_4 *}_{\lambda_4} (p_4^{\text{com}}) \mathfrak{v}^{\nu_4}(-\partial^2_4)  \\
		& \quad  \times \quad \,\langle \Omega | T\{F_{\mu_4 \nu_4}(x_4)F_{\mu_3 \nu_3}(x_3)F_{\mu_1 \nu_1}(x_1) F_{\mu_2 \nu_2}(x_2)\}| \Omega \rangle_{connected}  \\
		&  \quad \times \quad (-\overleftarrow{\partial_1^2}) \epsilon^{\mu_1 }_{\lambda_1}(p_1^{\text{com}}) \mathfrak{v}^{\nu_1}\, e^{i p_1^{\text{com}} x_1} \\
		&\quad \times \quad (-\overleftarrow{\partial_2^2}) \epsilon^{\mu_2 }_{\lambda_2}(p_2^{\text{com}}) \mathfrak{v}^{\nu_2}\, e^{i p_2^{\text{com}} x_2},
	\end{aligned}
\end{equation}
where the vector $\mathfrak{v} = \frac{2}{\sqrt{s}}(-1,0,0,0)$ is chosen so that 
\begin{equation}
	\mathfrak{v}^\mu {p_i}^{\text{com}}_{\mu} = 1 \quad \forall \quad i = 1,2,3,4
\end{equation}

\paragraph{Crossing using LSZ}
Consider the LSZ formula for the process $13 \rightarrow 2 4$ 
\begin{equation}
	\begin{aligned}
		\label{eq:1324_covariant_GI_LSZ}
		{p_1}_{\nu_1}{p_3}_{\nu_3} {p_2}_{\nu_2}{p_4}_{\nu_4}
		\; \cT_{13\rightarrow 24}{}_{\lambda_1, \lambda_3}^{\lambda_2, \lambda_4}&(p_1, p_3, p_2, p_4)	\\
		& =	\int d^4 x_1 \, d^4 x_2\, d^4 x_3 \,d^4 x_4 \\
		& \quad \times \quad	e^{-i p_2 x_2}\, \epsilon^{\mu_2 *}_{\lambda_2} (p_2) (-\partial^2_2)\\
		& \quad \times \quad e^{-i p_4 x_4}\,\epsilon^{\mu_4 *}_{\lambda_4} (p_4) (-\partial^2_4)  \\
		& \quad  \times \quad \,\langle \Omega | T\{F_{\mu_4 \nu_4}(x_4)F_{\mu_2 \nu_2}(x_2)F_{\mu_1 \nu_1}(x_1) F_{\mu_3 \nu_3}(x_3)\}| \Omega \rangle_{connected}  \\
		&  \quad \times \quad (-\overleftarrow{\partial_1^2}) \epsilon^{\mu_1 }_{\lambda_1}(p_1)\, e^{i p_1 x_1} \\
		&\quad \times \quad (-\overleftarrow{\partial_3^2}) \epsilon^{\mu_3 }_{\lambda_3}(p_3)\, e^{i p_3 x_3}.
	\end{aligned}
\end{equation}
A priori this formula and \eqref{eq:covariant_GI_LSZ} are both defined for positive energy momenta $p_i^0 \geq 0$. But suppose we can analytically continue the 
formulae to negative energy momenta as well, then evaluating \eqref{eq:1324_covariant_GI_LSZ} at $-p_2$ and $-p_3$ gives
\begin{equation}
	\begin{aligned}
		{p_1}_{\nu_1}{p_3}_{\nu_3} {p_2}_{\nu_2}{p_4}_{\nu_4}
		\; \cT_{13\rightarrow 24}{}_{\lambda_1, \lambda_3}^{\lambda_2, \lambda_4}&(p_1, -p_3, -p_2, p_4)	\\
		& =	\int d^4 x_1 \, d^4 x_2\, d^4 x_3 \,d^4 x_4 \\
		& \quad \times \quad	e^{i p_2 x_2}\, \epsilon^{\mu_2 *}_{\lambda_2} (-p_2) (-\partial^2_2)\\
		& \quad \times \quad e^{-i p_4 x_4}\,\epsilon^{\mu_4 *}_{\lambda_4} (p_4) (-\partial^2_4)  \\
		& \quad  \times \quad \,\langle \Omega | T\{F_{\mu_4 \nu_4}(x_4)F_{\mu_2 \nu_2}(x_2)F_{\mu_1 \nu_1}(x_1) F_{\mu_3 \nu_3}(x_3)\}| \Omega \rangle_{connected}  \\
		&  \quad \times \quad (-\overleftarrow{\partial_1^2}) \epsilon^{\mu_1 }_{\lambda_1}(p_1)\, e^{i p_1 x_1} \\
		&\quad \times \quad (-\overleftarrow{\partial_3^2}) \epsilon^{\mu_3 }_{\lambda_3}(-p_3)\, e^{-i p_3 x_3},
	\end{aligned}
\end{equation}
which looks very similar to \eqref{eq:covariant_GI_LSZ}, except for the correlator and the polarization vectors which are evaluated at negative energies.
Using the fact that bosonic operators commute we see that the correlation functions are equal
\begin{multline}
%	\label{eq:permutations}
	\langle \Omega | T\{F_{\mu_4 \nu_4}(x_4)F_{\mu_3 \nu_3}(x_3)F_{\mu_1 \mu_1}(x_1) F_{\mu_2 \nu_2}(x_2)\}| \Omega \rangle  =\\
	\langle \Omega | T\{F_{\mu_4 \nu_4}(x_4)F_{\mu_2 \nu_2}(x_2)F_{\mu_1 \nu_1}(x_1) F_{\mu_3 \nu_3}(x_3)\}| \Omega \rangle
\end{multline}
We now consider the negative energy polarization vectors $\epsilon^\mu_{\lambda}(-p)$. They are the analytic continuations of the positive energy polarizations, which were defined in \ref{eq:polarizations}. Since we deal with massless particles in this work, we choose the following analytic continuation for the momenta 
\begin{equation}
	p^0 \rightarrow -p^0 \quad, \myP \rightarrow -\myP, \quad \theta \rightarrow \theta, \quad \phi \rightarrow \phi
\end{equation}
which ensures that $p^\mu \rightarrow -p^\mu$. See \cite{Hebbar:2020ukp} for more details. Under this analytic continuation we see from the explicit form \ref{eq:polarizations} that
\begin{equation}
%	\label{eq:polarization_continuation_2}
	\epsilon^\mu_{\lambda}(-p) = \epsilon^\mu_{\lambda}(p) = \left(\epsilon^{\,\mu }_{-\lambda}(p)\right)^*
\end{equation}
Thus we have
\begin{equation}
	\begin{aligned}
		{p_1}_{\nu_1} {p_2}_{\nu_2} {p_3}_{\nu_3} {p_4}_{\nu_4}
	\; \cT_{12\rightarrow 34}{}_{\lambda_1, \lambda_2}^{\lambda_3, \lambda_4}&(p_1, p_2, p_3, p_4) \\
	&= {p_1}_{\nu_1}{p_3}_{\nu_3} {p_2}_{\nu_2}{p_4}_{\nu_4}
	\; \cT_{13\rightarrow 24}{}_{\lambda_1, -\lambda_3}^{-\lambda_2, \lambda_4}(p_1, -p_3, -p_2, p_4)\,,
	\end{aligned}
\end{equation}
which implies the crossing equation in any frame
\begin{equation}
	\label{eq:crossing_general_frame_LSZ}
		\cT_{12\rightarrow 34}{}_{\lambda_1, \lambda_2}^{\lambda_3, \lambda_4}(p_1, p_2, p_3, p_4) =
		\cT_{13\rightarrow 24}{}_{\lambda_1, -\lambda_3}^{-\lambda_2, \lambda_4}(p_1, -p_3, -p_2, p_4)\,.
\end{equation}

\paragraph{Crossing for COM frame amplitudes} Our prescription to arrive at the amplitude involved going to a special frame, the COM frame. But this presents a problem -  it is not possible to have both the processes in their COM frame in the above relation \eqref{eq:crossing_general_frame_LSZ}. For example putting say the  s channel amplitude on the right hand side in its COM frame would mean that the $t$-channel amplitude on the left hand side is in a frame different from its COM frame (which we called crossed frame). The way to deal with this, as was reviewed in \cite{Hebbar:2020ukp}, is to take a step back and first perform a Lorentz transformation and then analytically continue the amplitude, \emph{i.e.} cross the amplitude.  
More concretely, consider \eqref{eq:covariant_GI_LSZ} in the ($s$-channel) COM frame and perform the following (complexified) Lorentz transformation 
\begin{equation}
	\renewcommand\arraystretch{1.5}
	\Lambda = \begin{pmatrix}
		0 & \frac{i  \sqrt{-u}}{\sqrt{s}} & 0 & -\frac{i\sqrt{-t}}{\sqrt{s}} \\
		-\frac{\sqrt{-u}}{\sqrt{-t}} & 1 & 0 &\frac{\sqrt{-u}}{\sqrt{-t}}\\
		0 & 0 & 1 & 0 \\
		-\frac{i\sqrt{s}}{\sqrt{-t}} & \frac{i\sqrt{-u}}{\sqrt{s}} & 0 &- \frac{i u}{\sqrt{s}\sqrt{-t}}
	\end{pmatrix}
\end{equation}
 such that
\begin{equation}
	\begin{aligned}
		p_1^{\text{com}} &=\frac{ \sqrt{s}}{2}(1,0,0,1)                  && \tilde p_1 \equiv \Lambda p_1 = -\frac{ i\sqrt{-t}}{2}(1,0,0,1)  \\
		p_2^{\text{com}}  &=\frac{ \sqrt{s}}{2}(1,0,0,-1)                  \qquad\longrightarrow \qquad && \tilde p_2 \equiv \Lambda p_2 =-\frac{ i\sqrt{-t}}{2}\left(-1,\frac{2i\sqrt{-su}}{t},0,\frac{u-s}{t}\right) \\
		p_3^{\text{com}}  &=\frac{ \sqrt{s}}{2}\left(1,\frac{2\sqrt{t u}}{s},0,\frac{t-u}{s}\right) &&\tilde p_3 \equiv \Lambda p_3 =-\frac{ i\sqrt{-t}}{2}(-1,0,0,1) \\
		p_4^{\text{com}}  &=\frac{ \sqrt{s}}{2}\left(1,-\frac{2\sqrt{t u}}{s},0,\frac{u-t}{s}\right) &&\tilde p_4 \equiv \Lambda p_4 =-\frac{ i\sqrt{-t}}{2}\left(1,\frac{2i\sqrt{-su}}{t},0,\frac{u-s}{t}\right)
		\end{aligned}
\end{equation}
Note that we are still in the $s$-channel region $s > 0 $ and $t <0$. 
 For this choice of Lorentz transformation $\Lambda$, the photon polarization vectors also transform in the same way $
\epsilon^\mu_\lambda(\tilde p) = \epsilon^\mu_\lambda(\Lambda p)  = {\Lambda^\mu}_\nu \epsilon^\nu_\lambda(p)$ \footnote{In general we have $
	\epsilon^\mu_\lambda(\Lambda p) = e^{-i \lambda \omega}{\Lambda^\mu}_\nu \epsilon^\nu_\lambda(p)$. For a Lorentz transformation composed of boosts and rotations in the $x-z$ plane, the Wigner angle $\omega$ is $0$. See equation A.112 in \cite{Hebbar:2020ukp}. \label{footnote:wigner_angle_crossing}} (upto terms proportional to the momentum)
\begin{equation}
	\begin{aligned}
		\epsilon_\lambda(p_1)&=\frac{ 1}{\sqrt{2}}(0,1, i \lambda,0)                  && \epsilon_\lambda(\tilde p_1) = \Lambda \epsilon_\lambda(p_1) =\frac{ 1}{\sqrt{2}}(0,1, i \lambda,0) \\
		\epsilon_\lambda(p_2) &=\frac{ 1}{\sqrt{2}}(0,-1, i \lambda,0)                    \qquad\longrightarrow \qquad && \epsilon_\lambda(\tilde p_2) = \Lambda \epsilon_\lambda(p_2)=\frac{ 1}{\sqrt{2}}(0,\frac{s-u}{t}, i \lambda,\frac{2i\sqrt{-su}}{t})  \\
		\epsilon_\lambda(p_3)&=\frac{ 1}{\sqrt{2}}(0,\frac{t-u}{s}, i \lambda,-\frac{2\sqrt{tu}}{s})     &&\epsilon_\lambda(\tilde p_3) = \Lambda \epsilon_\lambda(p_3) =\frac{ 1}{\sqrt{2}}(0,-1, i \lambda,0)     \\
		\epsilon_\lambda(p_4) &=\frac{ 1}{\sqrt{2}}(0,\frac{u-t}{s}, i \lambda,\frac{2\sqrt{tu}}{s})   &&\epsilon_\lambda(\tilde p_4) = \Lambda \epsilon_\lambda(p_4) =\frac{ 1}{\sqrt{2}}(0,\frac{u-s}{t}, i \lambda,-\frac{2i\sqrt{-su}}{t})
	\end{aligned}
\end{equation}
Under this Lorentz transformation, the amplitude transforms as \eqref{TCovariance}\footnote{The little group phases $t_i=1$  since the Wigner angle is $0$. See footnote \ref{footnote:wigner_angle_crossing} above.}
\begin{equation}
	\begin{aligned}
		{\tilde {p_1}}_{\nu_1} { \tilde {p_2}}_{\nu_2} { \tilde {p_3}}_{\nu_3} { \tilde{ p_4}}_{\nu_4}
	\; \cT_{12\rightarrow 34}{}_{\lambda_1, \lambda_2}^{\lambda_3, \lambda_4}(s,t,u) =		{\tilde {p_1}}_{\nu_1}{ \tilde {p_2}}_{\nu_2}{ \tilde {p_3}}_{\nu_3} { \tilde {p_4}}_{\nu_4}
	\; \cT_{12\rightarrow 34}{}_{\lambda_1, \lambda_2}^{\lambda_3, \lambda_4}(\tilde{p}_1, \tilde {p}_2, \tilde{p}_3, \tilde{ p}_4)
	\end{aligned}
\end{equation}
Writing out the right hand side
\begin{equation}
	\begin{aligned}
		{\tilde {p_1}}_{\nu_1} { \tilde {p_2}}_{\nu_2} { \tilde{p_3}}_{\nu_3}&{ \tilde {p_4}}_{\nu_4}
		\; \cT_{12\rightarrow 34}{}_{\lambda_1, \lambda_2}^{\lambda_3, \lambda_4}(s,t,u)\\ & =	\int d^4 x_1 \, d^4 x_2\, d^4 x_3 \,d^4 x_4 \\
		& \quad \times \quad	e^{-i \tilde {p}_3 x_3}\, \epsilon^{\mu_3 *}_{\lambda_3} (\tilde {p}_3) (-\partial^2_3)\\
		& \quad \times \quad e^{-i \tilde {p}_4 x_4}\,\epsilon^{\mu_4 *}_{\lambda_4} (\tilde {p}_4) (-\partial^2_4)  \\
		& \quad  \times \quad \,\langle \Omega | T\{F_{\mu_4 \nu_4}(x_4)F_{\mu_3 \nu_3}(x_3)F_{\mu_1 \nu_1}(x_1) F_{\mu_2 \nu_2}(x_2)\}| \Omega \rangle_{connected}  \\
		&  \quad \times \quad (-\overleftarrow{\partial_1^2}) \epsilon^{\mu_1 }_{\lambda_1}(\tilde {p}_1)\, e^{i \tilde {p}_1 x_1} \\
		&\quad \times \quad (-\overleftarrow{\partial_2^2}) \epsilon^{\mu_2 }_{\lambda_2}(\tilde {p}_2)\, e^{i \tilde {p}_2 x_2},
	\end{aligned}
\end{equation}
 we now analytically continue to the $t$-channel region $t > 0$ and $s <0$ at fixed real $u$. Under this operation we have 
\begin{equation}
	\begin{aligned}
		\tilde {p}_1 &=-\frac{ i\sqrt{-t}}{2}(1,0,0,1)             && \tilde {p}_1^{\text{a.c.}} =\frac{ \sqrt{t}}{2}(1,0,0,1)  \\
		\tilde {p}_2 & =-\frac{ i\sqrt{-t}}{2}\left(-1,\frac{2i\sqrt{-su}}{t},0,\frac{u-s}{t}\right)               \qquad\longrightarrow \qquad && \tilde {p}_2^{\text{a.c.}} =-\frac{ \sqrt{t}}{2}\left(1,\frac{2\sqrt{s u}}{t},0,\frac{t-u}{s}\right) \\
		\tilde {p}_3 &=-\frac{ i\sqrt{-t}}{2}(-1,0,0,1) &&\tilde {p}_3^{\text{a.c.}} =-\frac{ \sqrt{t}}{2}(1,0,0,-1)  \\
		\tilde {p}_4 &=-\frac{ i\sqrt{-t}}{2}\left(1,\frac{2i\sqrt{-su}}{t},0,\frac{u-s}{t}\right) &&\tilde {p}_4^{\text{a.c.}} =\frac{ \sqrt{t}}{2}\left(1,-\frac{2\sqrt{s u}}{t},0,\frac{u-s}{t}\right)
	\end{aligned}
\end{equation}
which ensures that the momenta $P_1^{\overline{com}} = \tilde{p}_1^{\text{a.c.}}$, $P_3^{\overline{com}} = -\tilde{p}_3^{\text{a.c.}}$, $P_2^{\overline{com}}= -\tilde{p}_2^{\text{a.c.}}$ and $P_4^{\overline{com}}=\tilde{p}_4^{\text{a.c.}}$ are precisely in the $t$-channel COM frame. We also analytically continue the photon polarizations 
\begin{equation}
	\begin{aligned}
		\epsilon_\lambda(\tilde{p}_1)&=\frac{ 1}{\sqrt{2}}(0,1, i \lambda,0)                  && \epsilon_\lambda^{\text{a.c.}}(\tilde{p}_1) =\frac{ 1}{\sqrt{2}}(0,1, i \lambda,0) \\
			\epsilon_\lambda(\tilde{p}_2) &=\frac{ 1}{\sqrt{2}}(0,\frac{s-u}{t}, i \lambda,\frac{2i\sqrt{-su}}{t})                 \qquad\longrightarrow \qquad && 	\epsilon_\lambda^{\text{a.c.}}(\tilde{p}_2)=\frac{ 1}{\sqrt{2}}(0,\frac{s-u}{t}, i \lambda,-\frac{2\sqrt{su}}{t})  \\
			\epsilon_\lambda(\tilde{p}_3)&=\frac{ 1}{\sqrt{2}}(0,-1, i \lambda,0)      &&	\epsilon_\lambda^{\text{a.c.}}(\tilde{p}_3)=\frac{ 1}{\sqrt{2}}(0,-1, i \lambda,0)     \\
		\epsilon_\lambda(\tilde{p}_4) &=\frac{ 1}{\sqrt{2}}(0,\frac{u-s}{t}, i \lambda,-\frac{2i\sqrt{-su}}{t})   &&	\epsilon_\lambda^{\text{a.c.}}(\tilde{p}_4)=\frac{ 1}{\sqrt{2}}(0,\frac{u-s}{t}, i \lambda,\frac{2\sqrt{su}}{t}) \,.
	\end{aligned}
\end{equation}
Using $\epsilon^\mu_\lambda(-p) = \left(\epsilon^{\mu}_{-\lambda}(p)\right)^*$, we have 
\begin{equation}
	\begin{aligned}
		{\tilde {p_1}}_{\nu_1} { \tilde {p_2}}_{\nu_2} { \tilde{p_3}}_{\nu_3}&{ \tilde {p_4}}_{\nu_4}
		\; \cT_{12\rightarrow 34}{}_{\lambda_1, \lambda_2}^{\lambda_3, \lambda_4}(s,t,u)\\ & =	\int d^4 x_1 \, d^4 x_2\, d^4 x_3 \,d^4 x_4 \\
		& \quad \times \quad	e^{i P_3^{\overline{com}} x_3}\, \epsilon^{\mu_3 *}_{-\lambda_3} (P_3^{\overline{com}}) (-\partial^2_3)\\
		& \quad \times \quad e^{-i P_4^{\overline{com}} x_4}\,\epsilon^{\mu_4 *}_{\lambda_4} (P_4^{\overline{com}}) (-\partial^2_4)  \\
		& \quad  \times \quad \,\langle \Omega | T\{F_{\mu_4 \nu_4}(x_4)F_{\mu_3 \nu_3}(x_3)F_{\mu_1 \nu_1}(x_1) F_{\mu_2 \nu_2}(x_2)\}| \Omega \rangle_{connected}  \\
		&  \quad \times \quad (-\overleftarrow{\partial_1^2}) \epsilon^{\mu_1 }_{\lambda_1}(P_1^{\overline{com}})\, e^{i P_1^{\overline{com}}x_1} \\
		&\quad \times \quad (-\overleftarrow{\partial_2^2}) \epsilon^{\mu_2 }_{-\lambda_2}(P_2^{\overline{com}})\, e^{-i P_2^{\overline{com}} x_2},
	\end{aligned}
\end{equation}
the right hand side is nothing but the t channel amplitude evaluated in its COM frame with the values for its arguments given by
\begin{equation}
	\begin{aligned}
	\bar s = -(P_1^{\overline{com}}+P_3^{\overline{com}})^2 &= -(\tilde p_1 - \tilde p_3)^2 = t  \\
	\bar t = -(P_1^{\overline{com}}-P_2^{\overline{com}})^2 &= -(\tilde p_1 + \tilde p_2)^2 = s \,,
	\end{aligned}
\end{equation}
 and therefore we deduce that 
 \begin{equation}
 	 {P_1}^{\overline{com}}_{\nu_1} {P_2}^{\overline{com}}_{\nu_2}  {P_3}^{\overline{com}}_{\nu_3}{P_4}^{\overline{com}}_{\nu_4}
 	\; \cT_{12\rightarrow 34}{}_{\lambda_1, \lambda_2}^{\lambda_3, \lambda_4}(s,t,u)=
 	{P_1}^{\overline{com}}_{\nu_1} {P_2}^{\overline{com}}_{\nu_2}  {P_3}^{\overline{com}}_{\nu_3}{P_4}^{\overline{com}}_{\nu_4}
 	\; \cT_{13\rightarrow 24}{}_{\lambda_1, -\lambda_3}^{-\lambda_2, \lambda_4}(t,s,u)
 \end{equation}
and  we can finally extract the crossing relation by dotting on both sides with the vector $\bar {\mathfrak v} = \frac{2}{\sqrt{t} }(-1,0,0,0)$
\begin{equation}
	\begin{aligned}
		\cT_{12\rightarrow 34}{}_{\lambda_1, \lambda_2}^{\lambda_3, \lambda_4}(s,t,u) 
		= \cT_{13\rightarrow 24}{}_{\lambda_1, -\lambda_3}^{-\lambda_2, \lambda_4}(t,s,u)
	\end{aligned}
\end{equation} 
For the case at hand, i.e photon scattering, the two processes are the same since all the particles $1,2,3$ and $4$ are identical and the above $s-t$ crossing relation is a symmetry of the amplitude. Similarly one can repeat the process to establish the other crossing equations. 
\section{Asymptotic unitarity constraints}
As described in section \ref{sec:numerics}, we impose the unitarity constraints up to some $\Lmax$ and on a grid of $s$ values. In this appendix we describe how we supplement them with additional constraints by analyzing the unitarity equations as $\ell \rightarrow \infty $ and $s \rightarrow \infty $. In principle, these constraints should become redundant as we increase $\Lmax$ and choose finer grids in $s$. In practice, we find that adding in these asymptotic constraints improves convergence in $\Lmax$ and number of grid points.

\subsection{Large spin}
	\label{app:large_spin}
	 The first step is  to estimate the behaviour of partial waves at large spin. To this end, 
we follow the analysis of Appendix D.4 in \cite{Paulos:2017fhb}. We would like to use the Froissart-Gribov projection formula and write partial waves as contour integrals in the complex $z = \cos \theta$ plane. We define
\begin{equation}
	\begin{aligned}
		e^{\, \ell}_{\lambda \mu} (z) &= \frac{(-1)^{\lambda - \mu}}{2} \left[\Gamma(\ell+\lambda +1)\Gamma(\ell-\lambda +1)\Gamma(\ell+\mu +1)\Gamma(\ell-\mu +1)\right]^{\frac{1}{2}} \left(\frac{1+z}{2}\right)^{\frac{\lambda+\mu}{2}}\\
		&\times\left(\frac{1-z}{2}\right)^{-\frac{\lambda-\mu}{2}}
		 \left(\frac{z-1}{2}\right)^{-\ell -\mu - 1} \frac{1}{\Gamma(2\ell +2)} {}_2 F_1\left(\ell+\lambda+1, \ell +\mu +1, 2 \ell +2, \frac{2}{1-z}\right)~, 
	\end{aligned}
\end{equation} 
valid for $\lambda + \mu \geq 0$ and $\lambda - \mu \geq 0$. For other ranges of parameters, the function is defined by its symmetry properties
\begin{equation}
	e^{\, \ell}_{\lambda \mu} (z) = (-1)^{\lambda-\mu}e^{\, \ell}_{\mu \lambda} (z) = (-1)^{\lambda-\mu} e^{\, \ell}_{-\lambda, -\mu} (z) 
\end{equation}
This function has a branch cut in the complex $z$ plane between $-1$ and $1$ and its discontinuity there is the Wigner $d$ function:\footnote{In this respect, it's a generalization of the Legendre Q function, which obeys an analogous relation to the Legendre P polynomial.} \footnote{\label{footnote:e_function_singularities} In general, the Wigner $e$ function has additional singularities at $z = \pm 1$. The leading singular behaviour there is $(z+1)^{(\lambda+\mu)/2}$ and $(z-1)^{(\lambda-\mu)/2}$. In the case relevant for us, we have
	\begin{equation}
		\begin{aligned}
			e^{\ell}_{2,2}(z) & \sim  \frac{2 (-1)^{\ell}}{\ell(\ell+1)}\frac{1}{(z+1)} + \frac{12 (-1)^{\ell-1}}{(\ell-1)\ell(\ell+1)(\ell+2)}\frac{1}{(z+1)^2}\\
			e^{\ell}_{0,2}(z) &\sim \frac{2}{\sqrt{(\ell-1)\ell(\ell+1)(\ell+2)}} \frac{1}{(z+1)} + \frac{2 (-1)^\ell}{\sqrt{(\ell-1)\ell(\ell+1)(\ell+2)}}\frac{1}{(z-1)} 
		\end{aligned}
	\end{equation}
}
\begin{equation}
	\label{eq:discontinuity_wignerE}
	e^{\, \ell}_{\lambda \mu} (z + i \epsilon)-e^{\, \ell}_{\lambda \mu} (z - i \epsilon) = - i \pi d^{\, \ell}_{\lambda \mu} (z)  \qquad z  \in (-1,1)
\end{equation}
We now recall the definition of partial wave amplitudes
\begin{equation}
	{\cT^{\ell}}_{\lambda_1, \lambda_2}^{\lambda_3, \lambda_4}(s) = \int_{-1}^{+1} dz \; d^{\, \ell}_{\lambda_{12}\lambda_{34}} (z)\, \cT_{\lambda_1, \lambda_2}^{\lambda_3, \lambda_4}(s,z)
\end{equation}
At this point we would like to use \ref{eq:discontinuity_wignerE} to write the above equation as a contour integral in the $z$ plane. Scattering amplitudes have the following behaviour near $z = \pm 1$\footnote{see for example eq.(2.138) in \cite{Hebbar:2020ukp}.}
\begin{equation}
	\cT_{\lambda_1, \lambda_2}^{\lambda_3, \lambda_4}(s,z) = b_{\lambda_{12} \lambda_{34}}(z) \hat{\cT}_{\lambda_1, \lambda_2}^{\lambda_3, \lambda_4}(s,z),
\end{equation}
where we have defined the $b$ function 
\begin{equation}
	b_{\lambda \mu}(z) = \left(\frac{1+z}{2}\right)^{\frac{|\lambda + \mu|}{2}}\left(\frac{1-z}{2}\right)^{\frac{|\lambda - \mu|}{2}}
\end{equation}
and with the function $\hat \cT$ being regular near $z = \pm 1$.  This behaviour of the scattering amplitude precisely cancels the poles (more generally branch cuts) mentioned in footnote \ref{footnote:e_function_singularities}. 
We can now re-write the partial wave integral as a contour integral
\begin{equation}
	{\cT^{\ell}}_{\lambda_1, \lambda_2}^{\lambda_3, \lambda_4}(s) = \frac{1}{i \pi} \oint_C b_{\lambda_{12} \lambda_{34}}(z) e^{\, \ell}_{\lambda_{12} \lambda_{34}} (z) \hat \cT_{\lambda_1, \lambda_2}^{\lambda_3, \lambda_4}(s,z)
\end{equation}
with the contour $C$ circling the line segment $[-1,1]$ anti-clockwise. In fact, in the presence of massless particles, the contour should pass through $z=1$ and $z=-1$, as it will be clear below. 
Therefore for large enough $\ell$,\footnote{Note that $e_{\lambda \mu}^\ell \sim z^{-\ell}$ for large $|z|$.} we can open up the contour and drop the arcs at infinity, to arrive at the (generalized) Froissart-Gribov projection formula
\begin{multline}
	\label{eq:froissart_gribov_spinning}
	{\cT^{\ell}}_{\lambda_1, \lambda_2}^{\lambda_3, \lambda_4}(s) = \frac{1}{i \pi} \left( \int_{z_t}^{\infty} dz \; b_{\lambda_{12} \lambda_{34}}(z) \, e^{\, \ell}_{\lambda_{12} \lambda_{34}} (z) \text{Disc}_t \hat \cT_{\lambda_1, \lambda_2}^{\lambda_3, \lambda_4}(s,z) \right. \\
	\left. + \int_{-z_u}^{-\infty} dz \; b_{\lambda_{12} \lambda_{34}}(z) \, e^{\, \ell}_{\lambda_{12} \lambda_{34}} (z) \text{Disc}_u \hat \cT_{\lambda_1, \lambda_2}^{\lambda_3, \lambda_4}(s,z) \right)
\end{multline}
where the 1st term is due to the $t$ channel branch cut from $[z_t,\infty)$ and the 2nd term is due to the $u$ channel branch cut from $(-\infty, -z_u]$.

For theories with a mass gap, $z_t = 1 +  \frac{2t_0}{s-4m^2} > 1$ and $z_u =1 +  \frac{2u_0}{s-4m^2} > 1$  and due to the exponential decay in spin of the $e$ function, the partial wave amplitudes also have an exponential fall off in spin. For theories with massless particles this is not the case, because $z_t \rightarrow  1$ and $z_u \rightarrow 1$. 

We therefore consider the large $\ell$, $z \rightarrow 1^{+}$ limit of the hypergeometric function. Assuming $\lambda \ll \ell $, $\mu\ll \ell $ we find
\begin{equation}
	\begin{aligned}
		{}_2 F_1\left(\ell+\lambda+1, \ell +\mu +1, 2 \ell +2, \frac{2}{1-z}\right) &\approx  \frac{2 \, \Gamma(2\ell +2)}{\Gamma(\ell+\mu +1)\Gamma(\ell-\mu +1)} \left(\frac{z-1}{2}\right)^{\ell+1+\frac{\mu+\lambda}{2}}\\
		& \qquad \qquad \times \,K_{\lambda-\mu}\left(\sqrt{2 (z-1)}\, \ell\right).
	\end{aligned} 
\end{equation}
Hence we obtain an approximation for the $e$ function
\begin{equation}
	\begin{aligned}
		\label{eq:e_function_large_spin}
		e^{\, \ell}_{\lambda \mu} (z) &\approx (-1)^{\frac{\lambda - \mu}{2}}
		 K_{\lambda-\mu}\left(\sqrt{2 (z-1)}\, \ell\right),
	\end{aligned}
\end{equation}
valid for $\ell \gg 1$ and $z \rightarrow 1^+$ and $\lambda \ll \ell $, $\mu\ll \ell $.
For the other limit $\ell \gg 1$ and $z \rightarrow -1^-$, we use the relation
\begin{equation}
	\label{eq:e_function_parity}
	e^{\, \ell}_{\lambda \mu} (-z) = (-1)^{l-\lambda + 1}e^{\, \ell}_{\lambda, -\mu} (z)
\end{equation}
The other piece in \eqref{eq:froissart_gribov_spinning} is the discontinuity of the amplitude $\hat T 
\equiv 
\frac{T}{b}$. In our numerics we parametrize the amplitude as displayed in \eqref{eq:ansatz}. The prefactors, composed of $\chi_s$, $\chi_t$ and $\chi_u$, which were added for kinematical reasons, ensure the cancellation of the singularities at $z = \pm 1$. We therefore consider the $\myRho$ series:
\begin{equation}
	 \sum_{abc} \alpha_{abc} \myRho^a(s)\myRho^b(t)\myRho^c(u) 
\end{equation}
For such a series, the $t$-channel discontinuity comes from 
\begin{equation}
	\label{eq:rho_t_discontinuity}
	\myRho(t(s,z+i\epsilon))^b-\myRho(t(s,z-i\epsilon))^b  \approx 2ib \sqrt{2s}\sqrt{z-1}
\end{equation}
and the $u$-channel discontinuity comes from
\begin{equation}
	\label{eq:rho_u_discontinuity}
	\myRho(u(s,z-i\epsilon))^c-\myRho(u(s,z+i\epsilon))^c \approx 2ic \sqrt{2s}\sqrt{-z-1}
\end{equation}
By making a change of variable $z \rightarrow -z$ for the $u$ channel contribution and using the symmetry properties  \eqref{eq:e_function_parity} of the $e$ function and the $b$ function
\begin{equation}
	b_{\lambda \mu}(-z)  = b_{\lambda, -\mu}(z) 
\end{equation}
we can reduce all our computations to that of integrals of the following form:
\begin{equation}
	\int_{1}^{\infty} dz \left(1+z\right)^{m} \left(1-z\right)^{n}  \sqrt{z-1}  K_{a}\left(c\sqrt{z-1}\right)
\end{equation}
To perform this integral we make a change of variable $(z-1) = \xi^2$ and then expand in the variable $\xi$ to get the following type of integrals which are easily evaluated
\begin{equation} 
	\int_{0}^{\infty} d\xi \, \xi^b \, K_{a}\left(c\, \xi \right) = \frac{2^{b-1}}{c^{b+1}}\Gamma\left(\frac{b-a+1}{2}\right)\Gamma\left(\frac{b+a+1}{2}\right)
\end{equation}
Since $c \sim \ell$, the leading contribution comes from the lowest order term in the $\xi$ expansion. 
At leading order in  the large $\ell$ limit, we have\footnote{Recall that due to crossing symmetry $\Phi_3(s,t,u) = \Phi_1(u,t,s)$, which is why the same coefficients $\alpha^{(1)}_{abc}$ appear in the expansion for both the amplitudes, albeit with different orderings of the $a$, $b$ and $c$ indices.}

	\begin{equation}
		\label{eq:ansatz_large_spin_estimate}
		\begin{aligned}
			\Phi_1^\ell \equiv {\cT^{\ell}}_{+, +}^{+, +}(s) &\approx \frac{1}{\ell^3} \sum_{abc} \alpha^{(1)}_{abc} \left(b \,\myRho^c(-s) + (-1)^\ell  c \,\myRho^b(-s)\right)\sqrt{s}\, \myRho^a(s) \chi^2(s),\\
			\Phi_2^\ell \equiv {\cT^{\ell}}_{+, +}^{-, -}(s)  &\approx \frac{1}{\ell^3} \sum_{abc} \alpha^{(2)}_{abc} \left(b \,\myRho^c(-s) + (-1)^\ell  c \,\myRho^b(-s)\right)\sqrt{s}\, \myRho^a(s), \\
			\Phi_3^\ell \equiv {\cT^{\ell}}_{+, -}^{+, -}(s)  &\approx \frac{1}{\ell^3} \sum_{abc} \alpha^{(1)}_{cba} b \,\myRho^c(-s) \sqrt{s}\, \myRho^a(s) \chi^2(-s),\\
			%\Phi_4^\ell \equiv {T^{\ell}}_{+, -}^{-, +}(s) &\approx \frac{1}{\ell^3} \sum_{abc} \alpha^{(4)}_{abc} \left(-15\, b \,\myRho^c(-s) + (-1)^\ell  c \,\myRho^b(-s)\right)\sqrt{s}\, \myRho^a(s) \\
			\Phi_5^\ell \equiv {\cT^{\ell}}_{+, +}^{+, -}(s) &\approx \mathcal O(l^{-4}).
		\end{aligned}
	\end{equation} 	
We notice that the fifth amplitude $\Phi_5^\ell$ is sub-leading at large spin and therefore the unitarity condition \eqref{eq:unitarity_2} decomposes into two $1 \times 1$ conditions 
	\begin{equation}
		\begin{aligned}
			|1 + i(\Phi_1^\ell + \Phi_2^\ell)| \leq 1, \\
			|1 + i(2\Phi_3^\ell)| \leq 1 .
		\end{aligned}	
	\end{equation} 
	In addition we have from \eqref{eq:unitarity_1}
	
	\begin{equation}
		|1 + i(\Phi_1^\ell - \Phi_2^\ell)| \leq 1. \\
	\end{equation}
	Plugging in \eqref{eq:ansatz_large_spin_estimate} into the above unitarity equations we get the following large spin conditions:\footnote{$|1+ i \epsilon  f(s)| \leq 1 $ for $\epsilon \ll1$ implies that $\text{Im} f(s) \geq 0$. In our case the small parameter $\epsilon$ is $\frac{1}{\ell^3}$.}
	\begin{equation}
		\boxed{
			\sum_{abc}  \left( \alpha^{(1)}_{abc} \text{Im}(\myRho^a(s) \chi^2(s)) + \alpha^{(2)}_{abc} \text{Im}(\myRho^a(s))\right)b \sqrt{s} \, \,\myRho^c(-s) \geq 0,
		}
	\end{equation}
	
	\begin{equation}
		\boxed{
			\sum_{abc}  \left( \alpha^{(1)}_{abc} \text{Im}(\myRho^a(s) \chi^2(s)) - \alpha^{(2)}_{abc} \text{Im}(\myRho^a(s))\right)b \sqrt{s} \, \,\myRho^c(-s) \geq 0,
		}
	\end{equation}

	\begin{equation}
		\boxed{
			\sum_{abc}  \alpha^{(1)}_{cba} \text{Im}(\myRho^a(s))\,b \sqrt{s} \, \,\myRho^c(-s)  \chi^2(-s)\geq 0,
		}
	\end{equation}
valid for $s \geq 0$.

	\subsection{Large energy}
	\label{app:large_energy}
	We begin by considering the $s \rightarrow \infty$ expansion of a monomial term in the ansatz:
	\begin{equation}
		\begin{aligned}
			\myRho^a(s)\myRho^b(t)\myRho^c(u) \approx (-1)^{a+b+c} \Bigg[1 &- \frac{2}{\sqrt{s}} \left(i a + \frac{\sqrt{2}b}{\sqrt{1-z}} + \frac{\sqrt{2}c}{\sqrt{1+z}} \right) +i\frac{4 \sqrt{2} a}{s} \left( \frac{b}{\sqrt{1-z}} +  \frac{c}{\sqrt{1+z}} \right)   \\ 
			& +  \frac{1}{s} \left(-2a^2 + \frac{8bc}{\sqrt{1-z}\sqrt{1+z}} + \frac{4b^2}{1-z}+\frac{4c^2}{1+z} \right) + \ldots  
		\end{aligned}
	\end{equation}					
			In addition to the large energy expansion of the $\myRho$ series, we also need the large energy expansion of $\chi^2(s)$ , $\chi^2(u)$ and $\chi(t)\chi(u)$:
			
			\begin{equation}
				\begin{aligned}
					\chi^2(s) &\approx 9 - \frac{ 48 i }{\sqrt{s}}+ \ldots \\
					\chi^2(u) &\approx 9 - \frac{ 48 i }{\sqrt{s}} \frac{1}{\sqrt{1+z}} + \ldots \\
					\chi(s)\chi(t)\chi(u) &\approx 27 - \frac{72 i}{\sqrt{s}}  \left(1+\frac{\sqrt{2}}{\sqrt{1+z}} + \frac{\sqrt{2}}{\sqrt{1-z}} \right) + \ldots 
				\end{aligned}
			\end{equation}
			This leads to the following expansions of the ansatze at large energy:
			\begin{equation}
				\begin{aligned}
					\Phi_1 &\approx \sum_{abc} \alpha^{(1)}_{abc} (-1)^{a+b+c}\Bigg[9 - \frac{6}{\sqrt{s}}\left((8+3a)i +\frac{3 \sqrt{2} b}{\sqrt{1-z}}+\frac{3 \sqrt{2} c}{\sqrt{1+z}}\right) + \ldots \Bigg], \\
					\Phi_2 &\approx \sum_{abc} \alpha^{(2)}_{abc}	(-1)^{a+b+c} \Bigg[1 - \frac{2}{\sqrt{s}} \left(i a + \frac{\sqrt{2}b}{\sqrt{1-z}}  + \frac{\sqrt{2}c}{\sqrt{1+z}}\right) + \ldots \Bigg] ,\\
					\Phi_3 &\approx \sum_{abc} \alpha^{(1)}_{cba} (-1)^{a+b+c}\Bigg[9 - \frac{6}{\sqrt{s}}\left(3ia +\frac{3 \sqrt{2} b}{\sqrt{1-z}}+\frac{(8+3c) \sqrt{2} }{\sqrt{1+z}}\right) + \ldots \Bigg] ,\\
					\Phi_5 &\approx \sum_{abc} \alpha^{(5)}_{abc} (-1)^{a+b+c}\Bigg[27 - \frac{18}{\sqrt{s}}\left((4+3a)i +\frac{(4+3b) \sqrt{2} }{\sqrt{1-z}} +\frac{(4+3c) \sqrt{2} }{\sqrt{1+z}}\right) + \ldots \Bigg]~. 
				\end{aligned}
			\end{equation}
			We now work order by order in $\frac{1}{s}$. We begin with the leading order which is 
			\subsubsection*{O(1):}
			The unitarity condition on spin $0$ partial waves $|1 + i(\Phi^0_1 \pm \Phi^0_2 )| \leq 1$ implies that
			\begin{equation}
				\label{eq:large_energy_equation_1}
				\boxed{
					\sum_{abc} \alpha^{(1)}_{abc} (-1)^{a+b+c} = 0
				}~,
			\end{equation}
			and
			\begin{equation}
				\boxed{
					\sum_{abc} \alpha^{(2)}_{abc} (-1)^{a+b+c} = 0
				}~.
			\end{equation}
			Similarly, the unitarity condition on odd spin $\ell$ partial waves $|1+ 2 i (\Phi^\ell_3)| \leq 1$ implies that
			\begin{equation}
					\label{eq:large_energy_equation_2}
				\boxed{
					\sum_{abc} \alpha^{(1)}_{cba} (-1)^{a+b+c} = 0
				}~,
			\end{equation}
		which turns out to be equivalent to the condition \eqref{eq:large_energy_equation_1}.
			Finally the unitarity condition \eqref{eq:unitarity_2} simplifies at leading order due to the results we just derived above, and leads to
			\begin{equation}
				\boxed{
					\sum_{abc} \alpha^{(5)}_{abc} (-1)^{a+b+c} = 0
				}~.
			\end{equation}
			Effectively, the constraints above say that constant terms in the amplitudes should go to 0 as $s \rightarrow \infty $. We now consider the next order which is 
			
			\subsubsection*{O($s^{-1/2}$):}  From the unitarity condition on spin 0 partial waves $|1 + i(\Phi^0_1 \pm \Phi^0_2 )| \leq 1$, we get the two conditions:
			
			\begin{equation}
				\boxed{
					\sum_{abc}  (-1)^{a+b+c} \left((24+9a)\,\alpha_{abc}^{(1)} + a \, \alpha_{abc}^{(2)} \right) \leq 0
				}~,
			\end{equation}
			and
			\begin{equation}
				\boxed{
					\sum_{abc}  (-1)^{a+b+c} \left((24+9a)\,\alpha_{abc}^{(1)} - a \, \alpha_{abc}^{(2)} \right) \leq 0
				}~.
			\end{equation}
			Consider now the unitarity condition \eqref{eq:unitarity_2} for non-zero even spin $\ell$ partial waves. The imaginary part of the $\Phi^\ell_1$ and $\Phi^\ell_2$ is 0 and this immediately implies that the imaginary part of $\Phi_5$ must vanish at this order:
			
			\begin{equation}
				\boxed{
					\sum_{abc} \alpha^{(5)}_{abc} (-1)^{a+b+c} a = 0
				}~.
			\end{equation}
			In addition, it implies that $|1+2\phi^\ell_3| \leq 1$.\footnote{More precisely, it implies it only for even $\ell$, but we also have the same condition $|1+2\phi^\ell_3| \leq 1$ separately for odd $\ell$ from \eqref{eq:unitarity_3}.   }Since $\int dz  \,d^\ell_{22}(z) >  0$ for even $\ell$ and  $\int  dz  \, d^\ell_{22}(z) <  0$ for odd $\ell$, we get
			
			\begin{equation}
				\boxed{
					\sum_{abc} \alpha^{(1)}_{cba} (-1)^{a+b+c} a = 0
				}~.
			\end{equation}

\bibliographystyle{JHEP}
\bibliography{refs}

\end{document}